\documentclass[aps,prd,superscriptaddress,groupedaddress,nofootinbib,A4]{article}

\usepackage{jheppub}
\usepackage{subfigure}
\usepackage{multirow}
\usepackage{subfigure}
\usepackage{graphicx}
\usepackage{epsfig}
\usepackage{lscape}
\usepackage{rotating}

\title{Walking Technicolor in the light of $Z^{\prime}$ searches at the LHC}

\author[a,c]{Alexander Belyaev}
\author[a]{, Azaria Coupe}
\author[b]{, Mads Frandsen}
\author[c]{, Emmanuel Olaiya}
\author[c]{, Claire Shepherd-Themistocleous}

\affiliation[a]{School of Physics and Astronomy, University of Southampton, Highfield, Southampton SO17 1BJ, UK}
\affiliation[b]{High Energy Physics Center, University of Southern Denmark, Campusvej 55, DK-5230 Odense M, Denmark}
\affiliation[c]{Particle Physics Department, Rutherford Appleton Laboratory, Chilton, Didcot, Oxon OX11 0QX, UK}

\abstract{
We investigate the potential of the Large Hadron Collider (LHC)  to probe 
one of the most compelling Beyond the Standard Model (BSM) frameworks --- Walking Technicolor (WTC),
involving strong dynamics and having a slowly running (walking) new strong coupling.
For this purpose we use recent LHC Run2 data to explore the full  parameter space of the minimal WTC model using dilepton signatures from heavy neutral  $Z^{\prime}$ and $Z^{\prime\prime}$ resonances
predicted by the model.
This signature is  the most promising one for discovery of WTC at the LHC
for the low-intermediate values of the $\tilde g$ coupling -- one of the principle parameters of WTC.
We have demonstrated  complementarity of the dilepton signals from
both  resonances, have established  the most up-to-date limit on the WTC parameter space, and provided  projections for the the LHC potential to probe the  WTC parameter space 
at higher future luminosities  and upgraded energy.
We have explored the whole four-dimensional  parameter space of the model and have
found  the most conservative limit on the WTC scale $M_A$ above 3 TeV for the low values of 
$\tilde g$ which is significantly higher than
previous limits established by the LHC collaborations.
}

\makeindex

\begin{document}
\maketitle

\section{Introduction}
\label{sec:intro}

With the discovery of a Higgs boson at the LHC \cite{Chatrchyan:2012xdj,Aad:2012tfa} it has become not only possible, but imperative to discover the true origin of mass in the Universe. 
The traditional Standard Model (SM) Higgs mechanism of mass generation via spontaneous electroweak symmetry breaking (SEWSB) leads to the hierarchy problem, associated with the large fine tuning between the EWSB scale and the Planck mass. 
Several classes of Beyond the Standard Model (BSM) theories have been proposed to address the shortcomings of the SM,
and one of them is Technicolor which is based on  \textit{new strong dynamics}\cite{PhysRevD.20.2619,Weinberg:1975gm}. 
In Technicolor, EWSB is generated dynamically by the formation of a chiral condensate under the new strong dynamics, providing a natural scale for mass generation without fine tuning. Experimental bounds from  Electroweak Precision Data (EWPD) disfavour TC models with QCD-like dynamics~\cite{Peskin:1991sw}, so modern Technicolor models must have a modified strong coupling. 
Walking Technicolor (WTC)~\cite{King:1986zt,Chivukula:1988qr,King:1988kk,Appelquist:1988yp,Sundrum:1991rf,Lane:1991qh} and its recent developments~\cite{Sannino:2004qp,Dietrich:2006cm,Dietrich:2005jn,Ryttov:2007sr,Ryttov:2007cx,Sannino:2008ha,Foadi:2007ue} 
is a very compelling BSM candidate for the underlying theory of Nature.
It has a strong coupling $\alpha_{TC}$ with a very slowly running (``walking") regime between the TC energy scale and high energy Extended-TC scale. The lightest scalar resonance of WTC can be identified as the experimentally consistent Higgs boson, whose mass scale is naturally generated thus does not incur a hierarchy problem\cite{Foadi:2012bb,Belyaev:2013ida}. WTC also provides a rich phenomenology of composite spin-0 and multiple triplets of composite spin-1 resonances, making this a prime candidate for experimental particle physics searches.

Using LHC Run 1 dilepton data, the ATLAS Collaboration have interpreted experimental limits on a new heavy neutral resonance in the context of the  WTC parameter space in Ref.\cite{Aad:2014cka} and Ref.\cite{Aad:2015yza} using dilepton and $HV$
searches respectively. These WTC interpretations have been following the phenomenological exploration of WTC parameter space
performed in \cite{Belyaev:2008yj} for a 2-dimensional(2D) bench mark from the whole 4-dimensional(4D) parameter space of the model.

This study makes the next step in exploration of the LHC potential to test WTC.
First of all, we perform analysis in the full 4D parameter space of the model.
Secondly, we study the complementarity of the dilepton signals from
both  heavy neutral vector mesons of WTC and demonstrate its importance.
In this work we focus exclusively on Drell-Yan (DY) processes, and provide justification for the single peak analysis of current LHC constraints in the context of this model. 
Finally, we are establishing here the most up-to-date limit on WTC parameter space
from LHC dilepton searches for the whole 4D parameter space of the model,
and give projections for the the LHC potential to probe WTC parameter space 
at higher integrated luminosity in the future.

In Section \ref{sec:setup} we discuss the WTC model together with the constraints on its parameter space.  In Section \ref{sec:pheno} we explore the phenomenology of WTC and the LHC potential to probe the model. Finally in Section \ref{sec:conclusions} we summarise the results of this work, and comment on the future prospects for WTC exploration at the LHC.

\section{Minimal Walking Technicolor  Model}
\label{sec:setup}

Throughout this paper we focus on the global symmetry breaking pattern $SU(2)_{L}\times SU(2)_{R} \to SU(2)_{V}$. This pattern is realized by the Next to Minimal Walking Technicolor model (NMWT) \cite{Sannino:2004qp,Dietrich:2005jn,Belyaev:2008yj}, which features two Dirac fermions transforming in the 2-index symmetric representation of the technicolor gauge group $SU(3)$. 
However, any technicolor model must feature $SU(2)_{L}\times SU(2)_{R} \to SU(2)_{V}$ as a subgroup breaking pattern of the full symmetry breaking pattern $G\to H$. This is required to ensure mass generation for the $W$ and $Z$ bosons and to preserve an $SU(2)_{V}$ custodial symmetry in the new strong dynamics sector like that in the SM higgs sector. 

More generally, theories of composite dynamics with a Technicolor limit will feature this as a global symmetry breaking subpattern. Examples are composite Higgs and partially composite Higgs models \cite{Kaplan:1983fs,Kaplan:1983sm} with an underlying 4 dimensional realization,
%. such as those yielding an $SU(4)/Sp(4)$ coset 
e.g. \cite{Galloway:2010bp,Cacciapaglia:2014uja,Galloway:2016fuo,Agugliaro:2016clv,Alanne:2017rrs}. Another example is bosonic technicolor \cite{Simmons:1988fu,Dine:1990jd,Carone:2012cd,Alanne:2013dra}. In both partially composite Higgs models and in bosonic technicolor, the Higgs particle is a mixture of an elementary and composite scalar. Some aspects of how the spin-1 resonance phenomenology is affected by aligning the theory away from the Technicolor vacuum in composite Higgs and partially composite Higgs models are given in \cite{Franzosi:2016aoo,Galloway:2016fuo}. In general the mass scale set by the Goldstone boson decay constant of the strong interactions, $F_\pi$, is larger in composite Higgs models than in ordinary technicolor while it is smaller in the bosonic technicolor models. In partially composite Higgs models, the scale depends on the relative size of the elementary doublet vacuum expectation value $v$ and the vacuum alignment angle $\theta$.  In general $F_\pi$ in these different composite models is determined through a constraint of the form
\begin{align} 
v_{\rm EW}^2 = F_\pi^2 N_D \sin^2 \theta+ v^2
\end{align}
where $N_D$ is the number of electroweak doublet fermion families, $\theta=\pi/2$ corresponds to the technicolor vacuum and  $0<\theta<\pi/2$ to the composite Higgs vacuum. 
%the EW preserving vacuum at  $v$ is the vev of the fundamental scalar doublet. The pion decay constant of the strong interactions,  $F_\pi$, sets the mass scale of the composite resonances. 

In this study we restrict ourselves to the technicolor limit, which provides dynamical electroweak symmetry breaking, and a composite Higgs resonance, but requires a further extension to provide SM fermion masses. The composite Higgs resonance has been argued to be heavy in the Technicolor limit with respect to the electroweak scale, by analogy with  scalar resonances in QCD which are heavy compared to the QCD pion decay constant. However for composite sectors which are not a copy of QCD it is a non-perturbative problem to determine the lightest scalar mass. Both model computations \cite{Dietrich:2005jn,Kurachi:2006ej} and lattice simulations \cite{Fodor:2014pqa,Kuti:2014epa} of models like the NMWT model, which appear to be near the conformal window, have indicated the presence of a scalar  $0^{++}$ resonance that is much lighter than expected from simply scaling up scalar masses in QCD. The physics behind the origin of the fermion masses can also play a role in reducing this TC Higgs mass to the observed value at LHC and at the same time provide SM Higgs like couplings \cite{Foadi:2012bb,Belyaev:2013ida} to the SM particles. It is possible to probe the origin of the fermion masses, whether they are due to extended technicolor, fermion partial compositeness or a new elementary scalar, via the pseudo scalar sector of the theory, the analogues of the QCD $\eta$ and $\eta'$ resonances as discussed in \cite{DiVecchia:1980xq,Alanne:2016rpe}.

%has the group structure of $SU(2)_{L}\times SU(2)_{R}$, the simplest global chiral symmetry for the WTC class of theories. 
We follow the same prescription for constructing an effective theory of the underlying composite dynamics as in \cite{Appelquist:1999dq,Belyaev:2008yj}
by introducing composite spin-1 resonances transforming under the $SU(2)_{L}\times SU(2)_{R}$ global symmetry: Two new triplets of heavy spin-1 resonances are introduced at interaction eigenstate level  $A_{L/R}$ as gauge fields under $SU(2)_{L/R}$ respectively. The $SU(2)_L$ is gauged as $SU(2)_W$ such that the $A_{L}$ fields form a weak triplet analogous to the triplet in $W'$ models while the $A_{R}$ fields are  $SU(2)_W$ singlets.

%s this introduces a single new triplet and singlet vector meson field, $A_{L}$ and $A_{R}$ respectively.
Together with the Standard Model electroweak fields in the gauge eigenbasis, $\tilde{W}_{\mu}$ and $\tilde{B}_{\mu}$, we define chiral fields $C_{L/R \mu}$ 
%which include the left handed triplet fields and the right handed singlet fields respectively,
\begin{equation}
C_{L\mu}\equiv A_{L\mu}-\frac{g}{\tilde{g}}\tilde{W}_{\mu},
\hspace{2cm}
C_{R\mu}\equiv A_{R\mu}-\frac{g^{\prime}}{\tilde{g}}\tilde{B}_{\mu},
\end{equation}
where $g$, $g^{\prime}$ are the usual Standard Model EW coupling constants, and $\tilde{g}$ is the coupling constant of the NMWT gauge interactions. 
These fields transform homogeneously when the $A_{L/R}$ fields are introduced formally as gauge fields. 

The scalar composite Higgs resonance $H$ and the triplet of pions $\pi^a$ absorbed by the $W$ and $Z$ bosons are introduced as a bi-doublet field under the $SU(2)_{L/R}$ symmetries %Lagrangian also contains the Higgs self-interactions and its mixing with the EW and TC gauge fields within the 
described via the $2\times 2$ matrix M,
\begin{equation}
M=\frac{1}{\sqrt{2}}\big[v+H+2i\pi^{a}T^{a}], \quad M\to u_L M u_R^\dagger , \quad u_{L/R} \in SU(2)_{L/R}
\end{equation}
where $a=1,2,3$, $v=\mu/\sqrt{\lambda}$ is the vacuum expectation value associated with the breaking of the chiral symmetry, 
%$H$ is the composite Higgs
%\footnote{The composite Higgs appears as the lightest scalar mode of NMWT when the chiral symmetry is broken, and is a composite of techni-quarks}
%, $\pi^{a}$ are the Goldstone bosons created by the symmetry breaking, 
and $T^{a}$ are the generators of the $SU(2)$ groups, related to the Pauli Matrices by $T^{a}=\sigma^{a}/2$. The electroweak covariant derivative of $M$ is
\begin{equation}
D_{\mu}M = \partial_{\mu}M-ig\tilde{W}_{\mu}^{a}T^{a}M+ig^{\prime}M\tilde{B}_{\mu}T^{3}.
\end{equation}
With these definitions we write the low-energy effective Lagrangian of the model, up to dimension 4 operators as in \cite{Belyaev:2008yj}:
\begin{align}
\begin{split}
 \mathcal{L}_{boson} = & -\frac{1}{2}\textrm{Tr}\big[\tilde{W}_{\mu\nu}\tilde{W}^{\mu\nu}\big] - \frac{1}{4}\tilde{B}_{\mu\nu}\tilde{B}^{\mu\nu} - \frac{1}{2}\textrm{Tr}[F_{L\mu\nu}F_{L}^{\mu\nu}{+}F_{R\mu\nu}F_{R}^{\mu\nu}] \\
 & + m^{2}\textrm{Tr}[C_{L\mu}^{2}{+}C_{R\mu}^{2}]+\frac{1}{2}\textrm{Tr}[D_{\mu}MD^{\mu}M^{\dagger}]-\tilde{g}^{2}r_{2}\textrm{Tr}[C_{L\mu}MC_{R}^{\mu}M^{\dagger}] \\
 & - \frac{i\tilde{g}r_{3}}{4}\textrm{Tr}[C_{L\mu}(MD^{\mu}M^{\dagger}-D^{\mu}MM^{\dagger})+C_{R\mu}(M^{\dagger}D^{\mu}M-D^{\mu}M^{\dagger}M)] \\
 & + \frac{\tilde{g}^{2}s}{4}\textrm{Tr}[C_{L\mu}^{2}+C_{R\mu}^{2}]\textrm{Tr}[MM^{\dagger}]+\frac{\mu^{2}}{2}\textrm{Tr}[MM^{\dagger}]-\frac{\lambda}{4}\textrm{Tr}[MM^{\dagger}]^{2},
\end{split}
\label{eqn:bosonlagng}
\end{align}
where $\tilde{W}_{\mu\nu}$ and $\tilde{B}_{\mu\nu}$ are the SM electroweak field strength tensors, and $F_{L/R\mu\nu}$ are the field strength tensors corresponding to the vector meson fields.

%This Lagrangian also contains the Higgs self-interactions and its mixing with the EW and TC gauge fields within the $2\times 2$ matrix M,
%
%\begin{equation}
%M=\frac{1}{\sqrt{2}}\big[v+H+2i\pi^{a}T^{a}],
%\end{equation}
%
%where $a=1,2,3$, $v=\mu/\sqrt{\lambda}$ is the vacuum expectation value associated with the breaking of the chiral symmetry, $H$ is the composite Higgs%\footnote{The composite Higgs appears as the lightest scalar mode of NMWT when the chiral symmetry is broken, and is a composite of techni-quarks}
%, $\pi^{a}$ are the Goldstone bosons created by the symmetry breaking, and $T^{a}$ are the generators of the $SU(2)$ groups, related to the Pauli Matrices by $T^{a}=\sigma^{a}/2$. The Lagrangian also contains the covariant derivative of $M$,
%
%
%\begin{equation}
%D_{\mu}M = \partial_{\mu}M-ig\tilde{W}_{\mu}^{a}T^{a}M+ig^{\prime}M\tilde{B}_{\mu}T^{3}.
%\end{equation}

The global symmetry breaking pattern $SU(2)_{L}\times SU(2)_{R}\times U(1)_{V}\rightarrow SU(2)_{V}\times U(1)_{V}$ is triggered by the vev of $M$ and provides the 3 Goldstone degrees of freedom for the massive $W$ and $Z$ bosons. The heavy vector resonances, here introduced via the $A_{L/R}$ triplets, 
%which here are just added in according to global symmetries, 
can equivalently be treated as Higgs'ed
gauge fields of a 'hidden local symmetry' copy of the above global symmetry group \cite{Bando:1987br}, as discussed in \cite{Foadi:2007ue}. 
%which is a mirror to this gauge group, vector and axial vector triplets are formed respectively. 
The physical spectrum of the model then consist of the 2 triplets of spin-1 mesons which in the absence of electroweak interactions form a vector triplet $V$ under $SU(2)_V$ and the  axial-vector partner triplet $A$, analogous to the $\rho$ and $a_1$ vector mesons in QCD. 
%To understand the particle content directly from group theory arguments, consider that there is a doublet of techni-quarks under each of the $SU(2)_{L/R}$ groups. Chiral symmetry breaking leads to
%\begin{equation}
%2_{L} \otimes 2_{R} \otimes 1_{V} \rightarrow 3_{V}+1_{V},
%\end{equation}
In this study we focus on the two neutral resonance mass eigenstates which, in the presence of SM electroweak interactions, we for convenience refer to as $Z'$ and $Z''$ although these are distinct from sequential $Z'$ resonances.

%\textcolor{red}{Below here discuss a bit the differences with W' and ordinary Z' in terms of the symmetries and left right admixtures:}. 
%so the techni-quarks combine to form a vector-vector triplet, corresponding to the  $Z^{\prime}/W^{\pm\prime}$ particles ($\rho$-mesons). The associated mirror gauge group breaks similarly, producing an axial-vector triplet corresponding to the $Z^{\prime\prime}/W^{\pm\prime\prime}$ particles ($a$-mesons). Note that there is also a vector-singlet produced, however we do not consider this during the analysis of this model as it is decoupled from the theory, so does not mix with any other particles. 

The spin-1 sector of the Lagrangian in equation \ref{eqn:bosonlagng} contains five parameters, $m, \tilde{g}, r_2, r_3$ and $s$. The masses and decay constants of the vector and axial-vector resonances, in the limit of zero electroweak couplings, are given in terms of these parameters as
\begin{align}
\begin{split}
M_{V}^{2} & = m^{2}+\frac{\tilde{g}^{2}(s-r_{2})v^{2}}{4}, \quad F_{V}  = \frac{\sqrt{2}M_{V}}{\tilde{g}}, \\
M_{A}^{2} & = m^{2}+\frac{\tilde{g}^{2}(s+r_{2})v^{2}}{4}, \quad F_{A}  = \frac{\sqrt{2}M_{A}}{\tilde{g}}\chi,
\end{split}
\label{eqn:massvec-ax}
\end{align}

%where $M_{V}$ and $M_{A}$ are the masses of the vector and axial-vector mesons in the limit of zero electroweak coupling, respectively. The masses and Lagrangian coefficients can be combined to form decay constants for the vector and axial-vector mesons,
%
%\begin{align}
%\begin{split}
%F_{V} & = \frac{\sqrt{2}M_{V}}{\tilde{g}}, \\
%F_{A} & = \frac{\sqrt{2}M_{A}}{\tilde{g}}\chi, 
%\end{split}
%\label{eqn:decayvec-ax}
%\end{align}

where 

\begin{equation}
\chi \equiv 1 - \frac{v^{2}\tilde{g}^{2}r_{3}}{4M_{A}^{2}}.
\label{eqn:chi}
\end{equation}

The techni-pion decay constant $F_{\pi}$ may be expressed in terms of $F_{V}$ and $F_{A}$ as

\begin{equation}
F_{\pi}^{2} = (1+2\omega)F_{V}^{2}-F_{A}^{2},
\label{eqn:decayconsts}
\end{equation}
with 
\begin{equation}
\omega \equiv \frac{v^{2}\tilde{g}^{2}}{4M_{V}^{2}}(1-r_{3}+r_{2}).
\end{equation}
and with $F_{\pi}=246 \sqrt{N_D}$ GeV in technicolor models with $N_D$ families of technifermions and no elementary doublet scalars.  Here we assume $N_D=1$ as in the NMWT model.
%At this stage, there are five free parameters in the NMWT model, $M_{A}$, $\tilde{g}$, $s$, $r_{2}$ and $r_{3}$. 
We can now make use of the Weinberg Sum Rules (WSRs) \cite{PhysRevLett.18.507} to constrain the number of parameters in the effective model and connect them to 
%make the connection from the effective Lagrangian to 
the underlying fermionic dynamics. 

The assumed asymptotic freedom of the effective theory implies the 1st and 2nd WSRs respectively

\begin{equation}
\int_{0}^{\infty}ds\textrm{Im}\Pi_{LR}(s) = 0, \hspace{1.5cm}
\int_{0}^{\infty}ds s\textrm{Im}\Pi_{LR}(s) = 0.
\label{eqn:1stand2ndWSR}
\end{equation}
where $\Pi_{LR}(s) $ is the Lorentz invariant part of the LR correlation function:
\begin{equation}
\Pi^{a,b}_{\mu\nu LR}(q)= (q_{\mu}q_{\nu}-g_{\mu\nu}q^{2})\delta^{ab}\Pi_{LR}(q^{2}),
\end{equation} 
with
\begin{align}
i\Pi^{a,b}_{\mu\nu LR}(q) &= \int d^{4}x e^{iq\cdot x} [ \langle J^{a}_{\mu ,V}(x)J^{b}_{\nu ,V}(0) \rangle - \langle J^{a}_{\mu ,A}(x)J^{b}_{\nu ,A}(0) \rangle ].
\end{align}
Assuming that only the lowest spin-1 resonances $A,V$ saturate the WSRs, the vector and axial vector spectral densities are given in terms of the spin-1 masses and decay constants as
\begin{align}
\begin{split}
\textrm{Im}\Pi_{V}(s)&=\pi F_{V}^{2}\delta(s-M_{V}^{2}) \\
\textrm{Im}\Pi_{A}(s)&=\pi F_{\pi}^{2}\delta(s) + \pi F_{A}^{2}\delta(s-M_{A}^{2}),
\end{split}
\end{align}
%where $F_{V}(F_{A})$, $M_{V}(M_{A})$ are the decay constants and masses of the vector(axial) mesons respectively. 
%Applying these equations to 
The 1st WSR therefore implies that $\omega=0$ and
\begin{equation}
F_{V}^{2}-F_{A}^{2}=F_{\pi}^{2},
\end{equation} 
In terms of the Lagranian parameter this gives the relation 
%which we can use to constrain the NMWT parameter space by comparison of this WSR to equation \ref{eqn:decayconsts}. From this constraint $\omega=0$, we can write the relation
%
\begin{equation}
r_{2}=r_{3}-1,
\end{equation}

The second WSR is less dominated by the infrared dynamics than the first WSR as seen from Eq.~\ref{eqn:1stand2ndWSR}. We therefore allow for a modification of the second WSR encoded by the dimensionless parameter $a$ following \citep{Appelquist:1998xf} 
\begin{equation}
a\frac{8\pi^{2}}{d(R)}F_{\pi}^{4} = F_{V}^{2}M_{V}^{2}-F_{A}^{2}M_{A}^{2},
\label{fig:a-param}
\end{equation}
where $d(R)$ is the dimension of the gauge group representation of the underlying technifermions. The parameter $a$ measures the contribution of the underlying dynamics to the integral in  Eq.~\ref{eqn:1stand2ndWSR} from intermediate energies, above the confinement scale.

\begin{figure}[htb]
%\subfigure[]{\includegraphics[type=pdf,ext=.pdf,read=.pdf,width=0.5\textwidth]{figures/a-param-exclusions_psminus0.1_s0}}%
\subfigure[]{\includegraphics[type=pdf,ext=.pdf,read=.pdf,width=0.5\textwidth]{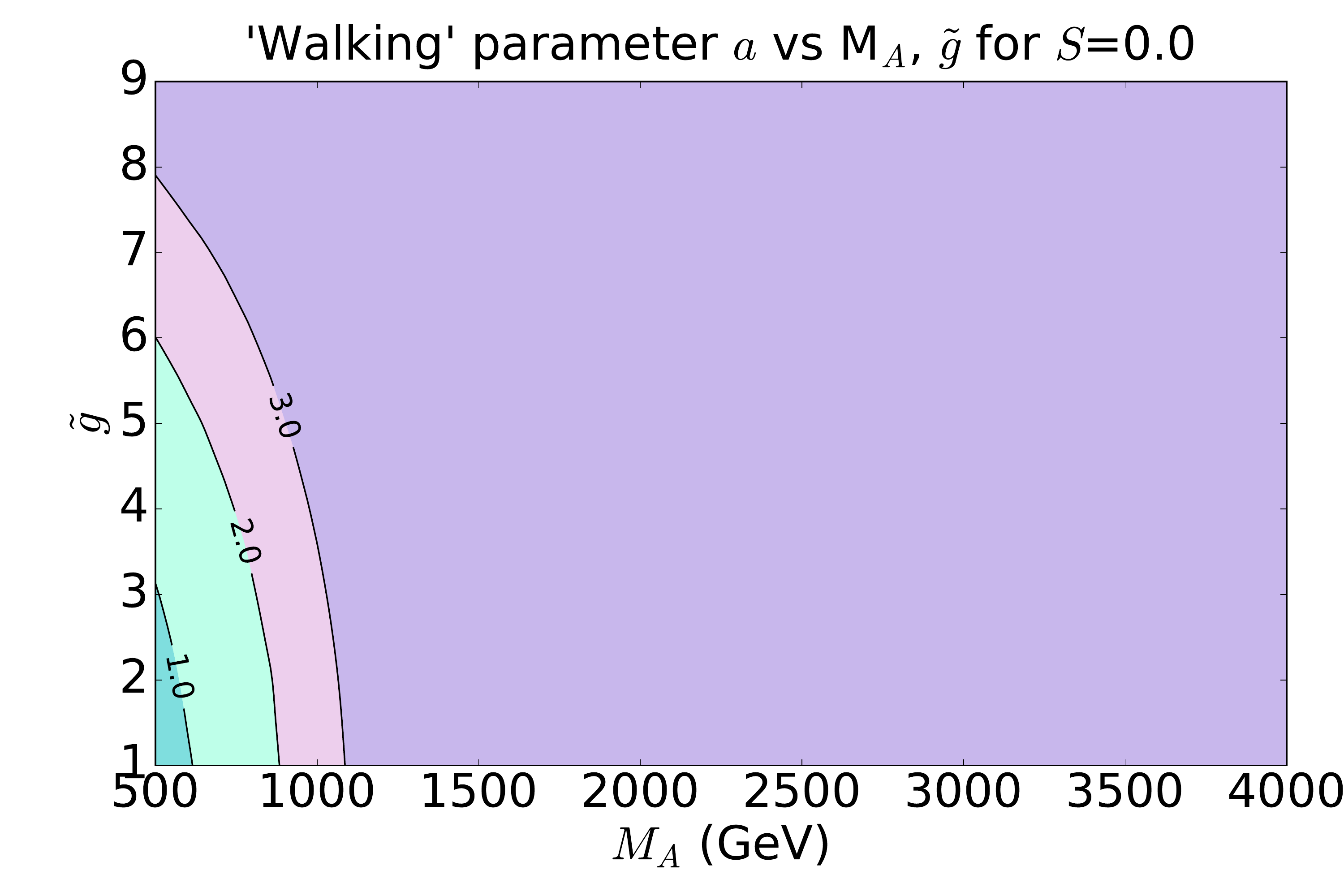}}%
\subfigure[]{\includegraphics[type=pdf,ext=.pdf,read=.pdf,width=0.5\textwidth]{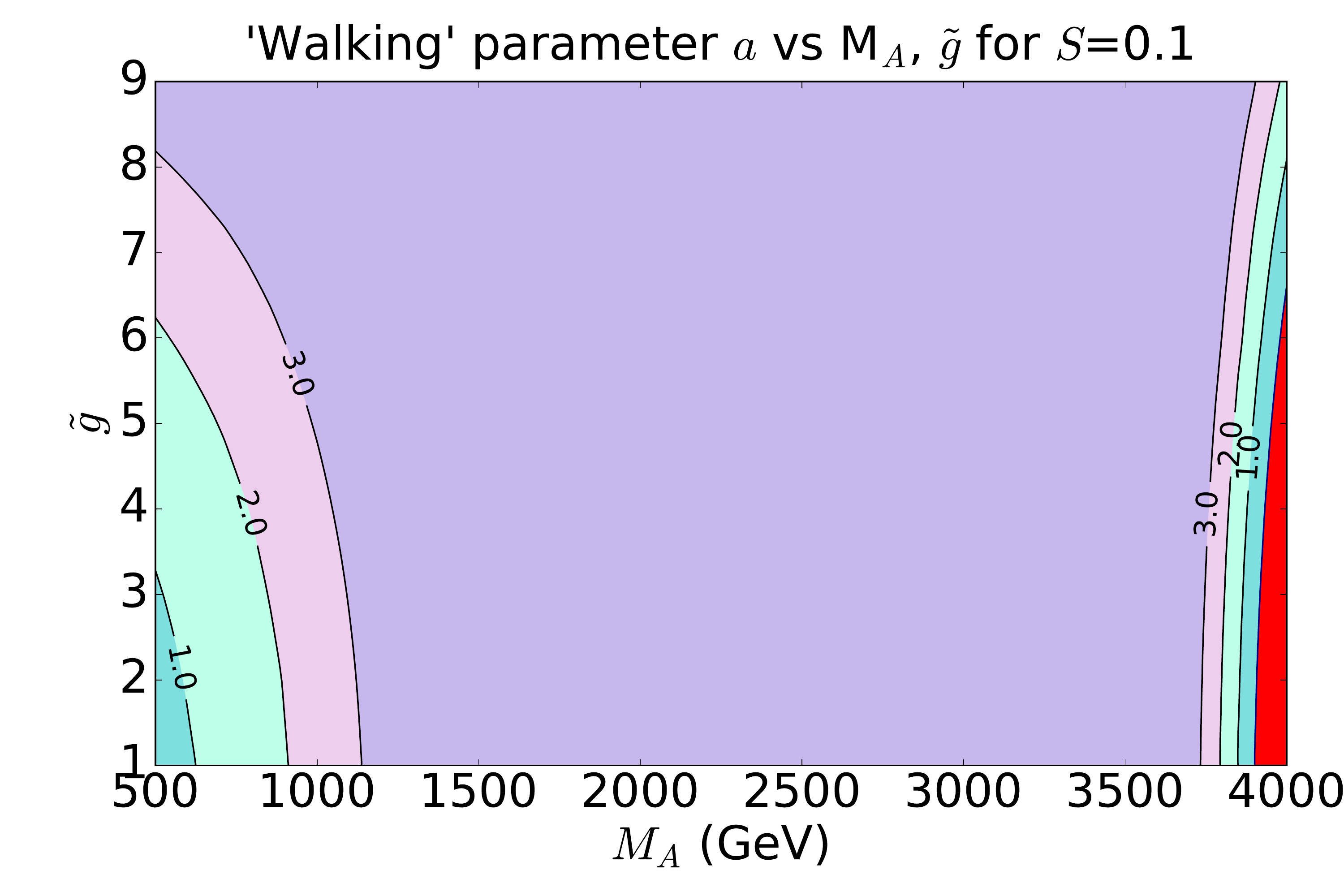}}
\subfigure[]{\includegraphics[type=pdf,ext=.pdf,read=.pdf,width=0.5\textwidth]{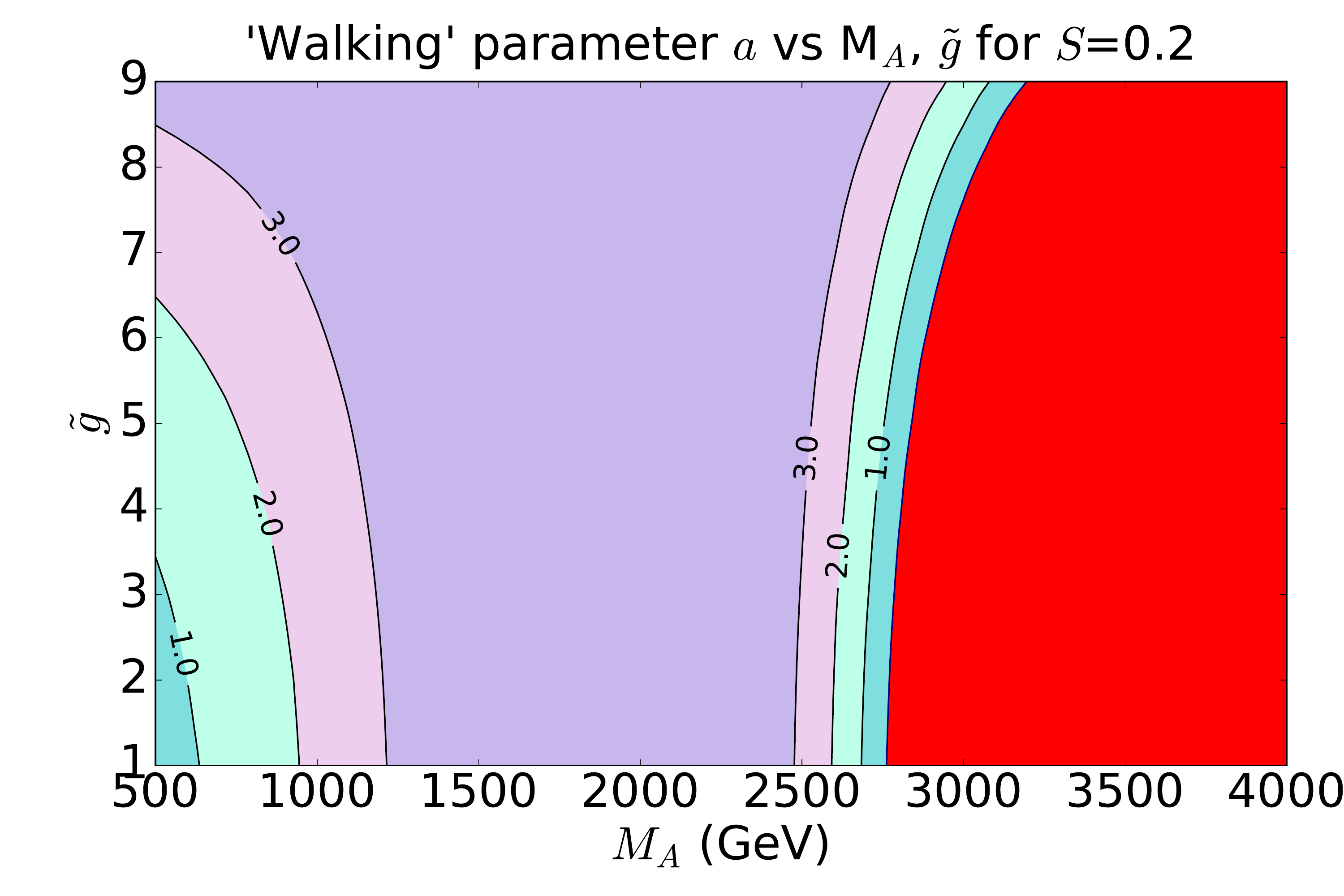}}%
\subfigure[]{\includegraphics[type=pdf,ext=.pdf,read=.pdf,width=0.5\textwidth]{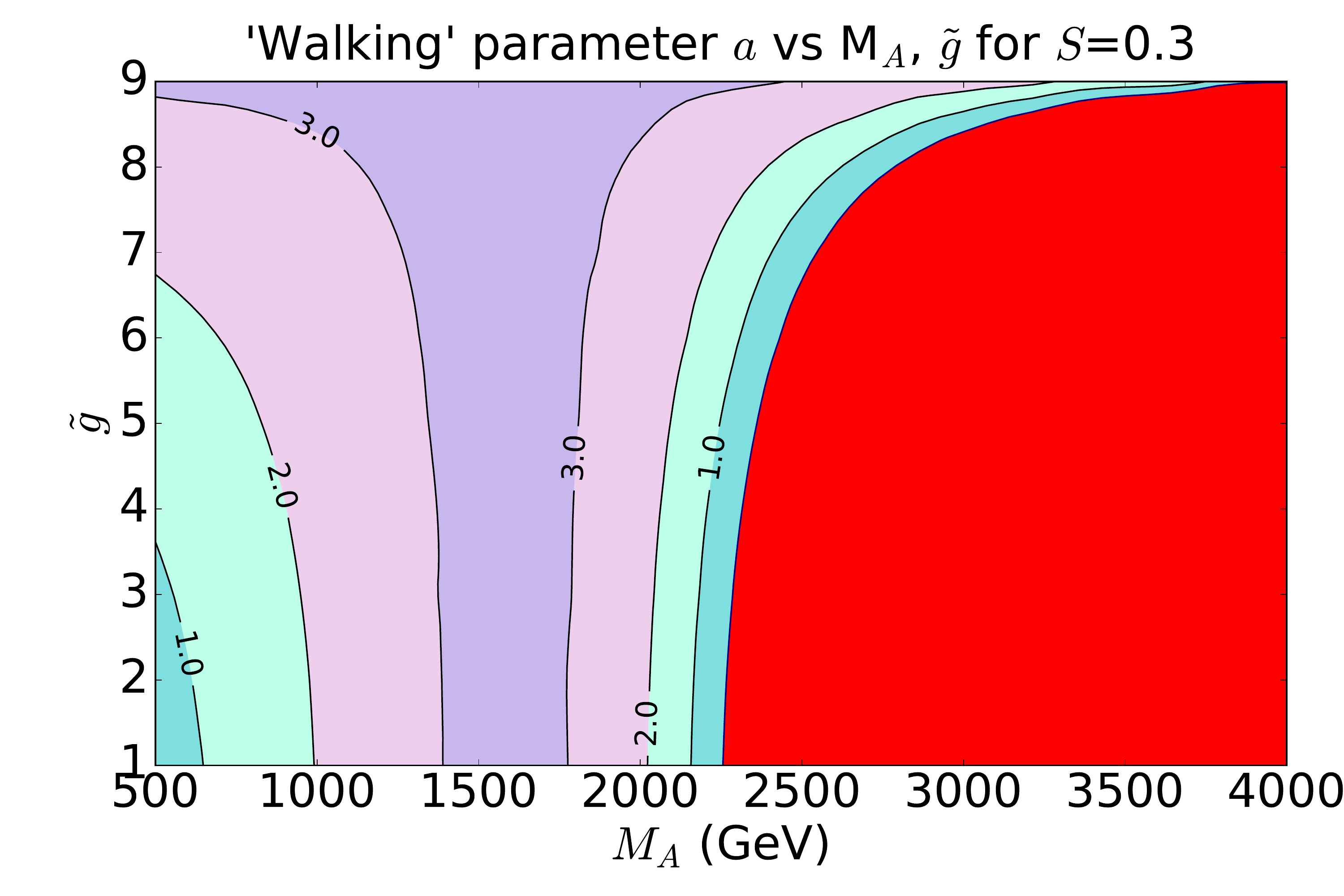}}
\caption{\label{fig:aparam} Contour levels for $a$ parameter in ($M_A$,$\tilde{g}$) NMWTC plane for various values of $S$ and fixed $s=0$. 
The red-shaded region corresponds to excluded  $a<0$ space.}
\end{figure}

Finally the electroweak Peskin-Takeuchi $S$ parameter is related to a zeroth Weinberg sum rule, 
% We can also make use of the WSR at order $\mathcal{O}(Q^{4})$, which is the definition of the  as defined in Ref\cite{Appelquist:1998xf},
\begin{equation}
S = 4\int_{0}^{\infty}\frac{ds}{s}\textrm{Im}\bar{\Pi}_{LR} = 4\pi\Bigg[\frac{F_{V}^{2}}{M_{V}^{2}}-\frac{F_{A}^{2}}{M_{A}^{2}}\Bigg].
\label{eqn:zerothwsr}
\end{equation}
Combining the first and second WSRs it follows that the $a$ parameter gives a negative contribution to axial-vector mass difference $M_A^2-M_V^2$ and a negative contribution to the $S$ parameter. We therefore expect $a$ to be positive in a near conformal theory yielding a smaller $S$-parameter and a more degenerate axial-vector mass spectum than in QCD. This is in line with e.g. model computations \cite{Kurachi:2006mu,Kurachi:2006ej} based on Schwinger-Dyson analysis, but here we take it as an assumption.

In QCD we expect $a\simeq 0$ while in near-conformal theories, where the coupling constant is assumed to be approximately constant in a region above the confinement scale, we expect $a>0$ \citep{Appelquist:1998xf} .

This allows us to trade one of the Lagrangian parameters for the $S$ parameter via the relation
\begin{equation}
S=\frac{8\pi}{\tilde{g}^{2}}(1-\chi^{2}).
\label{eqn:0thwsr-gt}
\end{equation}
We are left with a four dimensional parameter space that describes the model: $M_{A}, \tilde{g},  S, s.$
%\label{eqn:parameterspace}
%\end{equation}
The Lagrangian constant $s$ parametrises the interactions of the Technicolor spin-1 mesons with the Higgs sector (see \ref{eqn:bosonlagng}). Since we do not consider the composite Higgs phenomenology the only relevant effect of the $s$ parameter is on the branching ratio of the $Z'$ and $Z''$ states into dileptons. The branching ratios into dileptons are maximal for $s=0$ so we therefore restrict to this throughout. 
%therefore also affects the Since we do not consider these processes explicitly here we therefore set $s=0$.  
%which we do not consider e. Most of the interactions considered in this analysis do not involve the composite Higgs, so naively we set $s=0$ throughout. However, it will become %important to consider $s$ as a variable fourth parameter when we consider these Higgs processes.\AC{Edit this section to justify setting $s=0$ (minimal effect on dilepton branching, %exclusions would be stronger for $s>0$)}
%In this work, 
This leaves 3 relevant parameters
\begin{equation}
M_{A}, \hspace{1cm} \tilde{g}, \hspace{1cm} S.
\label{eqn:reducedparameterspace}
\end{equation}

We show the value of the $a$ parameter in the $M_A, \tilde{g}$ plane for different values of $S$ in Figure \ref{fig:aparam}. Restricting to positive values of $a$
%[\AB{Sasha: Mads, we need to explain and argue why do we choose positive a, at least shortly}]
we get an upper limit on the mass parameter $M_{A}$  which compliments the experimental limits we derive from dilepton searches.

\section{Phenomenology and LHC potential to probe WTC parameter space}
\label{sec:pheno}

In our analysis of heavy neutral spin-one resonances in the NMWT parameter space, we conduct a 3-dimensional scan over $M_{A}$, $\tilde{g}$ and $S$. The results  in this section are presented in the $M_{A}, \tilde{g}$ parameter space for discrete values of $S$ such as  $S=-0.1,0.0,\dots , 0.3$. The largest value, $S=0.3$
for the range we choose, is already disfavoured by EWPD~\cite{Baak:2014ora}, however we include it in this work for direct comparison to results of the previous work~\cite{Belyaev:2008yj}. The remaining limits of the scan over $S$ ensure that the tension with EWPD is minimised (for the zero $T$-paramter). In this section we present results at the benchmark $S=0.1$; fixed values of $S\neq 0.1$ are given in Appendix \ref{subsec:S-plots}.

There is an upper bound on $\tilde{g}$:

\begin{equation}
\tilde{g} < \sqrt{\frac{8\pi}{S}},
\label{eqn:gtlimit}
\end{equation}
which follows from Eq.~\ref{eqn:zerothwsr} and 
ensures that all physical quantities are real as we will see below.
For $S=0.3$, the biggest value of $S$ we consider here, the upper limit is $\tilde{g}= 9.15$. Therefore we present all results in the $M_{A}, \tilde{g}$ space with $\tilde{g}\leq 9$ to avoid  unphysical parameter space.

The phenomenology of the NMWT model is explored using the CalcHEP package \cite{Belyaev:2012qa} which allows to perform simple and robust analysis of tree-level collider events. The Lagrangian for NMWT was implemented using LanHEP \cite{Semenov:2002jw}, from which all interaction vertices are generated for use in CalcHEP. We focus on neutral heavy spin-one resonances in the Drell-Yan channel, with di-leptons signature. The mass spectra of the $Z^{\prime}/Z^{\prime\prime}$ are presented in section \ref{subsec:massspec}, the coupling strength of $Z^{\prime}/Z^{\prime\prime}$ vertices in section \ref{subsec:couplings}, followed by a discussion of the total widths and dilepton branching ratios in section \ref{subsec:width-and-br}, production and total cross sections for DY processes of $Z^{\prime}/Z^{\prime\prime}$ are given in section \ref{subsec:theory-cs}, section \ref{subsec:interference} explores the interference between the neutral resonances and discusses the validity of reinterpreting LHC constraints for the  NMWT model, and finally section \ref{sec:exclusion} explored the LHC potential to probe the WTC parameter space.
\subsection{Masses and couplings}

\subsubsection{Mass spectra}
\label{subsec:massspec}

Besides numerical analysis it is informative also to perform analytical one
as we do for some masses and couplings to understand the qualitative properties of the model
and the limits of the parameter space.
Diagonalising the neutral mixing matrix (see details in Appendix \ref{subsec:massmatrices}), we find the $Z^{\prime}/Z^{\prime\prime}$ masses to 2nd order in $\tilde{g}^{-1}$ take the form
\begin{eqnarray}
M_{Z^{\prime}}^{2} &=& M_{A}^{2}\left(1+{\frac{g_{1}^{2}+g_{2}^{2}}{\tilde{g}^{2}}}\chi^{2}\right),
\label{eqn:masses1}
\\
M_{Z^{\prime\prime}}^{2} &=&  
M_{A}^{2}\left(1+{\frac{g_{1}^{2}+g_{2}^{2}}{2\tilde{g}^{2}}}\right)
\left(\chi^2+\frac{\tilde{g}^{2}F_{\pi}^{2}}{2M_{A}^{2}}\right),
\label{eqn:masses2}
\end{eqnarray} 
where from equation \ref{eqn:0thwsr-gt} we express $\chi$ as
\begin{equation}
\chi = \sqrt{1-\frac{S\tilde{g}^{2}}{8\pi}},
\label{eqn:chi-function}
\end{equation}
and $U(1)_{Y}$ and $SU(2)_{L}$ couplings $g_{1}$, $g_{2}$ are functions of ($M_{A}$, $\tilde{g}$, $S$), see equations \ref{eqn:g1full}, \ref{eqn:g2full}. Both $g_{1}$ and $g_{2}$ have a very mild dependence on the model parameters\footnote{Variation in the couplings is less than $1\%$ level across the parameter space}.

\begin{figure}[htb]
\subfigure[]{\includegraphics[type=pdf,ext=.pdf,read=.pdf,width=0.5\textwidth]{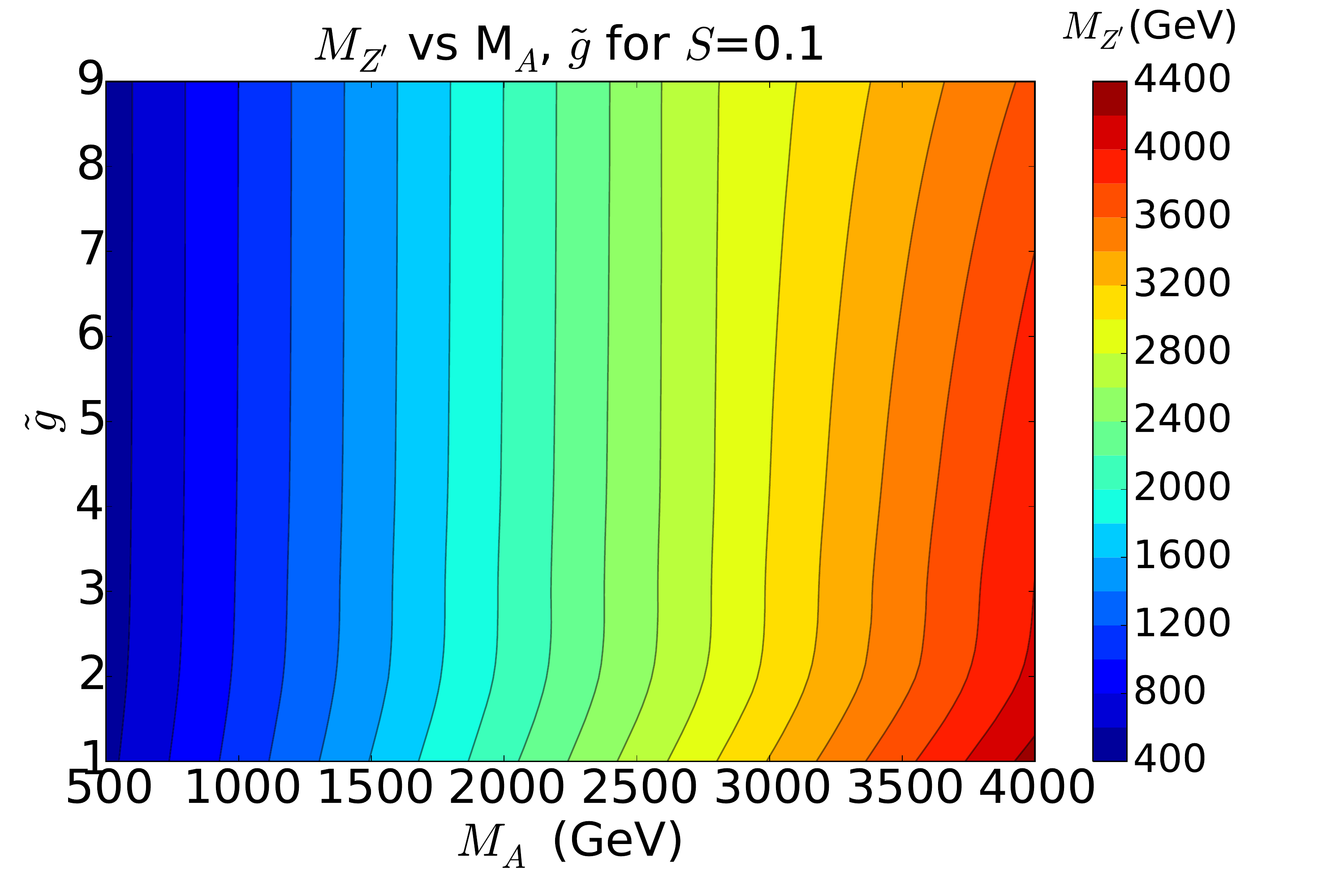}}%
%\subfigure[]{\includegraphics[type=pdf,ext=.pdf,read=.pdf,width=0.5\textwidth]{figures/Zpp_mass_ps0.1_s0}}\\
%\centering\subfigure[]{
\includegraphics[type=pdf,ext=.pdf,read=.pdf,width=0.5\textwidth]{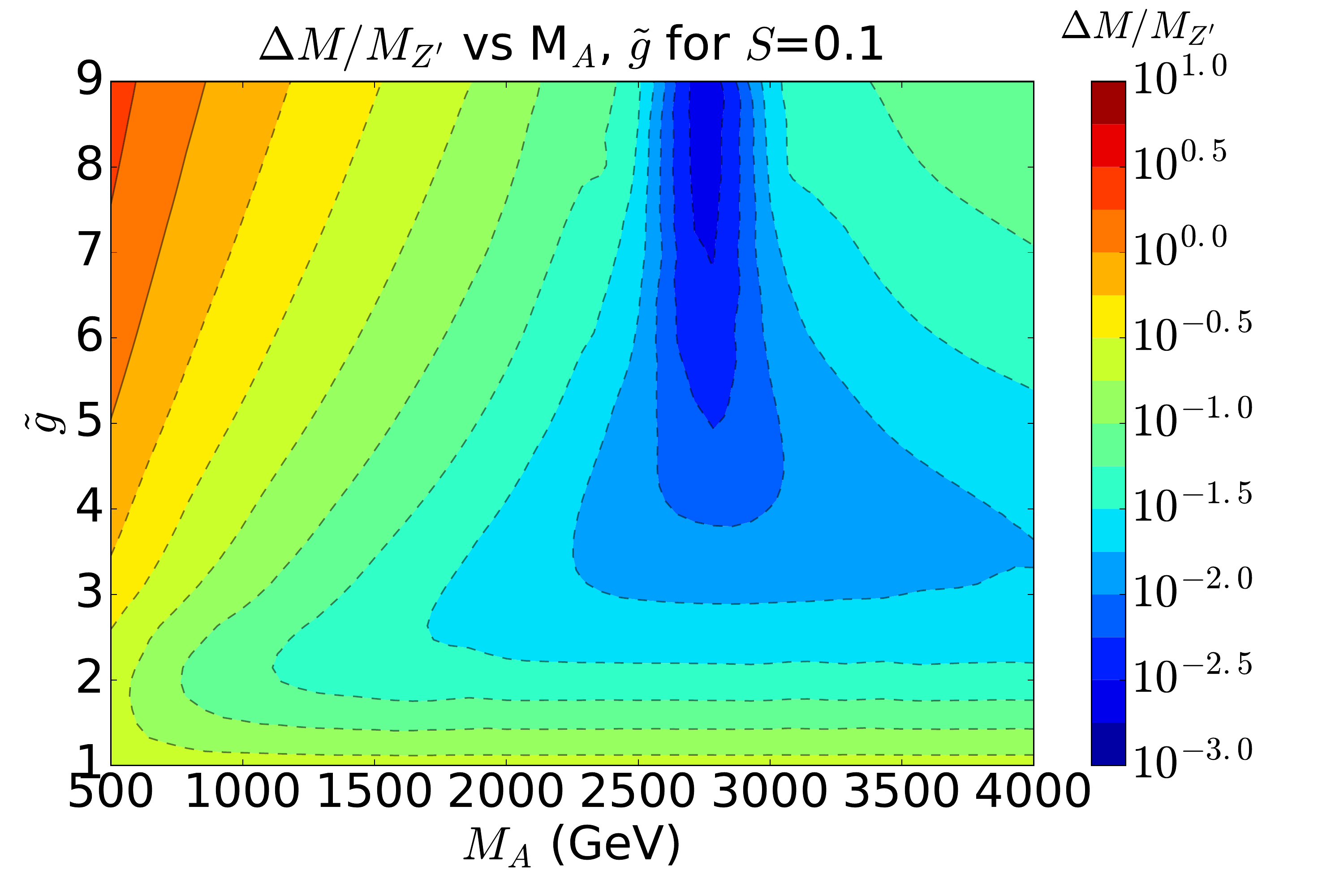}
%}%
\caption{\label{fig:masses} (a) $M_{Z^{\prime}}$(GeV) (b) $\Delta M/M_{Z^{\prime}}$ as a function of $M_A$, $\tilde{g}$, at benchmark values of $S=0.1$  and $s=0$}
\end{figure}

The mass spectrum of the $Z^{\prime}$ is shown in Figure \ref{fig:masses}a
and numerically presented in Table~\ref{tab:mass} where we also present the $Z''$ 
mass for the 3D grid in $(M_A,\tilde{g},S)$ space.
One can see that for $\tilde{g} \gtrsim 2$, $M_{Z^{\prime}}\simeq M_{A}^{2}$ 
as follows from Eq.~\ref{eqn:masses1}. 
In Figure~\ref{fig:masses}b we present the spectrum for the relative mass difference, $\Delta M/M_{Z'}$,
where $\Delta M=M_{Z''}-M_{Z'}$. One can see that $M_{Z^{\prime\prime}}$ behaviour is less trivial 
which reflects the 'competition' of $\tilde{g}$ and $F_\pi/M_A$ ratios in Eq.~\ref{eqn:masses2}.
For large  $M_{A}$ one can observe that  $Z^{\prime}$ starts to mildly depend on 
$\tilde{g}$.  This change in behaviour is due to a change of state of the $Z^{\prime}(Z^{\prime\prime})$ from mostly axial(vector) to mostly vector(axial)\cite{Belyaev:2008yj}. Figure \ref{fig:masses}b 
clearly reflects this mass inversion for $\tilde{g}>1$ at a fixed $M_{inv}=M_{A}$ which to 2nd order in $\tilde{g}^{-1}$  takes the form

\begin{equation}
M_{inv}^2 = \left(1+\frac{g_{1}^{2}+g_{2}^{2}}{\tilde{g}^{2}}\right)\frac{4\pi}{S}F_{\pi}^2.
\label{eqn:mass-inversion}
\end{equation}
 Using the benchmark $S=0.1$ the mass inversion occurs at $M_{A}=2760$GeV, we clearly observe this behaviour in Figure \ref{fig:masses}b.  

The mass splitting is large at low $M_{A}$, high $\tilde{g}$, opening new decay channels such as $Z^{\prime\prime}\rightarrow W^{+\prime}W^{-\prime}$. This is discussed further in section \ref{subsec:width-and-br}.

\begin{table}
\begin{center}
\begin{tabular}{cc|c|c|c|c|}
       \cline{3-6}
       & &     \multicolumn{4}{c|}{$M_{A}$(GeV)} \\ \hline
      \multicolumn{1}{|c|}{$S$} & \multicolumn{1}{|c|}{$\tilde{g}$} & 1000 & 1500 & 2000 & 2500 \\ \hline
      \multicolumn{1}{|c}{\multirow{5}{*}{-0.1}} &
      \multicolumn{1}{|c|}{1} & 1080(1339) & 1614(1984) & 2148(2639) & 2683(3296) \\ \cline{2-6}
      \multicolumn{1}{|c}{} &
      \multicolumn{1}{|c|}{3} & 1016(1163) & 1523(1640) & 2030(2138) & 2536(2643)  \\ \cline{2-6}
      \multicolumn{1}{|c}{} &
      \multicolumn{1}{|c|}{5} & 1006(1370) & 1509(1808) &2012(2283)  & 2515(2778) \\ \cline{2-6}
      \multicolumn{1}{|c}{} & 
      \multicolumn{1}{|c|}{7} & 1003(1642) & 1505(2049) & 2007(2510) & 2508(3001) \\ \cline{2-6}
      \multicolumn{1}{|c}{} &
      \multicolumn{1}{|c|}{9} & 1002(1947) & 1503(2334) & 2005(2788) & 2506(3280)  \\ \hline
      \multicolumn{1}{|c}{\multirow{5}{*}{0.1}} &
      \multicolumn{1}{|c|}{1} & 1078(1325) & 1610(1976) & 2144(2629) & 2678(3283) \\ \cline{2-6}
      \multicolumn{1}{|c}{} &
      \multicolumn{1}{|c|}{3} & 1015(1130) & 1520(1590) & 2023(2071) & 2522(2565) \\ \cline{2-6}
      \multicolumn{1}{|c}{} &
      \multicolumn{1}{|c|}{5} & 1005(1295) & 1507(1678) & 2010(2100) & 2511(2543) \\ \cline{2-6}
      \multicolumn{1}{|c}{} & 
      \multicolumn{1}{|c|}{7} & 1002(1518) & 1503(1821) & 2004(2175) & 2505(2560) \\ \cline{2-6}
      \multicolumn{1}{|c}{} &
      \multicolumn{1}{|c|}{9} & 1001(1773) & 1502(1998) & 2002(2277) & 2503(2591) \\ \hline
      \multicolumn{1}{|c}{\multirow{5}{*}{0.3}} &
      \multicolumn{1}{|c|}{1} & 1075(1320) & 1607(1968) & 2139(2618) & 2672(3270) \\ \cline{2-6}
      \multicolumn{1}{|c}{} &
      \multicolumn{1}{|c|}{3} & 1013(1097) & 1514(1541) & 1985(2034) & 2452(2540) \\ \cline{2-6}
      \multicolumn{1}{|c}{} &
      \multicolumn{1}{|c|}{5} & 1004(1215) & 1505(1537) & 1898(2008) & 2280(2510)  \\ \cline{2-6}
      \multicolumn{1}{|c}{} & 
      \multicolumn{1}{|c|}{7} & 1001(1382) & 1502(1560) & 1779(2002) & 2025(2503) \\ \cline{2-6}
      \multicolumn{1}{|c}{} &
      \multicolumn{1}{|c|}{9} & 1000(1580) & 1500(1593) & 1611(2000) & 1634(2500) \\ \hline   
\end{tabular}
\caption{\label{tab:mass} Masses of the neutral resonances at reference points in the $M_{A},\tilde{g},S$ parameter space, displayed in the format $M_{Z^{\prime}}$($M_{Z^{\prime\prime}})$ in GeV for each parameter space value}
\end{center}
\end{table}

\subsubsection{Couplings}
\label{subsec:couplings}

Here we explore analytic form of $Z^{\prime}$ and $Z^{\prime\prime}$  couplings to fermions. These are composed of elements of the neutral diagonalisation matrix $N_{ij}$ \cite{Foadi:2007ue}, details of the mixing matrix calculation are included in Appendix \ref{subsec:massmatrices}. 

For the vertices with fermions, the coupling strengths can be decomposed into left and right handed parts and
to 2nd order in $1/\tilde{g}$, 
$g_{Z'f\bar{f}}$ and $g_{Z''f\bar{f}}$ couplings
take the form:

\begin{equation}
g_{Z'f\bar{f}}^L = \frac{\chi}{2\sqrt{2}\tilde{g}}\left(-I_3 g_{2}^{2}+Y g_{1}^{2}\right), \hspace{1cm}
g_{Z'f\bar{f}}^R= \frac{\chi}{2\sqrt{2}\tilde{g}} q_f g_{1}^{2},
\label{eqn:zp-couplings}
\end{equation}

\begin{equation}
g_{Z''f\bar{f}}^L = \frac{1}{2\sqrt{2}\tilde{g}}\left(I_3 g_{2}^{2}+Y g_{1}^{2}\right), \hspace{1cm}
g_{Z''f\bar{f}}^R= \frac{1}{2\sqrt{2}\tilde{g}} q_f g_{1}^{2},
\label{eqn:zpp-couplings}
\end{equation}

where $I_3=\pm1/2$ is usual 3rd componet of the weak Isospin for up and down-femions respectively,
$Y=q_f-I_3$ is their hypercharge, and $q_f$ is the charge of the fermions.

The parameter dependence of the $Z^{\prime}$ and $Z^{\prime\prime}$ dilepton couplings are given as a ratio to the SM $g_{Zl^{+}l^{-}}$ in Figures \ref{eqn:zp-couplings} and \ref{eqn:zpp-couplings} respectively. Both L and R components of the $Z^{\prime}$ dilepton coupling increase as $\tilde{g}\rightarrow 1$, however as the coupling is diluted through the mixing effects between the gauge fields, $g_{Z^{\prime}l^{+}l^{-}}\geq g_{Zl^{+}l^{-}}$ is never realised.

Similarly, the L component of the $Z^{\prime\prime}$ dilepton coupling grows as $\tilde{g}\rightarrow 1$, however this is not the case for the R component. The R component is suppressed in comparison to the $Z^{\prime}$ as the mixing with the photon is smaller for $\gamma-Z''$ than $\gamma-Z'$; such mixing effects are discussed further in \ref{subsec:width-and-br}. 
%As there is no suppression by the factor of $\chi$, the $Z^{\prime\prime}$ dilepton coupling is stronger than that of the $Z^{\prime}$ in the low $\tilde{g}$ region, so we would expect constraints at low $\tilde{g}$ region of NMWT parameter space to be strongest from the $Z^{\prime\prime}$.
 
Again we see that the axial(vector) composition of the $Z^{\prime}(Z^{\prime\prime})$ affects both L and R coupling strengths, suppressing the coupling as the $Z^{\prime}(Z^{\prime\prime})$ becomes mostly vector(axial).

%%%%%%%%%% Ratio to SM of Z'/Z'' couplings to LH and RH components of e+e-

\begin{figure}[htb]
\subfigure[]{\includegraphics[type=pdf,ext=.pdf,read=.pdf,width=0.5\textwidth]{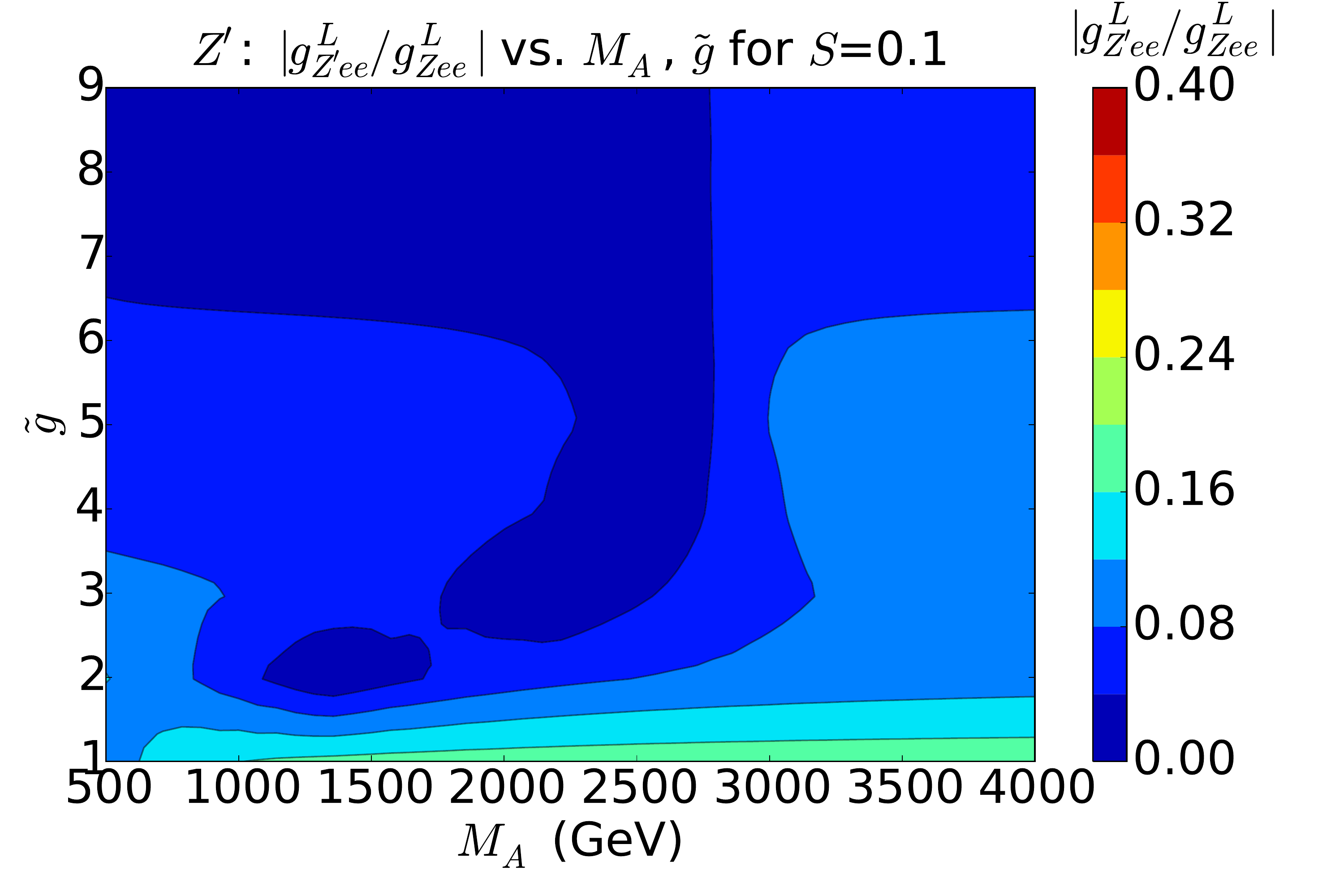}}%
\subfigure[]{\includegraphics[type=pdf,ext=.pdf,read=.pdf,width=0.5\textwidth]{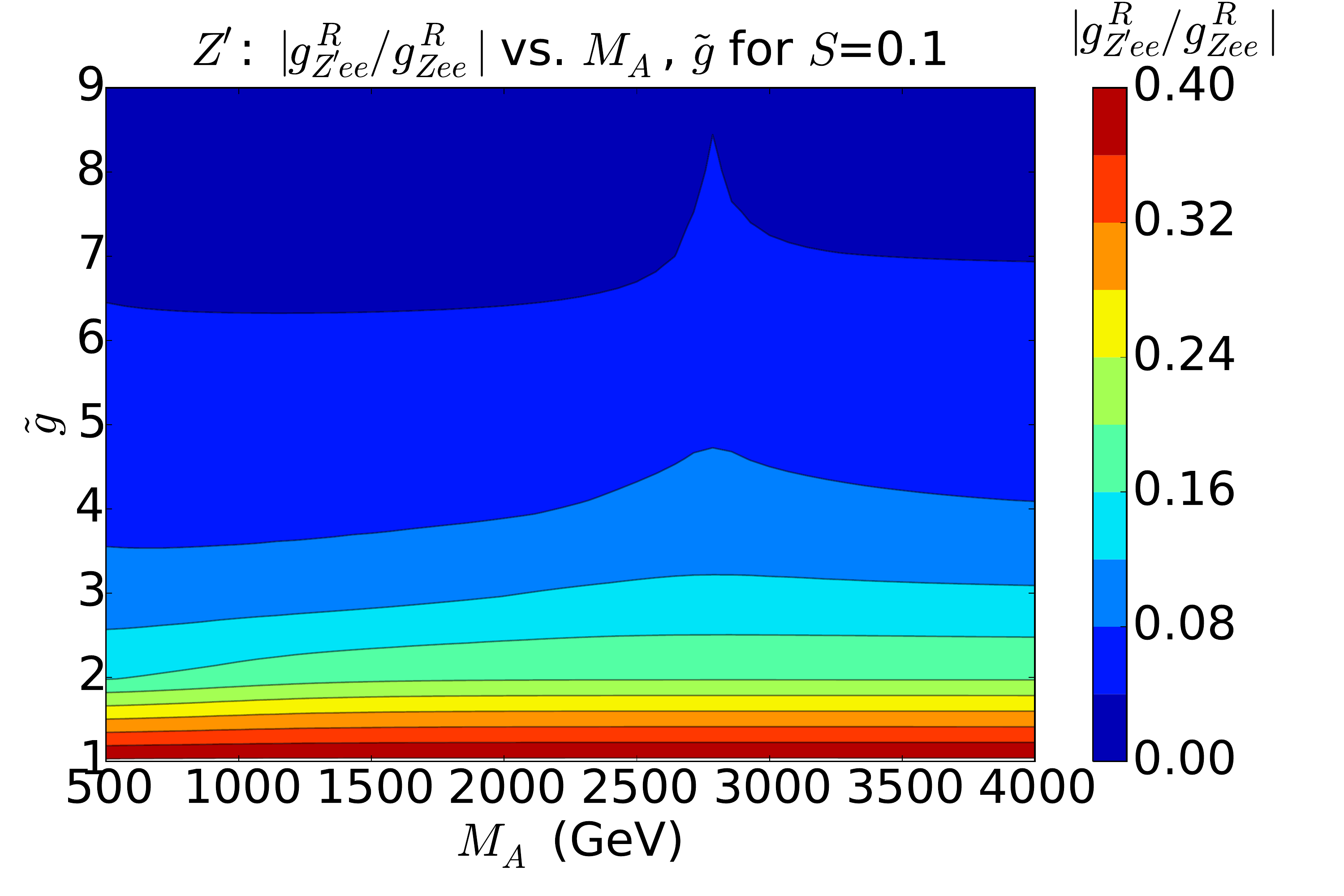}}
\caption{\label{fig:zp-coupling} Coupling of $Z^{\prime}$ to charged lepton pairs as a ratio to its SM equivalent separated into left and right handed components, (a) $\mid g_{Z^{\prime}l^{+}l^{-}}^{L}/g_{Zl^{+}l^{-}} \mid$, (b) $\mid g_{Z^{\prime}l^{+}l^{-}}^{R}/g_{Zl^{+}l^{-}} \mid$, as a function of $M_A$ and $\tilde{g}$ parameters at the benchmark values of $S=0.1$ and $s=0$}
\end{figure}

\begin{figure}[htb]
\subfigure[]{\includegraphics[type=pdf,ext=.pdf,read=.pdf,width=0.5\textwidth]{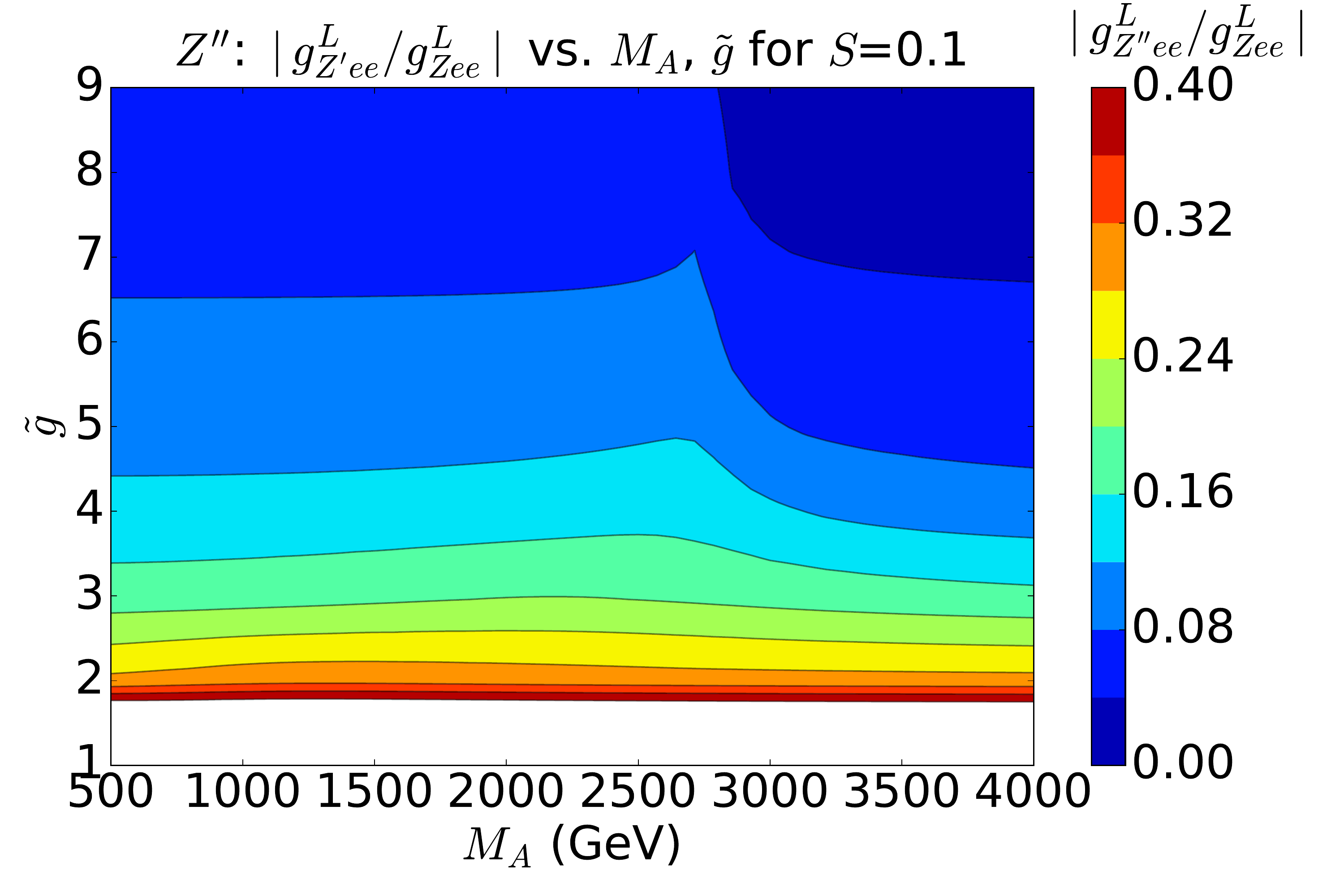}}%
\subfigure[]{\includegraphics[type=pdf,ext=.pdf,read=.pdf,width=0.5\textwidth]{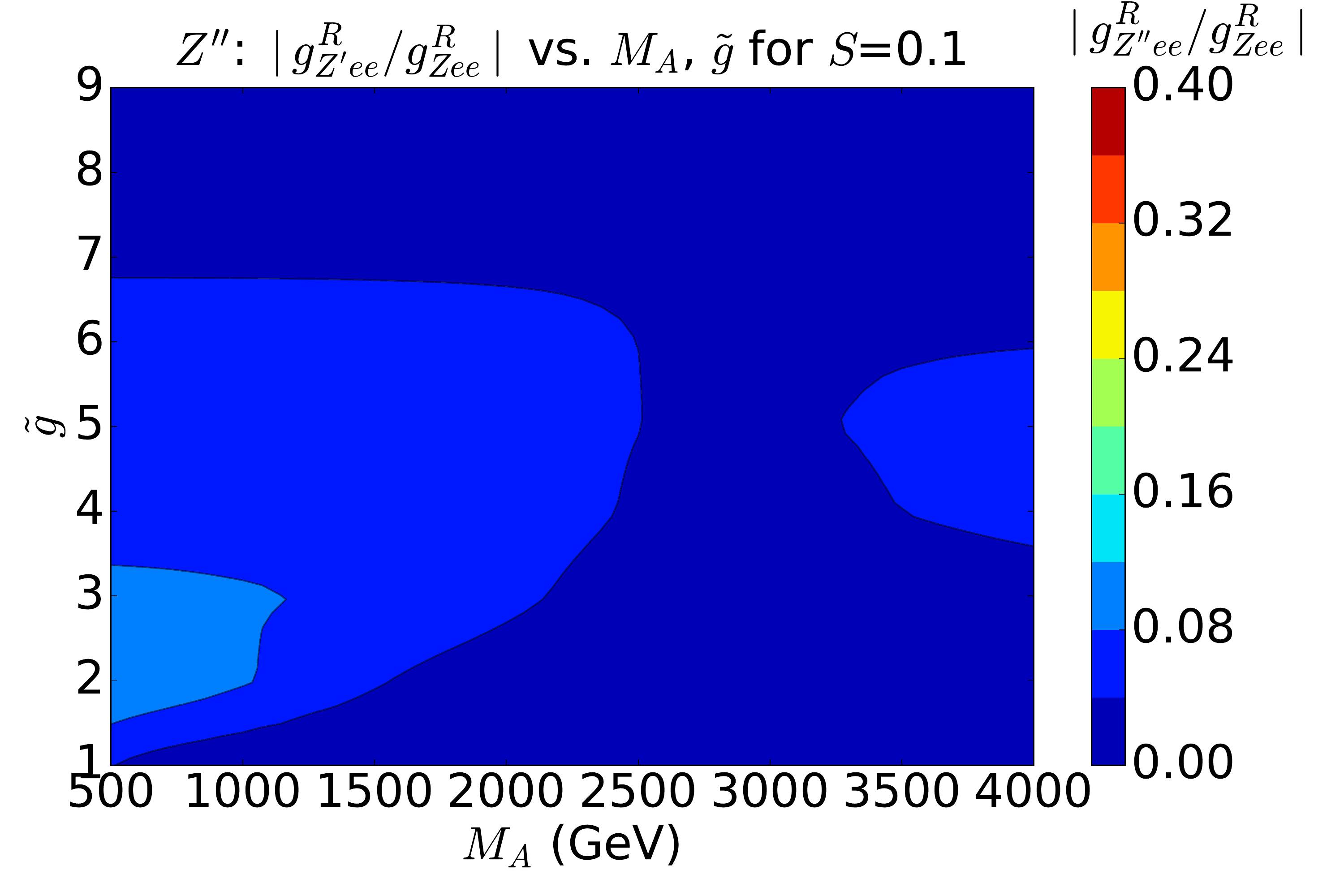}}
\caption{\label{fig:zpp-coupling} Coupling of $Z^{\prime\prime}$ to charged lepton pairs as a ratio to its SM equivalent separated into left and right handed components, (a) $\mid g_{Z^{\prime\prime}l^{+}l^{-}}^{L}/g_{Zl^{+}l^{-}} \mid$, (b) $\mid g_{Z^{\prime\prime}l^{+}l^{-}}^{R}/g_{Zl^{+}l^{-}} \mid$, as a function of $M_A$ and $\tilde{g}$ parameters at the benchmark values of $S=0.1$ and $s=0$}
\end{figure}

\subsection{Widths and branching ratios}
\label{subsec:width-and-br}

\begin{figure}[htb]
\subfigure[]{\includegraphics[type=pdf,ext=.pdf,read=.pdf,width=0.5\textwidth]{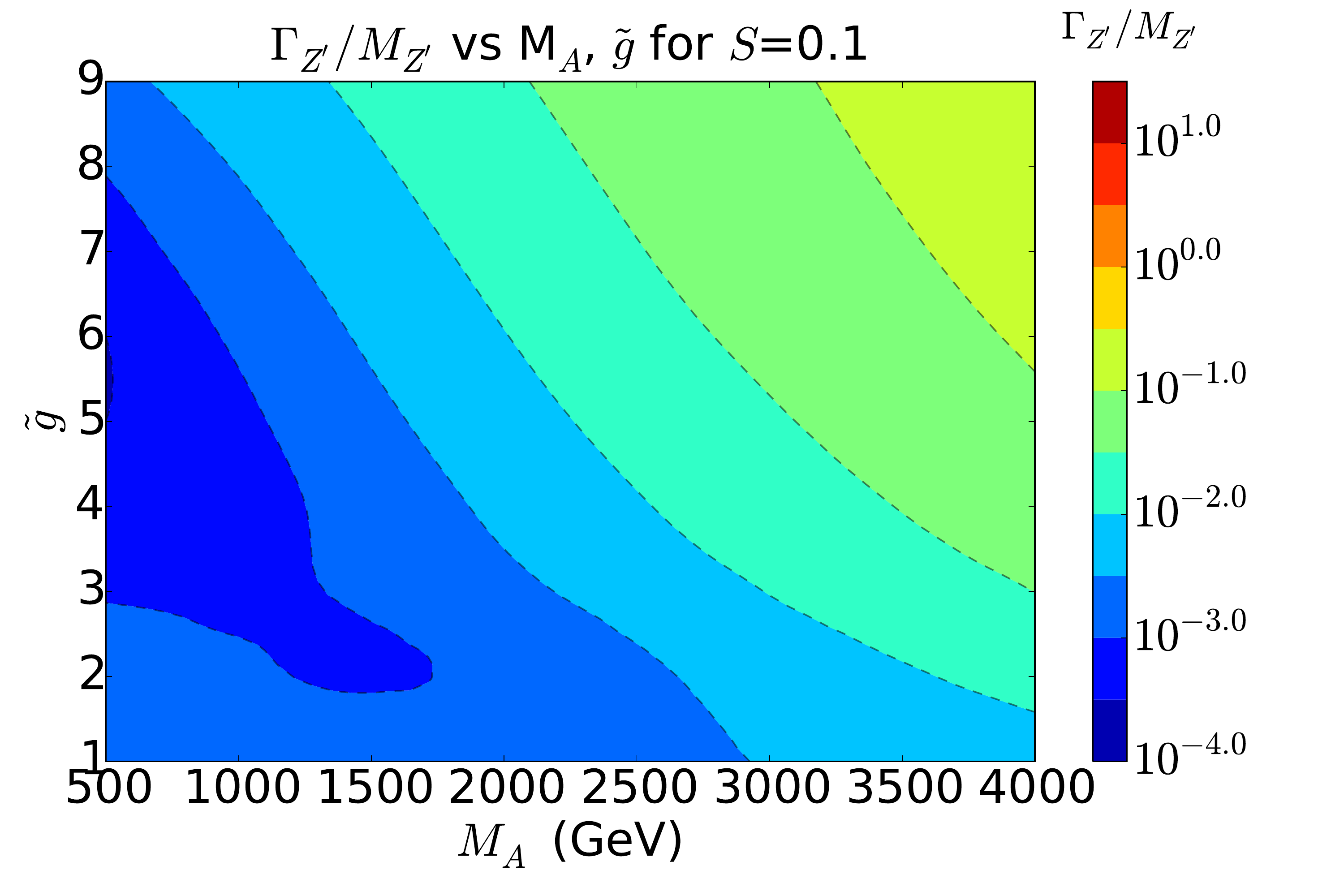}}%
\subfigure[]{\includegraphics[type=pdf,ext=.pdf,read=.pdf,width=0.5\textwidth]{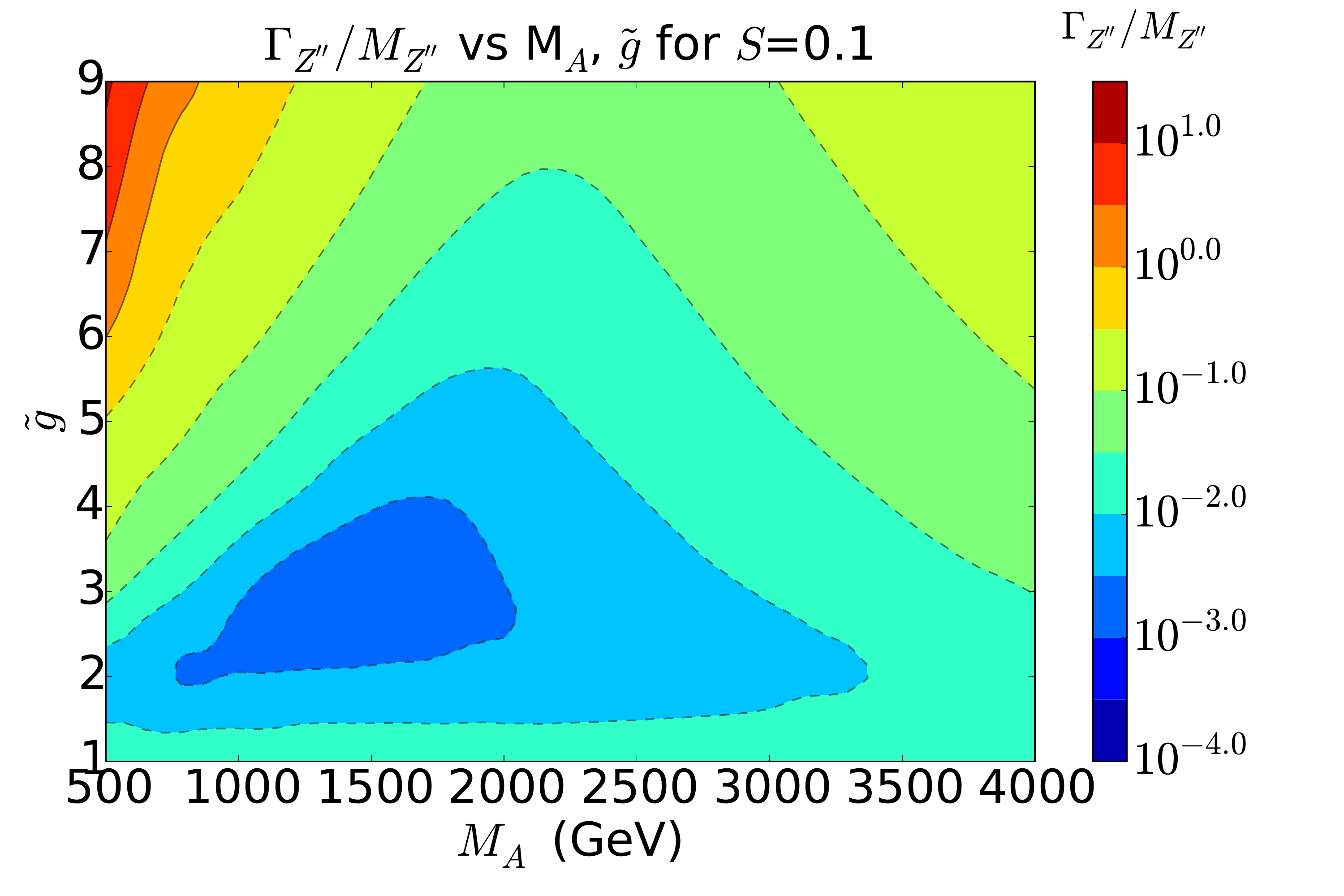}}
\caption{\label{fig:width2mass} (a) $\Gamma_{Z^{\prime}}/M_{Z^{\prime}}$, (b) $\Gamma_{Z^{\prime\prime}}/M_{Z^{\prime\prime}}$ as a function of $M_A$ and $\tilde{g}$ parameters at benchmark values of $S=0.1$ and $s=0$}
\end{figure}

The width-to-mass ratio $\Gamma/M$ for $Z^{\prime}$ and $Z^{\prime\prime}$ is shown in Fig.~\ref{fig:width2mass}.

\begin{figure}[htb]
\subfigure[]{\includegraphics[type=pdf,ext=.pdf,read=.pdf,width=0.5\textwidth]{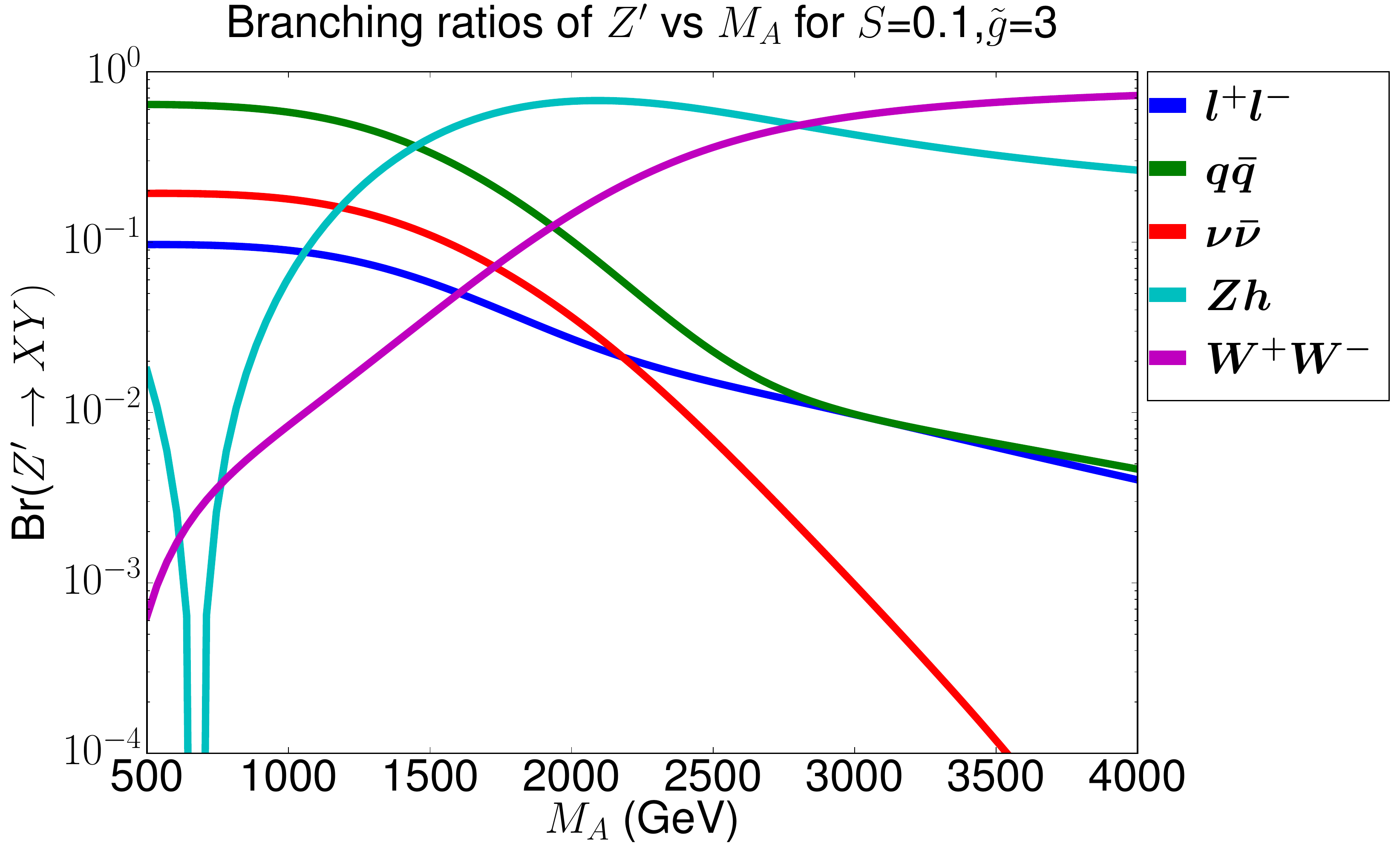}}%
\subfigure[]{\includegraphics[type=pdf,ext=.pdf,read=.pdf,width=0.5\textwidth]{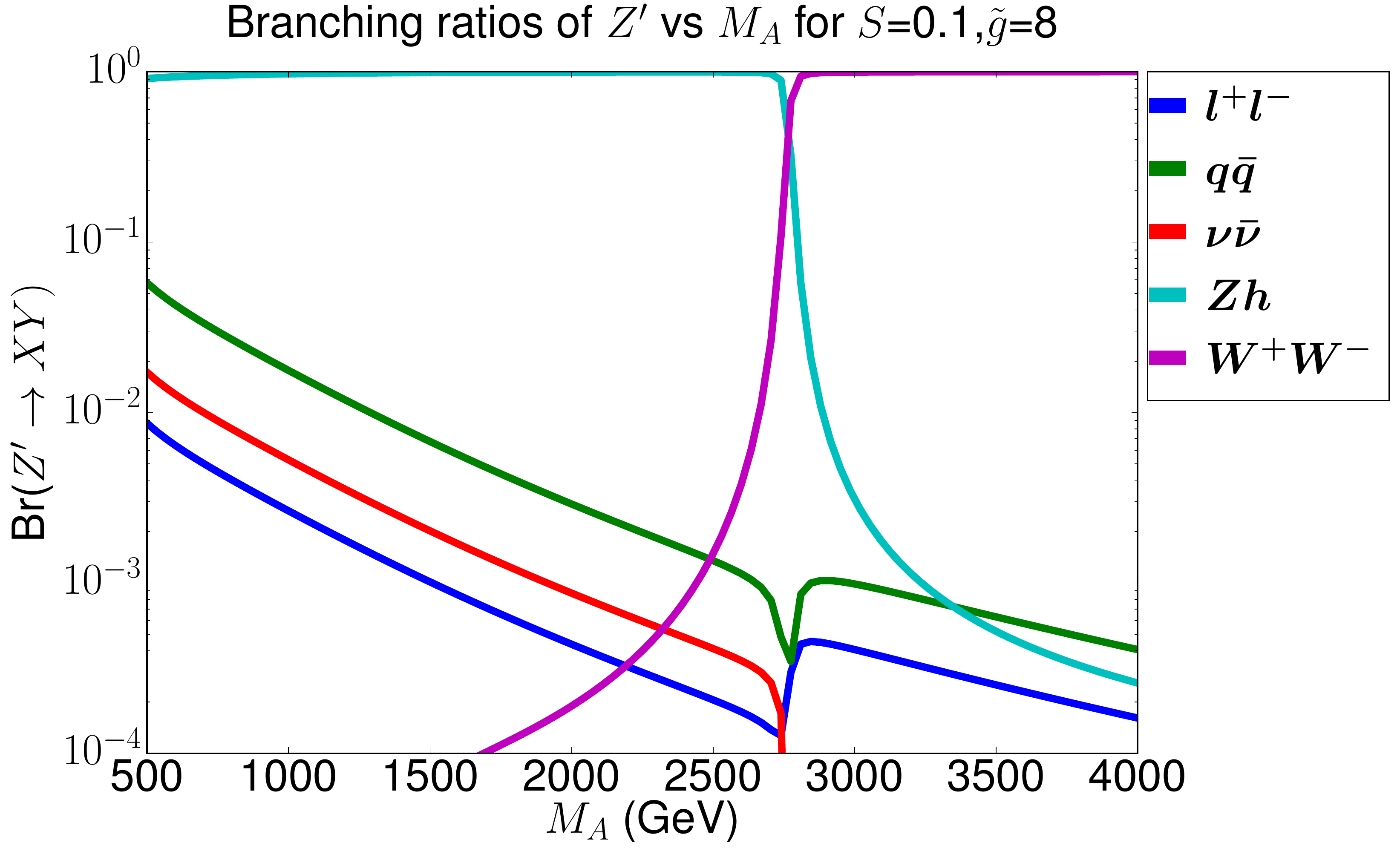}}\\
\subfigure[]{\includegraphics[type=pdf,ext=.pdf,read=.pdf,width=0.5\textwidth]{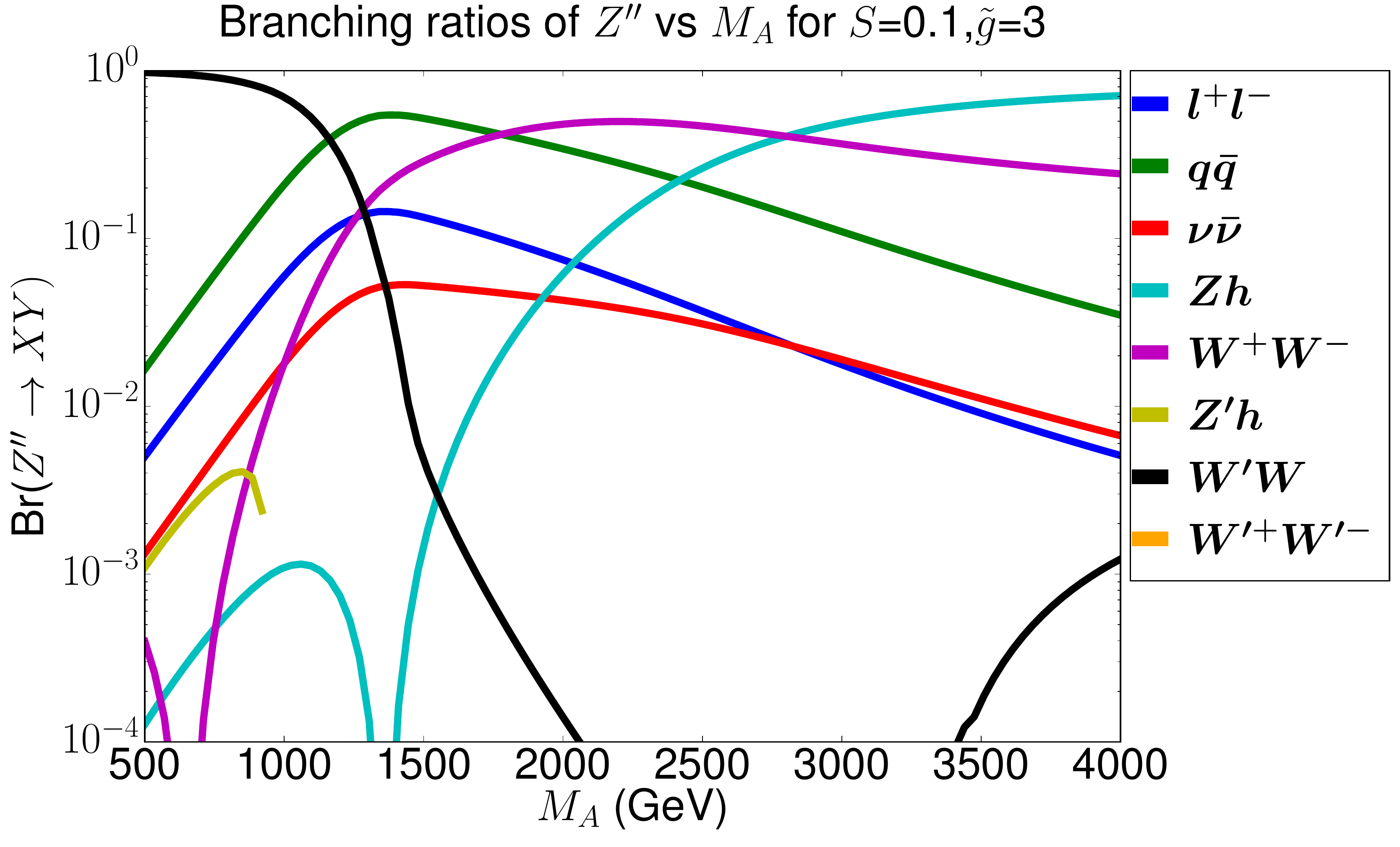}}%
\subfigure[]{\includegraphics[type=pdf,ext=.pdf,read=.pdf,width=0.5\textwidth]{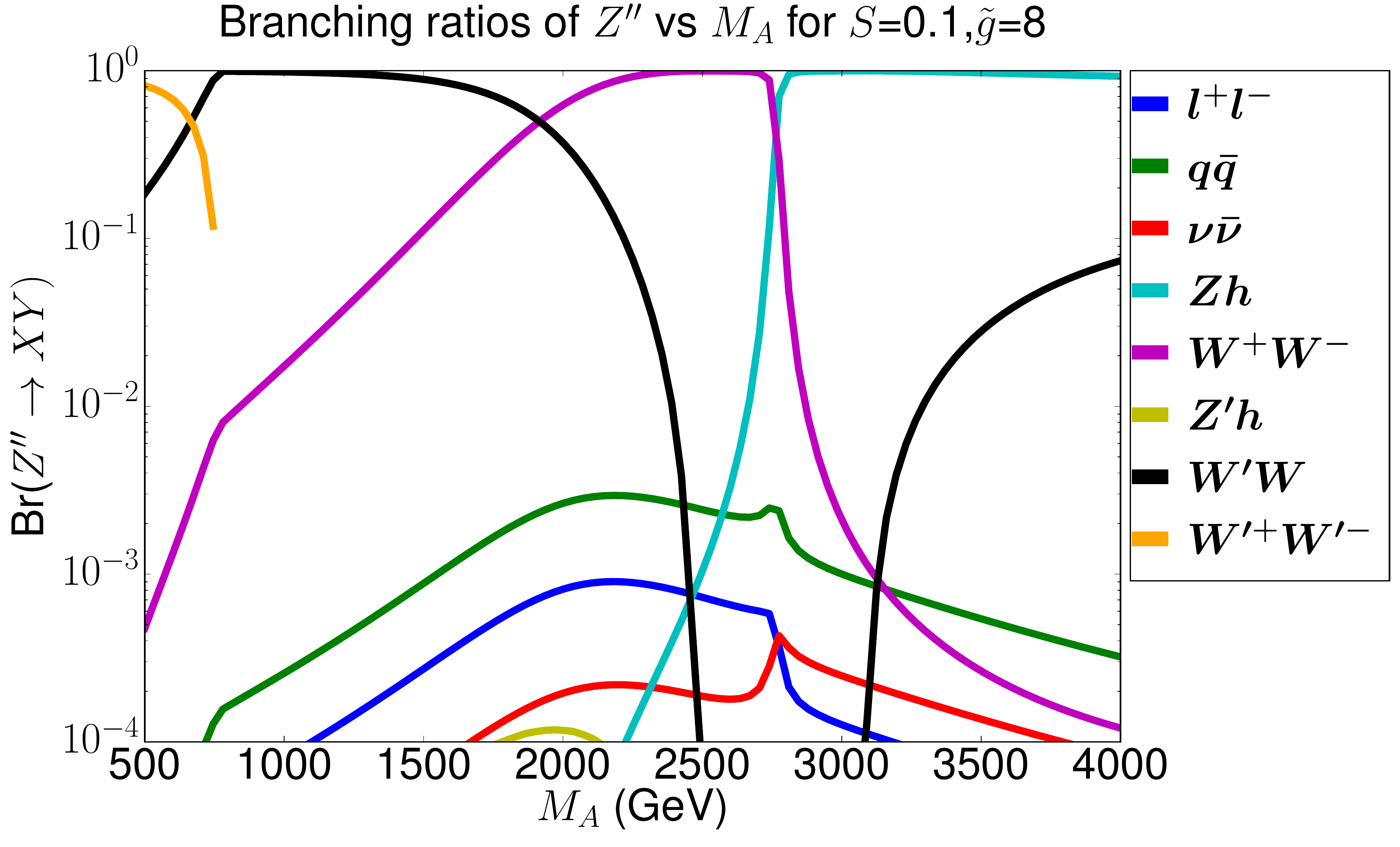}}
\vspace*{-0.5cm}
\caption{\label{fig:mzp-allbr} $Br(Z')$(a,b) and $Br(Z'')$(c,d) for all decay channels  as a function of $M_A$ at the fixed values of (a)$\tilde{g}=3$, (b)$\tilde{g}=8$, at benchmark values of $S=0.1$ and $s=0$}
\end{figure}

One can see that  ${Z^{\prime}}$ is  generically narrow in the whole parameter space --
the $\Gamma/M$ is always below 10\%. One should also note that for large values of 
$\tilde{g}$  and $M_A<M_{inv}$ the main contribution to the width is coming 
from $Z'\to ZH$ decay as one can see from Fig.~\ref{fig:mzp-allbr}(a,b)
where we present  $Br(Z')$(a,b) and $Br(Z'')$(c,d) for all decay channels  as a function of $M_A$ at the fixed values of (a,c)$\tilde{g}=3$, (b,d)$\tilde{g}=8$, at benchmark values of $S=0.1$ and $s=0$.
This happens because of  the following asymptotic of $g_{Z'ZH}$ coupling at large $\tilde{g}$,
\begin{equation}
\label{eq:ZPZH}
g_{Z'ZH} = -{\frac{\tilde{g}^2 v}{16}}\sqrt{\frac{(g_2^2 + g_1^2) S}{\pi} },
\end{equation}
which makes $\Gamma(Z')$ increase with the increase of   $\tilde{g}$.
One can see the numerical results confirming this effect in Table~\ref{tab:width},
where we present $\Gamma(Z')$ and  $\Gamma(Z'')$ for the 3D grid in $(M_A,\tilde{g},S)$ space.

\begin{table}[htb]
\begin{center}
\begin{tabular}{cc|c|c|c|c|}
       \cline{3-6}
       & &     \multicolumn{4}{c|}{$M_{A}$(GeV)} \\ \hline
      \multicolumn{1}{|c|}{$S$} & \multicolumn{1}{|c|}{$\tilde{g}$} & 1000 & 1500 & 2000 & 2500 \\ \hline
      \multicolumn{1}{|c}{\multirow{5}{*}{-0.1}} &
      \multicolumn{1}{|c|}{1} & 2.91(35.28) & 4.54(52.92) & 6.68(72.28) & 9.76(94.34) \\ \cline{2-6}
      \multicolumn{1}{|c}{} &
      \multicolumn{1}{|c|}{3} & 1.29(10.79) & 2.92(7.73) & 7.20(12.28) & 17.39(24.99)  \\ \cline{2-6}
      \multicolumn{1}{|c}{} &
      \multicolumn{1}{|c|}{5} & 1.37(180.97) & 5.10(117.65) & 16.44(110.57)  & 44.28(143.36) \\ \cline{2-6}
      \multicolumn{1}{|c}{} & 
      \multicolumn{1}{|c|}{7} & 2.89(932.69) & 11.15(691.70) & 35.46(648.36) & 93.58(742.68) \\ \cline{2-6}
      \multicolumn{1}{|c}{} &
      \multicolumn{1}{|c|}{9} & 6.75(3028.96) &  23.56(2435.70) & 69.88(2375.84) & 176.01(2685.93)  \\ \hline  %%%%%% fill in from here
      \multicolumn{1}{|c}{\multirow{5}{*}{0.1}} &
      \multicolumn{1}{|c|}{1} & 2.72(33.70) & 4.02(48.98) & 5.50(64.11) & 7.50(79.44) \\ \cline{2-6}
      \multicolumn{1}{|c}{} &
      \multicolumn{1}{|c|}{3} & 0.88(4.13) & 1.80(2.69) & 4.74(6.40) & 12.93(15.07) \\ \cline{2-6}
      \multicolumn{1}{|c}{} &
      \multicolumn{1}{|c|}{5} & 0.79(76.29) & 3.60(19.00) & 12.85(14.75) & 36.46(36.86) \\ \cline{2-6}
      \multicolumn{1}{|c}{} & 
      \multicolumn{1}{|c|}{7} & 1.99(350.34) & 8.64(109.07) & 28.30(46.82) & 75.39(76.16) \\ \cline{2-6}
      \multicolumn{1}{|c}{} &
      \multicolumn{1}{|c|}{9} & 5.66(899.79) & 19.44(328.60) & 55.33(124.77) & 134.68(135.22) \\ \hline
      \multicolumn{1}{|c}{\multirow{5}{*}{0.3}} &
      \multicolumn{1}{|c|}{1} &  2.70(32.48) & 4.62(47.28) & 8.91(64.77) & 19.03(90.61) \\ \cline{2-6}
      \multicolumn{1}{|c}{} &
      \multicolumn{1}{|c|}{3} & 1.87(2.75) & 9.37(10.55) & 34.98(37.18) & 99.34(107.84) \\ \cline{2-6}
      \multicolumn{1}{|c}{} &
      \multicolumn{1}{|c|}{5} & 5.53(30.22) & 27.87(27.69) & 79.15(97.60) & 197.15(288.29) \\ \cline{2-6}
      \multicolumn{1}{|c}{} & 
      \multicolumn{1}{|c|}{7} & 18.16(108.87) & 64.34(59.34) & 113.87(195.62) & 217.11(580.30) \\ \cline{2-6}
      \multicolumn{1}{|c}{} &
      \multicolumn{1}{|c|}{9} & 72.97(125.19) & 160.17(109.98) & 116.31(318.94) & 124.76(617.72) \\ \hline   
\end{tabular}
\caption{\label{tab:width} Widths of the neutral resonances in the $M_{A},\tilde{g},S$ parameter space, displayed in the format $\Gamma_{Z^{\prime}}$($\Gamma_{Z^{\prime\prime}})$ in GeV for each parameter space value}
\end{center}
\end{table}

For $M_A>M_{inv}$, $Z'$ ``switches" its properties from  pseudo-vector to vector, and its
width is enhanced then by the  $Z'\to W+W-$ decay for  large  $\tilde{g}$ 
with the respective $g_{Z'WW}$ coupling proportional to $\tilde{g}$.
In the region of low values of  $\tilde{g}$ and not so large values of $M_A$ the contribution from
$Z'\to f\bar{f}$ also play an important role. This also happen for small values of $S\simeq 0$
as one can see from Figs.~\ref{fig:mzp-allbr-gt3-with-s} and \ref{fig:mzp-allbr-gt8-with-s}
in Appendix, where we present additional plots for $S=-0.1,0,0.2$ and $0.3$.
From Fig.~\ref{fig:width2mass} one can see that the picture of the width-to-mass ratio for $Z''$
is qualitatively different from the one for $Z'$:
though the  $\Gamma/M$ is also  below 10\% for $\tilde{g}\lesssim 5$,
for bigger values of $\tilde{g}$ the $\Gamma/M$ becomes very large especially in the small $M_A$ region
where  $\tilde{g}$-enhanced $Z''\to W'W'$ decay opens for vector $Z''$, or for large values of $M_A$
where $\tilde{g}$-enhanced $Z''\to ZH$ decay opens for pseudo-vector $Z''$(see Fig.~\ref{fig:mzp-allbr}
as well as analogous Fig.~\ref{fig:mzp-allbr-gt8-with-s}  and Fig.~\ref{fig:mzp-allbr-gt8-with-s} from Appendix~\ref{appendix:br}).
In this region  $Z^{\prime\prime}$ does not contribute to the dilepton signature at the LHC and therefore this region can be safely explored and interpreted using  $Z^{\prime}$ dilepton signature at the LHC.

\begin{figure}[htb]
\subfigure[]{\includegraphics[type=pdf,ext=.pdf,read=.pdf,width=0.5\textwidth]{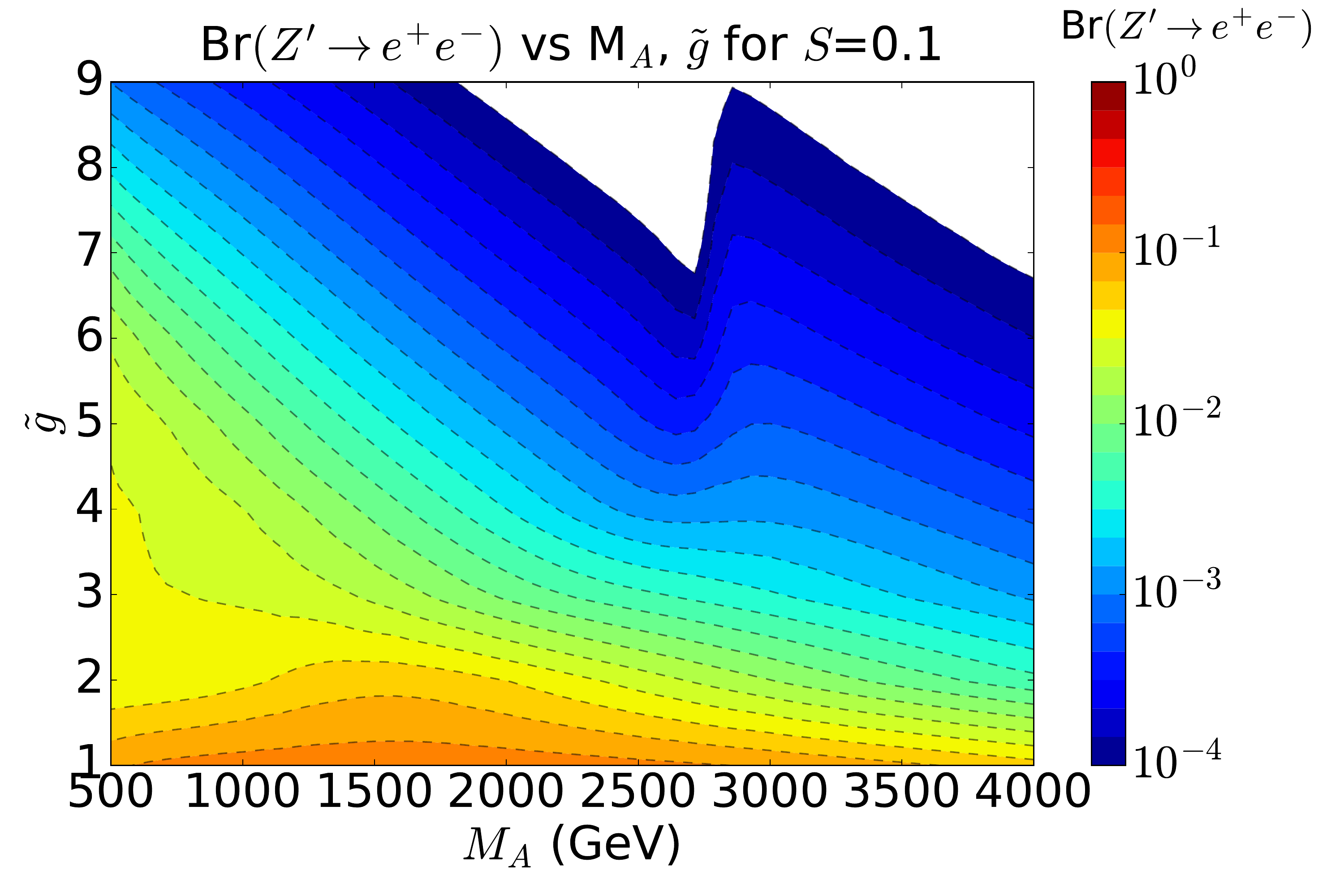}}%
\subfigure[]{\includegraphics[type=pdf,ext=.pdf,read=.pdf,width=0.5\textwidth]{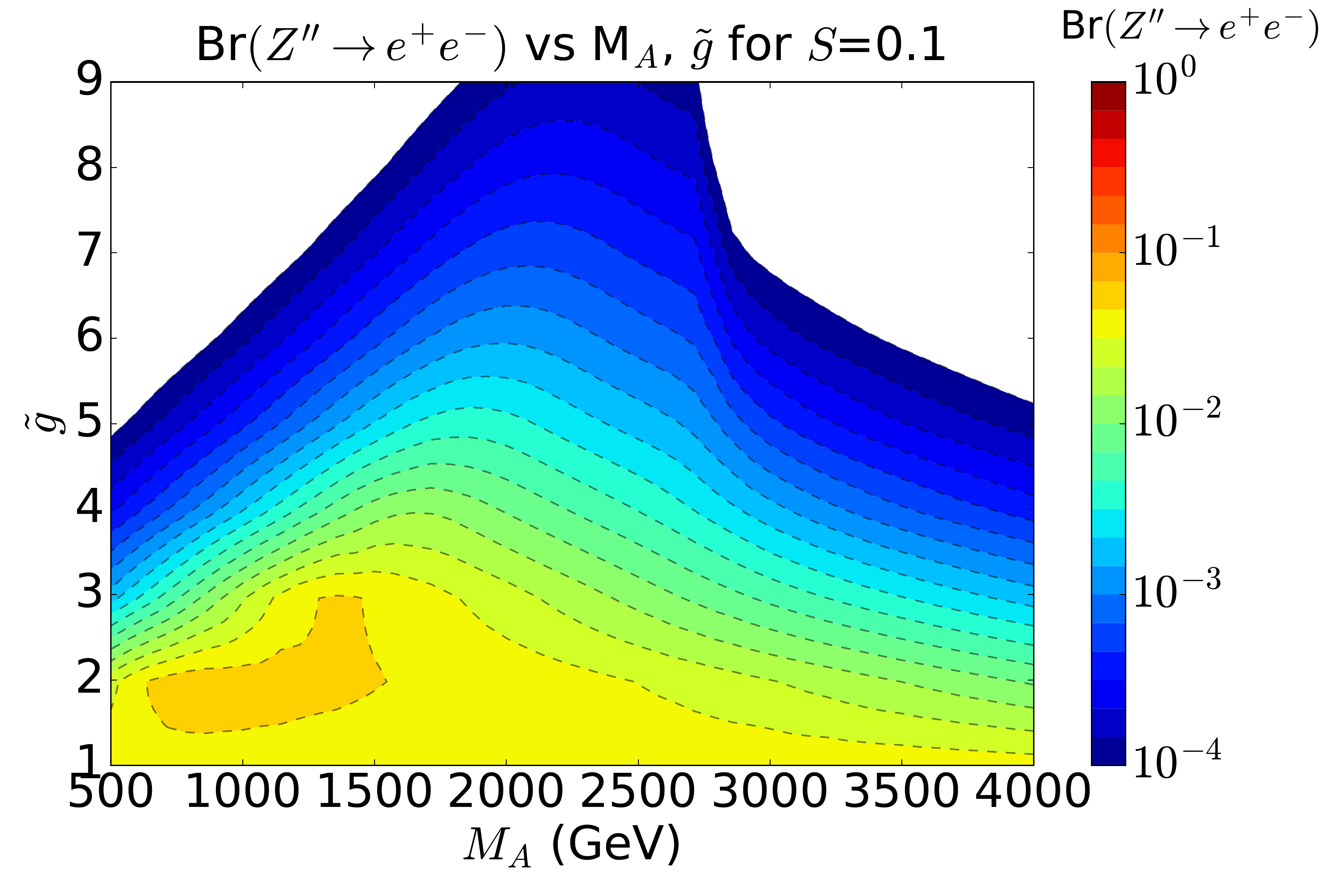}}
\vspace*{-0.5cm}
\caption{\label{fig:br-ee} (a)$Br(Z^{\prime}\rightarrow e^{+}e^{-})$, (b)$Br(Z^{\prime\prime}\rightarrow e^{+}e^{-})$  as a function of $M_A$ and $\tilde{g}$ parameters at benchmark values of $S=0.1$ and $s=0$}
\end{figure}
% Table of di-electron branchings

\begin{table}[htb]
\begin{center}
\begin{tabular}{cc|c|c|c|c|}
       \cline{3-6}
       & &     \multicolumn{4}{c|}{$M_{A}$(GeV)} \\ \hline
      \multicolumn{1}{|c|}{$S$} & \multicolumn{1}{|c|}{$\tilde{g}$} & 1000 & 1500 & 2000 & 2500 \\ \hline
      \multicolumn{1}{|c}{\multirow{5}{*}{-0.1}} &
      \multicolumn{1}{|c|}{1} & 10.941(3.963) & 10.759(3.873) & 9.854(3.749) & 8.467(3.576) \\ \cline{2-6}
      \multicolumn{1}{|c}{} &
      \multicolumn{1}{|c|}{3} & 2.226(0.782) & 1.377(1.572) & 0.704(1.313) & 0.350(0.807)  \\ \cline{2-6}
      \multicolumn{1}{|c}{} &
      \multicolumn{1}{|c|}{5} & 0.827(0.019) & 0.327(0.038) & 0.134(0.052)  & 0.061(0.049) \\ \cline{2-6}
      \multicolumn{1}{|c}{} & 
      \multicolumn{1}{|c|}{7} & 0.217(0.002) & 0.083(0.004) & 0.035(0.005) & 0.016(0.005) \\ \cline{2-6}
      \multicolumn{1}{|c}{} &
      \multicolumn{1}{|c|}{9} & 0.062(0.000) & 0.026(0.001) & 0.012(0.001) & 0.006(0.001)  \\ \hline
      \multicolumn{1}{|c}{\multirow{5}{*}{0.1}} &
      \multicolumn{1}{|c|}{1} & 11.788(4.080) & 12.280(4.112) & 12.084(4.154) & 11.119(4.174) \\ \cline{2-6}
      \multicolumn{1}{|c}{} &
      \multicolumn{1}{|c|}{3} & 2.986(1.991) & 1.930(4.455) & 0.903(2.487) & 0.502(1.229) \\ \cline{2-6}
      \multicolumn{1}{|c}{} &
      \multicolumn{1}{|c|}{5} & 1.171(0.042) & 0.373(0.220) & 0.133(0.360) & 0.050(0.183) \\ \cline{2-6}
      \multicolumn{1}{|c}{} & 
      \multicolumn{1}{|c|}{7} & 0.211(0.005) & 0.072(0.021) & 0.029(0.058) & 0.013(0.043) \\ \cline{2-6}
      \multicolumn{1}{|c}{} &
      \multicolumn{1}{|c|}{9} & 0.038(0.001) & 0.016(0.005) & 0.008(0.014) & 0.004(0.015) \\ \hline
      \multicolumn{1}{|c}{\multirow{5}{*}{0.3}} &
      \multicolumn{1}{|c|}{1} & 11.988(4.162) & 10.784(4.186) & 7.532(4.040) & 4.429(3.595) \\ \cline{2-6}
      \multicolumn{1}{|c}{} &
      \multicolumn{1}{|c|}{3} & 1.255(2.910) & 0.356(1.077) & 0.301(0.233) & 0.147(0.085) \\ \cline{2-6}
      \multicolumn{1}{|c}{} &
      \multicolumn{1}{|c|}{5} & 0.129(0.099) & 0.033(0.142) & 0.058(0.016) & 0.028(0.006)  \\ \cline{2-6}
      \multicolumn{1}{|c}{} & 
      \multicolumn{1}{|c|}{7} & 0.012(0.016) & 0.005(0.033) & 0.019(0.002) & 0.012(0.001) \\ \cline{2-6}
      \multicolumn{1}{|c}{} &
      \multicolumn{1}{|c|}{9} & 0.000(0.009) & 0.000(0.011) & 0.010(0.000) & 0.010(0.000) \\ \hline   
\end{tabular}
\caption{\label{tab:branching} Di-electron branching fraction of $Z^{\prime}$, $Z^{\prime\prime}$ in the $M_{A},\tilde{g},S$ parameter space, displayed in the format $Br(Z^{\prime}\rightarrow e^{+}e^{-})$($Br(Z^{\prime\prime}\rightarrow e^{+}e^{-}))$ in $\%$.}
\end{center}
\end{table}

Let us take a closer look at dilepton signature and the respective $Z'$ and $Z''$ branching ratios
in 2D ($M_A$, $\tilde{g}$) parameter space, presented in Fig.~\ref{fig:br-ee}
for S=0.1 and Table~\ref{tab:branching}
presenting numerical values for $Br(Z'\to e^+e^-)$ and  $Br(Z''\to e^+e^-)$
for 3D grid in $(M_A,\tilde{g},S)$ space.
Besides an expected $1/\tilde{g}$ suppression, in Fig.~\ref{fig:br-ee} one
can observe that for low values of $\tilde{g}$ both $Br(Z'\to \ell^+\ell^-)$ and  $Br(Z''\to \ell^+\ell^-)$
are enhanced above the $3\%$ value corresponding to  $Br(Z\to \ell^+\ell^-)$ in SM.
One can see from Table~\ref{tab:branching} that, for example, for $M_A=1500$~GeV,
$S=0.1$ and $\tilde g=1$ $Br(Z'\to e^+e^-)\simeq 12.3\%$ which is about 4 times bigger than the
SM value. This enhancement is related to a quite subtle effect which does not follow from
Eq.~\ref{eqn:zp-couplings} which is valid for intermediate-large values of  $\tilde g$;
for  $\tilde g \simeq 1$ one can check numerically that photon-$Z'$ mixing is enhanced,
while $Z-Z'$ is suppressed, which leads to a relative suppression
of $Br(Z'\to \nu\nu)$ and $Br(Z'\to q_d\bar{q}_d)$ with respect to
$Br(Z'\to \ell^+\ell^-)$ and $Br(Z'\to q_u\bar{q}_u)$.

Talking about all other decay channels, which actually define
$Br(Z'\to e^+e^-)$ and  $Br(Z''\to e^+e^-)$, there are four more decays:
 $q\bar{q}$, $\nu\bar\nu$, $VV$ and $Vh$ channels
 as one can see from see Fig.~\ref{fig:mzp-allbr} 
(as well as analogous Fig.~\ref{fig:mzp-allbr-gt3-with-s}  and Fig.~\ref{fig:mzp-allbr-gt8-with-s}  from Appendix~\ref{appendix:br}) which are already mentioned above.
Besides the dominant role of $WW$ and $ZH$ channels for large value of $\tilde g$ one should note
dips in $Z'$ and $Z''$ branchings into these channels occurring for small-intermediate values of $\tilde g$.
This happens because the respective $Z^{\prime(\prime\prime)}WW$ and $Z^{\prime(\prime\prime)}ZH$ couplings change the sign around these dips,
such that at the dips the respective branchings go to zero.
The reason for this is the cancellation occurring because of the contribution from  several 
different terms to these couplings -- from gauge kinetic terms as well as from $r_2$ and $r_3$ terms
from the Lagrangian defined by Eq.~\ref{eqn:bosonlagng}. One should note that in case of such cancellation
and absence of $ZH$ signal, which has been explored by the ATLAS collaboration to probe WTC parameter space~\cite{Aad:2015yza},
the role of  dilepton searches in probing WTC parameter space becomes especially appealing
as a crucial complementary  channel.

\subsection{Cross sections}
\label{subsec:theory-cs}

\begin{figure}[htb]
\subfigure[]{\includegraphics[width=0.5\textwidth]{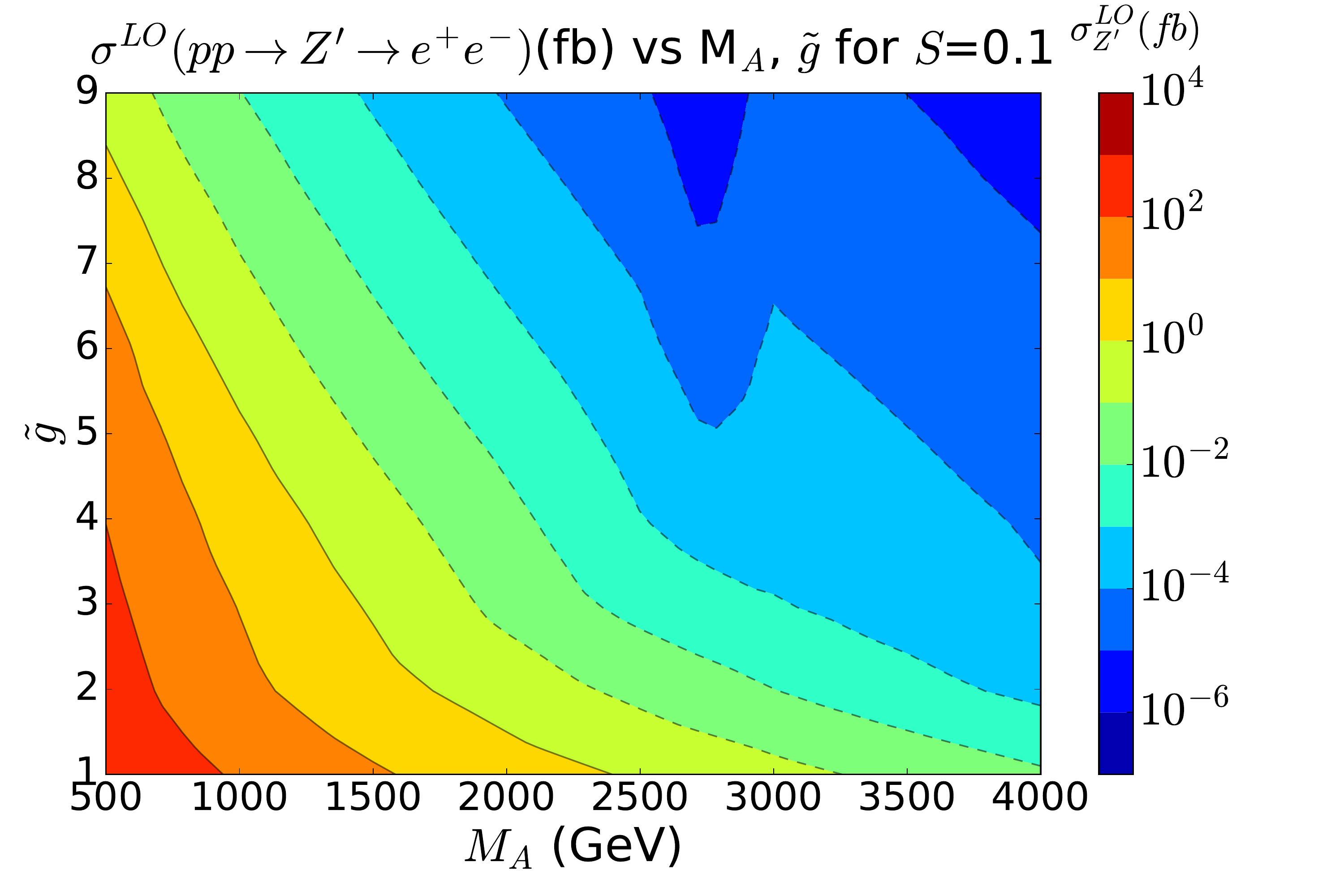}}%
\subfigure[]{\includegraphics[width=0.5\textwidth]{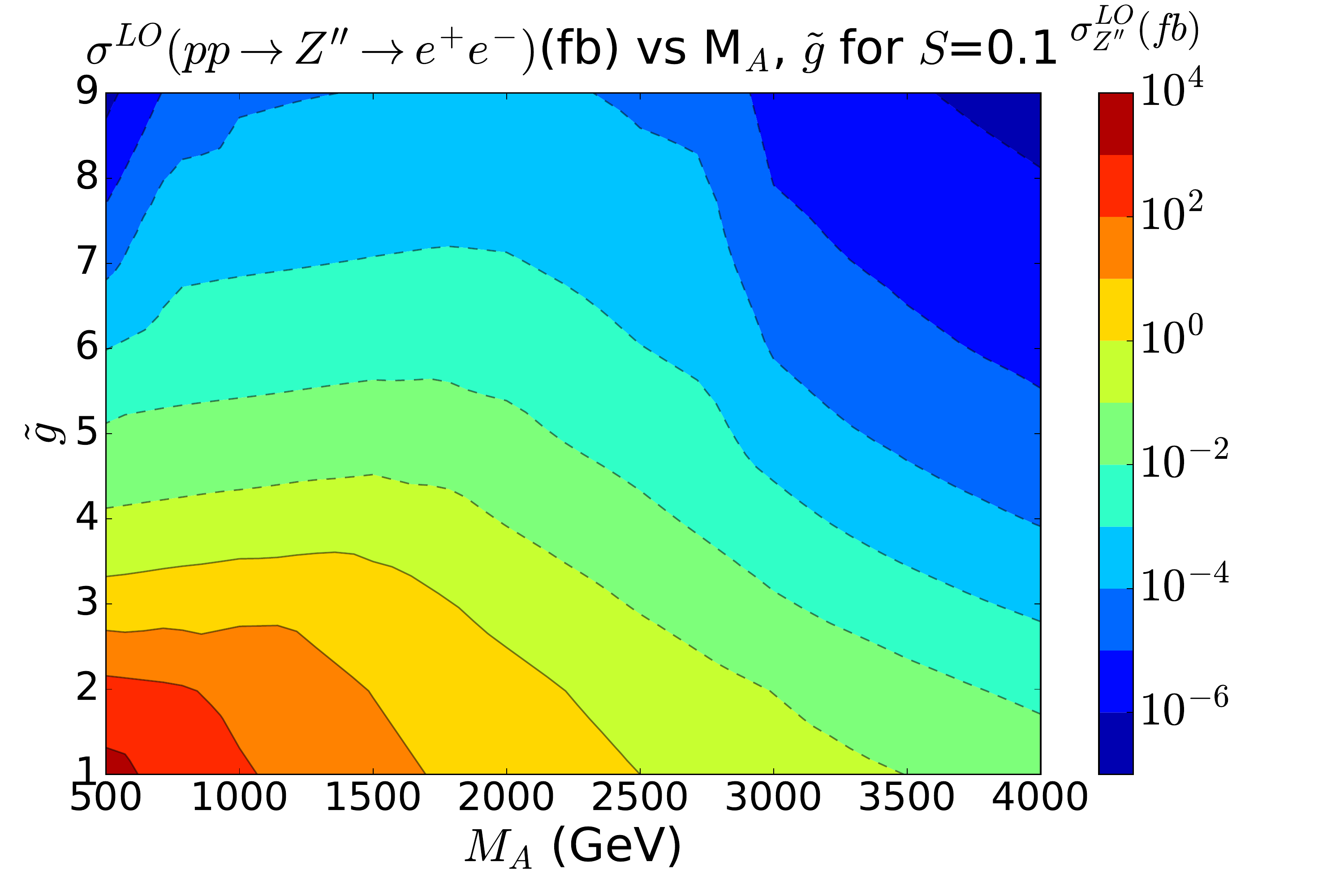}}\\
\caption{\label{fig:theory-cs} (a)$\sigma^{LO}(pp\rightarrow Z^{\prime}\rightarrow e^{+}e^{-})$ (fb)(b)$\sigma^{LO}(pp\rightarrow Z^{\prime\prime}\rightarrow e^{+}e^{-})$ (fb) at $\sqrt{s}=13$TeV as a function of $M_A$, $\tilde{g}$ at benchmark values of $S=0.1$ and $s=0$}
\end{figure}

Both $Z'$ and $Z''$ can be resonantly produced in a DY process, giving rise to dilepton
signatures. The cross section rates are 
%eventually 
directly related to
$Z'$ and $Z''$ coupling to fermions and dilepton branching ratios discussed earlier.
Production cross sections for LO  DY  $pp\to Z'/Z''\to e^+ e^-$ processes for LHC@13TeV
are presented in Fig.~\ref{fig:theory-cs} 
as contour levels of the cross section in $(M_A,\tilde g)$ space for S=0.1
(see also Fig.~\ref{fig:zp-theory-cs-with-s} and Fig.~\ref{fig:zpp-theory-cs-with-s}
for analogous results for different values of $S$ in the Appendix~\ref{appendix:cs})
as well as in Table~\ref{tab:cross} as a numerical results for 3D $(M_A,\tilde{g},S)$
grid.
Cross sections are calculated using CalcHEP \cite{Belyaev:2012qa} via the High Energy Physics Model Database  HEPMDB \cite{hepmdb}, linked to the IRIDIS4 supercomputer. 
The PDF set used is NNPDF23 LO \texttt{as\char`_0130\char`_QED}%as_0130_QED 
\cite{Ball:2013hta}, and the  QCD scale $Q$ is set to be the dilepton invariant mass, $Q=M(e^{+}e^{-})$.
The cross section has been evaluated in the narrow width approximation (NWA)
to be consistent with the latest CMS limit~\cite{Sirunyan:2018exx} which we use
for the interpretation of our signal as we discuss below.
In the experimental CMS paper the cross section for $Z'$ models was calculated in a
mass window of $\pm5\%\sqrt{s}$ at  the resonance mass, following the prescription of Ref.~\cite{Accomando:2013sfa}
where it was checked that for this cut the cross section is close to the one from the NWA to within 10\%.
To account for NNLO QCD effects in our analysis below, the LO cross sections are multiplied by a mass-dependent  K-factor
which was found using  WZPROD program~\cite{Hamberg:1990np,vanNeerven:1991gh,ZWPROD} which we have modified
to evaluate the cross sections for $Z'$ and $W'$ resonances and linked to LHAPDF6 library~\cite{Buckley:2014ana}
as described in Ref.~\cite{Accomando:2010fz}. The resulting NNLO  K-factors are presented in Table \ref{tab:k-factors}.

From Fig.~\ref{fig:theory-cs} one can observe for $Z^{\prime}$ and $Z^{\prime\prime}$ DY cross sections
an expected $1/\tilde g$ suppression discussed above as well as eventual PDF suppression with the 
increase of the mass of the resonances. Also, one should make an important remark that in the large mass region
for low-intermediate values of $\tilde g$ the signal from the $Z^{\prime\prime}$ is higher than the one from 
the  $Z^{\prime}$.
This highlights the complementarity between the two resonances, indicating that the $Z^{\prime}$ and $Z^{\prime\prime}$ DY processes will exclude different areas of the parameter space. This motivates our study of both resonances in conjunction, as we will exclude a greater portion of the parameter space with combined searches.

%\AC{Replacing with production cross sections to enforce narrow width for resonance - $Z''$ replaced, waiting for $Z'$ to run}
\begin{table}[htb!]
\begin{center}
\begin{tabular}{cc|c|c|c|c|}
       \cline{3-6}
       & &     \multicolumn{4}{c|}{$M_{A}$(GeV)} \\ \hline
      \multicolumn{1}{|c|}{$S$} & \multicolumn{1}{|c|}{$\tilde{g}$} & 1000 & 1500 & 2000 & 2500 \\ \hline
      \multicolumn{1}{|c}{\multirow{5}{*}{-0.1}} &
      \multicolumn{1}{|c|}{1} & 6.37$\times 10^{2}$(3.08$\times 10^{3}$) & 1.03$\times 10^{2}$(4.29$\times 10^{2}$) & 23.7(83.5) & 6.54(19.1) \\ \cline{2-6}
      \multicolumn{1}{|c}{} &
      \multicolumn{1}{|c|}{3} & 3.37$\times 10^{2}$(2.39$\times 10^{2}$) & 49.6(52.2) & 10.4(14.1) & 2.66(4.31)  \\ \cline{2-6}
      \multicolumn{1}{|c}{} &
      \multicolumn{1}{|c|}{5} & 1.43$\times 10^{2}$(37.1) & 22.9(9.83) & 5.29(2.84)  & 1.47(0.89) \\ \cline{2-6}
      \multicolumn{1}{|c}{} & 
      \multicolumn{1}{|c|}{7} & 80.2(7.89) & 13.0(2.54) & 3.03(0.81) & 0.85(0.26) \\ \cline{2-6}
      \multicolumn{1}{|c}{} &
      \multicolumn{1}{|c|}{9} & 53.9(2.00) &  8.78(0.74) & 2.05(0.25) & 0.58(8.59$\times 10^{-2}$)  \\ \hline  
      \multicolumn{1}{|c}{\multirow{5}{*}{0.1}} &
      \multicolumn{1}{|c|}{1} & 6.39$\times 10^{2}$(3.10$\times 10^{3}$) & 1.04$\times 10^{2}$(4.34$\times 10^{2}$) & 24.0(84.7) & 6.64(19.5) \\ \cline{2-6} 
      \multicolumn{1}{|c}{} &
      \multicolumn{1}{|c|}{3} & 3.06$\times 10^{2}$(2.72$\times 10^{2}$) & 39.8(65.3) & 5.79(20.0) & 0.96(6.50) \\ \cline{2-6}
      \multicolumn{1}{|c}{} &
      \multicolumn{1}{|c|}{5} & 1.17$\times 10^{2}$(47.7) & 18.5(14.4) & 4.03(4.72) & 0.81(1.89) \\ \cline{2-6}
      \multicolumn{1}{|c}{} & 
      \multicolumn{1}{|c|}{7} & 54.0(11.5) & 8.75(4.70) & 2.01(1.85) & 0.52(0.76) \\ \cline{2-6}
      \multicolumn{1}{|c}{} &
      \multicolumn{1}{|c|}{9} & 27.7(3.22) & 4.50(1.73) & 1.05(0.85) & 0.29(0.41) \\ \hline
      \multicolumn{1}{|c}{\multirow{5}{*}{0.3}} &
      \multicolumn{1}{|c|}{1} &  6.43$\times 10^{2}$(3.12$\times 10^{3}$) & 1.05$\times 10^{2}$(4.40$\times 10^{2}$) & 24.3(85.8) & 6.75(19.8) \\ \cline{2-6}
      \multicolumn{1}{|c}{} &
      \multicolumn{1}{|c|}{3} & 2.70$\times 10^{2}$(3.15$\times 10^{2}$) & 16.1(93.9) & 8.68(19.0) & 3.47(4.70) \\ \cline{2-6}
      \multicolumn{1}{|c}{} &
      \multicolumn{1}{|c|}{5} & 90.4(63.2) & 11.8(24.1) & 6.98(3.82) & 2.64(1.04) \\ \cline{2-6}
      \multicolumn{1}{|c}{} & 
      \multicolumn{1}{|c|}{7} & 27.9(17.6) & 4.30(10.2) & 5.18(1.09) & 2.64(0.31) \\ \cline{2-6}
      \multicolumn{1}{|c}{} &
      \multicolumn{1}{|c|}{9} & 1.35(5.65) & 0.22(5.43) & 5.13(5.22$\times 10^{-2}$) & 4.79(1.47$\times 10^{-2}$) \\ \hline   
\end{tabular}
\end{center}
\caption{\label{tab:cross} Cross section $\sigma(pp\rightarrow Z^{\prime}/Z^{\prime\prime})$ at LO in the $M_{A},\tilde{g},S$ parameter space at $\sqrt{s}=13$TeV, displayed in the format $\sigma_{Z^{\prime}}$($\sigma_{Z^{\prime\prime}})$ in fb for each parameter space value}
\end{table}

\clearpage
\begin{table}[htb]
\begin{center}
\begin{tabular}{|c|c|}
\hline
$M_{Z^{\prime}}$ (GeV) & K$_{NNLO}$ \\ \hline
500 &   1.35 \\
600 &	1.36 \\
700 &	1.36 \\
800 &	1.37 \\
900 &	1.38 \\
1000 &	1.39 \\
1100 &	1.39 \\
1200 &	1.40 \\
1300 & 	1.40 \\
1400 &	1.41 \\
\hline
\end{tabular}
\begin{tabular}{|c|c|}
\hline
$M_{Z^{\prime}}$ (GeV) & K$_{NNLO}$ \\ \hline
1500 &	1.41 \\
1600 &	1.41 \\
1700 &	1.42 \\
1800 &	1.42 \\
1900 &	1.42 \\
2000 &	1.41 \\
2100 &	1.41 \\
2200 &	1.41 \\
2300 &	1.41 \\
2400 &	1.40 \\
\hline
\end{tabular}
\begin{tabular}{|c|c|}
\hline
$M_{Z^{\prime}}$ (GeV) & K$_{NNLO}$ \\ \hline
2500 &	1.40 \\
2600 &	1.39 \\
2700 &	1.39 \\
2800 &	1.38 \\
2900 &	1.37 \\
3000 &	1.36 \\
3100 &	1.35 \\
3200 &	1.34 \\
3300 &	1.33 \\
3400 &	1.32 \\
\hline
\end{tabular}
\begin{tabular}{|c|c|}
\hline
$M_{Z^{\prime}}$ (GeV) & K$_{NNLO}$ \\ \hline
3500 &	1.31 \\
3600 &	1.30 \\
3700 &	1.29 \\
3800 &	1.28 \\
3900 &	1.26 \\
4000 &	1.25 \\
4100 &	1.24 \\
4200 &	1.22 \\
4300 &	1.21 \\
4400 &	1.19 \\
%4500 &	1.18 \\
\hline
\end{tabular}
\end{center}
\caption{\label{tab:k-factors}K-factors for NNLO QCD corrections to Drell-Yan cross sections at $\sqrt{s}=13$TeV 
evaluated with the help of the modified ZWPROD program as described in the text, using  NNPDF23 LO \texttt{as\char`_0130\char`_QED} 
and NNPDF23 NNLO \texttt{as\char`_0119\char`_QED}\cite{Ball:2012cx} PDFs for LO and NNLO cross sections respectively.}
\end{table}

\subsection{$Z^{\prime}/Z^{\prime\prime}$ interference and validity of the re-interpretation of the LHC limits \label{subsec:interference}}

Following our results in the previous section,
we  explore the interference between the $Z'$ and $Z''$ boson which gives rise to the di-lepton signature.
This is an important point for our study since  we aim to re-interpret the LHC limits based on
a {\it single} resonance search in the di-lepton channel. Besides interference, the validity of such an interpretation also depends on
how well these resonances are separated, their relative contribution to the signal and their width-to-mass ratio.

\begin{figure}[htb]
\subfigure[]{\hspace*{-0.5cm}\includegraphics[width=0.55\textwidth]{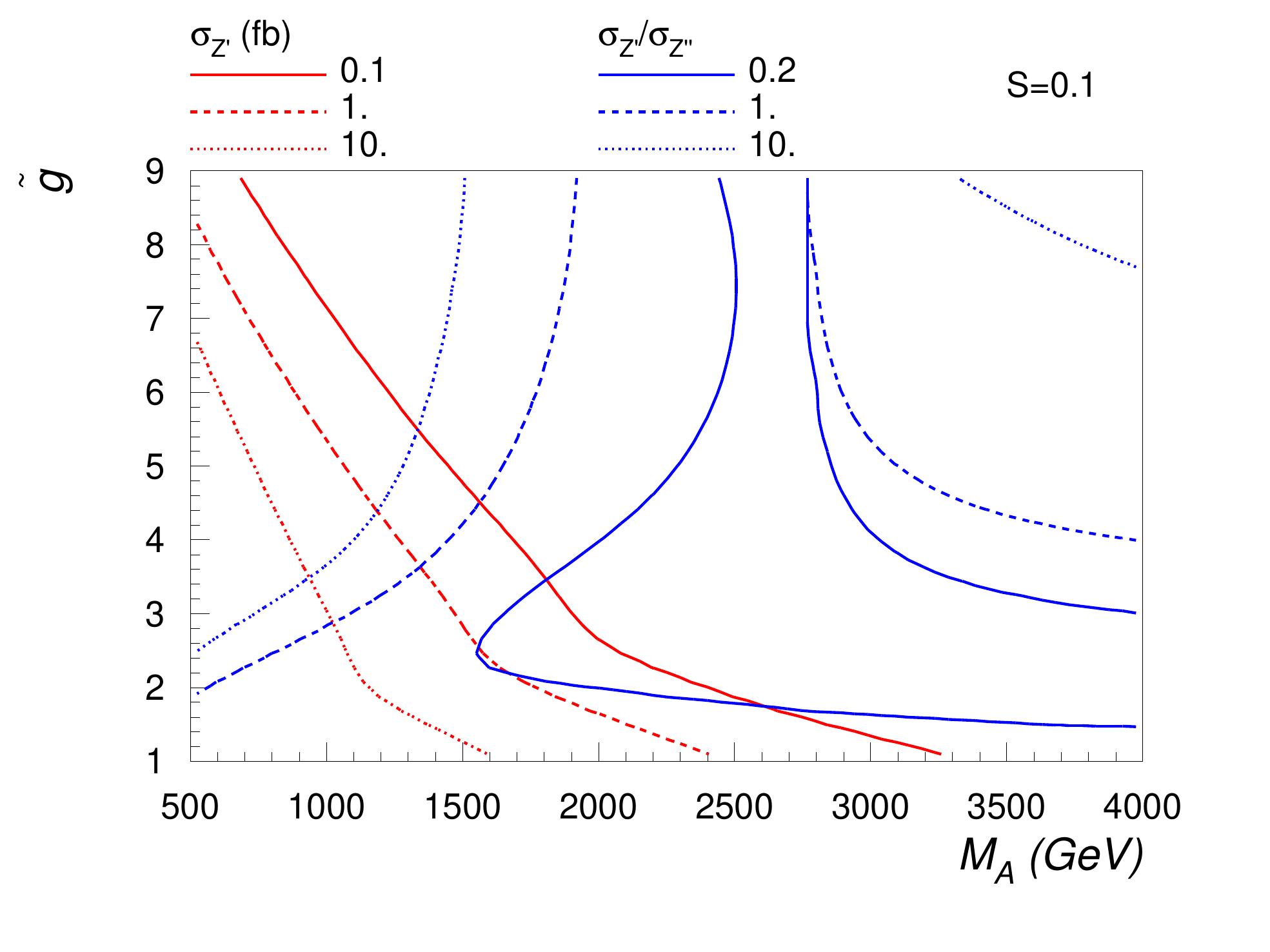}}\hspace*{-0.8cm}%
\subfigure[]{\includegraphics[width=0.55\textwidth]{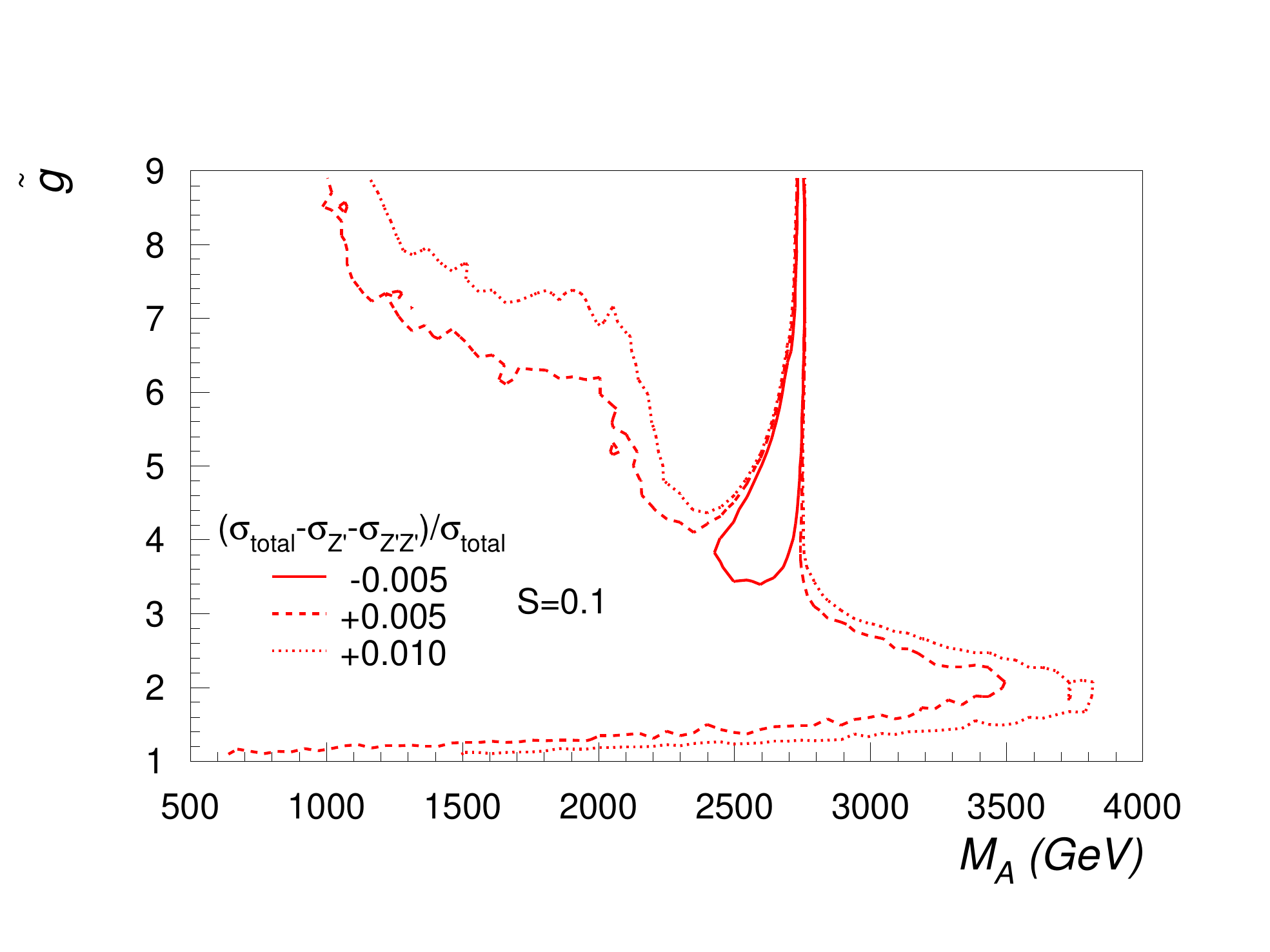}}\hspace*{-0.5cm}\\
\caption{\label{fig:interf}
Left(a): the contour levels for $pp \to Z'\to e^+e^-$ production cross section at the LHC@13TeV
as well as relative ratio of di-lepton rates for $Z'$  vs $Z''$ production for S=0.1.
Right(b):  the interference between $Z'$ and $Z''$ contributing to di-lepton signature
from $pp \to Z'/Z'' \to \ell^+\ell^-$ process.
}
\end{figure}

In Fig.~\ref{fig:interf}(a) the contour levels for $pp \to Z'\to e^+e^-$ production cross section at the LHC@13TeV
as well as relative ratio of di-lepton rates for $Z'$  vs $Z''$ production for S=0.1.
As in the recent experimental CMS paper, the cross section for $Z'$ and $Z''$ was evaluated using finite width and
mass window of $\pm5\%\sqrt{s}$ at  the resonance mass to correctly estimate the size of the $Z'/Z''$interference.
Qualitatively the picture is similar for other values of S-parameter.
First of all one can notice that with a luminosity of roughly 40 fb$^{-1}$
for which the limits on di-lepton resonances are publicly available, one can expect a di-lepton cross section of the order of 0.1 fb,
which translates to $M_A$ of about 3 TeV for low $\tilde{g}$ values.
As we will see in the following section this rough estimation agrees with an accurate limit we establish later in our paper.
Second, one can clearly see that the role of $Z''$ becomes important and even dominant for $M_A$ above 1.5 TeV
and $\tilde{g}$ below about 4. 
Fig.~\ref{fig:interf}(b) presents the interference between $Z'$ and $Z''$ contributing to di-lepton signature. One can see 
that the interference is at the percent level and can be safely neglected. This is an important condition for interpretation
of the LHC limits on single resonance search. Taking this into account and the fact  that
the  $Z'$ contribution to di-lepton signature is dominant, in the region of small $M_A<1$ TeV
we conclude that one can use  LHC limits for di-lepton single resonance searches.
Using similar logic, one can see that in the region of intermediate and large $M_A>1.5$ TeV where $M_{Z''}$ contribution to di-lepton signature is dominant one can use  LHC limits for single resonance di-lepton searches in the case of $Z''$.

Finally, in the intermediate region of $M_A$ between 1 and 1.5 TeV when di-lepton signals from  $Z'$ and $Z''$ are comparable,
well separated in mass (above 10\%) recalling  $Z'-Z''$ mass difference from Fig.~\ref{fig:masses}
and their width-to-mass ratio is small (few percent) (Fig.~\ref{fig:width2mass}),
the LHC  limits can be applied {\it separately} to $Z'$ or $Z''$ signatures.
Therefore in the whole parameter space of interest (with $\sigma (pp \to Z'/Z'' \to \ell^+\ell^-)\simeq 0.1$ fb)
one can use the signal either from $Z'$ or $Z''$ to best probe the model parameter space. This procedure sets the strategy which we use in the following section.
The statistical combination of signatures from both resonances is outside of the scope of this paper
since it requires also the change the procedure in setting the limit at the experimental level.

\subsection{Probing Technicolor parameter space at the LHC}
\label{sec:exclusion}

\subsubsection{The setup for the LHC limits}
The CMS $Z^{\prime}$ dielectron  13TeV limits~\cite{Sirunyan:2018exx} which we use for the interpretation of the WTC parameter space 
are expressed as $R_{\sigma}$ = $\sigma(pp\rightarrow Z^{\prime}\rightarrow e^{+}e^{-})/\sigma(pp\rightarrow Z\rightarrow e^{+}e^{-})$, which is the ratio of the cross section for dielectron production through a $Z^{\prime}$ boson to the cross section for dielectron production through a Z boson. The limits are expressed as a ratio in order to remove the dependency on the theoretical prediction of the Z boson cross section and correlated experimental uncertainties.

To reproduce these limits, a simulated dataset of the CMS mass distribution is generated using a background probability density function:

\begin{equation}
m^{\kappa}e^{\alpha + \beta m + \gamma m^{2} + \delta m^{3} + \epsilon m^{4}}
\end{equation}

%Would like to state that actual values of \kappa, \alpha, \beta, \gamma, \delta$ and $\epsilon but they have not been published

where $\kappa, \alpha, \beta, \gamma, \delta$ and $\epsilon$ are function parameters. This probability density function was used by to describe the dielectron mass background distribution, where the background is predominantly Drell-Yan dielectron events.  A simulated CMS dataset is obtained by normalising the Z boson region (60 $<$ $m_{ee}$ $<$ 120 GeV) in simulation to data. The total number of data events corresponding to a given integrated luminosity is $N_{Lumi}$. Using the above probability density function we generate hundreds of datasets, each with a total number of events which is a Poisson fluctuation on $N_{Lumi}$. For each dataset we step through mass values and set a 95$\%$ confidence level (CL) limit on $R_{\sigma}$. The limits are set using a Bayesian method with an unbinned extended likelihood function. Using both the signal and background probability density functions, the likelihood distribution is calculated as a function of the number of signal events for a given mass. The 95$\%$ CL upper limit on the number of signal events $N_{95}$ for a given mass is taken to be the value such that integrating the likelihood from 0 to $N_{95}$ is 0.95 of the total likelihood integral. This number $N_{95}$ is converted to a limit on the ratio of cross sections by dividing by the total number of acceptance and efficiency corrected Z bosons, the signal acceptance and efficiency. At each mass point, a limit is calculated for each of the hundreds of simulated datasets. Using the limits computed from each simulated dataset, the median $95\%$ CL limit and the one and two sigma standard deviations on the $95\%$ CL limit for each mass point can be calculated. The signal probability distribution used in the likelihood is a convolution of a Breit-Wigner function and a Gaussian function with exponential tails to either side. The limits are calculated in a mass window of $\pm$ 6 times the signal width, with this window being symmetrically enlarged until there is a minimum of 100 events in it.

To generate 14TeV dataset limits, the above procedure is repeated but the background probability density function is multiplied by an NNPDF scale factor to convert the 13TeV background distribution into a 14TeV distribution. In this work  the PDF set NNPDF LO \texttt{as\char`_0130\char`_QED} is applied.

\subsubsection{LHC potential to probe Walking Technicolor Parameter Space}
\label{subsec:13tev-exclusions}

With the set up described above we have evaluated limits on the NMWT parameter space according to Run 2 at CMS. We use the 95$\%$ CL observed limit on $\sigma(pp\rightarrow Z^{\prime}\rightarrow e^{+}e^{-})/\sigma(pp\rightarrow Z\rightarrow e^{+}e^{-})$ at $\sqrt{s}=13$TeV based on a dataset of integrated luminosity $36$fb$^{-1}$ \cite{Sirunyan:2018exx}. 
\begin{figure}[htb]
\subfigure[]{\includegraphics[width=0.5\textwidth]{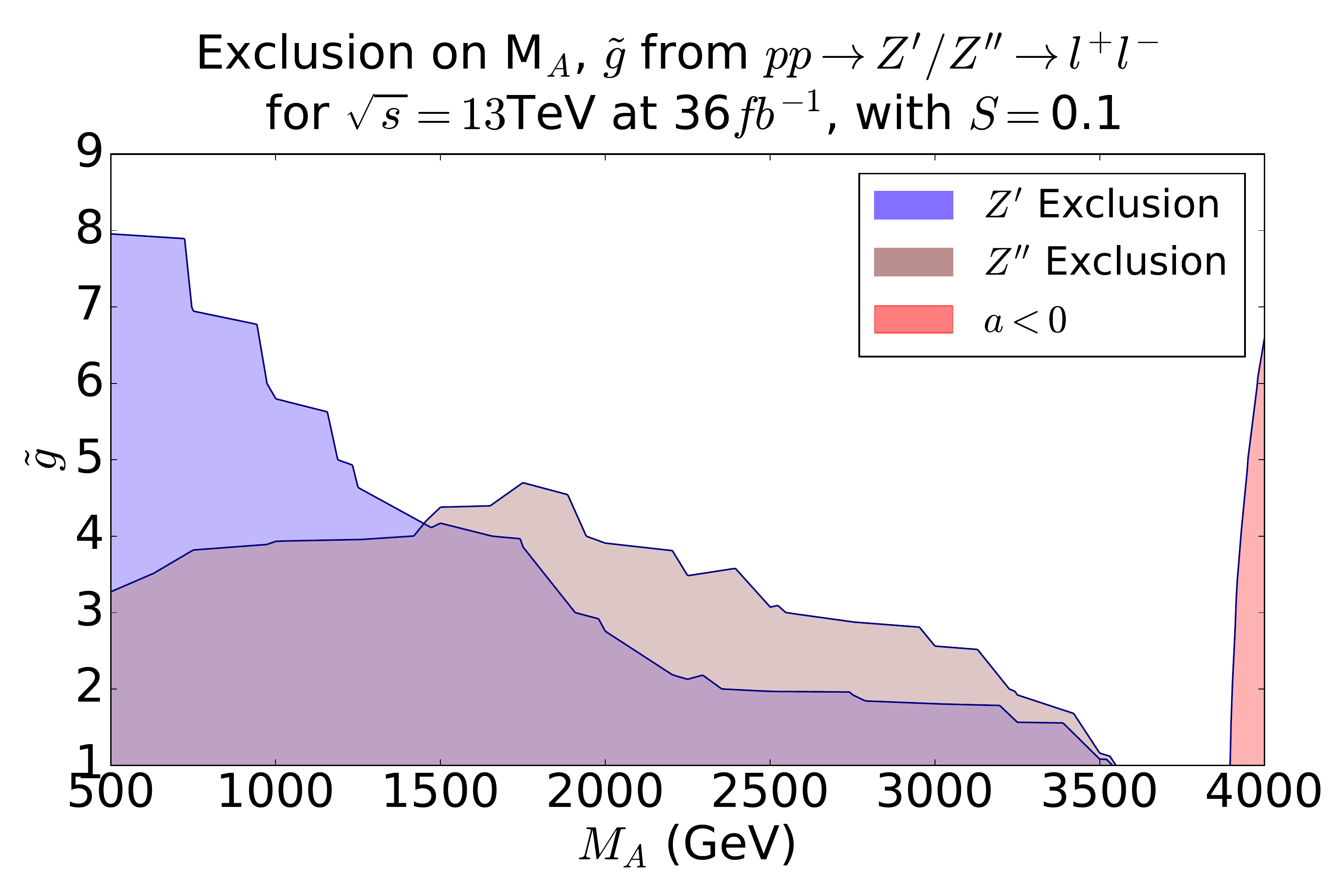}}%
\subfigure[]{\includegraphics[width=0.5\textwidth]{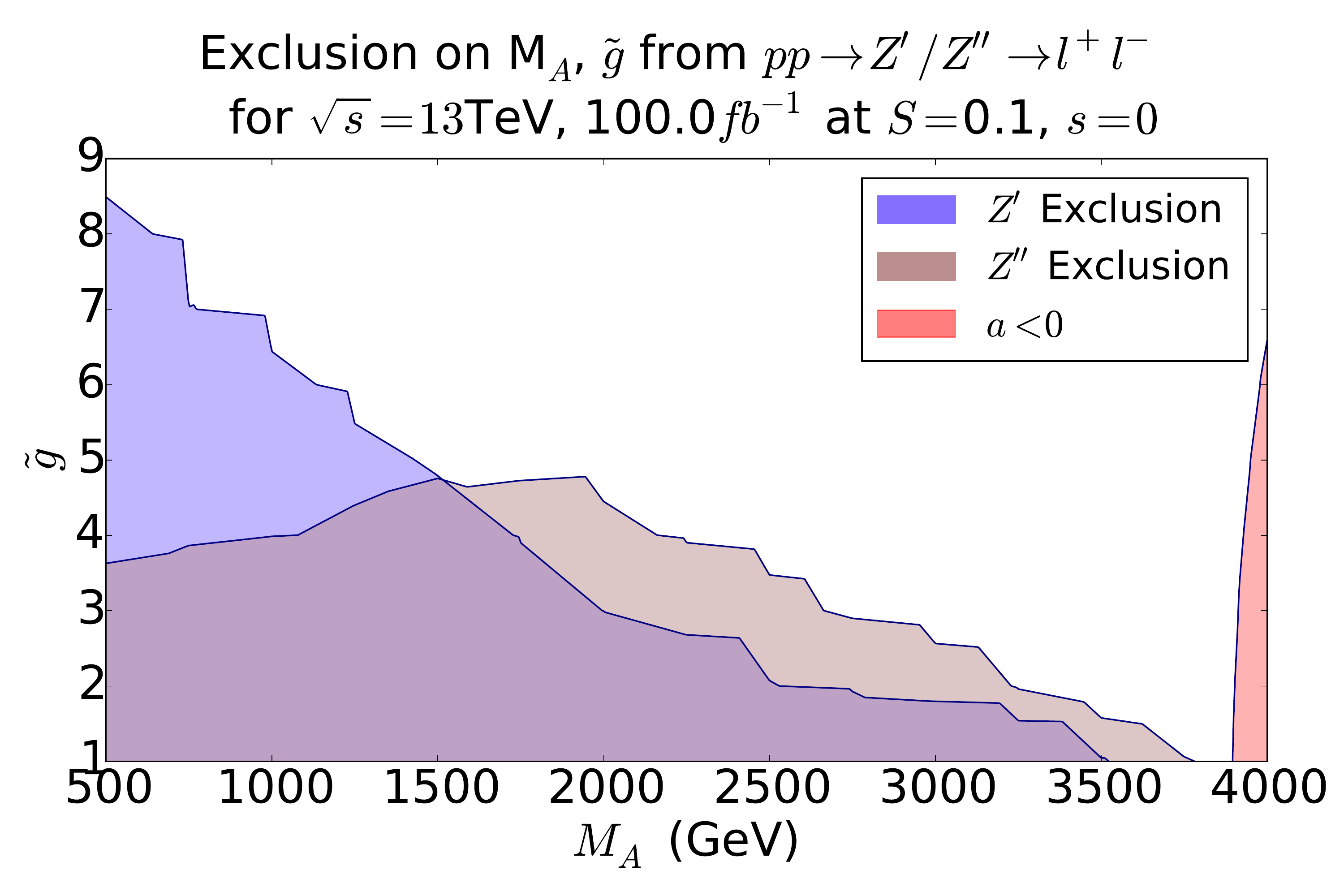}}\\
\subfigure[]{\includegraphics[width=0.5\textwidth]{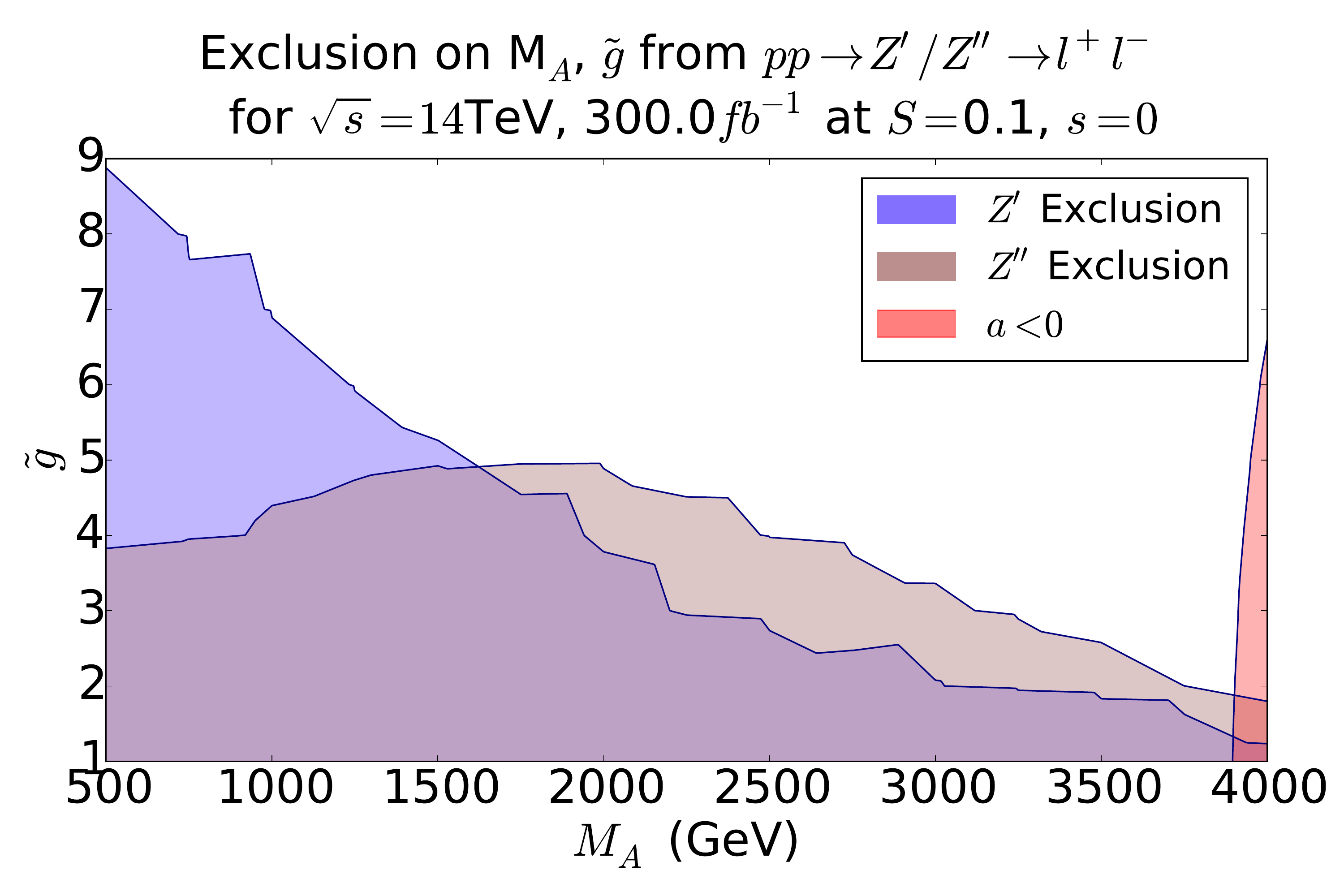}}%
\subfigure[]{\includegraphics[width=0.5\textwidth]{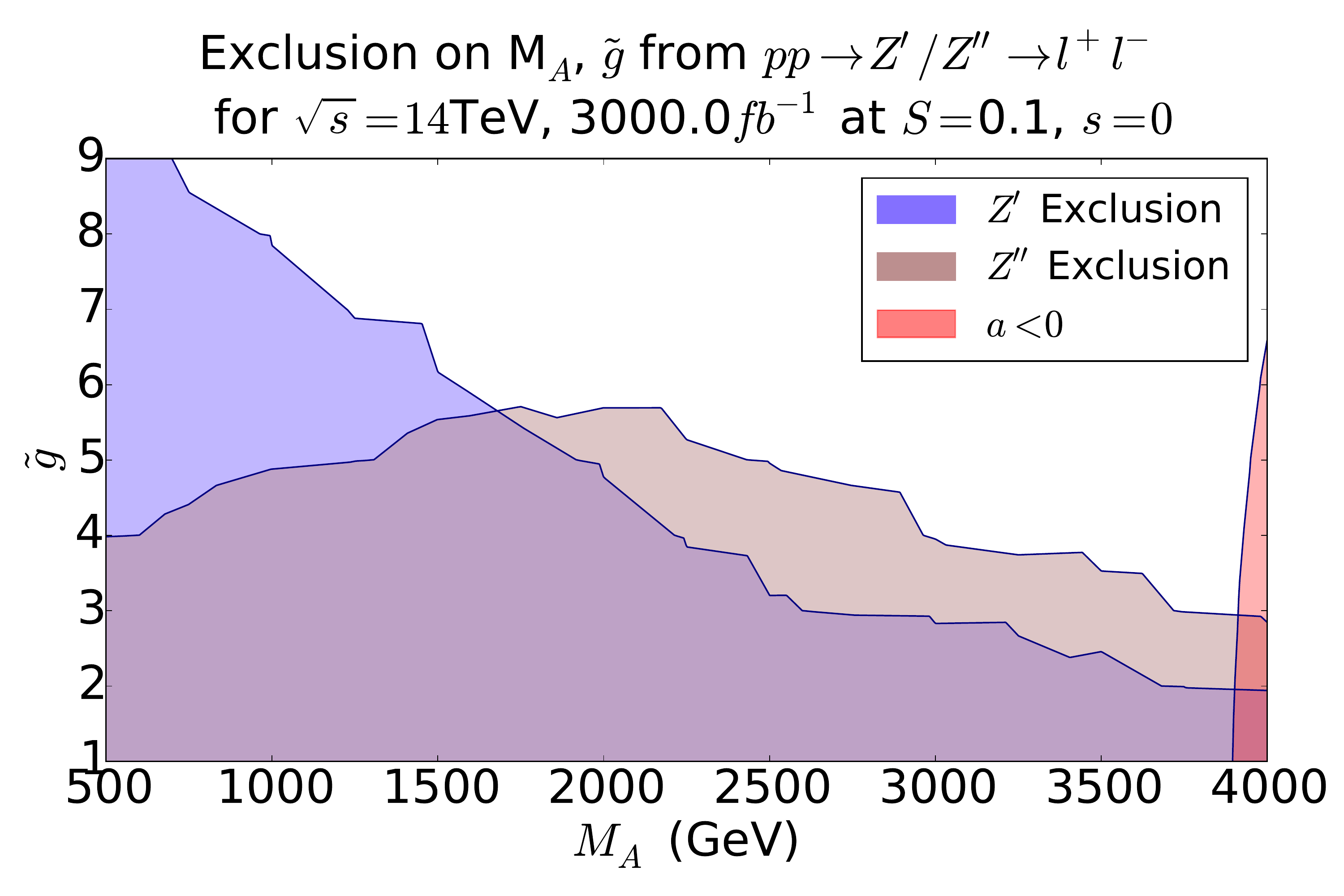}}
\caption{\label{fig:13-14-tev-exclusions} Exclusion of the $M_A$-$\tilde{g}$ parameter space from $Z^{\prime}$ and $Z^{\prime\prime}$ DY processes at $\sqrt{s}=13$TeV and luminosity of $36$fb$^{-1}$(a); Predicted exclusion regions for the NMWT parameter space at (a) $\sqrt{s}=13$TeV and $\mathcal{L}=100fb^{-1}$, (b)$\sqrt{s}=14$TeV and $\mathcal{L}=300fb^{-1}$, (c) $\sqrt{s}=14$TeV and $\mathcal{L}=3000fb^{-1}$}
\end{figure}

The SM DY cross section at NNLO is given to be $\sigma(pp\rightarrow Z/\gamma^{*}\rightarrow e^{+}e^{-})=1.928$nb, which we use to convert the ratio of cross sections to a limit on $\sigma(pp\rightarrow Z^{\prime}\rightarrow e^{+}e^{-})$. This limit is then projected onto the $(M_{A}, \tilde{g})$ plane and compared to the  signal cross sections for $Z^{\prime}$ and $Z^{\prime\prime}$ which we have evaluated at NNLO level.
Figure \ref{fig:13-14-tev-exclusions}a presents the NMWT parameter space in the $(M_A,\tilde g)$ plane
for S=0.1 which is already excluded with the recent CMS results.
One can observe an important complementarity of $Z'$ and $Z''$; as was expected from the plots with cross sections,
$Z''$ extends the coverage of the LHC in large $\tilde g$ and $M_A$ region.
Analogous exclusion plots for different values of $S$ are presented 
in Fig.~\ref{fig:exclusions-s-minus0.1}a, Fig.~\ref{fig:exclusions-s-0}a,  Fig.~\ref{fig:exclusions-s-0.2}a and Fig.~\ref{fig:exclusions-s-0.3}a for $S=-0.1,0.0,0.2$ and  $0.3$ respectively.

We have also found the projected LHC limit for higher integrated luminosities.
To do this we have  simulated the SM DY background and have obtain an expected limit for $36fb^{-1}$,
confirming to within a few $\%$ the CMS expected limits using the method described in the previous section for the sake of its validation.
Then we have obtained analogous expected limits for 
$100$fb$^{-1}$ at $\sqrt{s}=13$TeV as well as for  $300$fb$^{-1}$ and $3000$fb$^{-1}$ at $\sqrt{s}=14$TeV. We follow the CMS limit setting procedure except for mass points with less than 10 events where we set limits using Poisson statistics. The excluded regions of the $M_{A}, \tilde{g}$ parameter space are shown in Figure \ref{fig:13-14-tev-exclusions}b, c, d respectively.
Analogous exclusion plots for different values of $S$ are presented 
in Fig.~\ref{fig:exclusions-s-minus0.1}, Fig.~\ref{fig:exclusions-s-0},  Fig.~\ref{fig:exclusions-s-0.2} and Fig.~\ref{fig:exclusions-s-0.3}  for $S=-0.1,0.0,0.2$ and  $0.3$ respectively.

Already at $100$fb$^{-1}$ the excluded region visibly increases in $M_{A}$ and $\tilde{g}$ for both $Z^{\prime}$ and $Z^{\prime\prime}$ resonances. For example for small values of $\tilde g$ it increases 
for $M_A$ from 3.5 TeV to about 3.8 TeV.
Figure \ref{fig:13-14-tev-exclusions} also shows the theoretical upper limit on $M_{A}$ imposed by the $a$ parameter (see section \ref{sec:setup}). Requiring $a>0$ and combining it with the current or projected 
experimental limits one gets the full picture of the surviving parameter space.

With the beam energy increase to $\sqrt{s}=14$TeV and total integrated luminosity 300 fb$^{-1}$ or more
the entire range of $M_{A}$ that we explore is excluded in the region of $\tilde{g}<2$, and the predictions for the final high-luminosity run of the LHC (Figure \ref{fig:13-14-tev-exclusions}d) increase the exclusions in both the $M_{A}$ and $\tilde{g}$ directions ruling out the whole parameter space for $\tilde{g}<3$.

\begin{figure}[htb]
\subfigure[]{\includegraphics[width=0.5\textwidth]{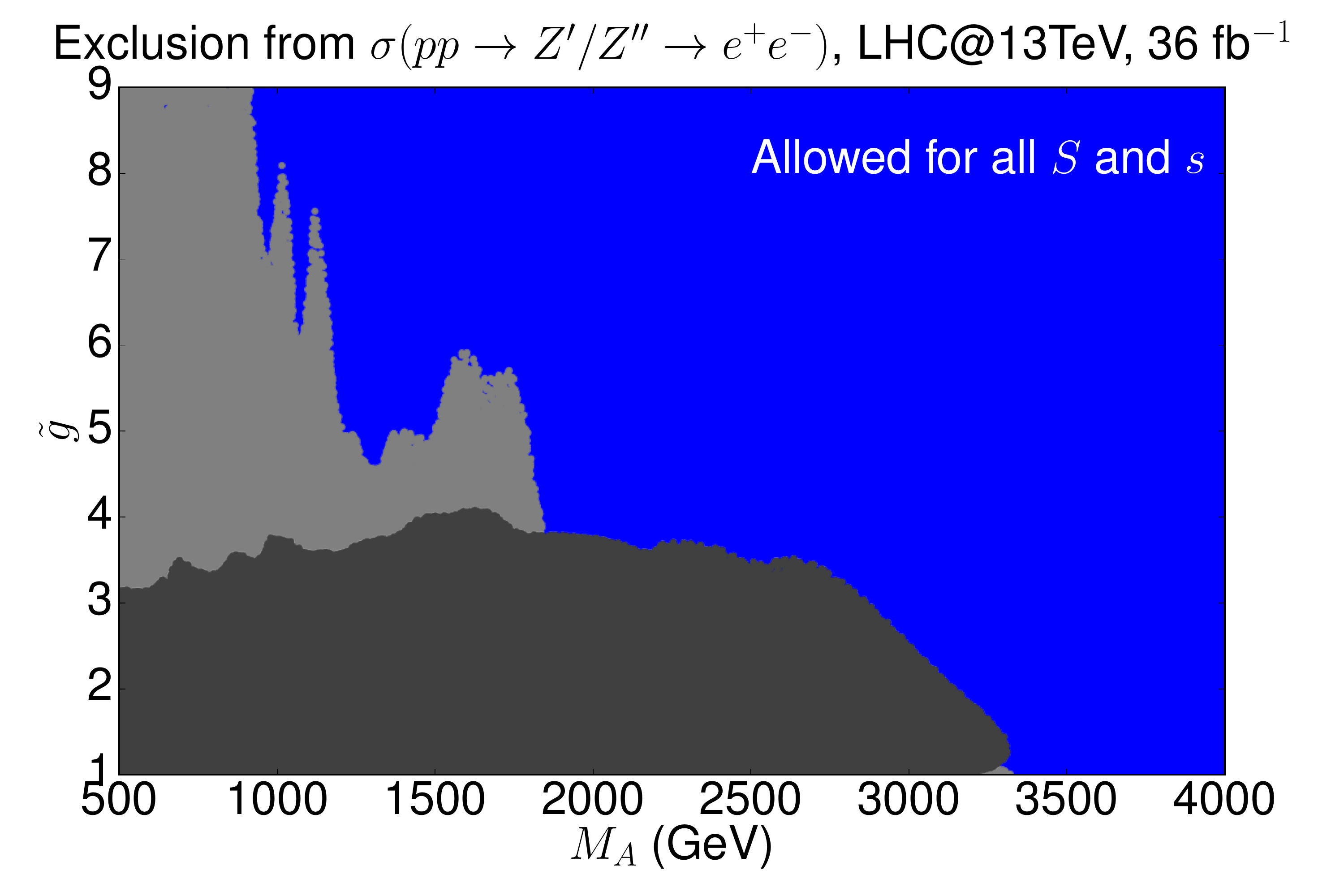}}%
\subfigure[]{\includegraphics[width=0.5\textwidth]{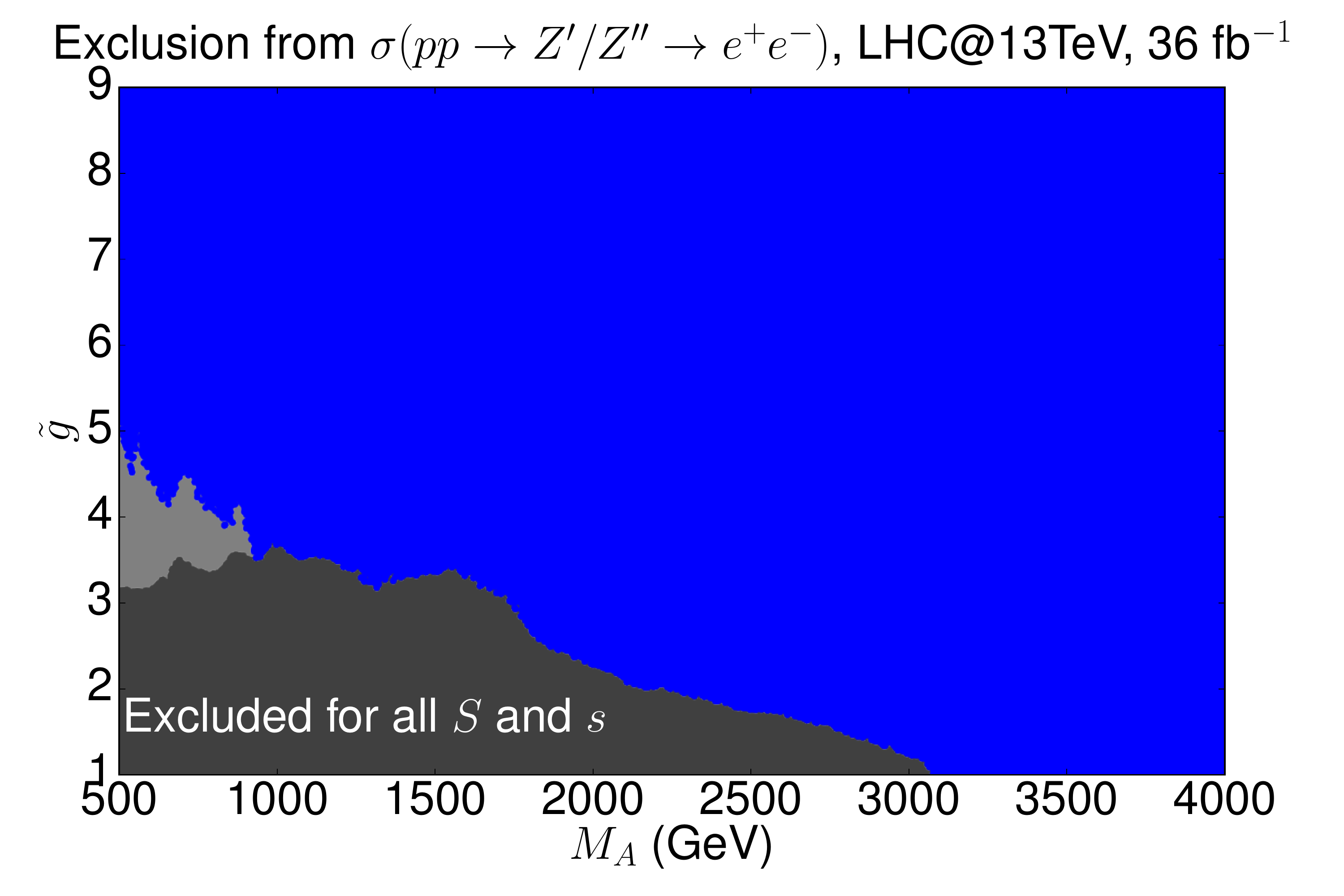}}
\caption{\label{fig:AA/AE} Projections on $M_{A}, \tilde{g}$ parameter space of theoretical DY $Z^{\prime}/Z^{\prime\prime}$ cross section showing the allowed for all $S$ and $s$ region (a), excluded for all $S$ and $s$ region (b) for the current CMS exclusion for LHC@13TeV and 36 fb$^{-1}$ integrated luminosity. Blue points are allowed, light grey points are excluded by the $Z^{\prime}$, and dark grey points are excluded by the $Z^{\prime\prime}$. }
\end{figure}

To see the picture of the LHC sensitivity to the {\it whole} NMWT parameter space
we have performed a scan of the full 4D ($M_A,\tilde g, S,s$) parameter space with  $\sim 1$x$10^{7}$  random points.
In Fig.~\ref{fig:AA/AE} we present the projection of this scan into ($M_{A},\tilde{g}$) plane,
with $S$ and $s$ range $(-0.1,0.3)$ and $(-1,1)$ respectively for LHC@13TeV and 36 fb$^{-1}$ integrated luminosity.
In Fig.~\ref{fig:AA/AE}a we overlayed  the excluded points from $Z'$ or $Z''$ signals on top of the allowed points 
to show the $(M_{A},\tilde{g})$ parameter space which is allowed for all values of $S$ and $s$ parameters,
while in  Fig.~\ref{fig:AA/AE}b we overlayed  the allowed points on top of the excluded  points 
to show the $(M_{A},\tilde{g})$ parameter space which is excluded for  all values of $S$ and $s$ parameters.
The excluded points from the $Z^{\prime\prime}$ cross section (dark grey) are layered on top of those excluded by the $Z^{\prime}$ cross section (light grey). 
It is important to stress that the most conservative limit on $M_A$ (parameter space which is excluded for all values of $S$ and $s$ parameters) is about 3.1 TeV for low values of $\tilde g$ and that this limit  is significantly higher
(by about 1 TeV) than
previous limits established by the ATLAS collaboration in Refs.~\cite{Aad:2015yza,Aad:2014cka}
for $S=0.3, s=0$ benchmark in ($M_{A},\tilde{g},S,s$) plane which actually gives one of the most optimistic limits
for NMWT.

\section{Conclusions}
\label{sec:conclusions}

Walking Technicolor remains one of the most appealing BSM theories involving strong dynamics.
In this study we have fully explored the 4D parameter space of WTC using dilepton signatures
from $Z^{\prime}/Z^{\prime\prime}$ production and decay at the LHC.
This signature is  the most promising one for discovery of WTC at the LHC
for the low-intermediate values of the $\tilde g$ parameter.

We have studied the complementarity of the dilepton signals from
both  heavy neutral vector resonances and have demonstrated its importance.
As a result, we  have established  the most up-to-date limit on the WTC parameter space
and provided  projections for the the LHC potential to probe WTC parameter space 
at higher future luminosity and upgraded energy.

Our  results on the LHC potential to probe WTC parameter space are presented in Fig.~\ref{fig:exclusions-s-minus0.1},\ref{fig:exclusions-s-0},\ref{fig:13-14-tev-exclusions},\ref{fig:exclusions-s-0.2}
and \ref{fig:exclusions-s-0.3} for the $(M_{A}$, $\tilde{g})$ plane  for $S=-0.1,0.0,0.1,0.2$ and 0.3 respectively,
which gives a clear idea how the properties of the model and the respective LHC reach 
depend on the value of the $S$ parameter. This extends the results found previously for just 
$S=0.3$ which is not quite motivated in light of present EWPD.
Moreover, as another new element of the exploration of WTC, 
we have provided an analytic description for features such as the $Z^{\prime}/Z^{\prime\prime}$ masses,   the mass inversion $M_{inv}$ point as well as some couplings  in our paper.
We have also presented all these properties in the form of figures in the $(M_{A}$, $\tilde{g})$
plane and 3D $(M_{A}$, $\tilde{g}$, $S)$ tables  for clear insight into the model behaviour, and for direct comparison with prior works.  
We have discussed the theoretical upper limit on $M_{A}$ from the requirement of ``walking" dynamics, and in combination with the exclusions from experiment we have found the strongest constraints on WTC to date. The predicted exclusions indicate  that within the scope of the LHC, the low $\tilde{g}$ regions of the WTC parameter space can be closed completely. 

We have explored the effect of the $S$ and $s$ parameters on the WTC exclusions
using a very detailed scan of the 4D  parameter space
and establishing the current LHC limit in this 4D space which we present in 
Fig.~\ref{fig:AA/AE}.  The results we have found reflect 
the most conservative limit on $M_A$ around 3.1 TeV, which for low values of $\tilde g$ is significantly higher (by about 1 TeV) than
previous limits established by the ATLAS collaboration in Refs.~\cite{Aad:2015yza,Aad:2014cka}
for the most optimistic benchmark with $S=0.3$.
The complete 4D scan also indicates the important influence  of the value of the $S$-parameter 
on the dilepton signal rate, while the  $s$ parameter has little effect on the rate of the dilepton signal
but could be important for the complementary  $VV$ and $VH$ signatures.

Besides $Z'$ and $Z''$ complementarity for the exploration of the dilepton signal in the low-intermediate $\tilde g$ region, it is important to note  the complementarity 
for the $VV$ and $VH$ signatures which would  allow us to probe the large values of $\tilde g$.
This is the subject of the upcoming study~\cite{WW-ZH-AB-AC}.

\section*{Acknowledgements}
The authors acknowledge the use of the IRIDIS High Performance Computing Facility, and
associated support services at the University of Southampton, in the completion of this work.
AB acknowledges partial  support from the STFC grant ST/L000296/1.
AB also thanks the NExT Institute, Royal Society Leverhulme Trust Senior Research Fellowship LT140094, 
Royal Society Internationl Exchange grant IE150682 and
Soton-FAPESP grant.
AB acknowledges partial support from the InvisiblesPlus RISE from the European
Union Horizon 2020 research and innovation programme under the Marie Sklodowska-Curie grant
agreement No 690575.
AC acknowledges the University of Southampton for support under the Mayflower Scholarship PhD programme.
AC acknowledges partial support from SEPnet under the GRADnet Scholarship award.  

\newpage\bibliography{bib}
\bibliographystyle{JHEP}

\newpage\appendix

\section{Appendix}
\label{sec:appendix}

\subsection{Mass Matrices in NMWT}
\label{subsec:massmatrices}

We calculate $N_{ij}$ by diagonalising the bosonic mixing matrices \ref{eqn:matrixlanhep}, perturbatively calculating the eigenvalues and eigenvectors of the matrices that diagonalise $\mathcal{M}_{C}^{2}$ and $\mathcal{M}_{N}^{2}$ order by order in $1/\tilde{g}$. Details of the calculation are presented in here, with the results for $C_{ij}$ and $N_{ij}$ to 2nd order in $1/\tilde{g}$\footnote{Each of these $C_{ij}$ and $N_{ij}$ represent the mixing of the vector boson/meson states, e.g $N_{24}$ represents a mixed $Z-Z^{\prime\prime}$ state, and components with $i=j$ represent mixing of a gauge field with itself}. At 0th order, the eigenvalues for the $\gamma, Z$ are degenerate and $m_{\gamma}^{2},m_{Z}^{2}=0$, so the eigenvectors cannot be uniquely defined at this stage. To resolve this degeneracy we introduce a generic parameter $x$ which is fixed at 2nd order to be $x=g_{2}/g_{1}$.

From the covariant derivative terms of the effective bosonic Lagrangian \ref{eqn:bosonlagng}, we  construct the mixing matrices that diagonalise to give physical masses for the vector bosons. The Lagrangian of the vector bosons in the mass eigenbasis is

\begin{equation}
\mathcal{L}_{mass} = \begin{pmatrix}\tilde{W}^{-}_{\mu} & A_{L\mu}^{-} & A_{R\mu}^{-}\end{pmatrix}  \mathcal{M}_{C}^{2} \begin{pmatrix}
\tilde{W}^{+\mu} \\
 A_{L}^{+\mu} \\
 A_{R}^{+\mu}
\end{pmatrix} + \frac{1}{2}\begin{pmatrix}\tilde{B}_{\mu} & \tilde{W}^{0}_{\mu} & A_{L\mu}^{0} & A_{R\mu}^{0}\end{pmatrix} \mathcal{M}_{N}^{2} 
\begin{pmatrix}
\tilde{B}^{\mu} \\
\tilde{W}^{0\mu} \\
A_{L}^{0\mu} \\
A_{R}^{0\mu}\end{pmatrix},
\end{equation}

where these mass matrices for the charged and neutral bosons are

\begin{equation}
\mathcal{M}_{C}^{2} = 
\begin{pmatrix}
\frac{g_{2}^{2}}{\tilde{g}^{2}}M_{V}^{2}(1+\omega) &
-\frac{g_{2}}{\sqrt{2}\tilde{g}}M_{A}^{2}\chi &
-\frac{g_{2}}{\sqrt{2}\tilde{g}}M_{V}^{2} \\
-\frac{g_{2}}{\sqrt{2}\tilde{g}}M_{A}^{2}\chi &
M_{A}^{2} &
0 \\
-\frac{g_{2}}{\sqrt{2}\tilde{g}}M_{V}^{2} &
0 &
M_{V}^{2}
\end{pmatrix},
\end{equation}

\begin{equation}
\mathcal{M}_{N}^{2} =  
\begin{pmatrix}
\frac{g_{1}^{2}}{\tilde{g}^{2}}M_{V}^{2}(1+\omega) &
-\frac{g_{1}g_{2}}{\tilde{g}^{2}}M_{V}^{2}\omega &
\frac{g_{1}}{\sqrt{2}\tilde{g}}M_{A}^{2}\chi &
-\frac{g_{1}}{\sqrt{2}\tilde{g}}M_{V}^{2} \\
-\frac{g_{1}g_{2}}{\tilde{g}^{2}}M_{V}^{2}\omega &
\frac{g_{2}^{2}}{\tilde{g}^{2}}M_{V}^{2}(1+\omega) &
-\frac{g_{2}}{\sqrt{2}\tilde{g}}M_{A}^{2}\chi &
-\frac{g_{2}}{\sqrt{2}\tilde{g}}M_{V}^{2} \\
\frac{g_{1}}{\sqrt{2}\tilde{g}}M_{A}^{2}\chi &
-\frac{g_{2}}{\sqrt{2}\tilde{g}}M_{A}^{2}\chi &
M_{A}^{2} &
0 \\
-\frac{g_{1}}{\sqrt{2}\tilde{g}}M_{V}^{2} &
-\frac{g_{2}}{\sqrt{2}\tilde{g}}M_{V}^{2} &
0 &
M_{V}^{2}
\end{pmatrix}.
\label{eqn:matrixlanhep}
\end{equation}

In order to perform the analytic diagonalisation of these matrices, we perform an expansion in $1/\tilde{g}$ and calculate the eigenvectors and eigenvalues of the matrix order by order. Rephrasing the $\chi$ and $M_{V}^{2}$ parameters such that

\begin{align*}
&M_{V}^{2} = \frac{F_{\pi}^{2}\tilde{g}^{2}}{2}+M_{A}^{2}\chi^{2}, \\
&\chi = \sqrt{1-\frac{S\tilde{g}^{2}}{8\pi}}, 
\end{align*}

we can rewrite these matrices in terms of the parameters of the model that we have used in this paper. Further to this, from the WSRs \cite{PhysRevLett.18.507} we set $\omega=0$ and fix $F_{\pi}=246$GeV, so the mass matrices are written entirely from the free parameters, $M_{A}$, $\tilde{g}$ and $S$. 

Consider the diagonalisation of the neutral matrix $\mathcal{M}_{N}^{2}$ \ref{eqn:matrixlanhep}. From the logic above we see that this can be written as

\small

\begin{equation}
\hspace{-20mm}
\mathcal{M}_{N}^{2} =  
\begin{pmatrix}
\frac{1}{8}g_{1}^{2}\Bigg(M_{A}^{2}(\frac{8}{\tilde{g}^{2}}{-}\frac{S}{\pi}){+}4F_{\pi}^{2}\Bigg) &
0 &
\frac{g_{1}M_{A}^{2}}{4\tilde{g}}\sqrt{8{-}\frac{\tilde{g}^{2}S}{\pi}}&
\frac{g_{1}}{8\pi\sqrt{2}\tilde{g}}(M_{A}^{2}(\tilde{g}^{2}S{-}8\pi){-}4\tilde{g}^{2}\pi F_{\pi}^{2}) \\ 
0 &
\frac{1}{8}g_{2}^{2}\Bigg(M_{A}^{2}\Bigg(\frac{8}{\tilde{g}^{2}}{-}\frac{S}{\pi}\Bigg){+}4F_{\pi}^{2}\Bigg) &
-\frac{g_{2}M_{A}^{2}}{4\tilde{g}}\sqrt{8{-}\frac{\tilde{g}^{2}S}{\pi}} &
\frac{g_{2}}{8\pi\sqrt{2}\tilde{g}}(M_{A}^{2}(\tilde{g}^{2}S{-}8\pi){-}4\tilde{g}^{2}\pi F_{\pi}^{2}) \\ 
\frac{g_{1}}{8\pi\sqrt{2}\tilde{g}}(M_{A}^{2}(\tilde{g}^{2}S{-}8\pi){-}4\tilde{g}^{2}\pi F_{\pi}^{2}) & 
-\frac{g_{2}M_{A}^{2}}{4\tilde{g}}\sqrt{8{-}\frac{\tilde{g}^{2}S}{\pi}} & 
M_{A}^{2} &
0 \\
\frac{g_{1}}{8\pi\sqrt{2}\tilde{g}}(M_{A}^{2}(\tilde{g}^{2}S{-}8\pi){-}4\tilde{g}^{2}\pi F_{\pi}^{2})  &
\frac{g_{2}}{8\pi\sqrt{2}\tilde{g}}(M_{A}^{2}(\tilde{g}^{2}S{-}8\pi){-}4\tilde{g}^{2}\pi F_{\pi}^{2})&
0 &
M_{A}^{2}(1{-}\frac{\tilde{g}^{2}S}{8\pi})+\frac{1}{2}\tilde{g}^{2}F_{\pi}^{2}
\end{pmatrix}.
\label{eqn:matrix-rewrite}
\end{equation}
\normalsize

To expand in powers of $1/\tilde{g}$ we can rewrite the independent $M_{A}$, $S$ and $F_{\pi}$ parameters in terms of $\tilde{g}$ and dependent parameters of the model.  As stated above, in the regime of large $\tilde{g}$,  $M_{A}^{2}$ is dominated by the $r_{2}$ term of equation \ref{eqn:massvec-ax}, however it is not obvious to see that in the case of small $\tilde{g}$, the $m^{2}$ term dominates. We can determine the scaling of $m^{2}$ from the 1st WSR and the definition of the pion decay constant in NMWT. From equation \ref{eqn:decayconsts} we see that $F_{\pi}^{2}$ can be written in terms of $M_{A}^{2}/\tilde{g}^{2}$. In the low $\tilde{g}$ regime this would lead to $F_{\pi}^{2}\propto m^{2}/\tilde{g}^{2}$, so one would na{\"i}vely expect $F_{\pi}\propto 1/\tilde{g}$. However, $F_{\pi}$ is fixed to avoid deviations from the 1st WSR, so $m^{2}$ \textbf{must} itself scale with $\tilde{g}^{2}$. Finally, from equation \ref{eqn:zerothwsr} we see that $S$ can be written in terms of $\tilde{g}^{-2}$.

At leading order in $1/\tilde{g}$, the mass squared terms for the neutral bosons are 

\begin{equation}
M_{\gamma}^{2} = 0, \hspace{3mm} M_{Z}^{2}=0, \hspace{3mm} M_{Z^{\prime}}^{2}= M_{A}^{2}, \hspace{3mm} M_{Z^{\prime\prime}}^{2}=M_{A}^{2}(1-\frac{\tilde{g}^{2}S}{8\pi})+\frac{1}{2}\tilde{g}^{2}F_{\pi}^{2}.
\label{eqn:LOmasses}
\end{equation}

As there are two degenerate eigenvalues$=0$, we must define the eigenvectors at 0th order with a generic term $x$ which is fixed only at 2nd order in the $1/\tilde{g}$ expansion. The 0th order eigenvectors are then

\begin{equation}
\bar{v}_{0} = 
\begin{pmatrix}
\frac{x}{\sqrt{1+x^{2}}} &
\frac{1}{\sqrt{1+x^{2}}} &
0 &
0 \\
\frac{1}{\sqrt{1+x^{2}}} &
-\frac{x}{\sqrt{1+x^{2}}} &
0 &
0 \\
0 &
0 &
1 &
0 \\
 0 &
0 &
0 &
1 
\end{pmatrix}.
\label{eqn:evecs0}
\end{equation}

We can now construct the higher order corrections order by order. To calculate the 1st order corrections, we consider the eigenvalue equation

\begin{equation}
M\bar{v}=\lambda\bar{v}
\label{eqn:evaleqn}
\end{equation}

where $M=M_{0}+M_{1}+M_{2}+\dots$ is the mixing matrix, $\bar{v}=\bar{v}_{0}+\bar{v}_{1}+\bar{v}_{2}+\dots$ are the eigenvectors of $M$, and $\lambda=\lambda_{0}+\lambda_{1}+\lambda_{2}+\dots$ are the eigenvalues of $M$. At first order we have

\begin{align*}
&(M_{0}+M_{1})(\bar{v}_{0}+\bar{v}_{1})=(\lambda_{0}+\lambda_{1})(\bar{v}_{0}+\bar{v}_{1}) \\
&M_{0}\bar{v}_{1}+M_{1}\bar{v}_{0}+M_{1}\bar{v}_{1} = \lambda_{0}\bar{v}_{1}+\lambda_{1}\bar{v}_{0}+\lambda_{1}\bar{v}_{1} \\
&\lambda_{1}=\bar{v}_{0}^{T}(M_{0}-\lambda_{0})\bar{v}+\bar{v}_{0}^{T}M_{1}\bar{v}_{0} \\
&\lambda_{1}=\bar{v}_{0}^{T}M_{1}\bar{v}_{0},
\end{align*}

where we have used the 0th order eigenvalue equation $M_{0}\bar{v}_{0}=\lambda_{0}\bar{v}_{0}$ to remove 0th order terms, and have discarded terms of order $>1$. 

We can immediately see that the 1st order eigenvalues are $\lambda_{1}^{i}=0$ for all $i=1,\dots,4$, as $\mathcal{M}_{N}^{2}$ does not have any diagonal components at order $1/\tilde{g}$. We do not expect to see corrections to the squared masses of the vector bosons at odd order in $1/\tilde{g}$ as then we would find mass terms dependent on fractional powers in the coupling. The eigenvectors will contribute to the 2nd order mass corrections, and in terms of model parameters and the unknown $x$ we find

\begin{equation}
\bar{v}_{1} = 
\begin{pmatrix}
0 &
0 &
\frac{g_{2}-g_{1}x}{4\tilde{g}\sqrt{1+x^{2}}}\sqrt{8-\frac{\tilde{g}^{2}S}{\pi}} &
\frac{g_{2}+g_{1}x}{\tilde{g}\sqrt{2}\sqrt{1+x^{2}}} \\
0 & 
0 & 
-\frac{g_{1}+g_{2}x}{4\tilde{g}\sqrt{1+x^{2}}}\sqrt{8-\frac{\tilde{g}^{2}S}{\pi}} &
\frac{g_{1}-g_{2}x}{\tilde{g}\sqrt{2}\sqrt{1+x^{2}}} \\
\frac{g_{1}}{4\tilde{g}}\sqrt{8-\frac{\tilde{g}^{2}S}{\pi}} &
-\frac{g_{2}}{4\tilde{g}}\sqrt{8-\frac{\tilde{g}^{2}S}{\pi}} &
0 & 
0 \\
-\frac{g_{1}}{\sqrt{2}\tilde{g}} &
-\frac{g_{2}}{\sqrt{2}\tilde{g}} &
0 &
0 
\end{pmatrix}.
\end{equation}

To find the 2nd order eigenvalues, we follow the same procedure as above, and keeping only 2nd order terms we find

\begin{equation}
\lambda_{2} = \bar{v}_{0}^{T}M_{1}\bar{v}_{1}+\bar{v}_{0}^{T}M_{2}\bar{v}_{0}-\bar{v}_{0}^{T}\lambda_{1}\bar{v}_{1},
\end{equation}

where we use the fact that $\lambda_{1}=0$ to reduce this to 

\begin{equation}
\lambda_{2} = \bar{v}_{0}^{T}M_{1}\bar{v}_{1}+\bar{v}_{0}^{T}M_{2}\bar{v}_{0}.
\label{eqn:ord2-evecs}
\end{equation}

At this order we can now fix $x$, which turns out to be $x = g_{2}/g_{1}$, and we arrive at the 2nd order corrections to the neutral vector boson masses; 

\begin{align}
&M_{\gamma}^{2} = 0, \hspace{3mm} M_{Z}^{2}=\frac{1}{4}(g_{1}^{2}+g_{2}^{2})F_{\pi}^{2},\\
 M_{Z^{\prime}}^{2}= \frac{g_{1}^{2}+g_{2}^{2}}{16\pi\tilde{g}^{2}}M_{A}^{2}(8\pi-\tilde{g}^{2}S), &\hspace{3mm} M_{Z^{\prime\prime}}^{2}=\frac{g_{1}^{2}+g_{2}^{2}+2\tilde{g}^{2}}{16\pi\tilde{g}^{2}}(M_{A}^{2}(8\pi-\tilde{g}^{2}S)+4\pi\tilde{g}^{2}F_{\pi}^{2}).
\label{eqn:NNLOmasses}
\end{align}

Finally, the rotation matrices $\mathcal{C}$ and $\mathcal{N}$ can be constructed from the transpose of the sum of 0th, 1st and 2nd order eigenvectors;

\begin{equation}
\mathcal{N} = 
\begin{pmatrix}
\frac{g_{2}}{\sqrt{g_{1}^{2}+g_{2}^{2}}} &
\frac{g_{1}}{\sqrt{g_{1}^{2}+g_{2}^{2}}} &
\frac{g_{1}\chi}{\sqrt{2}\tilde{g}} &
-\frac{g_{1}}{\sqrt{2}\tilde{g}} \\
\frac{g_{1}}{\sqrt{g_{1}^{2}+g_{2}^{2}}} & 
-\frac{g_{2}}{\sqrt{g_{1}^{2}+g_{2}^{2}}} & 
-\frac{g_{2}\chi}{\sqrt{2}\tilde{g}}&
-\frac{g_{2}}{\sqrt{2}\tilde{g}}\\
0 &
-\frac{\sqrt{g_{1}^{2}+g_{2}^{2}}\chi}{\sqrt{2}\tilde{g}} &
1 & 
-\frac{(g_{1}-g_{2})(g_{1}+g_{2})(2M_{A}^{2}\chi^{2}+\tilde{g}^{2}F_{\pi}^{2})\chi}{\tilde{g}^{2}M_{A}^{2}(4\chi^{2}-1)+2\tilde{g}^{4}F_{\pi}^{2}} \\
\frac{\sqrt{2}g_{1}g_{2}}{\sqrt{g_{1}^{2}+g_{2}^{2}}\tilde{g}} &
\frac{(g_{1}-g_{2})(g_{1}+g_{2})}{\sqrt{2}\tilde{g}\sqrt{g_{1}^{2}+g_{2}^{2}}} &
\frac{4(g_{1}-g_{2})(g_{1}+g_{2})M_{A}^{2}\chi}{2\tilde{g}^{2}M_{A}^{2}(\chi^{2}-4)+\tilde{g}^{4}F_{\pi}^{2}} &
1 
\end{pmatrix},
\end{equation}

\begin{equation}
\mathcal{C} = 
\begin{pmatrix}
1 &
-\frac{g_{2}\chi}{\sqrt{2}\tilde{g}} &
-\frac{g_{2}}{\sqrt{2}\tilde{g}} \\
\frac{g_{2}\chi}{\sqrt{2}\tilde{g}} &
1 &
\frac{g_{2}\chi}{\sqrt{2}\tilde{g}}\Bigg(1+\frac{2M_{A}^{2}}{2M_{A}^{2}(3\chi^{2}-1)+3\tilde{g}^{2}F_{\pi}^{2}} \Bigg) \\
\frac{g_{2}}{\sqrt{2}\tilde{g}} &
-\frac{3g_{2}^{2}M_{A}^{2}\chi}{2\tilde{g}^{2}M_{A}^{2}(\chi^{2}-3)+\tilde{g}^{4}F_{\pi}^{2}} &
1
\end{pmatrix},
\end{equation}

where $\mathcal{N}$ and $\mathcal{C}$ diagonalise the neutral and charged mass matrices respectively.
It is the elements of these rotation matrices that comprise the vector boson couplings in NMWT, as discussed in section \ref{subsec:couplings}. 

\subsection{Dependent Parameters in terms of $S, M_{A},\tilde{g},s$}

From the equations defined in section \ref{sec:setup}, we  derive expressions for all of the dependent parameters of NMWT in terms of its 4 independent parameters. Begin by constructing simultaneous equations for $v$ and $r_{2}$ parameters, the first of which comes from rearranging equation \ref{eqn:chi},

\begin{equation}
v^{2}(r_{2}+1) = \frac{4M_{A}^{2}}{\tilde{g}^{2}}(1-\chi),
\label{eqn:vr2eq1}
\end{equation}

and the second from equation \ref{eqn:massvec-ax},

\begin{align}
M_{V}^{2}-M_{A}^{2} &= m^{2}-m^{2}+\frac{\tilde{g}^{2}(s-r_{2})v^{2}}{4}-\frac{\tilde{g}^{2}(s+r_{2})v^{2}}{4} \\
r_{2}v^{2} &= \frac{2}{\tilde{g}^{2}}(M_{A}^{2}-M_{V}^{2}).
\label{eqn:vr2eq2}
\end{align}

The we resolve $v$ by subtracting equation \ref{eqn:vr2eq2} from equation \ref{eqn:vr2eq1}, and substituting the definitions of $\chi$ and $M_{V}$ from equations \ref{eqn:0thwsr-gt} and \ref{eqn:vr2eq1} respectively:

\begin{align}
v^{2} &= \frac{1}{\sqrt{2}G_{F}}+\frac{4M_{A}^{2}}{\tilde{g}^{2}}\Bigg(1-\frac{\tilde{g}^{2}S}{16\pi}-\sqrt{1-\frac{\tilde{g}^{2}S}{8\pi}}\Bigg).
\label{eqn:veqn}
\end{align}

Then we substitute this into equation \ref{eqn:vr2eq2} to find $r_{2}$,

\begin{equation}
r_{2} = \frac{\tilde{g}^{2}(G_{F}M_{A}^{2}S-2\sqrt{2}\pi)}{\tilde{g}^{2}(2\sqrt{2}\pi-G_{F}M_{A}^{2}S)-4G_{F}M_{A}^{2}(\sqrt{2\pi}\sqrt{8\pi-\tilde{g}^{2}S}-4\pi)}.
\label{eqn:r2eqn}
\end{equation}

Now we  find $m$ in terms of the NMWT parameters from equation \ref{eqn:massvec-ax},

\begin{align}
m^{2} &= M_{A}^{2}-\frac{\tilde{g}^{2}(s+r_{2})v^{2}}{4}  \\
m^{2} &= \frac{\tilde{g}^{2}(1-s)(2\sqrt{2}\pi-G_{F}M_{A}^{2}S)+4G_{F}M_{A}^{2}(4\pi(1-s)+s\sqrt{2\pi}\sqrt{8\pi-\tilde{g}^{2}S})}{16\pi G_{F}}
\end{align}

Then  we  relate the Fermi constant $G_{F}$ to the model parameters by finding the form of $F_{\pi}$. We combine the definitions of $F_{V}$ and $F_{A}$ with the 1st WSR,

\begin{equation}
F_{\pi}^{2}=\frac{2M_{V}^{2}}{\tilde{g}^{2}}-\frac{2M_{A}^{2}}{\tilde{g}^{2}}\chi^{2},
\end{equation}

and substitute the definition of $\chi$ from equation \ref{eqn:chi},

\begin{equation}
F_{\pi}^{2}=\frac{2}{\tilde{g}}\Bigg(\frac{v^{2}\tilde{g}^{2}r_{3}}{2}-\frac{v^{4}\tilde{g}^{4}r_{3}^{2}}{16M_{A}^{2}}-(r_{3}-1)\frac{v^{2}\tilde{g}^{2}}{2}\Bigg).
\end{equation}

Finally, we arrive at the expression for $F_{\pi}$,

\begin{equation}
F_{\pi}^{2}=v^{2}\Bigg(1-\frac{v^{2}\tilde{g}^{2}r_{3}^{2}}{8M_{A}^{2}}\Bigg).
\end{equation}

\subsection{Solving for EW couplings}

The other important quantities to derive analytic formulae for are the EW equivalent couplings $g_{1}$ and $g_{2}$ in terms of the independent parameters. These couplings can be derived as roots of the characteristic equation for the $Z$ boson eigenvalue, i.e we can solve the equation $\det[\mathcal{M}_{N}^{2}-M_{Z}^{2}]=0$. Taking the absolute values of the roots, we find two solutions to this equation which correspond to the couplings $g_{2}$ and $g_{1}$ respectively,

\begin{equation}
g2 = \tilde{g}\sqrt{\frac{(\tilde{g}^{2}-2e^{2})abM_{Z}^{2}+\sqrt{abM_{Z}^{2}(2e^{2}M_{Z}^{2}+\tilde{g}^{2}b)(a(\tilde{g}^{2}M_{Z}^{2}-2e^{2}b)+2e^{2}M_{A}^{4}\chi^{2})}}{M_{V}^{2}a(4e^{2}+\tilde{g}^{2}(b-M_{Z}^{2}))-M_{A}^{4}\chi^{2}(2e^{2}M_{Z}^{2}+\tilde{g}^{2}b))}}
\label{eqn:g2full}
\end{equation}

\begin{equation}
g1 = \tilde{g}\sqrt{\frac{(\tilde{g}^{2}-2e^{2})abM_{Z}^{2}-\sqrt{abM_{Z}^{2}(2e^{2}M_{Z}^{2}+\tilde{g}^{2}b)(a(\tilde{g}^{2}M_{Z}^{2}-2e^{2}b)+2e^{2}M_{A}^{4}\chi^{2})}}{M_{V}^{2}a(4e^{2}+\tilde{g}^{2}(b-M_{Z}^{2}))-M_{A}^{4}\chi^{2}(2e^{2}M_{Z}^{2}+\tilde{g}^{2}b))}}
\label{eqn:g1full}
\end{equation}

where $a=(M_{A}^{2}-M_{Z}^{2})$, $b=(M_{V}^{2}-M_{Z}^{2})$, and we have not replaced $M_{V}$ and $\chi$, as they are purely functions of the independent parameters and not of either $g_{1}$ or $g_{2}$.

%As with $G_{F}$, the coupling constant $e$ and  $M_{Z}$, are fixed. In terms of the denominator, $g_{1}$ and $g_{2}$ are identical, and taking into account the relative magnitudes of the three particle masses, essentially the denominator boils down to $\sqrt{M_{V}^{2}-M_{A}^{2}\chi^{2}}$. 

\subsection{Effect of $S$ on $Z^{\prime}/Z^{\prime\prime}$ properties}
\label{subsec:S-plots}

Here we provide the additional figures and information relevant to the phenomenological study presented in this paper. Throughout the paper we have chosen $S=0.1$ and $s=0$ as the benchmark parameter space values, the effect of varying $S$ is discussed here. As $s$ is the Lagrangian parameter that quantifies Higgs interactions with the WTC gauge bosons, we continue to assume $s=0$ throughout.

\subsubsection{Mass Spectra}

\begin{figure}[htb]
\subfigure[]{\includegraphics[type=pdf,ext=.pdf,read=.pdf,width=0.5\textwidth]{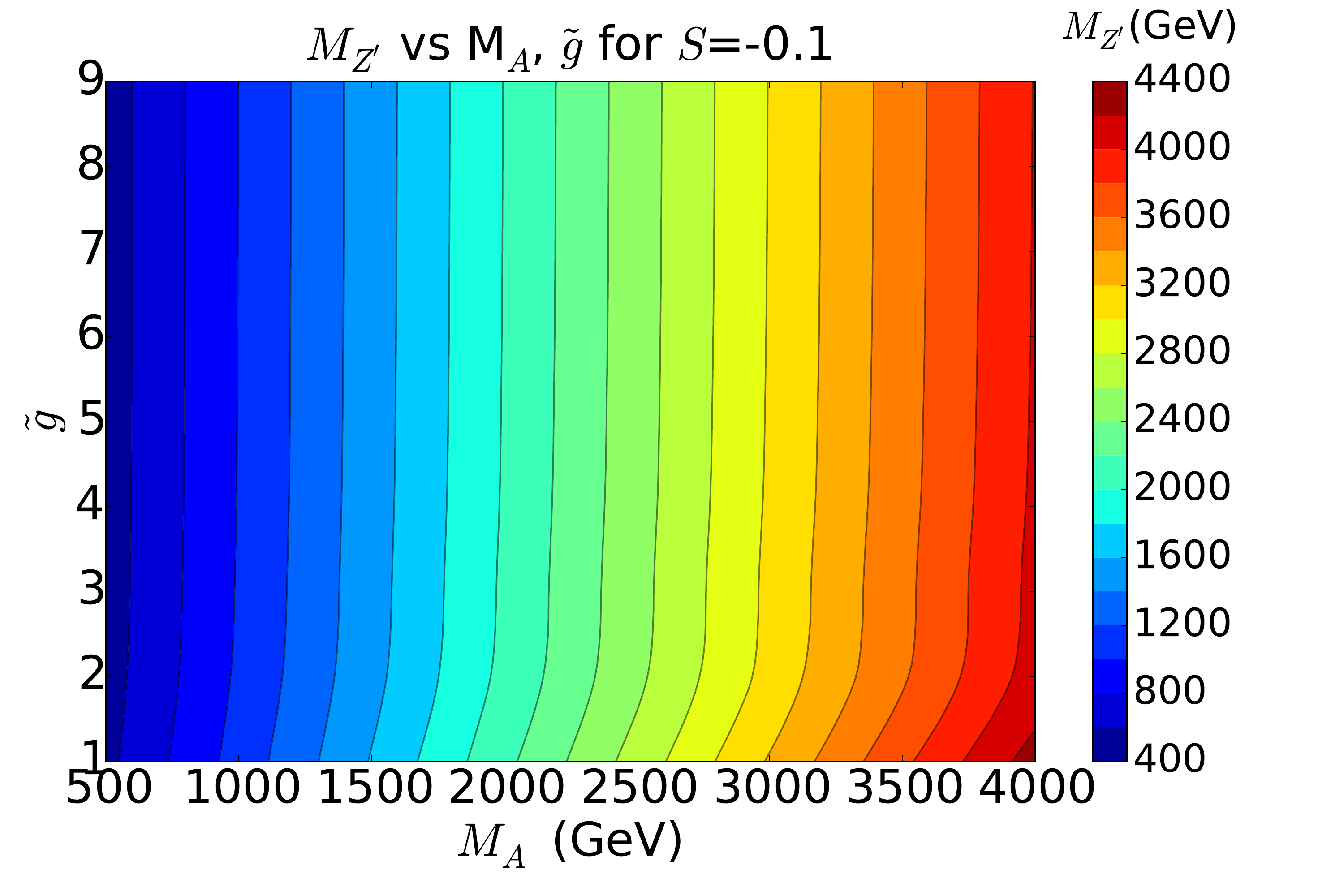}}%
\subfigure[]{\includegraphics[type=pdf,ext=.pdf,read=.pdf,width=0.5\textwidth]{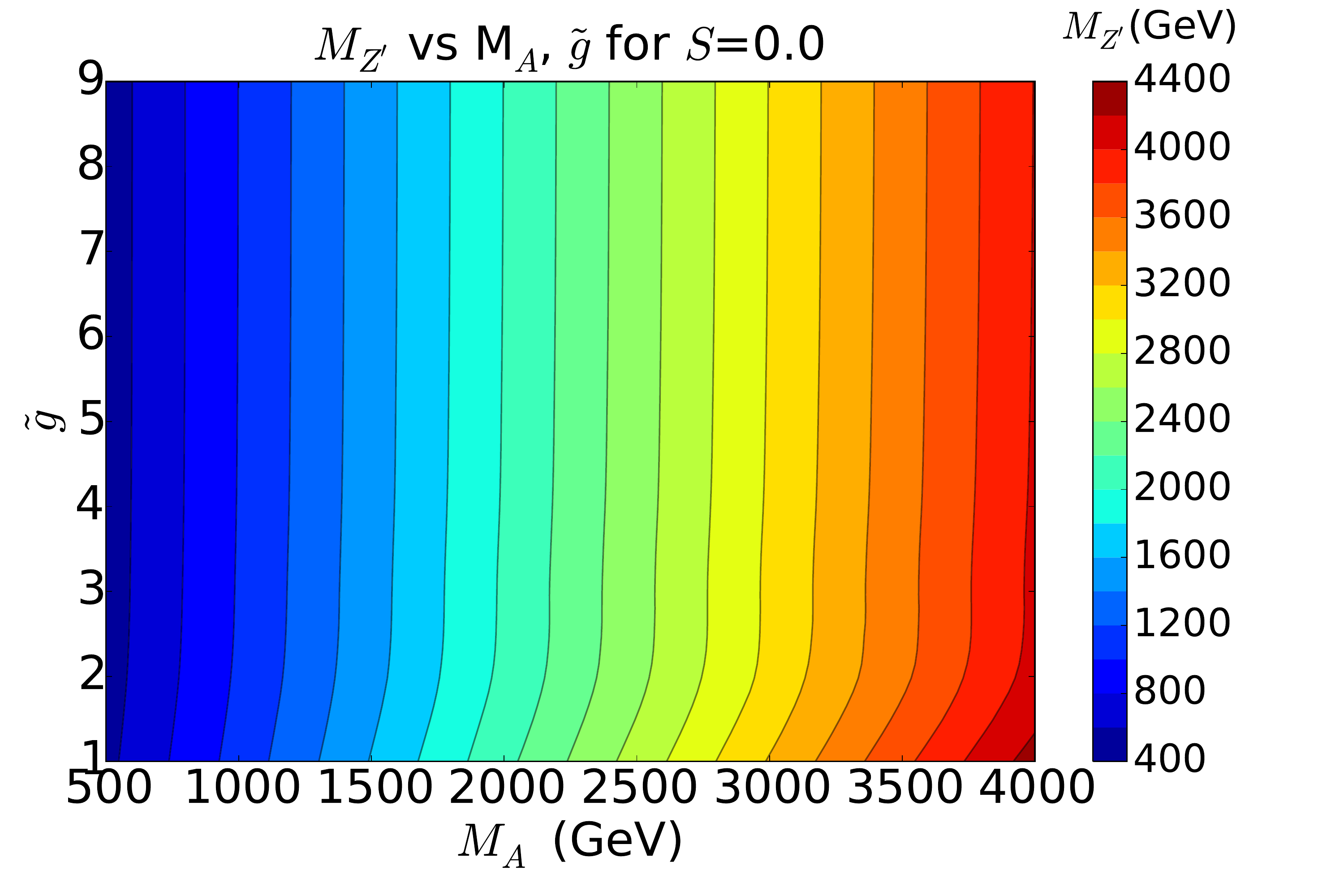}}\\
\subfigure[]{\includegraphics[type=pdf,ext=.pdf,read=.pdf,width=0.5\textwidth]{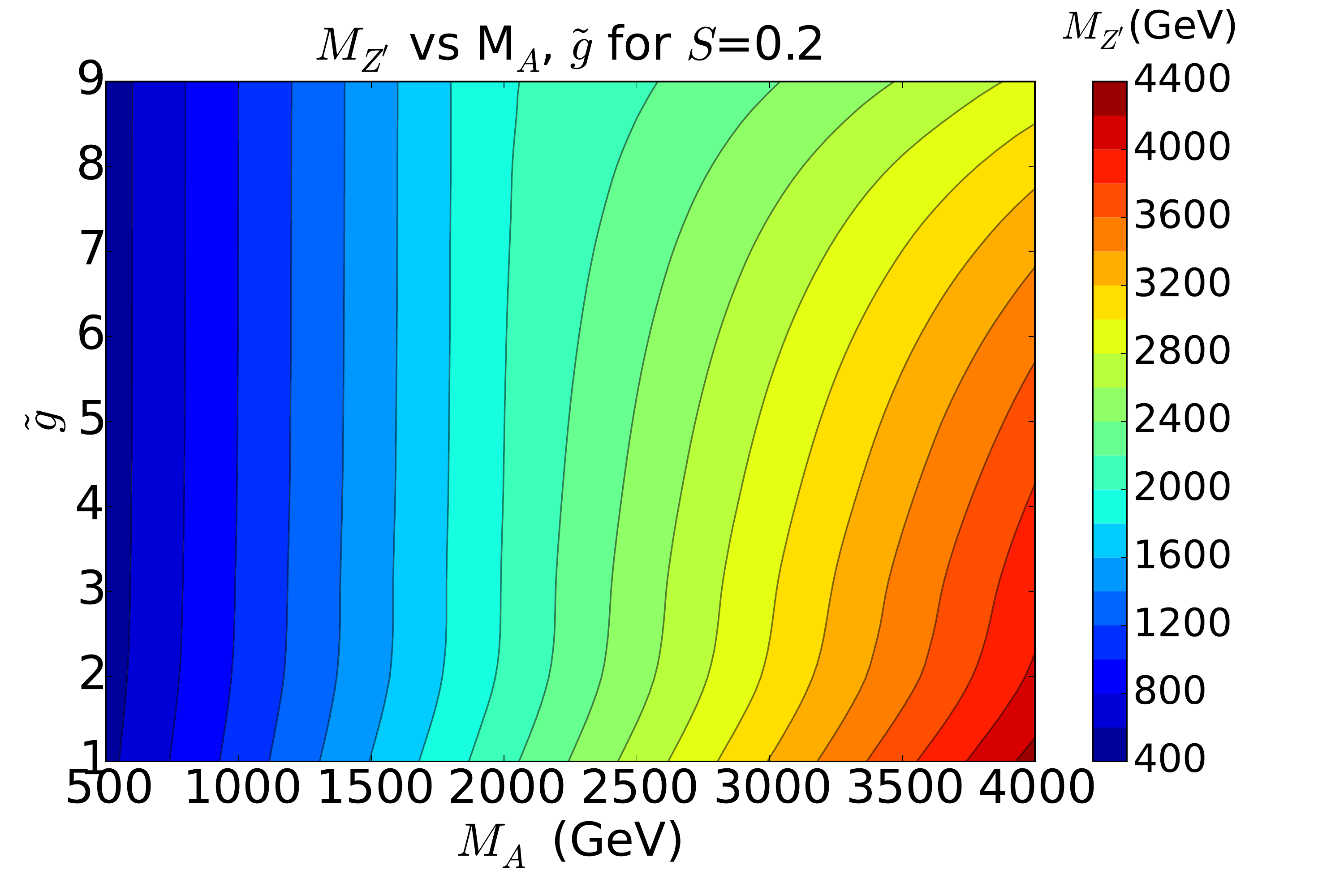}}%
\subfigure[]{\includegraphics[type=pdf,ext=.pdf,read=.pdf,width=0.5\textwidth]{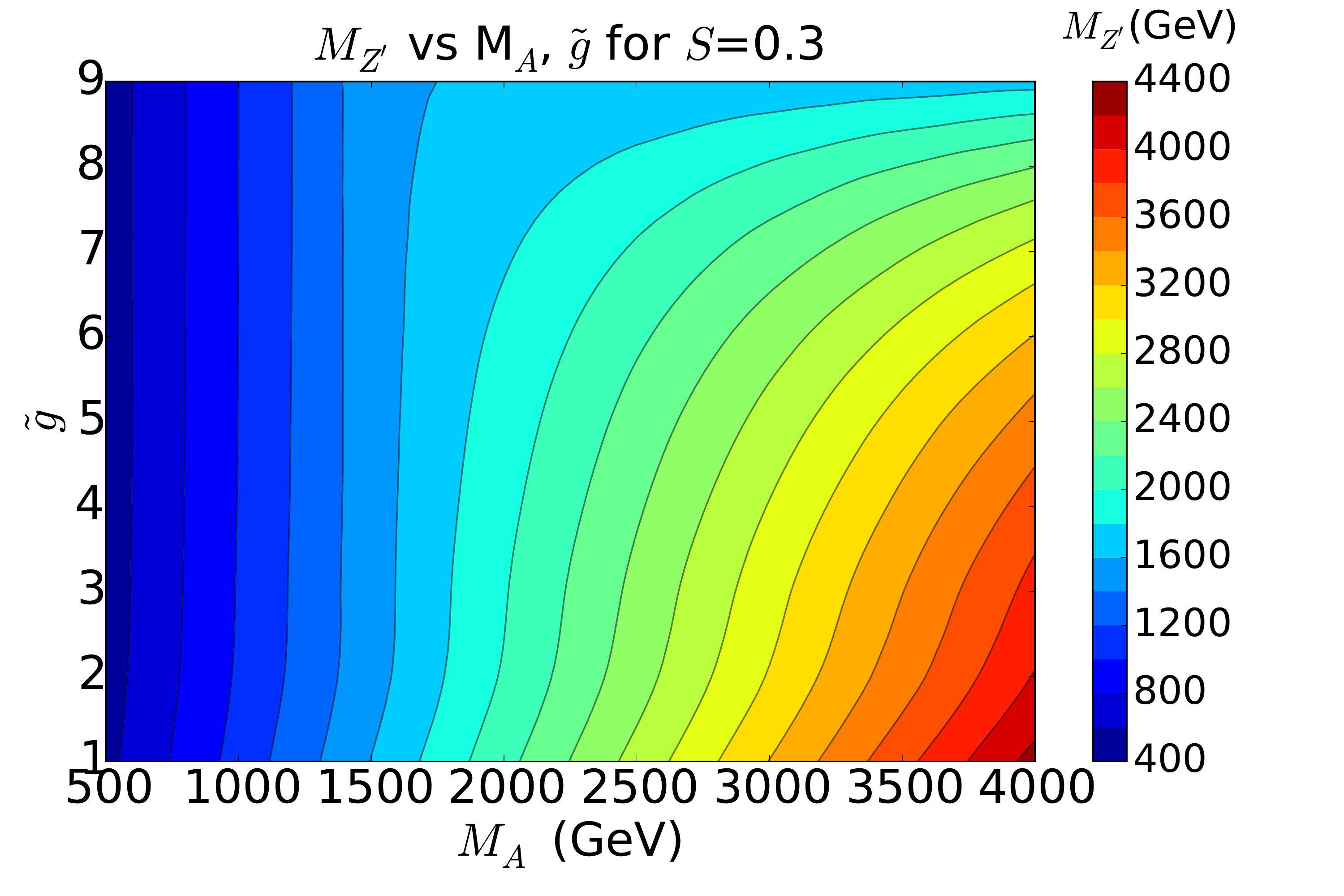}}
\caption{\label{fig:mzp-mass-with-s} $M_{Z^{\prime}}$(GeV) as a function of $M_A$ and $\tilde{g}$ parameters for the fixed values of  
$S=-0.1$ (a), $S=0.0$ (b), $S=0.2$ (c), $S=0.3$ (d) respectively, and $s=0$ throughout}
\end{figure}

\begin{figure}[htb]
\subfigure[]{\includegraphics[type=pdf,ext=.pdf,read=.pdf,width=0.5\textwidth]{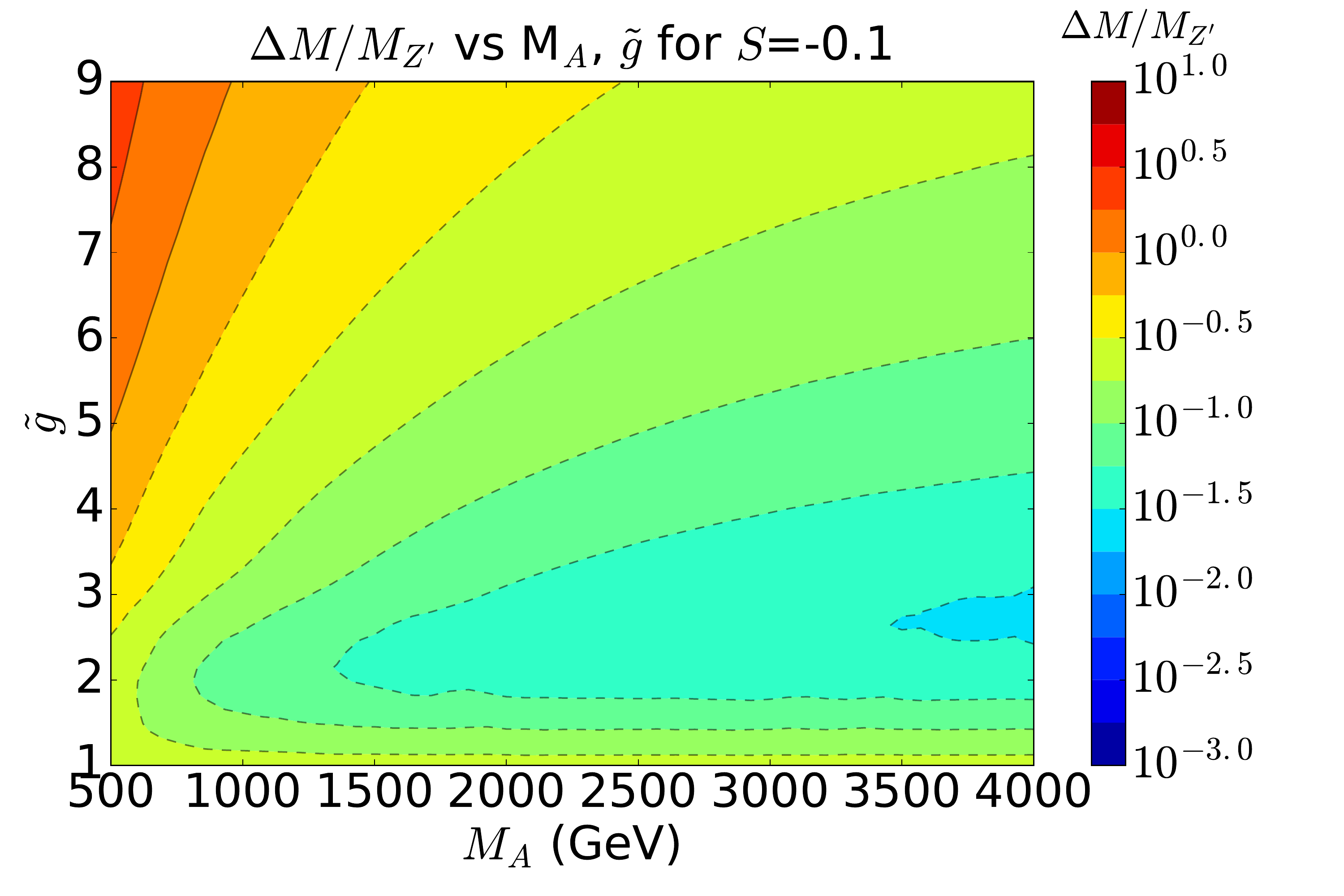}}%
\subfigure[]{\includegraphics[type=pdf,ext=.pdf,read=.pdf,width=0.5\textwidth]{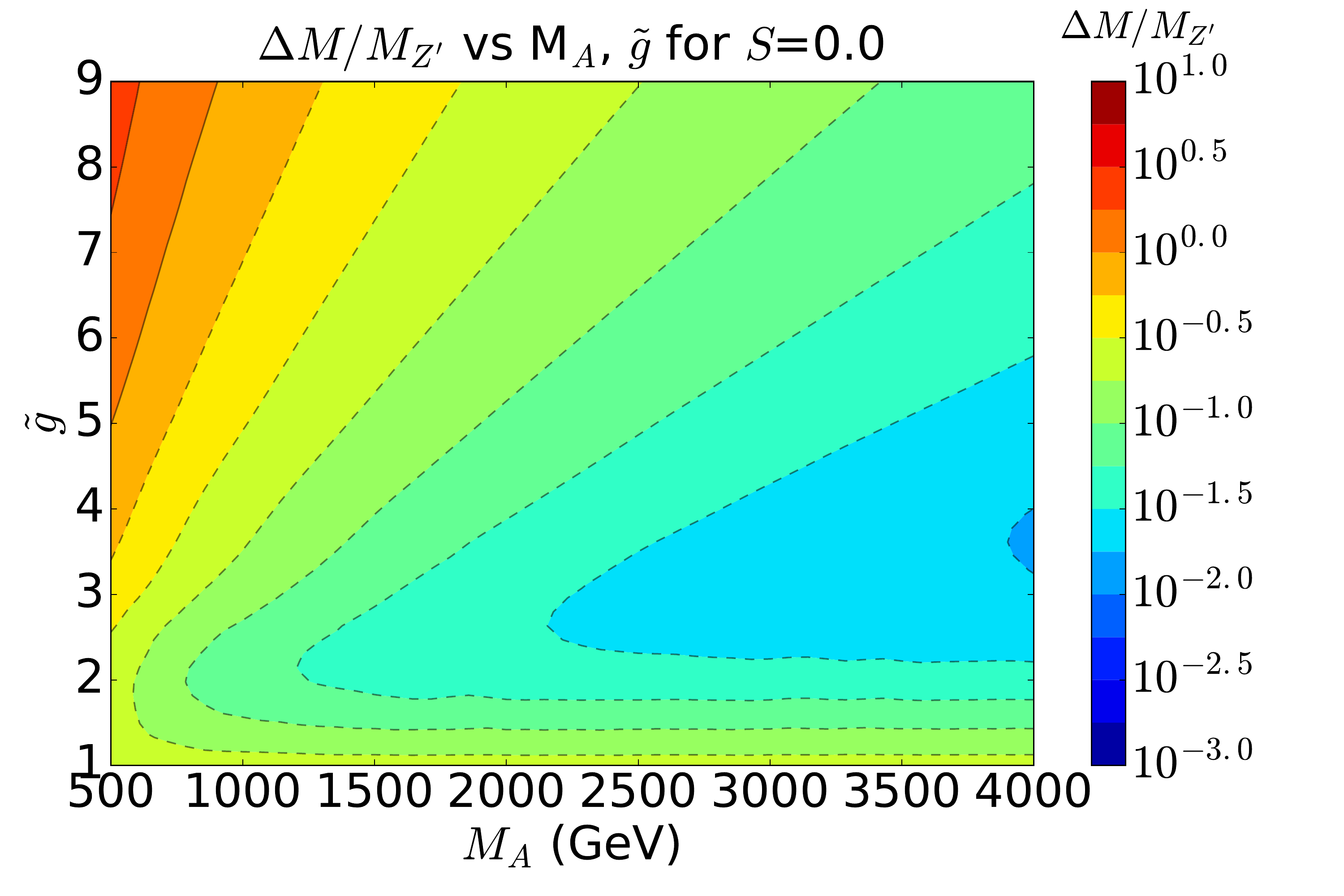}}\\
\subfigure[]{\includegraphics[type=pdf,ext=.pdf,read=.pdf,width=0.5\textwidth]{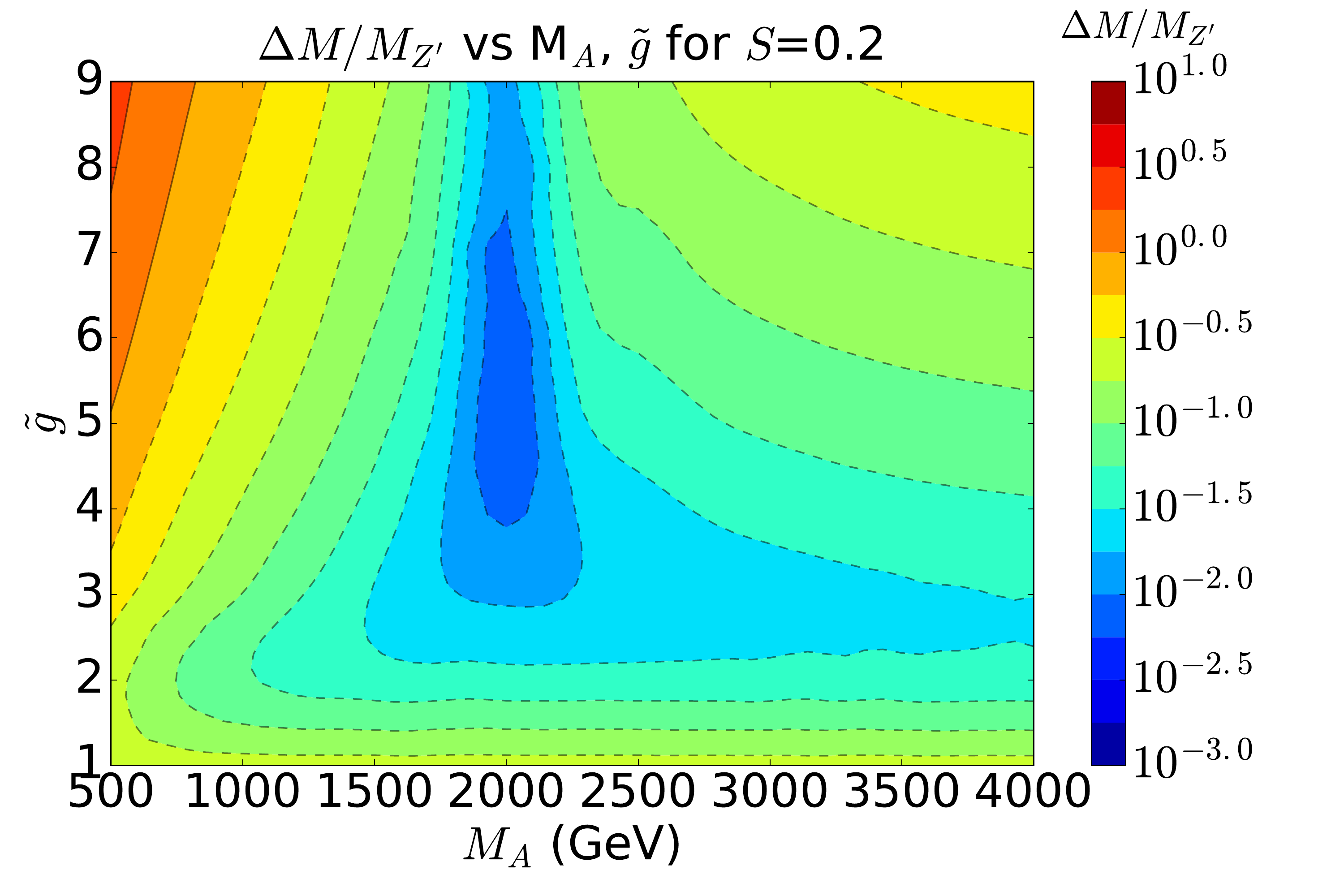}}%
\subfigure[]{\includegraphics[type=pdf,ext=.pdf,read=.pdf,width=0.5\textwidth]{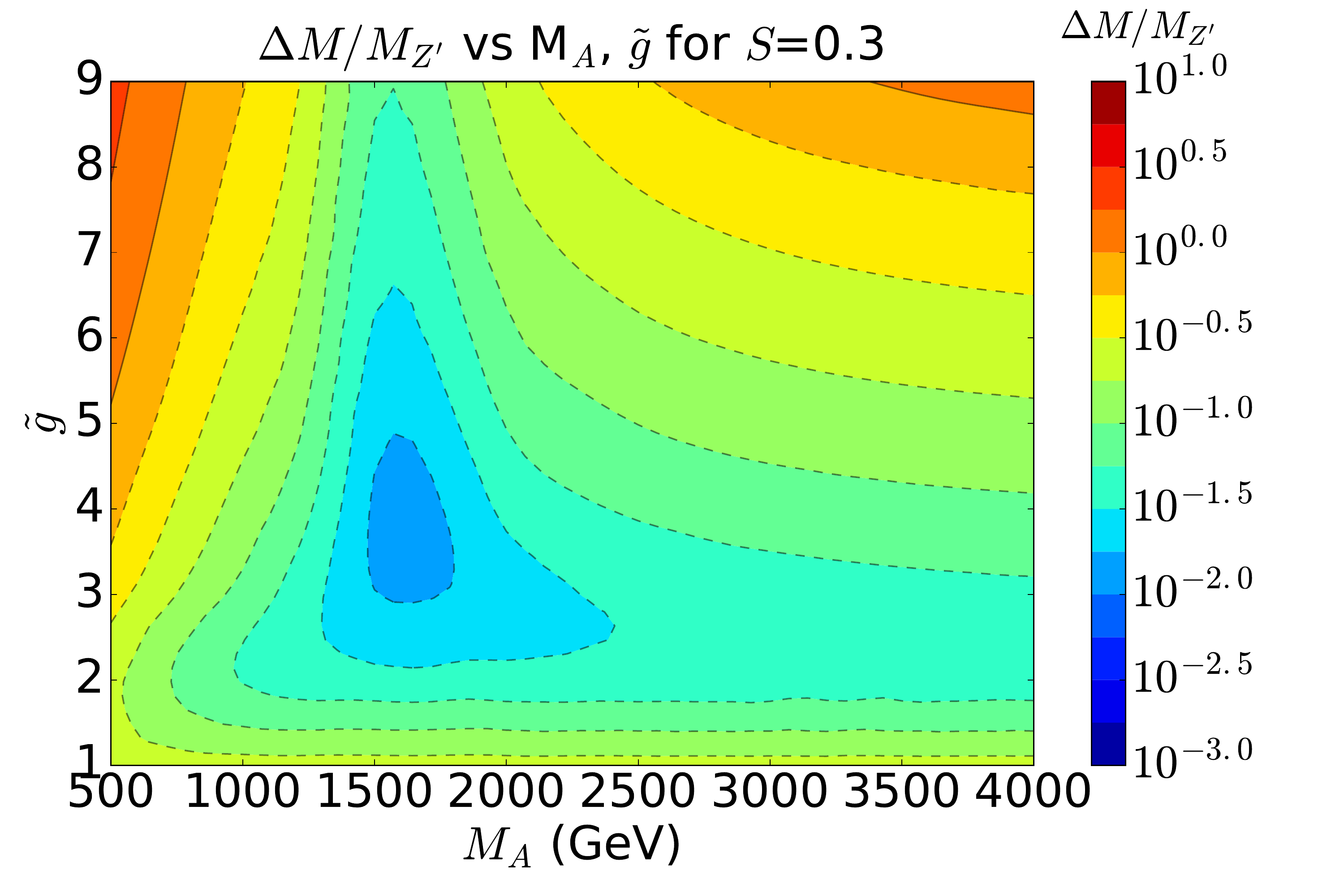}}
\caption{\label{fig:mass-ratio-rel-with-s} $\Delta M/M_{Z^{\prime}}$ as a function of $M_A$ and $\tilde{g}$ parameters for the fixed values of  
$S=-0.1$ (a), $S=0.0$ (b), $S=0.2$ (c), $S=0.3$ (d) respectively, and $s=0$ throughout}
\end{figure}

Figures~\ref{fig:mzp-mass-with-s} and \ref{fig:mass-ratio-rel-with-s} present $M_{Z^{\prime}}$ and $\Delta M/M_{Z^{\prime}}$ respectively for different values of $S$. The main feature to note is the mass inversion $M_{inv}$ defined by Eq.(\ref{eqn:mass-inversion}) such that
$M_{inv}^2\propto 1/S$.
The inversion point with $\Delta \simeq 0$ can be seen in Fig~\ref{fig:mass-ratio-rel-with-s} 
where the $Z^{\prime}$ is axial-vector below the inversion point 
and vector above it. One can observe the inversion only for large values of 
$S=0.2$ and $0.3$ for the  $M_A$ around 2 and 1.6 TeV respectively according to the  Eq.(\ref{eqn:mass-inversion}).

\clearpage

\subsubsection{Couplings}

\begin{figure}[htb]
\subfigure[]{\includegraphics[type=pdf,ext=.pdf,read=.pdf,width=0.5\textwidth]{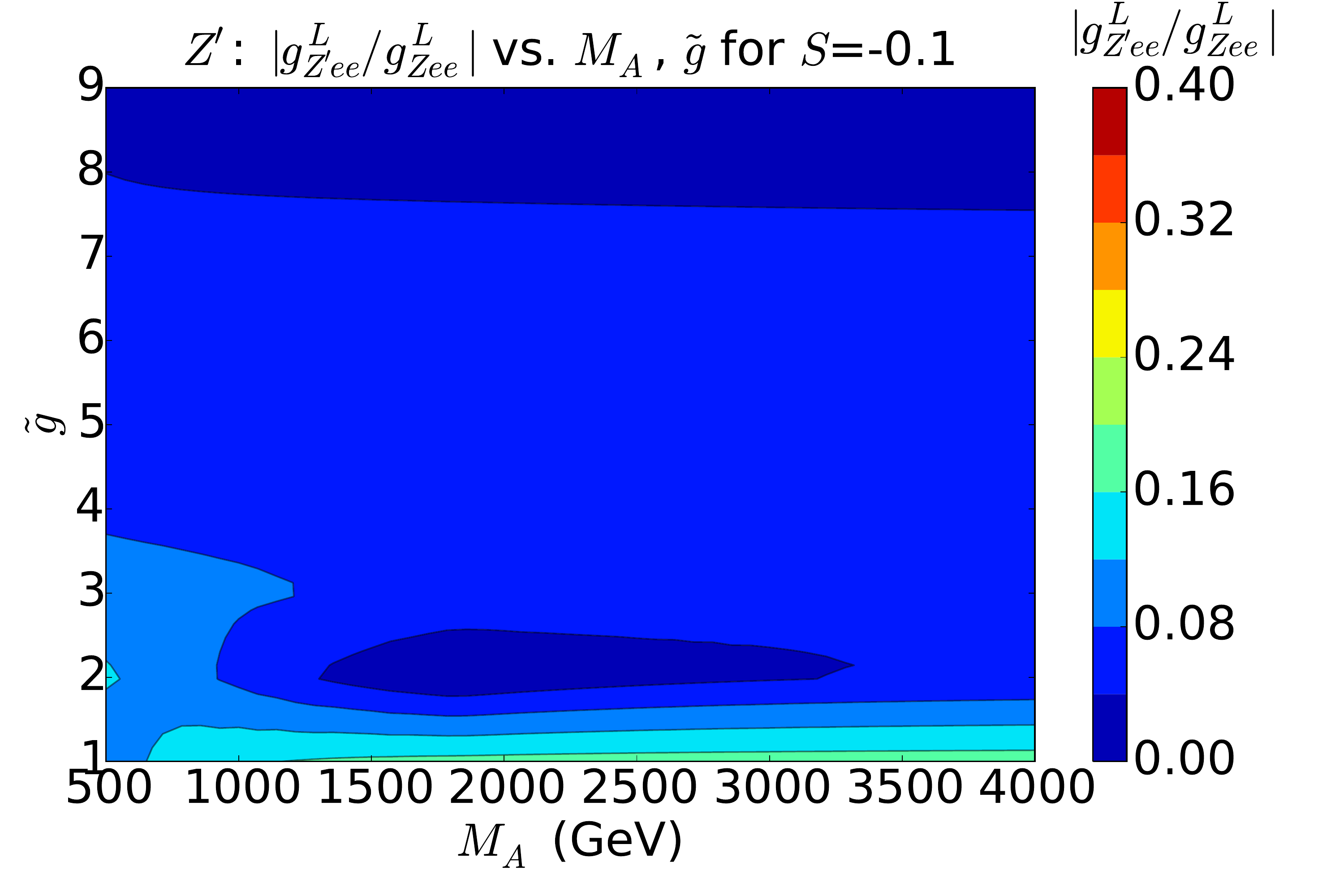}}%
\subfigure[]{\includegraphics[type=pdf,ext=.pdf,read=.pdf,width=0.5\textwidth]{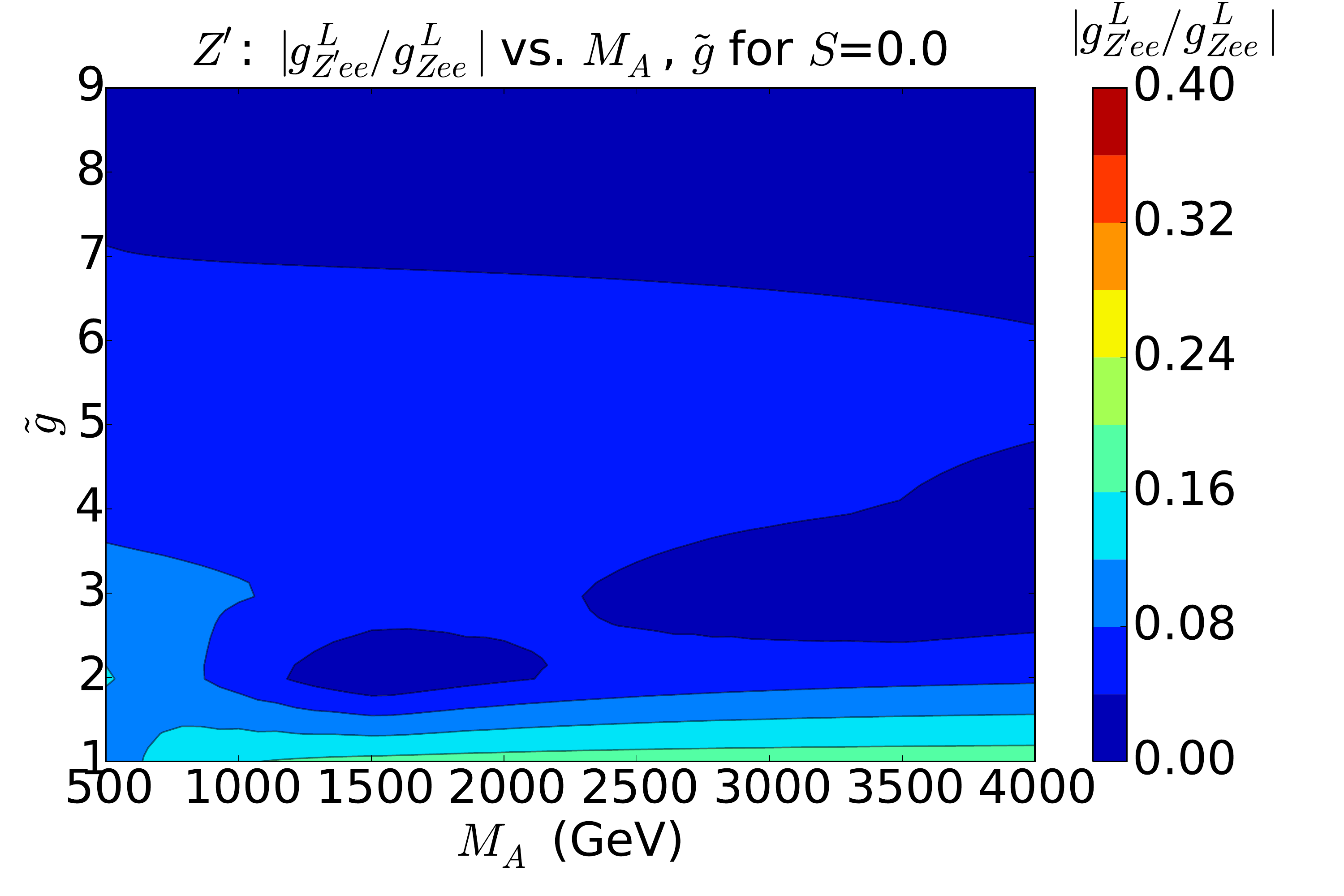}}\\
\subfigure[]{\includegraphics[type=pdf,ext=.pdf,read=.pdf,width=0.5\textwidth]{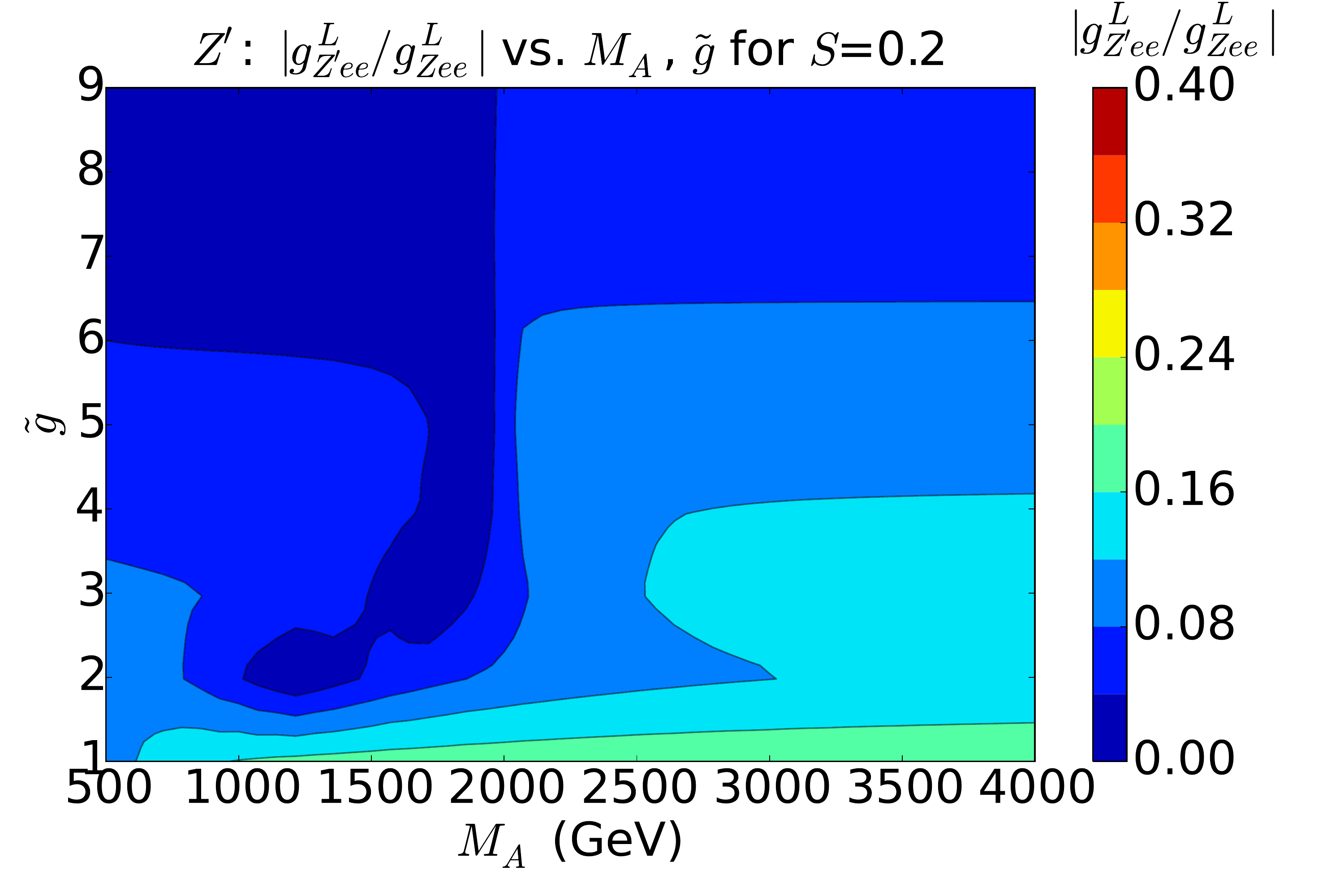}}%
\subfigure[]{\includegraphics[type=pdf,ext=.pdf,read=.pdf,width=0.5\textwidth]{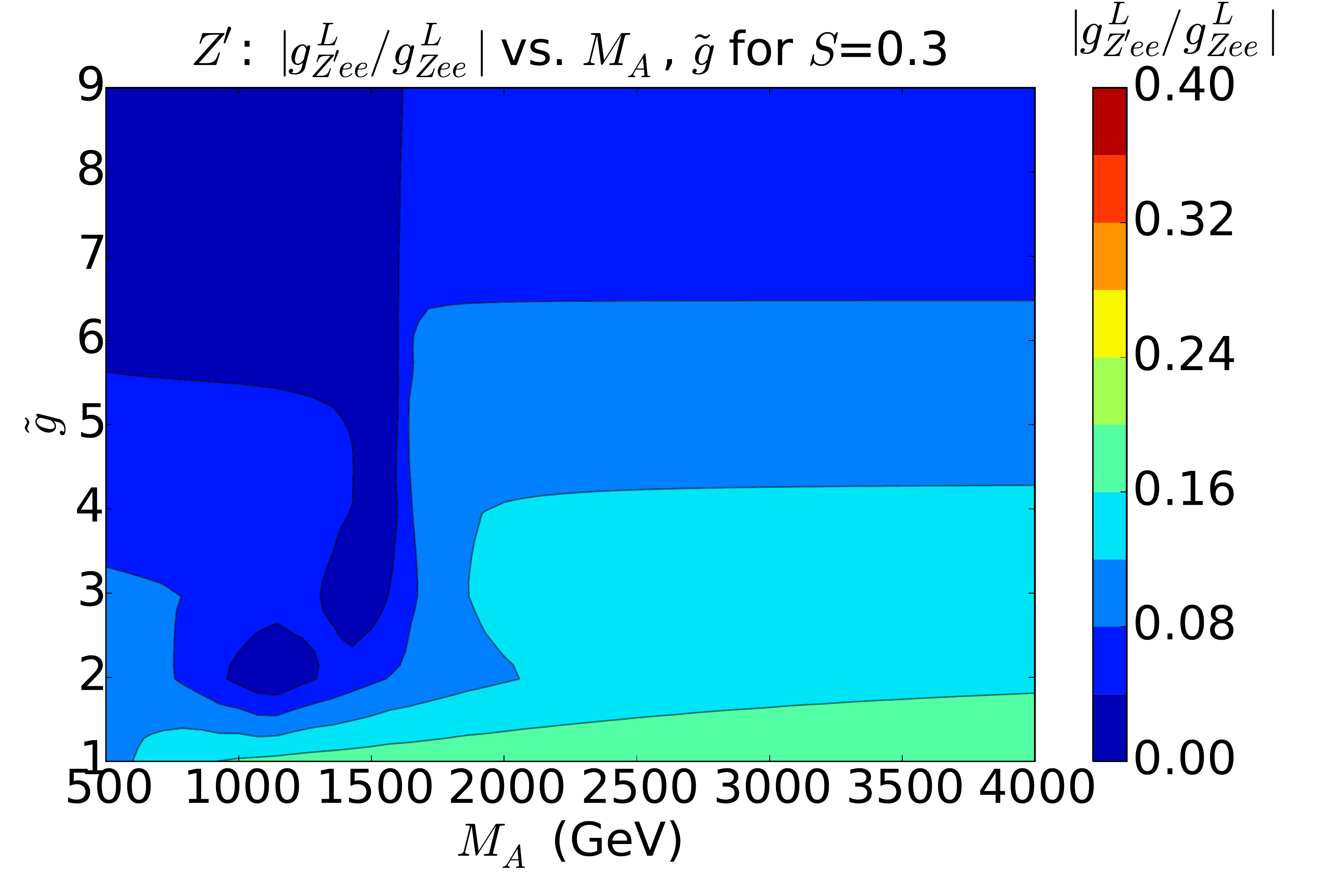}}
\caption{\label{fig:zp-LHcoupling-with-s} Left handed component of the coupling of $Z^{\prime}$ to charged lepton pairs as a ratio to its SM equivalent, $\mid g_{Z^{\prime}l^{+}l^{-}}/g_{Zl^{+}l^{-}} \mid$, as a function of $M_A$ and $\tilde{g}$ parameters for the fixed values of  
$S=-0.1$ (a), $S=0.0$ (b), $S=0.2$ (c), $S=0.3$ (d) respectively}
\end{figure}

\begin{figure}[htb]
\subfigure[]{\includegraphics[type=pdf,ext=.pdf,read=.pdf,width=0.5\textwidth]{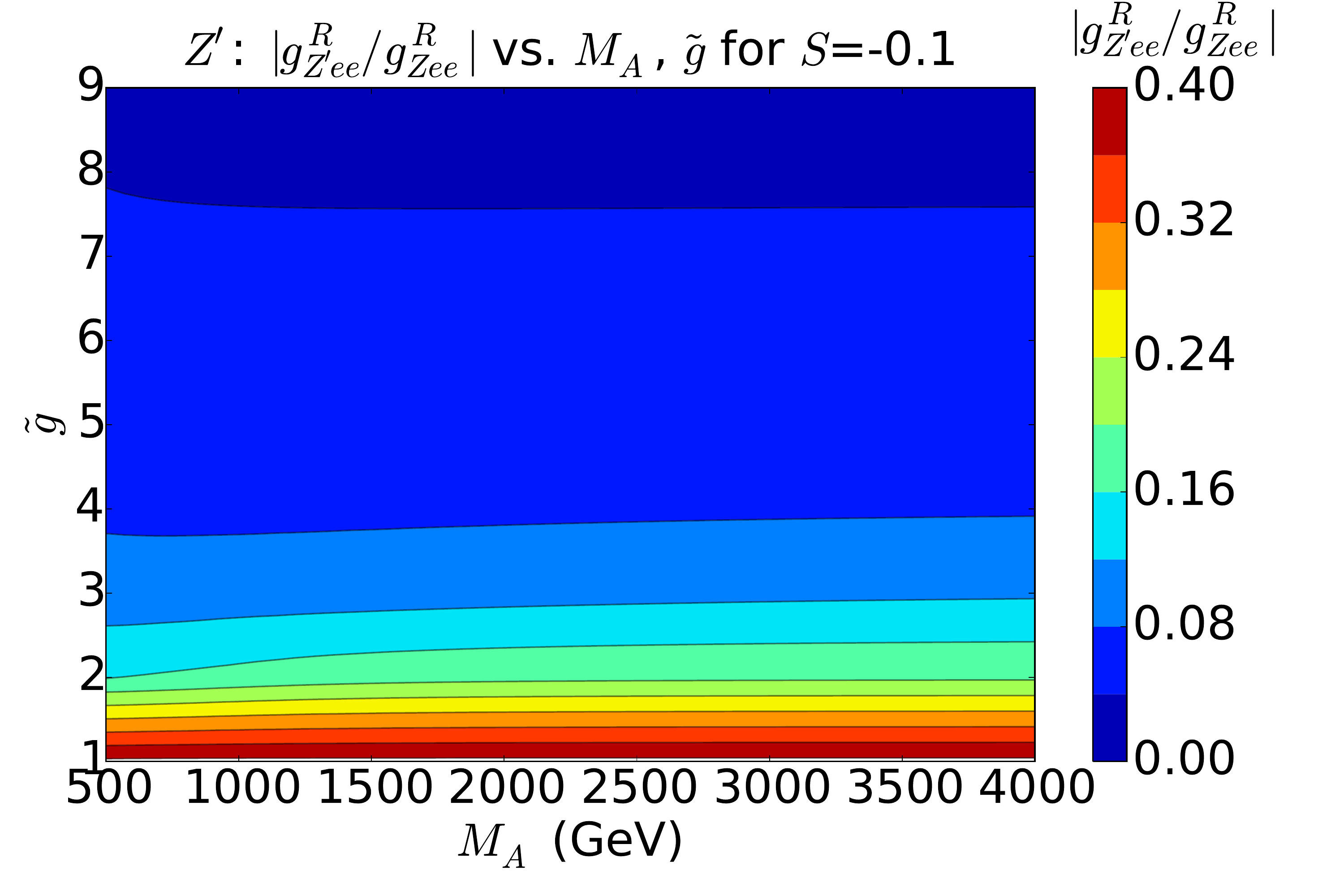}}%
\subfigure[]{\includegraphics[type=pdf,ext=.pdf,read=.pdf,width=0.5\textwidth]{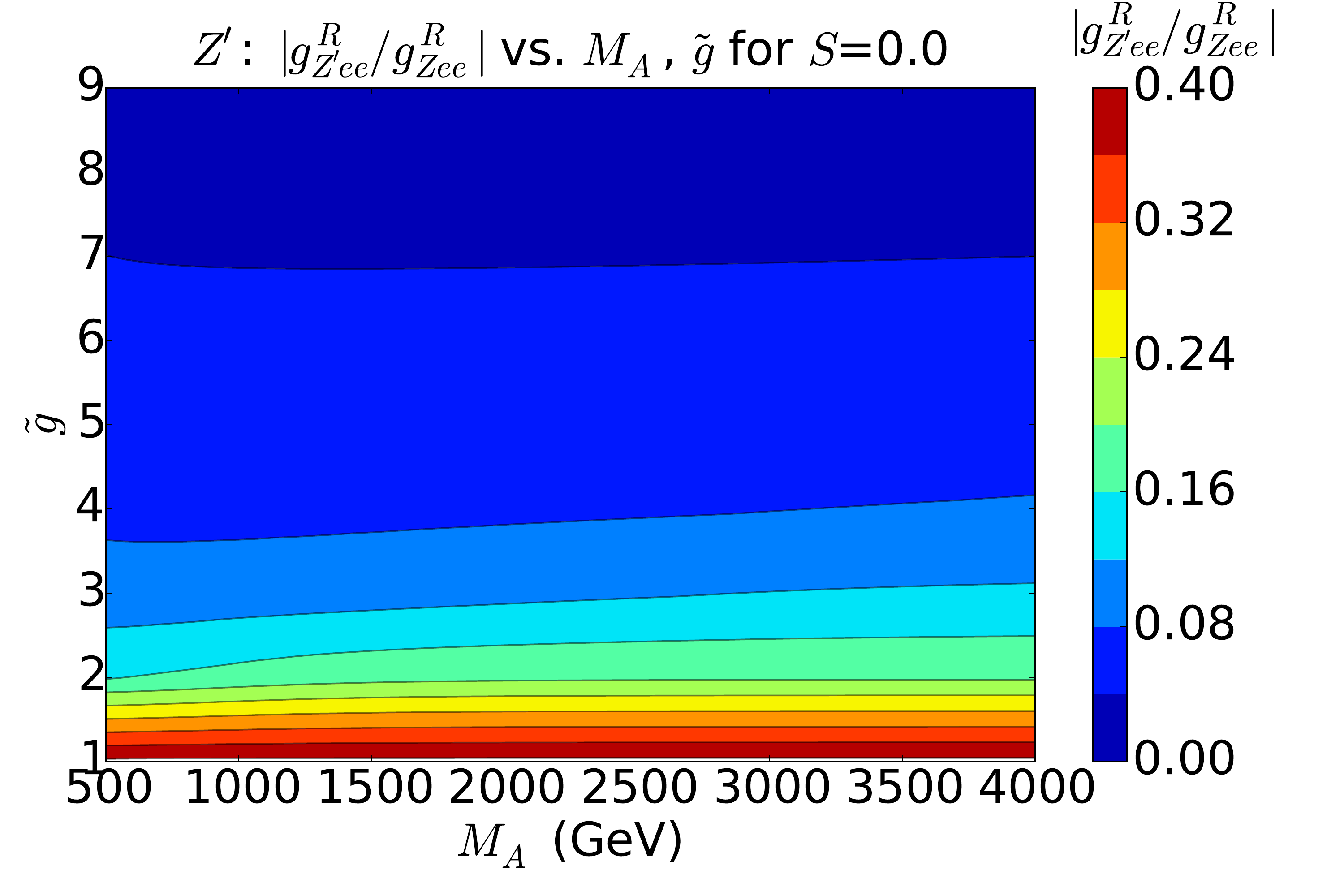}}\\
\subfigure[]{\includegraphics[type=pdf,ext=.pdf,read=.pdf,width=0.5\textwidth]{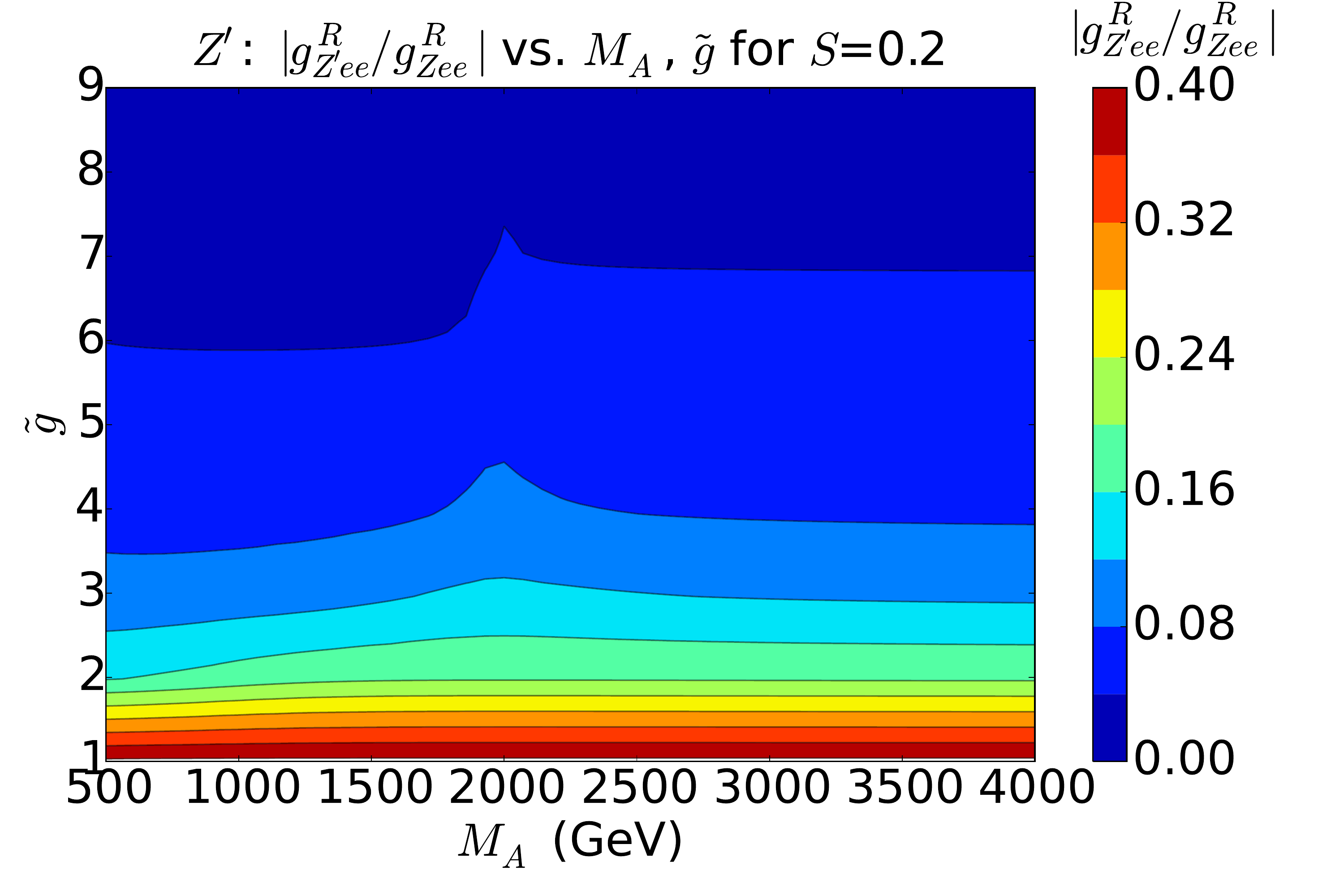}}%
\subfigure[]{\includegraphics[type=pdf,ext=.pdf,read=.pdf,width=0.5\textwidth]{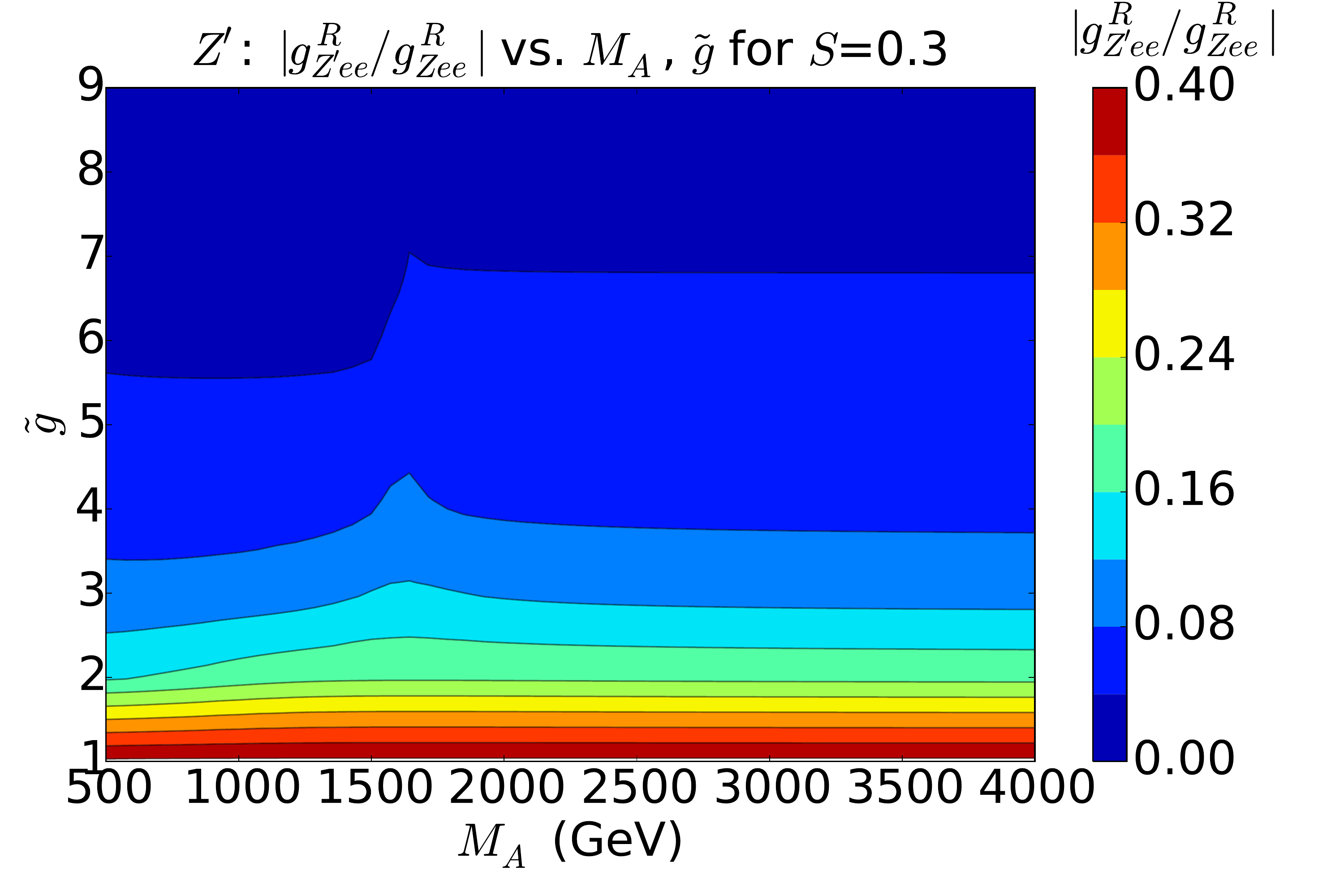}}
\caption{\label{fig:zp-RHcoupling-with-s} Right handed component of the coupling of $Z^{\prime}$ to charged lepton pairs as a ratio to its SM equivalent, $\mid g_{Z^{\prime}l^{+}l^{-}}/g_{Zl^{+}l^{-}} \mid$, as a function of $M_A$ and $\tilde{g}$ parameters for the fixed values of  
$S=-0.1$ (a), $S=0.0$ (b), $S=0.2$ (c), $S=0.3$ (d) respectively}
\end{figure}

\begin{figure}[htb]
\subfigure[]{\includegraphics[type=pdf,ext=.pdf,read=.pdf,width=0.5\textwidth]{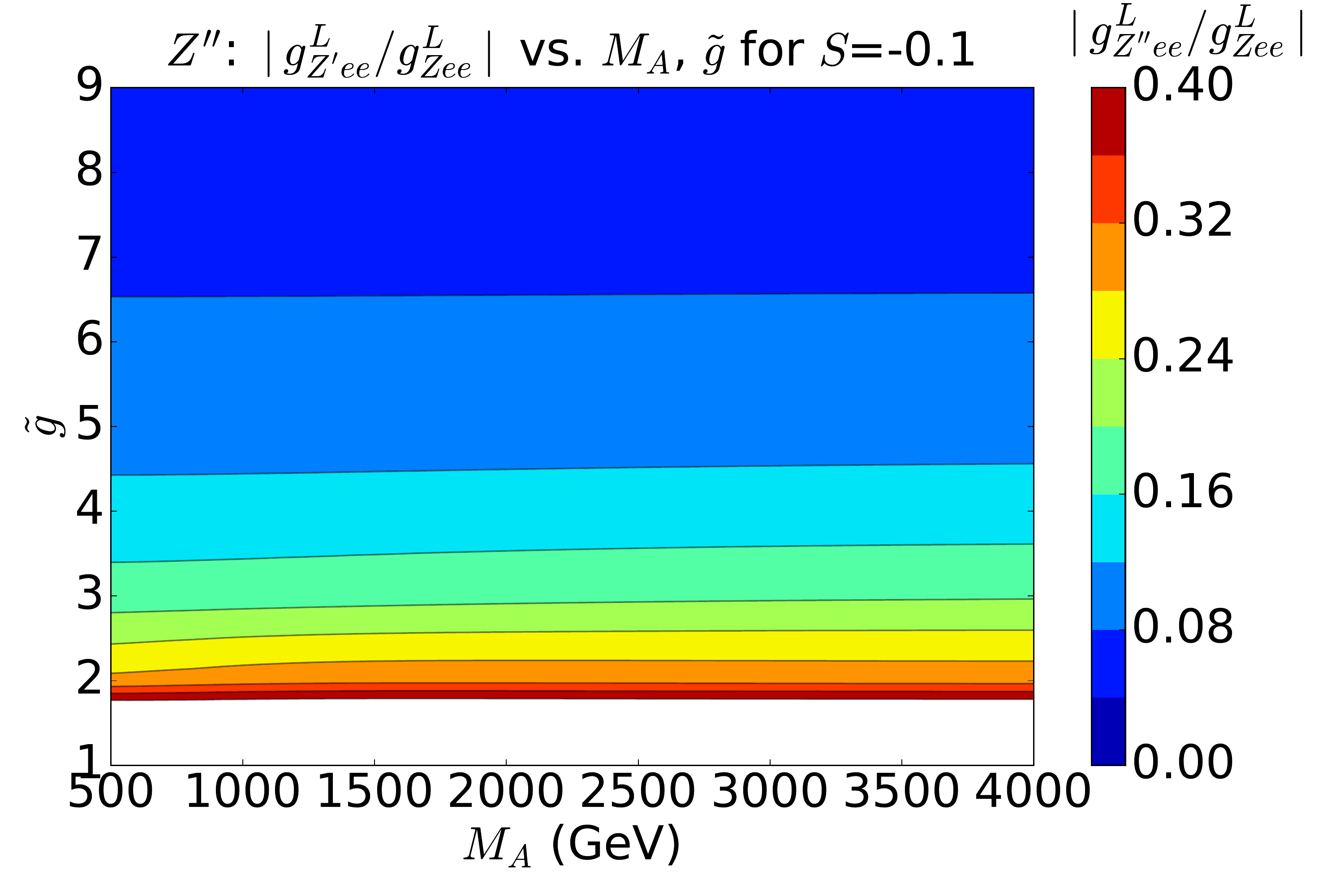}}%
\subfigure[]{\includegraphics[type=pdf,ext=.pdf,read=.pdf,width=0.5\textwidth]{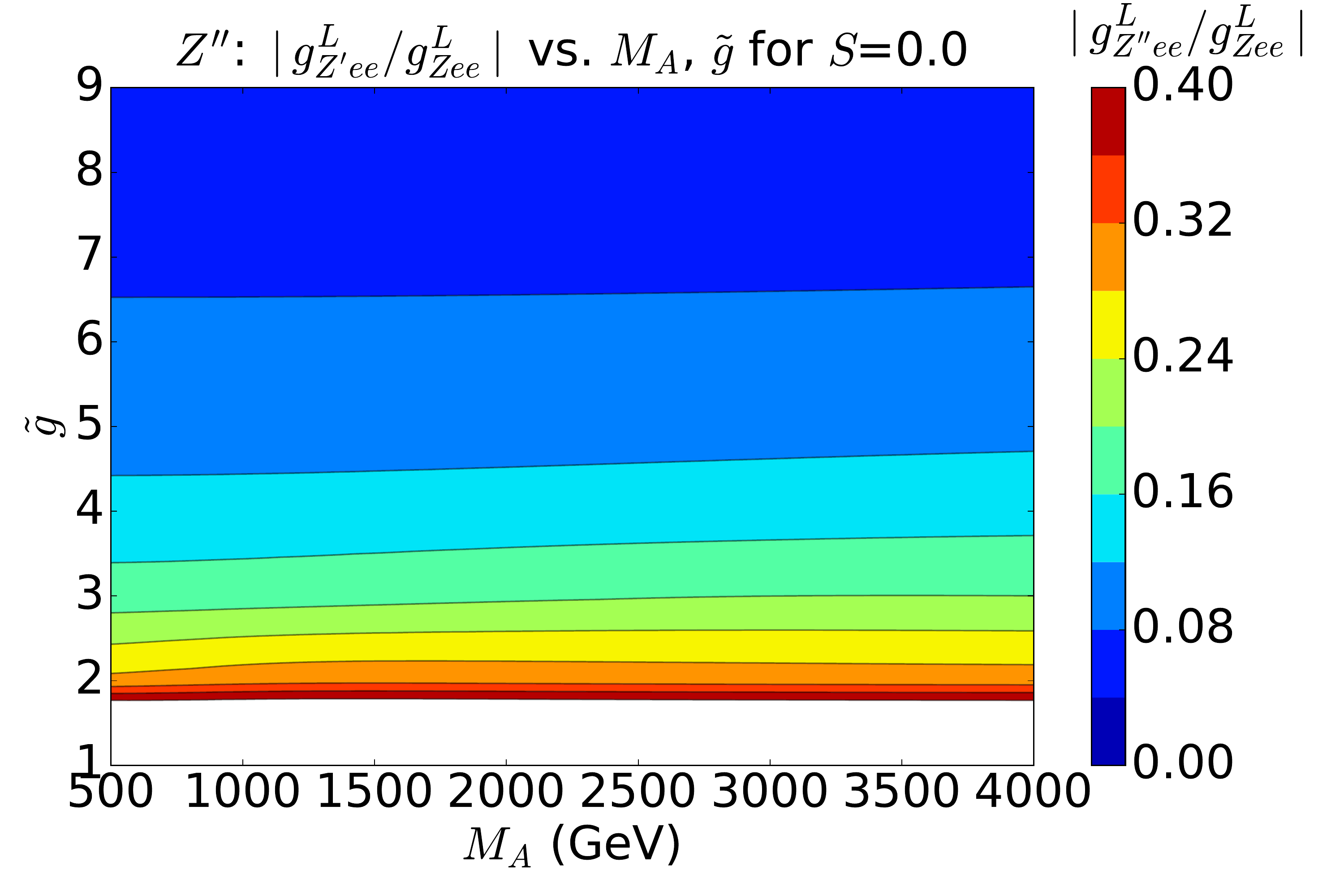}}\\
\subfigure[]{\includegraphics[type=pdf,ext=.pdf,read=.pdf,width=0.5\textwidth]{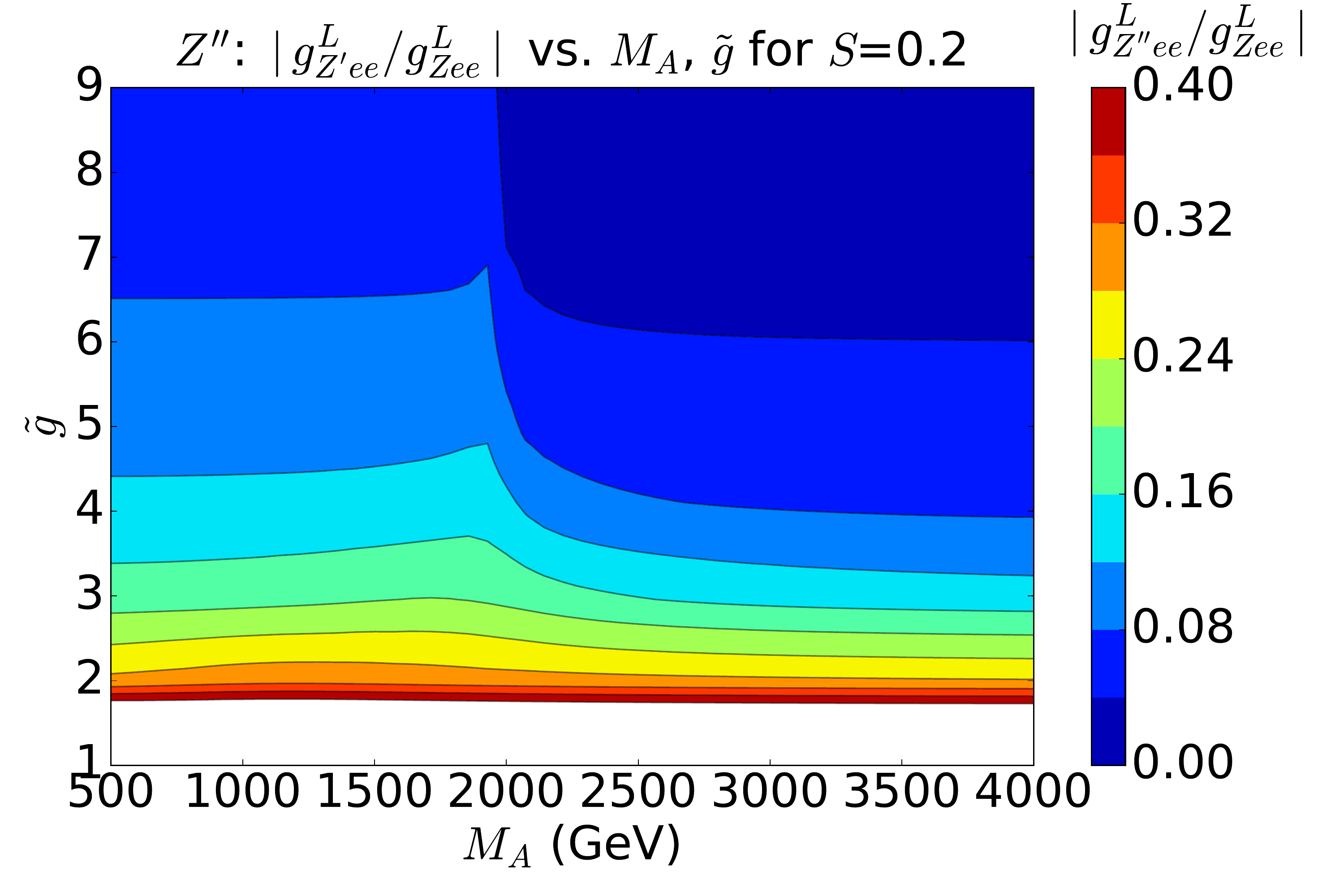}}%
\subfigure[]{\includegraphics[type=pdf,ext=.pdf,read=.pdf,width=0.5\textwidth]{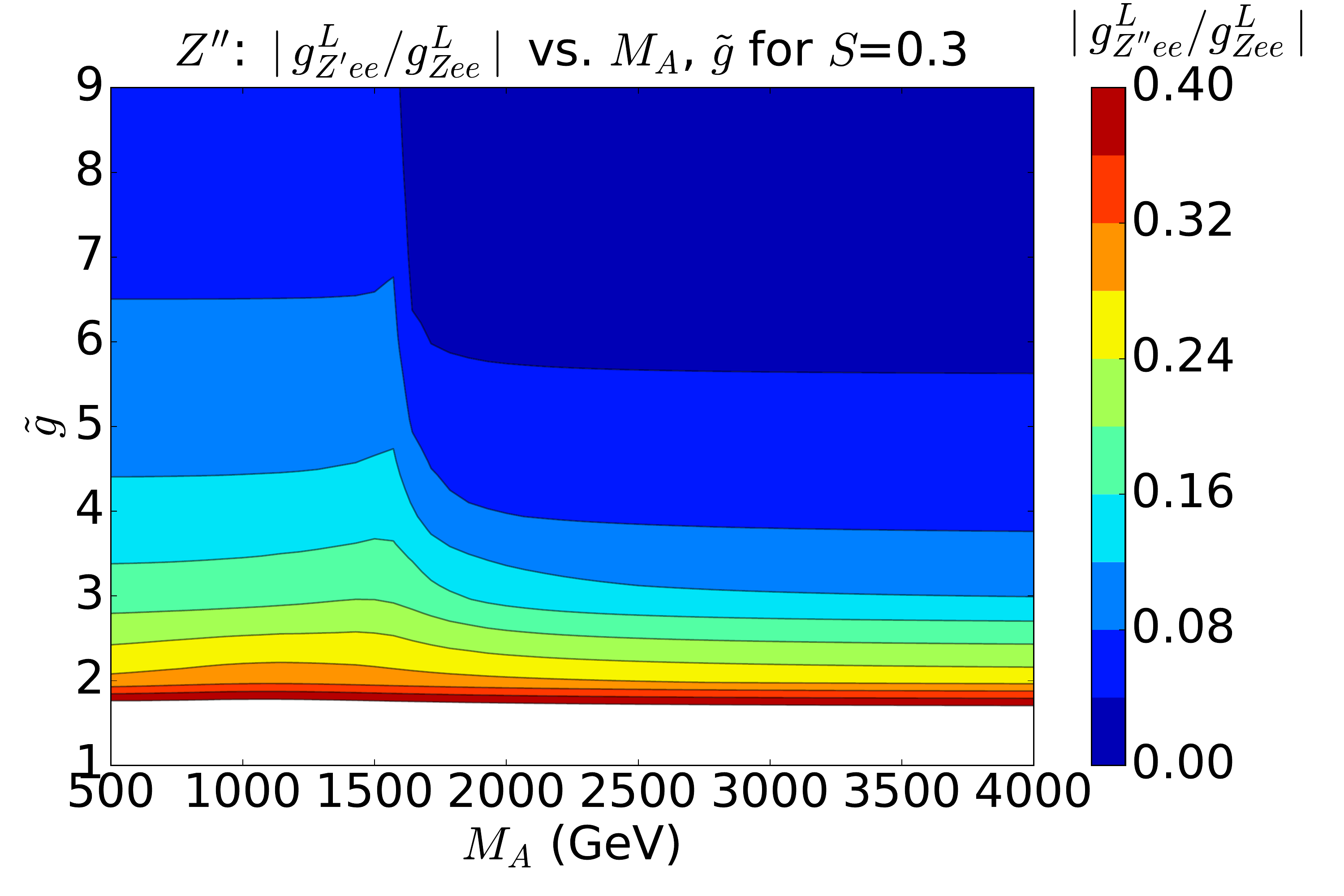}}
\caption{\label{fig:zpp-LHcoupling-with-s} Left handed component of the coupling of $Z^{\prime\prime}$ to charged lepton pairs as a ratio to its SM equivalent, $\mid g_{Z^{\prime\prime}l^{+}l^{-}}/g_{Zl^{+}l^{-}}\mid$, as a function of $M_A$ and $\tilde{g}$ parameters for the fixed values of  
$S=-0.1$ (a), $S=0.0$ (b), $S=0.2$ (c), $S=0.3$ (d) respectively}
\end{figure}

\begin{figure}[htb]
\subfigure[]{\includegraphics[type=pdf,ext=.pdf,read=.pdf,width=0.5\textwidth]{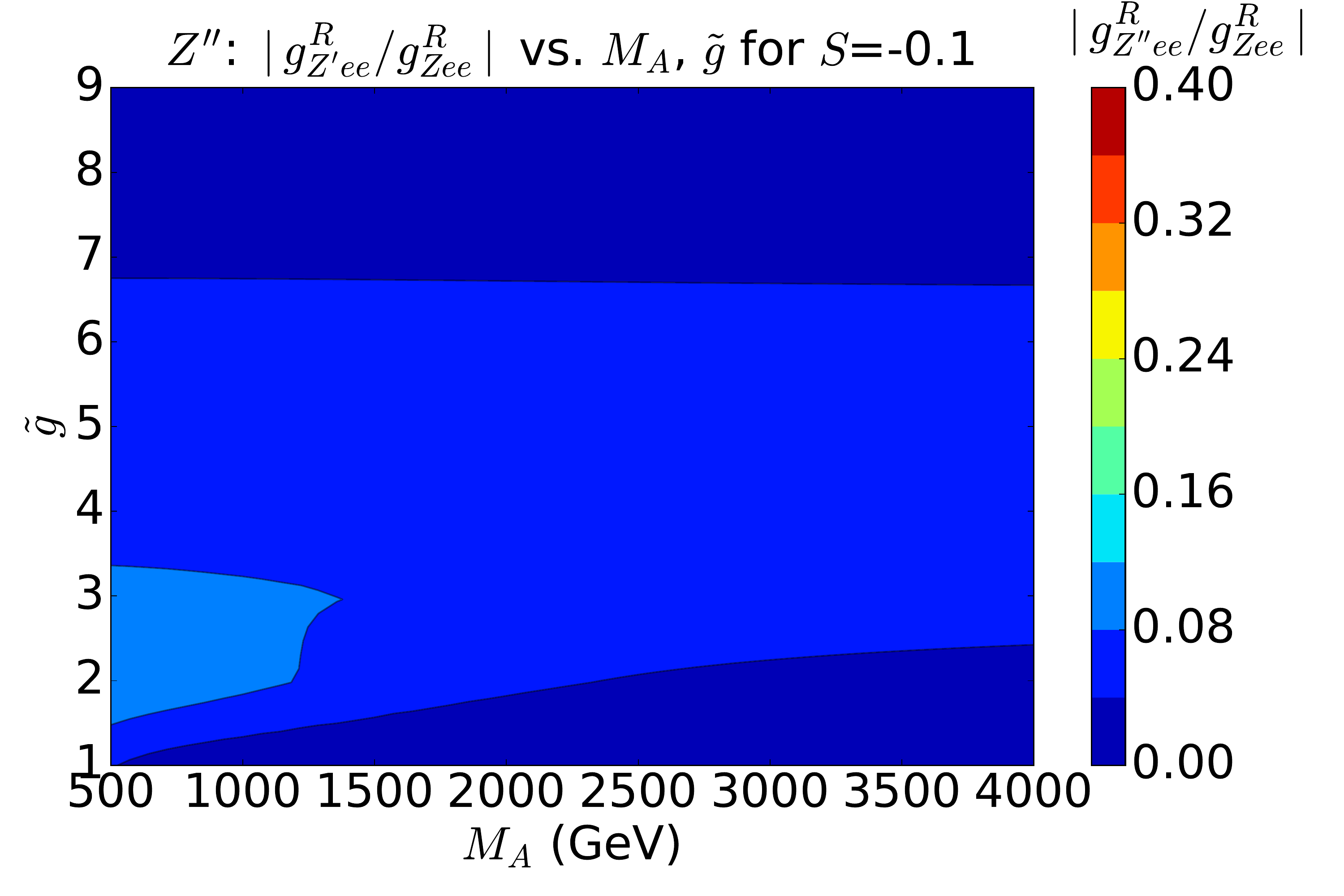}}%
\subfigure[]{\includegraphics[type=pdf,ext=.pdf,read=.pdf,width=0.5\textwidth]{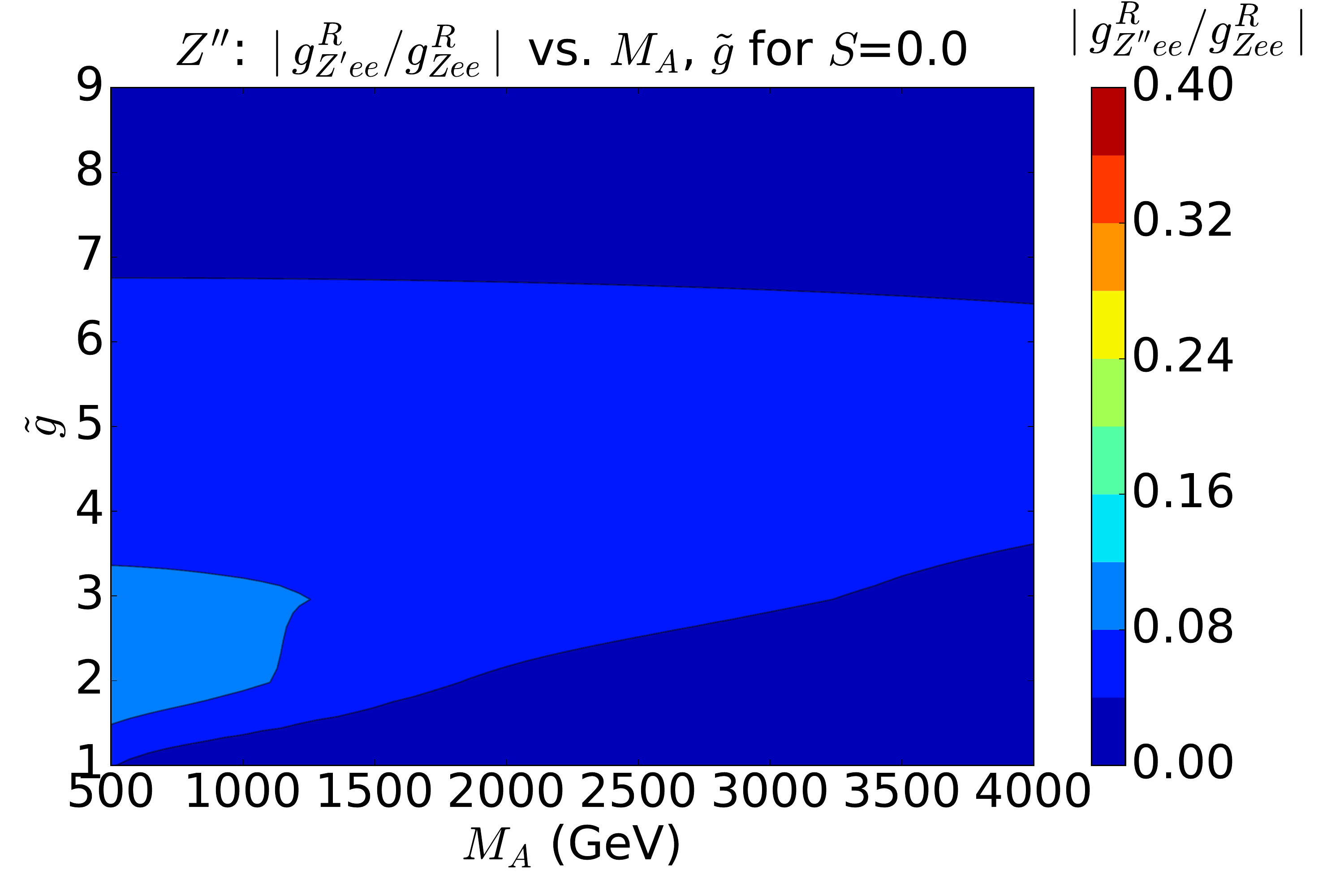}}\\
\subfigure[]{\includegraphics[type=pdf,ext=.pdf,read=.pdf,width=0.5\textwidth]{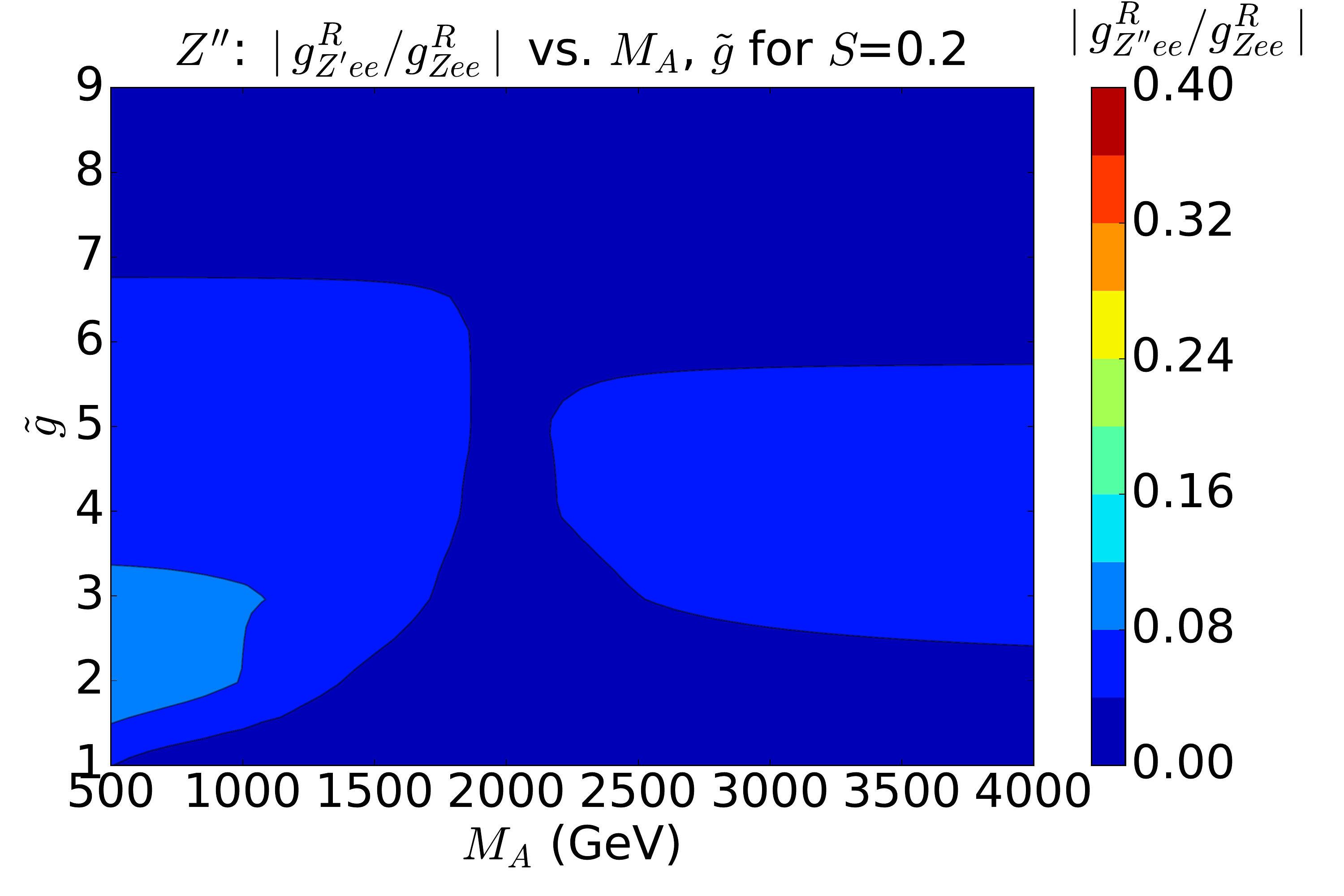}}%
\subfigure[]{\includegraphics[type=pdf,ext=.pdf,read=.pdf,width=0.5\textwidth]{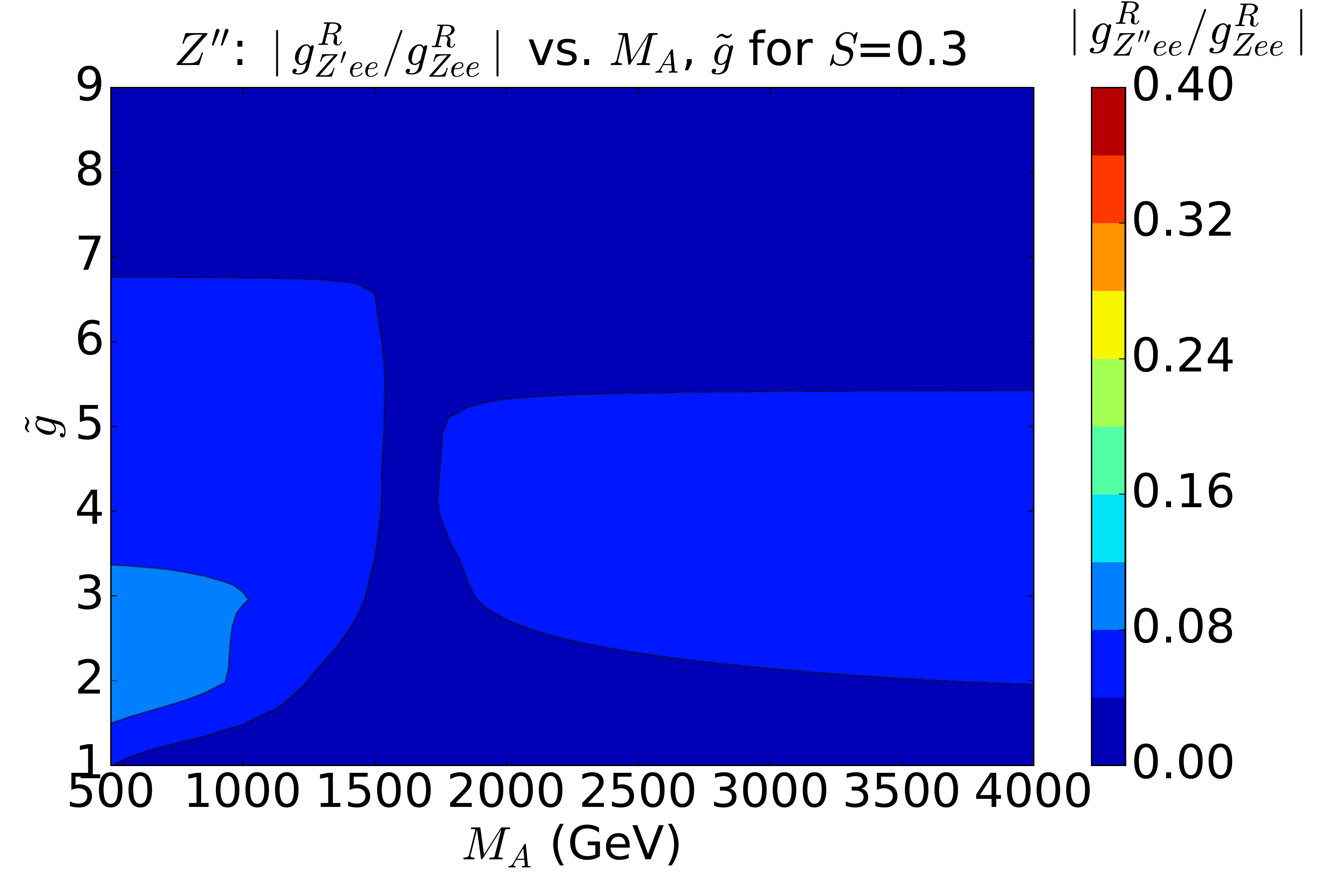}}
\caption{\label{fig:zpp-RHcoupling-with-s} Right handed component of the coupling of $Z^{\prime\prime}$ to charged lepton pairs as a ratio to its SM equivalent, $\mid g_{Z^{\prime\prime}l^{+}l^{-}}/g_{Zl^{+}l^{-}}\mid$, as a function of $M_A$ and $\tilde{g}$ parameters for the fixed values of  
$S=-0.1$ (a), $S=0.0$ (b), $S=0.2$ (c), $S=0.3$ (d) respectively}
\end{figure}

In Figures \ref{fig:zp-LHcoupling-with-s}-\ref{fig:zp-RHcoupling-with-s} and
Figures \ref{fig:zpp-LHcoupling-with-s}-\ref{fig:zpp-RHcoupling-with-s}
we present the  L-R components of the dilepton couplings for the $Z^{\prime}$ and $Z^{\prime\prime}$, respectively, for different values of $S$. These are analogous to the couplings presented in section \ref{subsec:couplings}, where the analytic form for the coupling components are also presented. The $S$ dependence of these couplings is implicit in $\chi$, $g_{1}$, and $g_{2}$, and the effect on the parameter space dependence for varying $S$ is presented here.

\clearpage

\subsubsection{Widths and branching ratios}
\label{appendix:br}

\begin{figure}[htb]
\subfigure[]{\includegraphics[type=pdf,ext=.pdf,read=.pdf,width=0.5\textwidth]{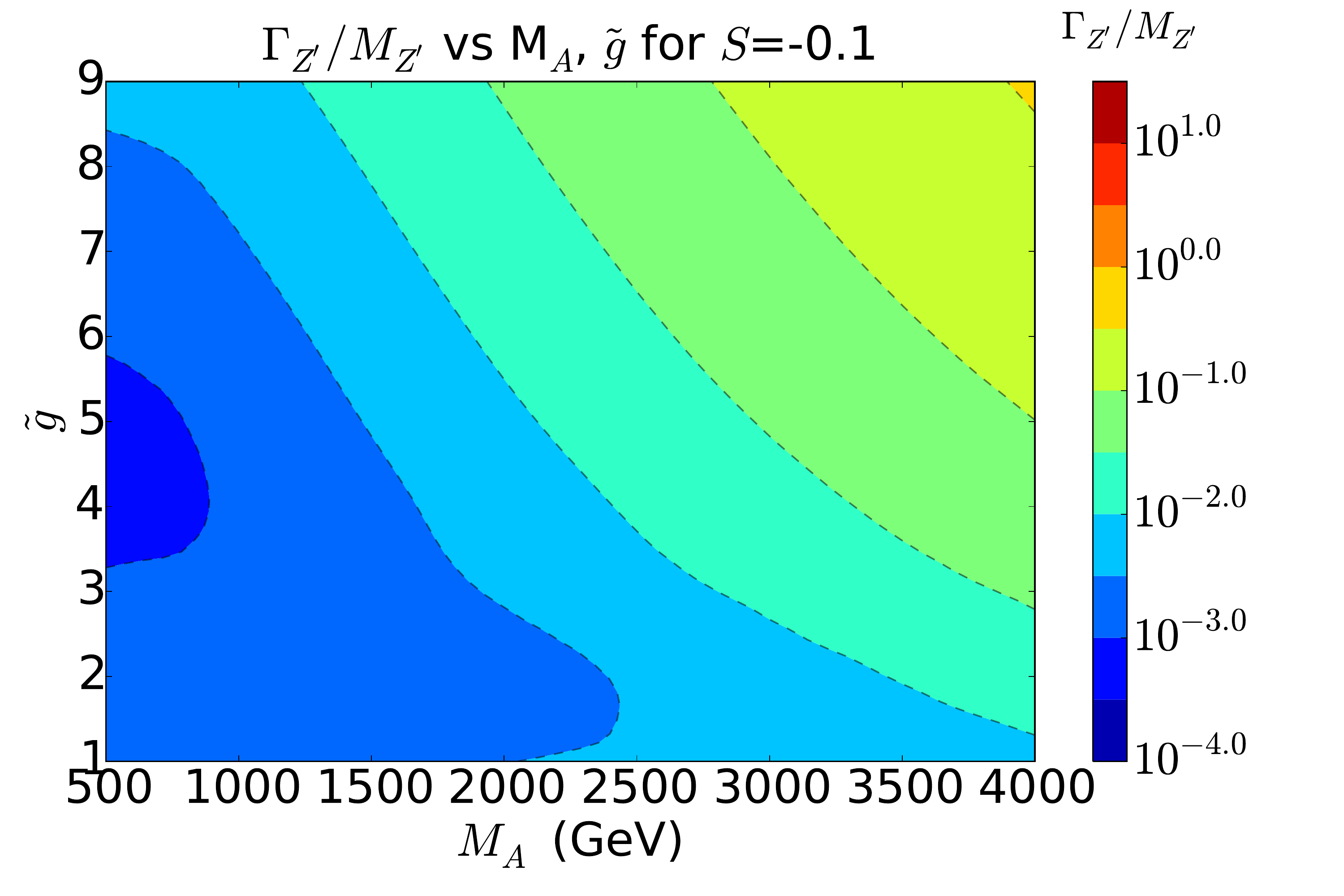}}%
\subfigure[]{\includegraphics[type=pdf,ext=.pdf,read=.pdf,width=0.5\textwidth]{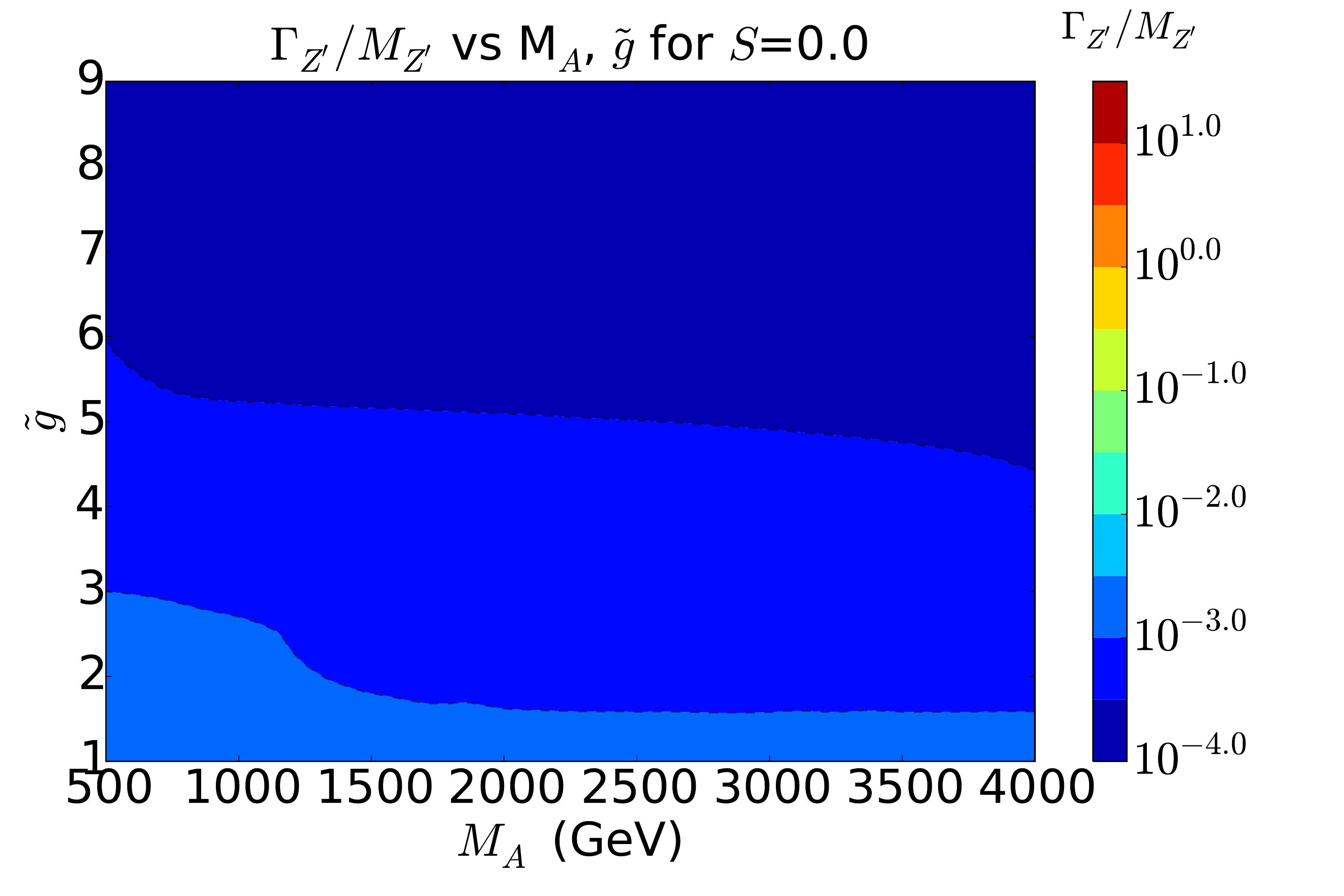}}\\
\subfigure[]{\includegraphics[type=pdf,ext=.pdf,read=.pdf,width=0.5\textwidth]{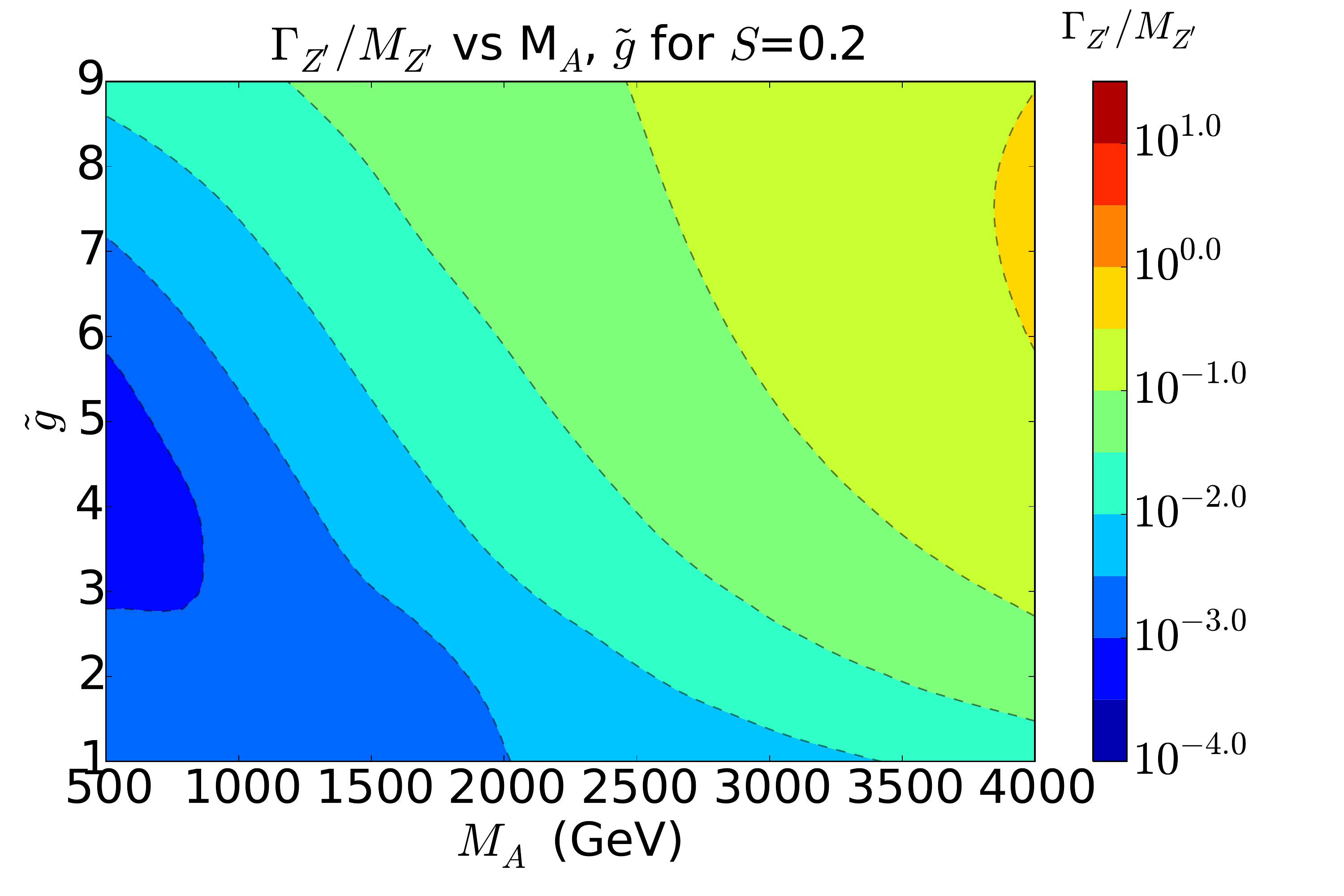}}%
\subfigure[]{\includegraphics[type=pdf,ext=.pdf,read=.pdf,width=0.5\textwidth]{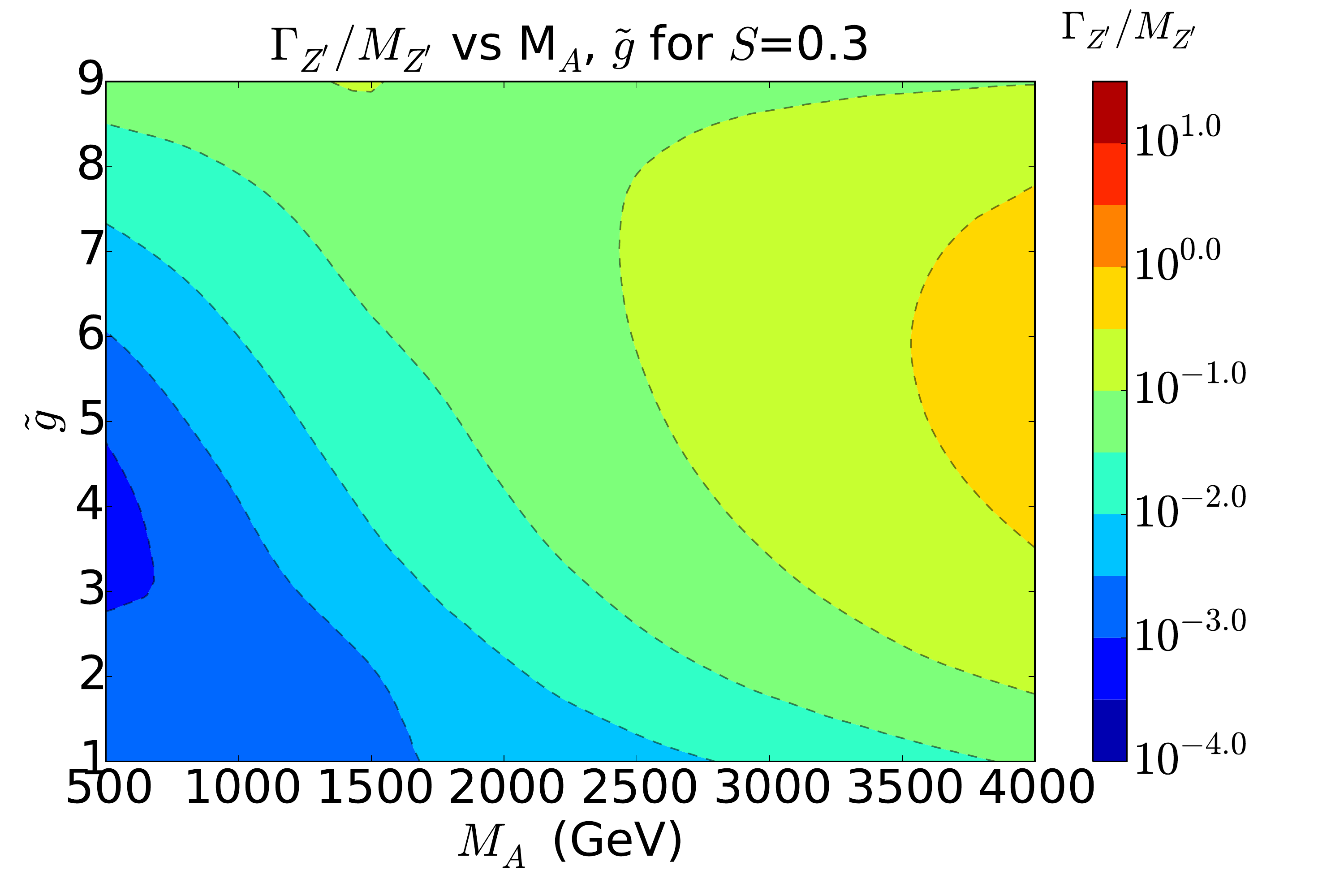}}
\caption{\label{fig:mzp-width2mass-with-s} $\Gamma_{Z^{\prime}}/M_{Z^{\prime}}$ as a function of $M_A$ and $\tilde{g}$ parameters for the fixed values of  
$S=-0.1$ (a), $S=0.0$ (b), $S=0.2$ (c), $S=0.3$ (d) respectively}
\end{figure}

\begin{figure}[htb]
\subfigure[]{\includegraphics[type=pdf,ext=.pdf,read=.pdf,width=0.5\textwidth]{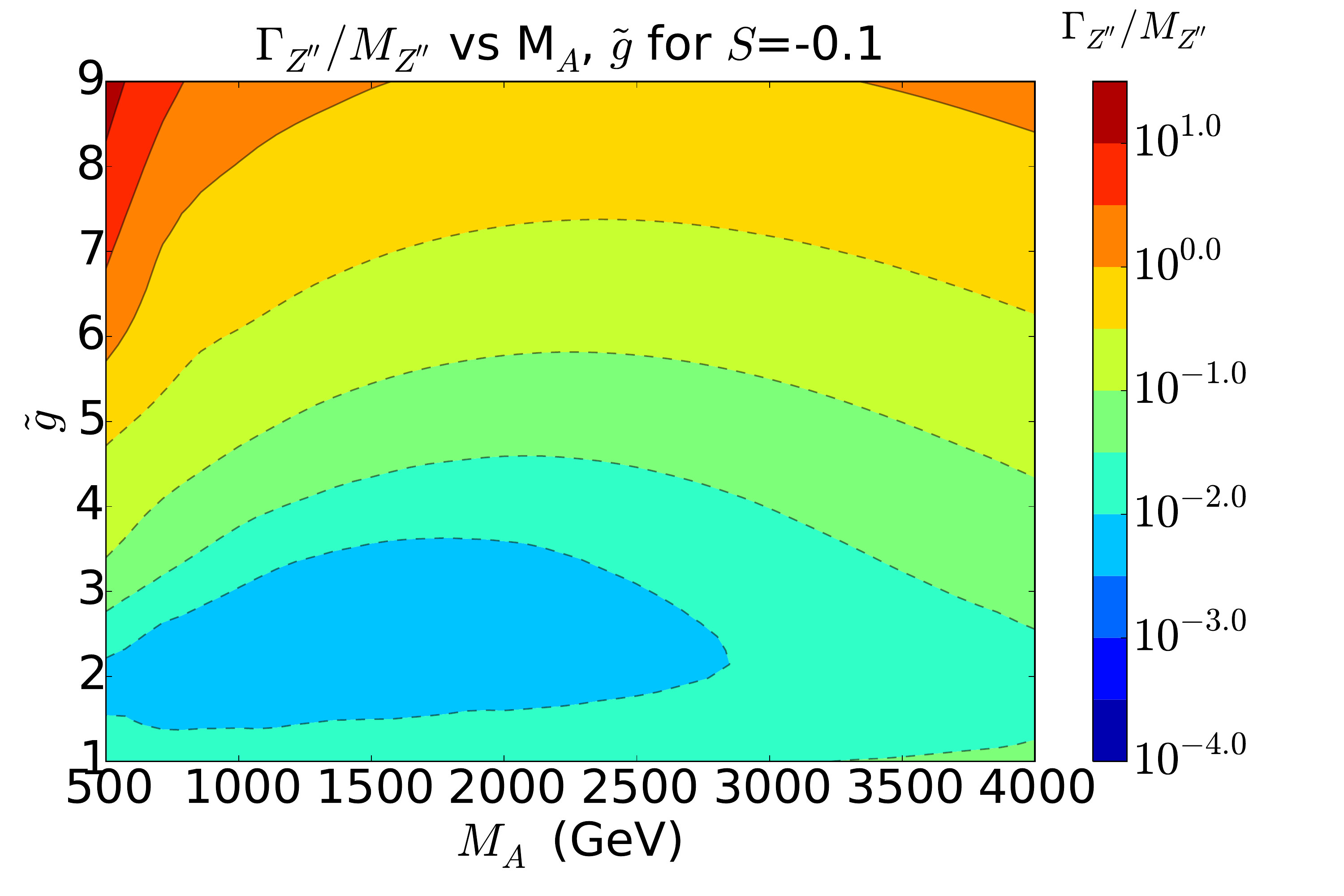}}%
\subfigure[]{\includegraphics[type=pdf,ext=.pdf,read=.pdf,width=0.5\textwidth]{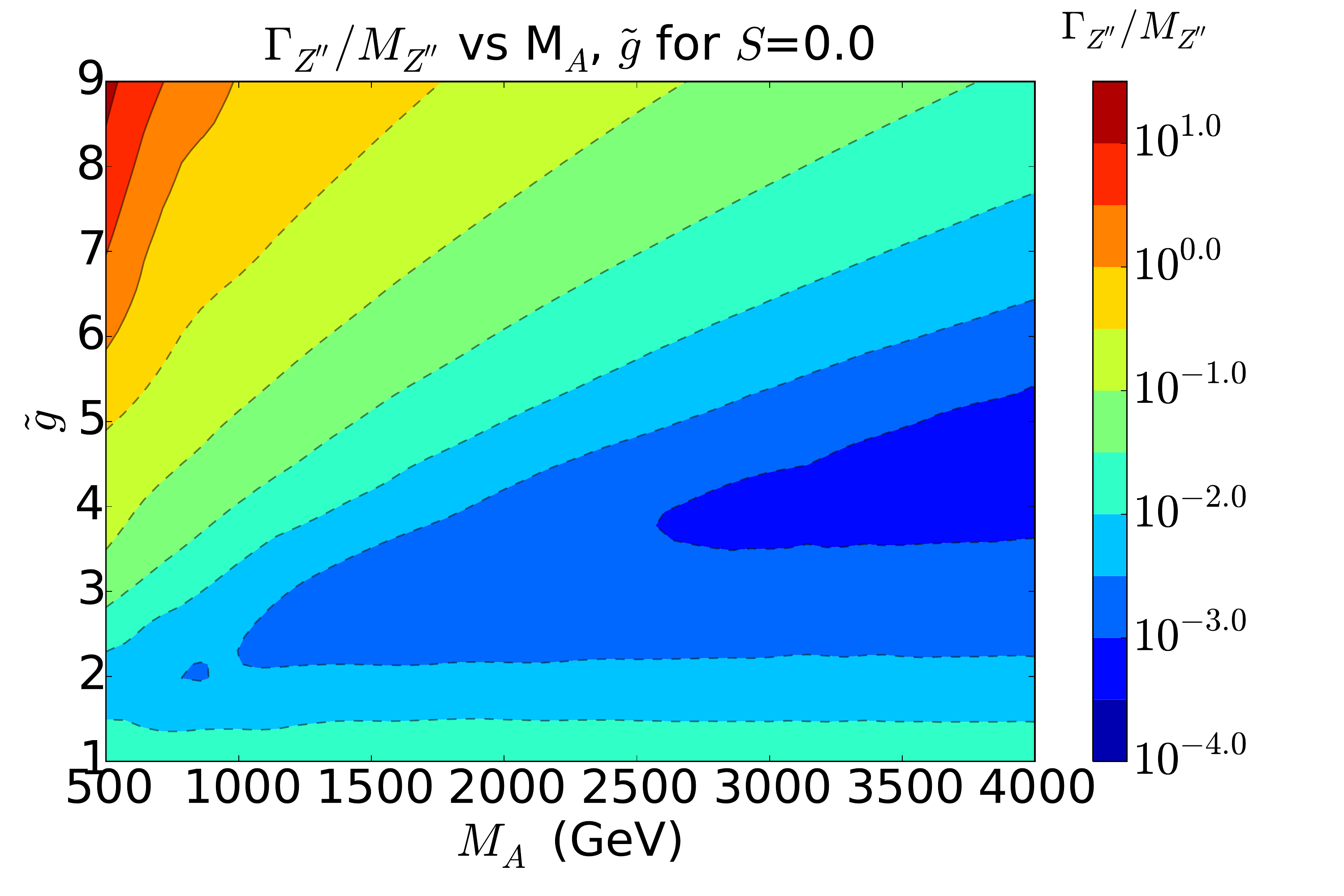}}\\
\subfigure[]{\includegraphics[type=pdf,ext=.pdf,read=.pdf,width=0.5\textwidth]{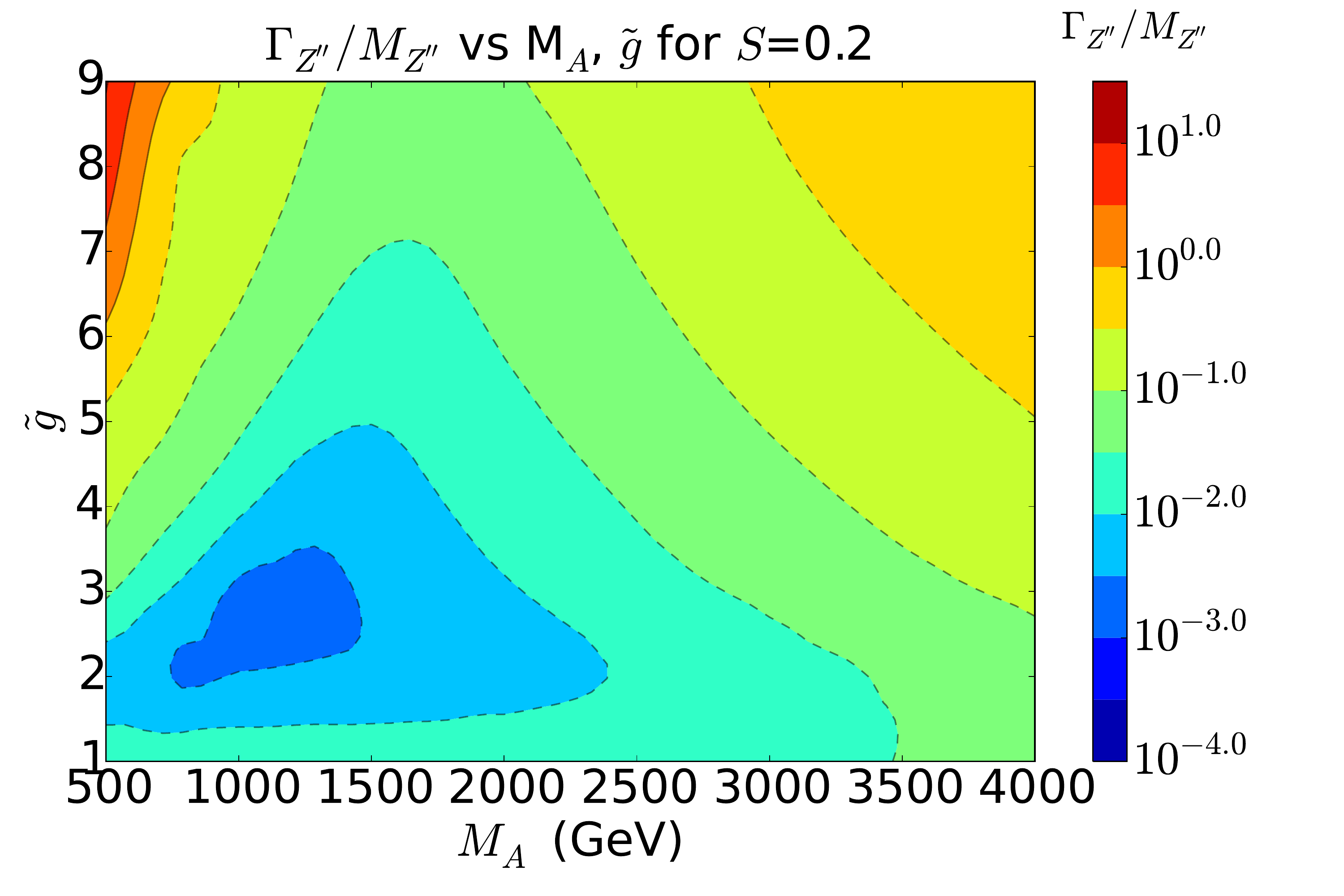}}%
\subfigure[]{\includegraphics[type=pdf,ext=.pdf,read=.pdf,width=0.5\textwidth]{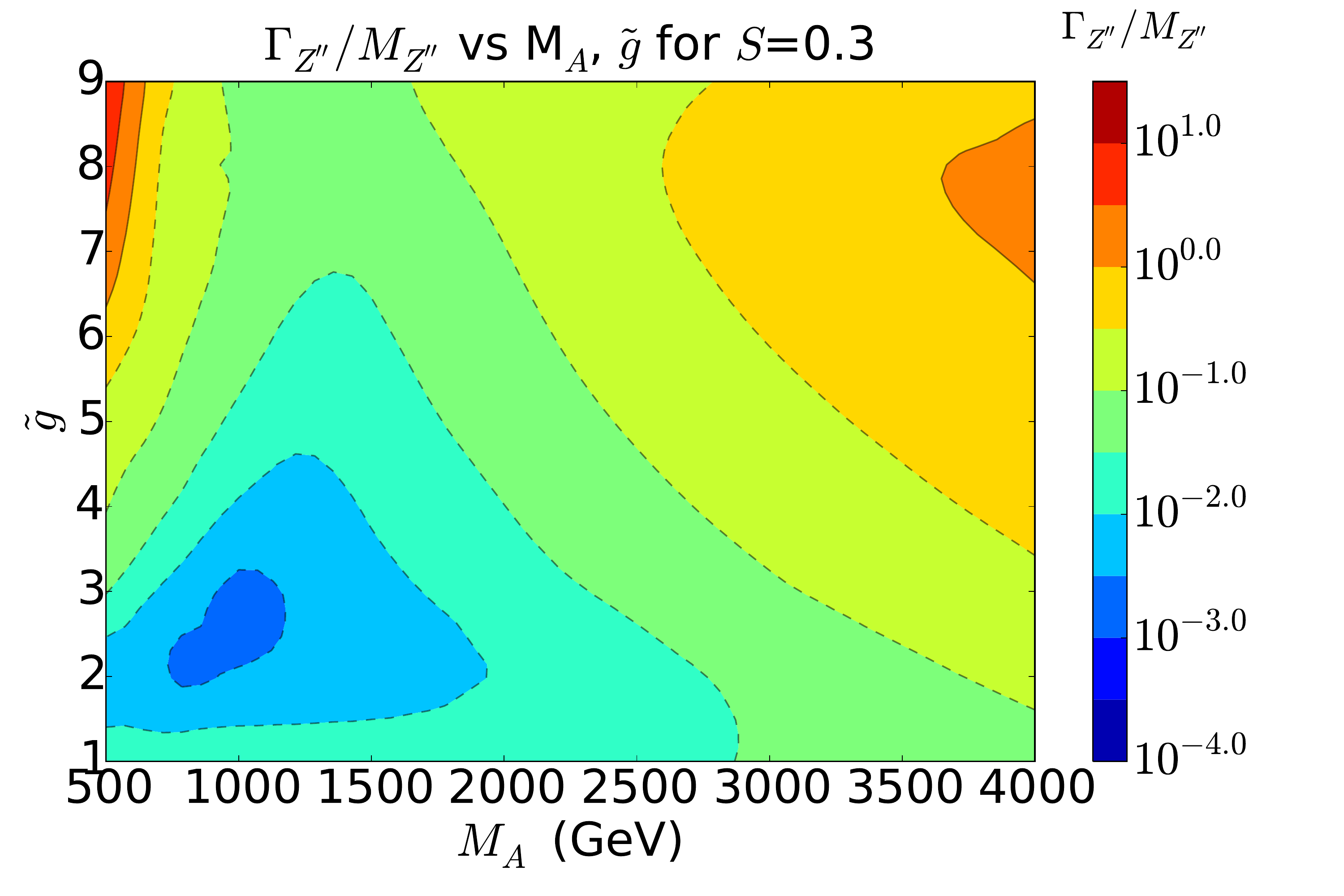}}
\caption{\label{fig:mzpp-width2mass-with-s} $\Gamma_{Z^{\prime\prime}}/M_{Z^{\prime\prime}}$ as a function of $M_A$ and $\tilde{g}$ parameters for the fixed values of  
$S=-0.1$ (a), $S=0.0$ (b), $S=0.2$ (c),  $S=0.3$ (d) respectively}
\end{figure}

The width to mass ratio for $Z^{\prime}$ and $Z^{\prime\prime}$for different $S$ are shown in Figures \ref{fig:mzp-width2mass-with-s} and \ref{fig:mzpp-width2mass-with-s}. The widths largely show similar behaviour to those at the benchmark value of $S=0.1$ (Figure \ref{fig:width2mass}), with the exception of $S=0$. At $S=0$, the $Z^{\prime}$ width to mass ratio is very small (less than $\%$ level), so the $Z^{\prime}$ resonance is always narrow at this $S$. The $Z^{\prime\prime}$ also has a narrower width for much of the parameter space at $S=0$, however the region of $\Gamma_{Z^{\prime\prime}}\geq M_{Z^{\prime\prime}}$ nevertheless appears in the region with low $M_A$ and high 
$\tilde{g}$.

% Width/Mass ratio

\begin{figure}[htb]
\subfigure[]{\includegraphics[type=pdf,ext=.pdf,read=.pdf,width=0.5\textwidth]{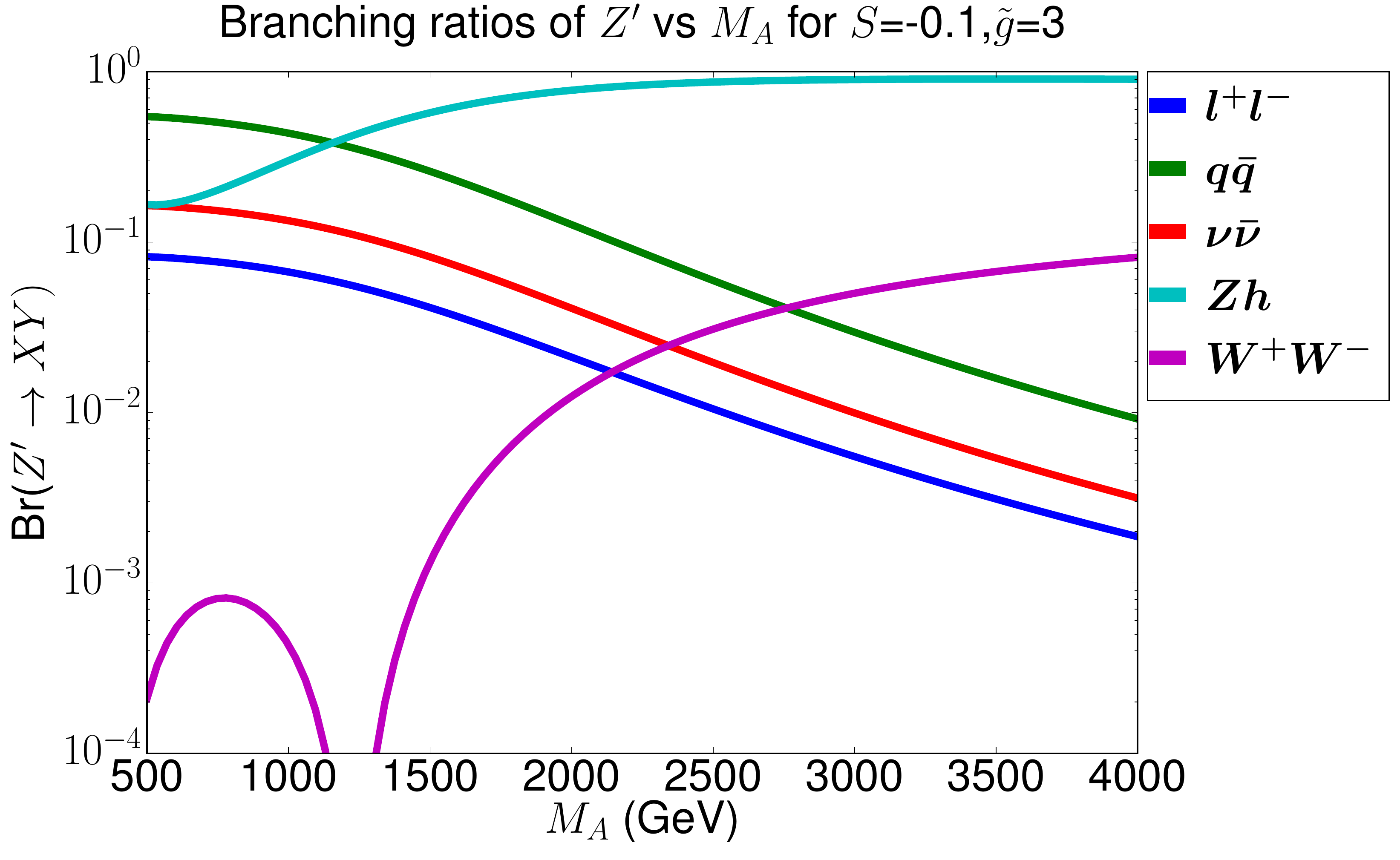}}%
\subfigure[]{\includegraphics[type=pdf,ext=.pdf,read=.pdf,width=0.5\textwidth]{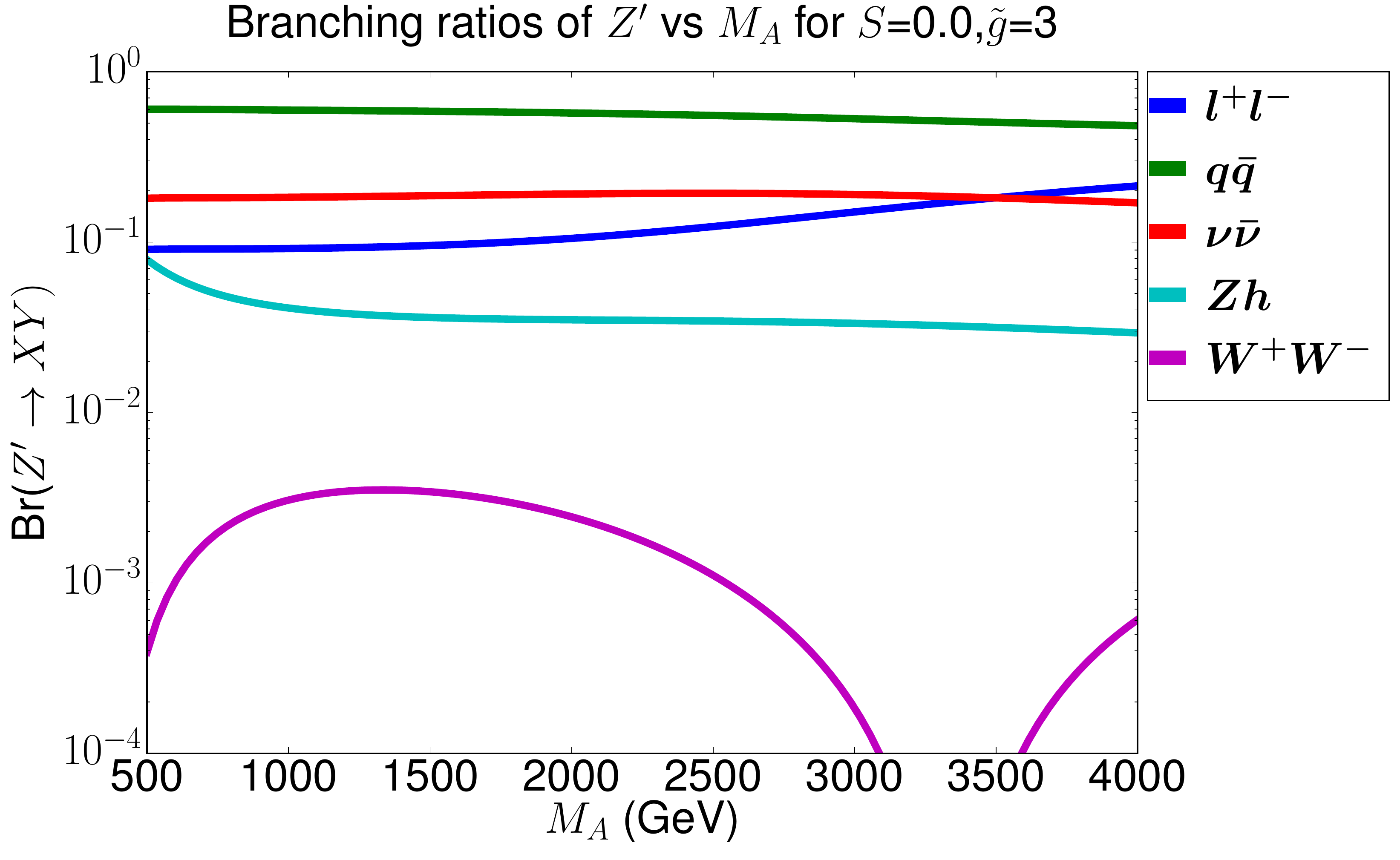}}\\
\subfigure[]{\includegraphics[type=pdf,ext=.pdf,read=.pdf,width=0.5\textwidth]{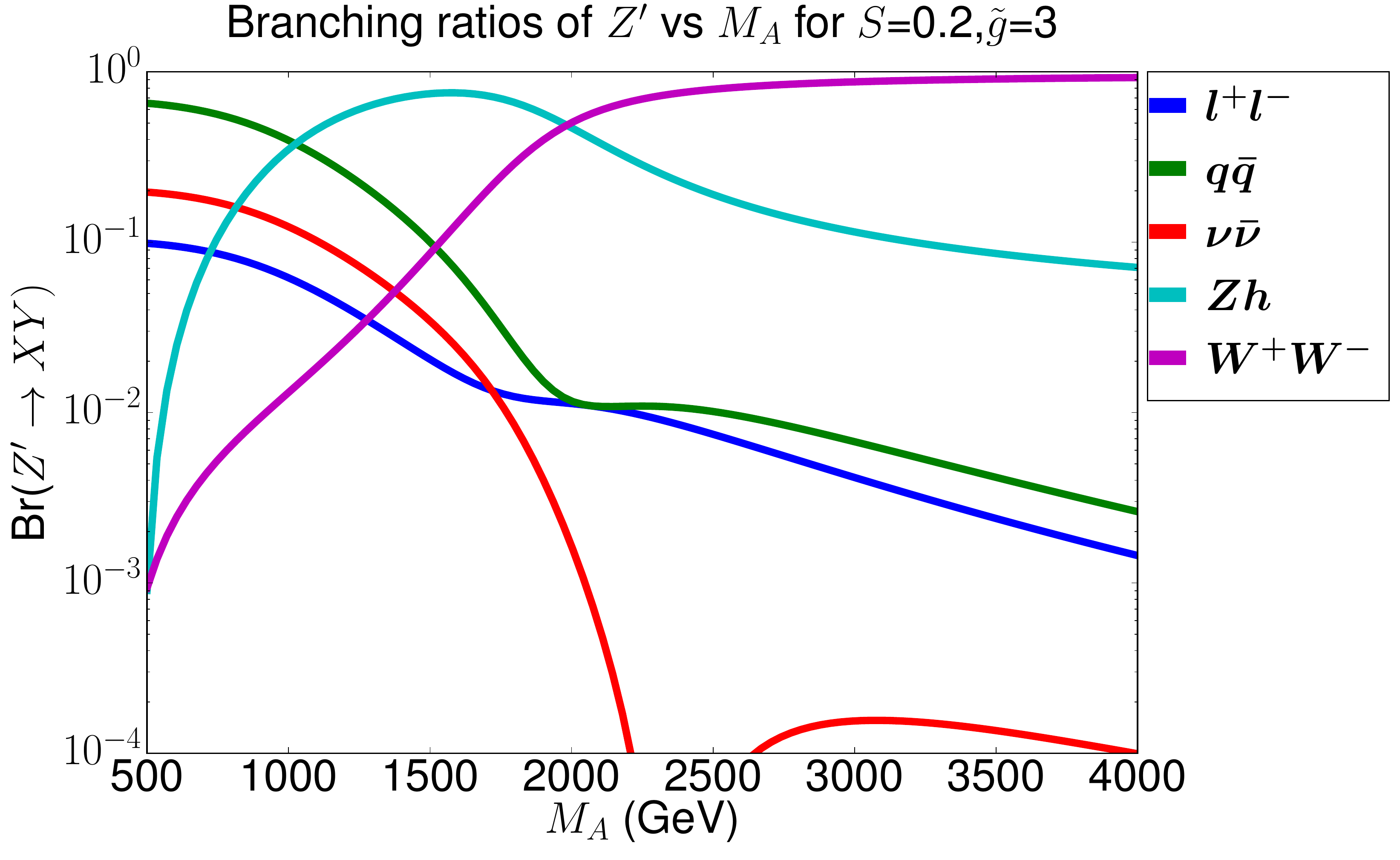}}%
\subfigure[]{\includegraphics[type=pdf,ext=.pdf,read=.pdf,width=0.5\textwidth]{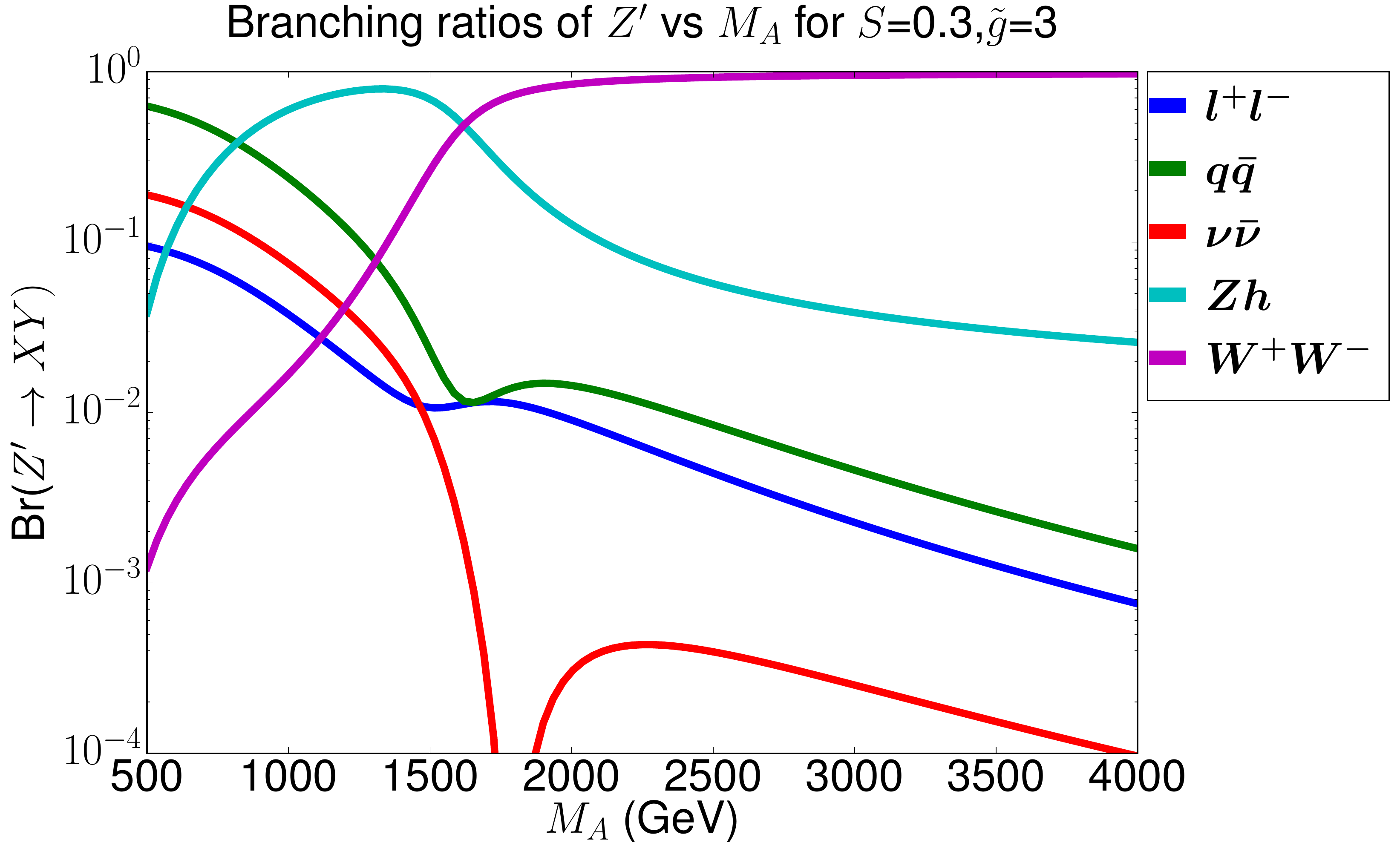}}
\caption{\label{fig:mzp-allbr-gt3-with-s} 
$Br(Z')$ for all decay channels  as a function of $M_A$ at  fixed value of 
$\tilde{g}=3$ for
$S=-0.1$ (a), $S=0.0$ (b), $S=0.2$ (c),  $S=0.3$ (d) respectively}
\end{figure}

\begin{figure}[htb]
\subfigure[]{\includegraphics[type=pdf,ext=.pdf,read=.pdf,width=0.5\textwidth]{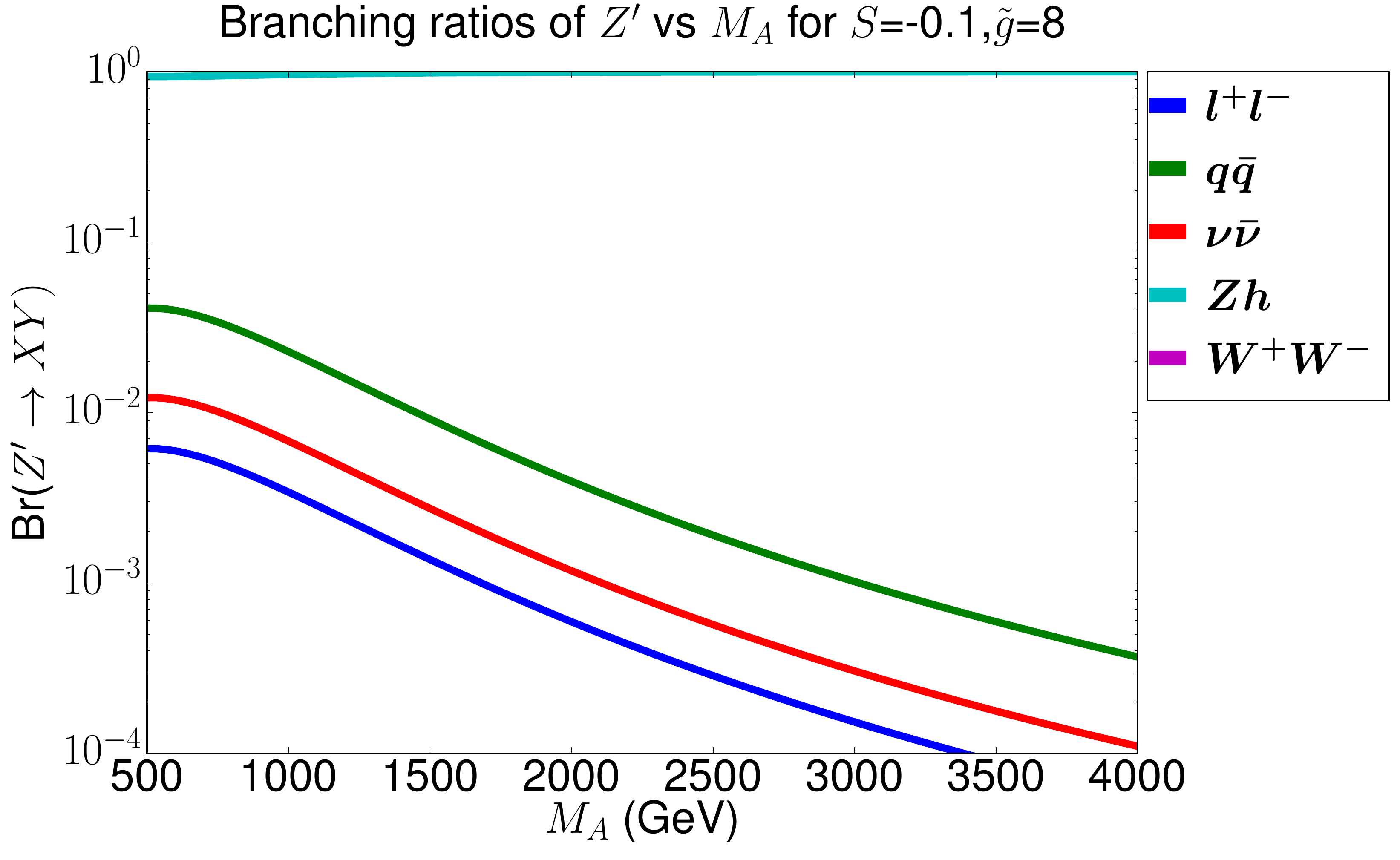}}%
\subfigure[]{\includegraphics[type=pdf,ext=.pdf,read=.pdf,width=0.5\textwidth]{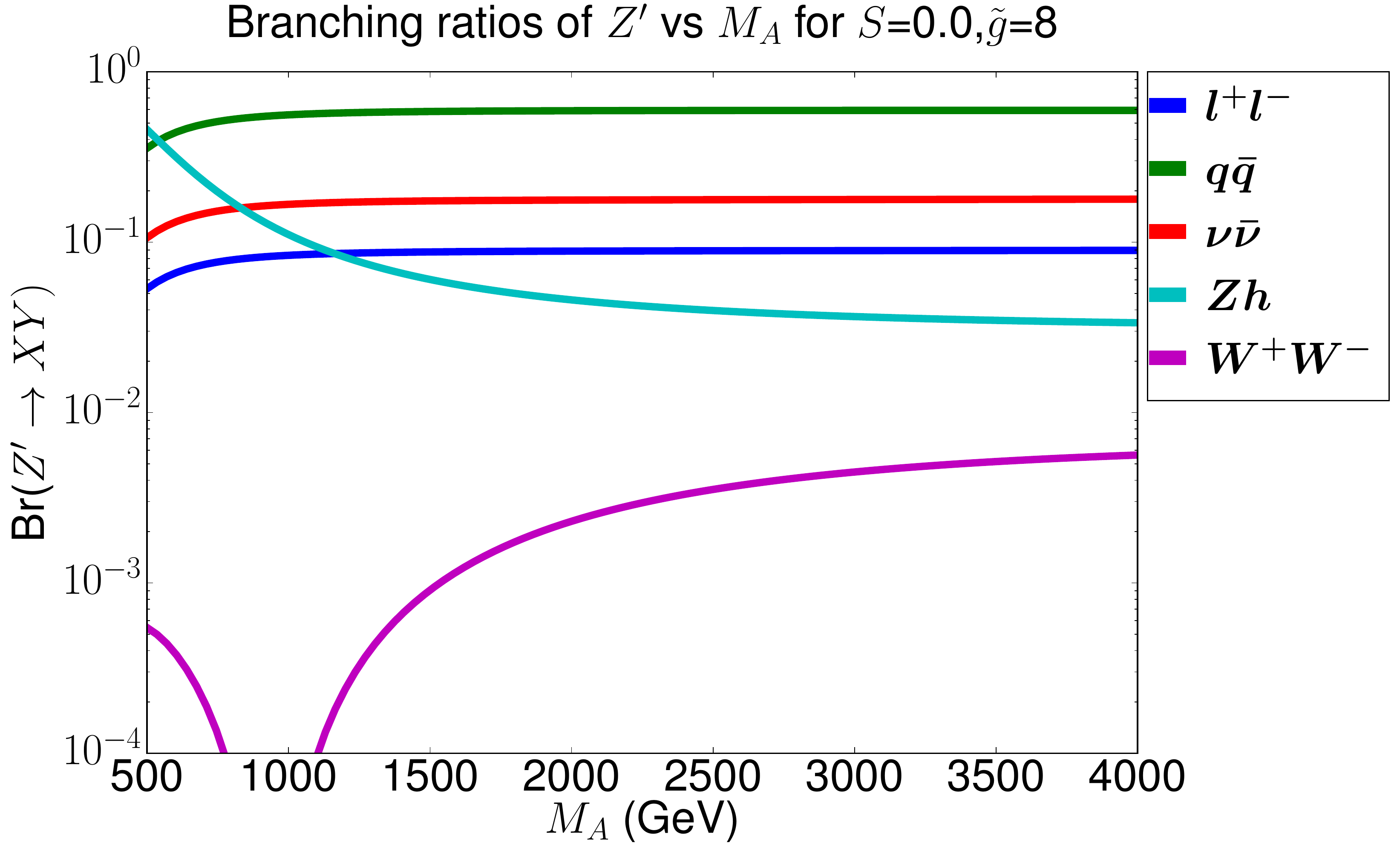}}\\
\subfigure[]{\includegraphics[type=pdf,ext=.pdf,read=.pdf,width=0.5\textwidth]{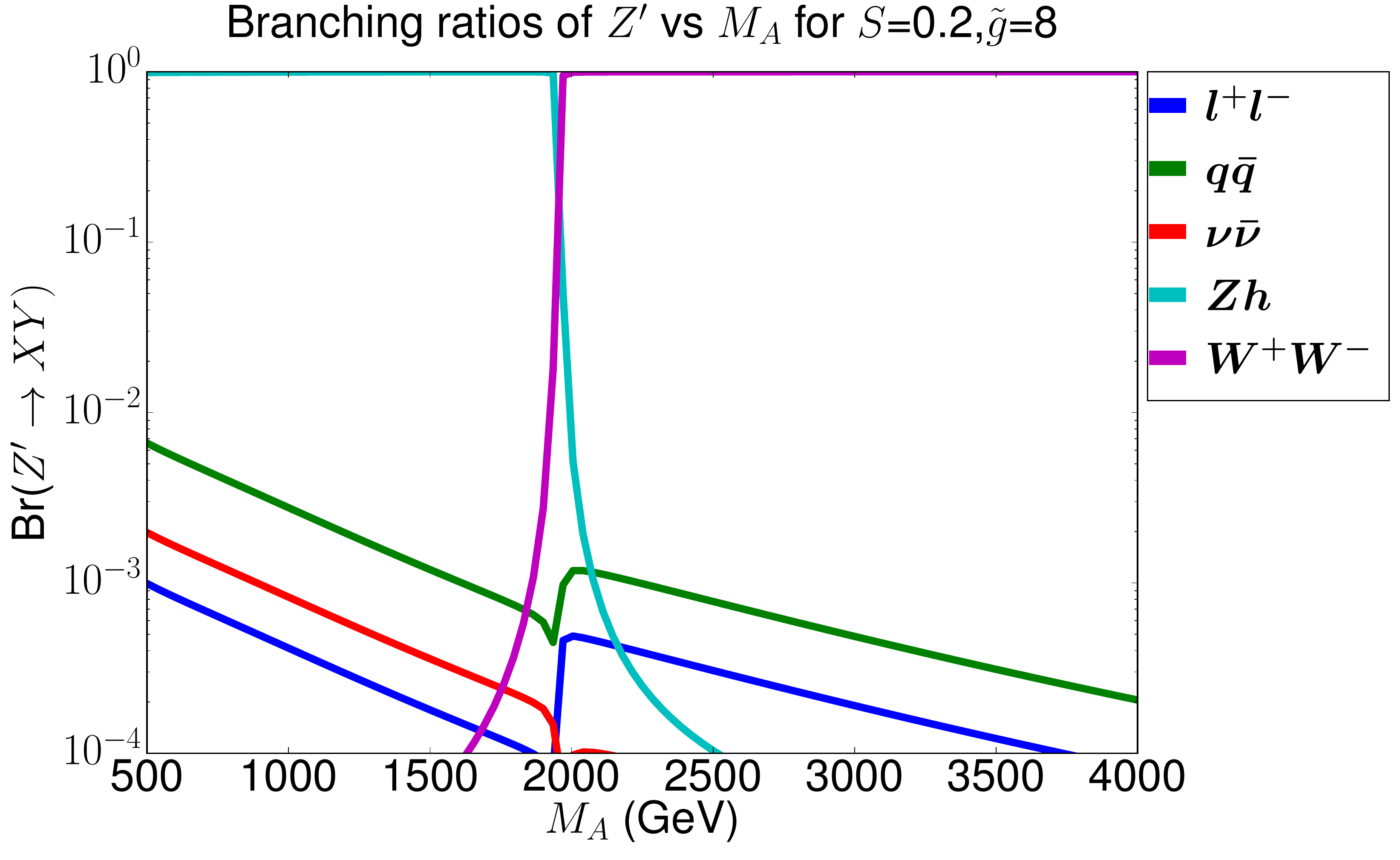}}%
\subfigure[]{\includegraphics[type=pdf,ext=.pdf,read=.pdf,width=0.5\textwidth]{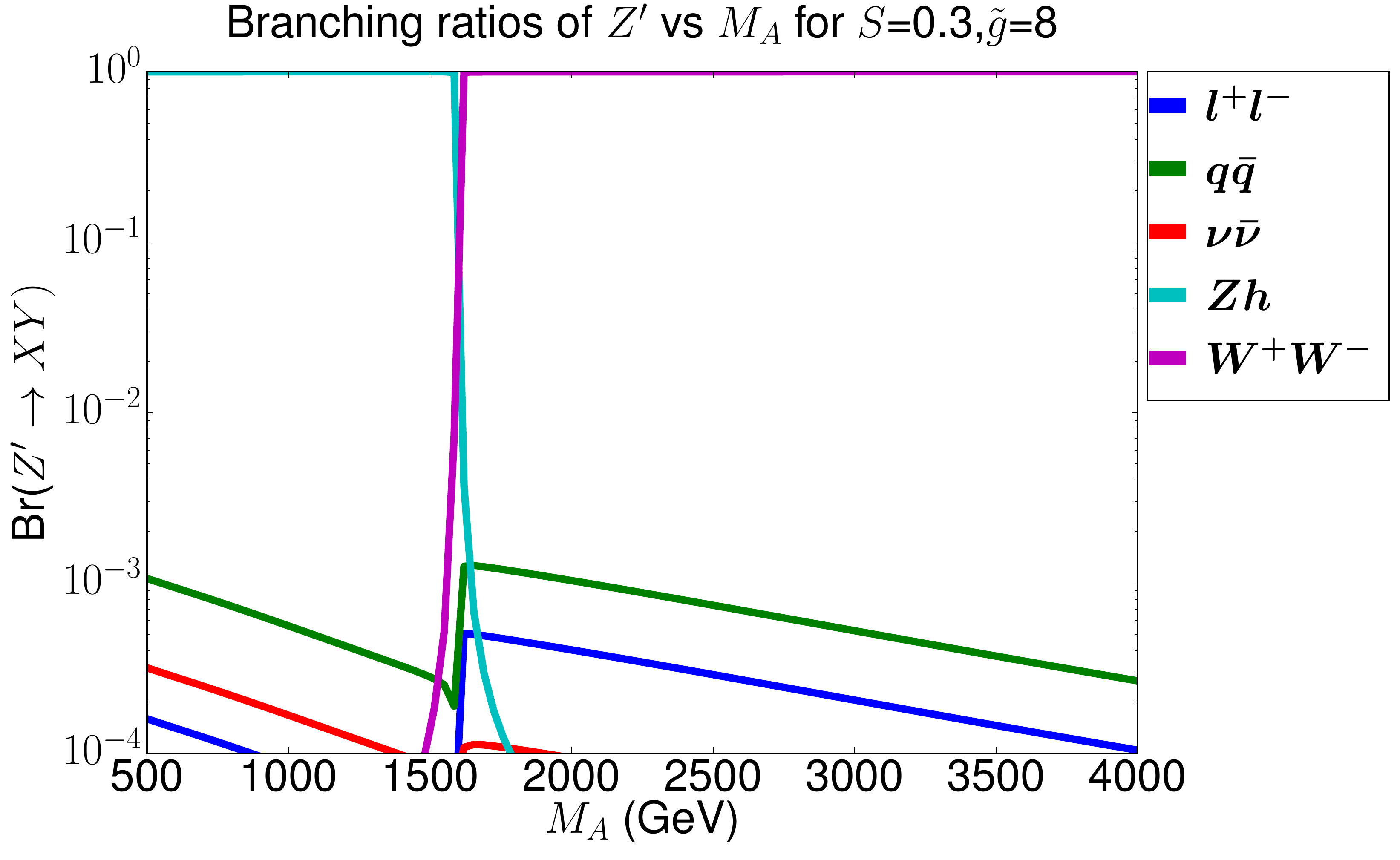}}
\caption{\label{fig:mzp-allbr-gt8-with-s} $Br(Z')$ for all decay channels  as a function of $M_A$ at  fixed value of 
$\tilde{g}=8$ for
$S=-0.1$ (a), $S=0.0$ (b), $S=0.2$ (c),  $S=0.3$ (d) respectively
$S=-0.1$ (a), $S=0.0$ (b), $S=0.2$ (c),  $S=0.3$ (d) respectively}
\end{figure}

\begin{figure}[htb]
\subfigure[]{\includegraphics[type=pdf,ext=.pdf,read=.pdf,width=0.5\textwidth]{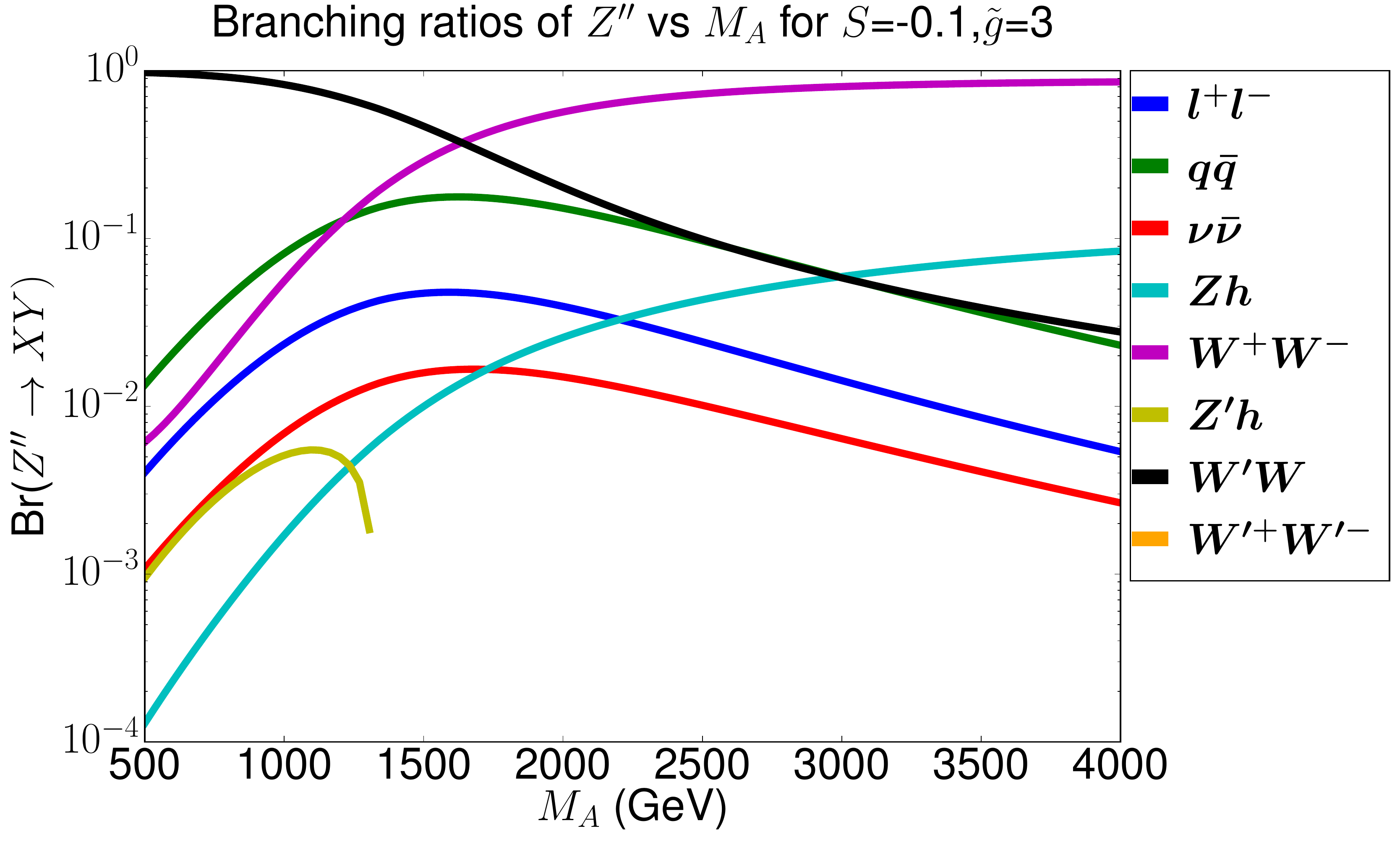}}%
\subfigure[]{\includegraphics[type=pdf,ext=.pdf,read=.pdf,width=0.5\textwidth]{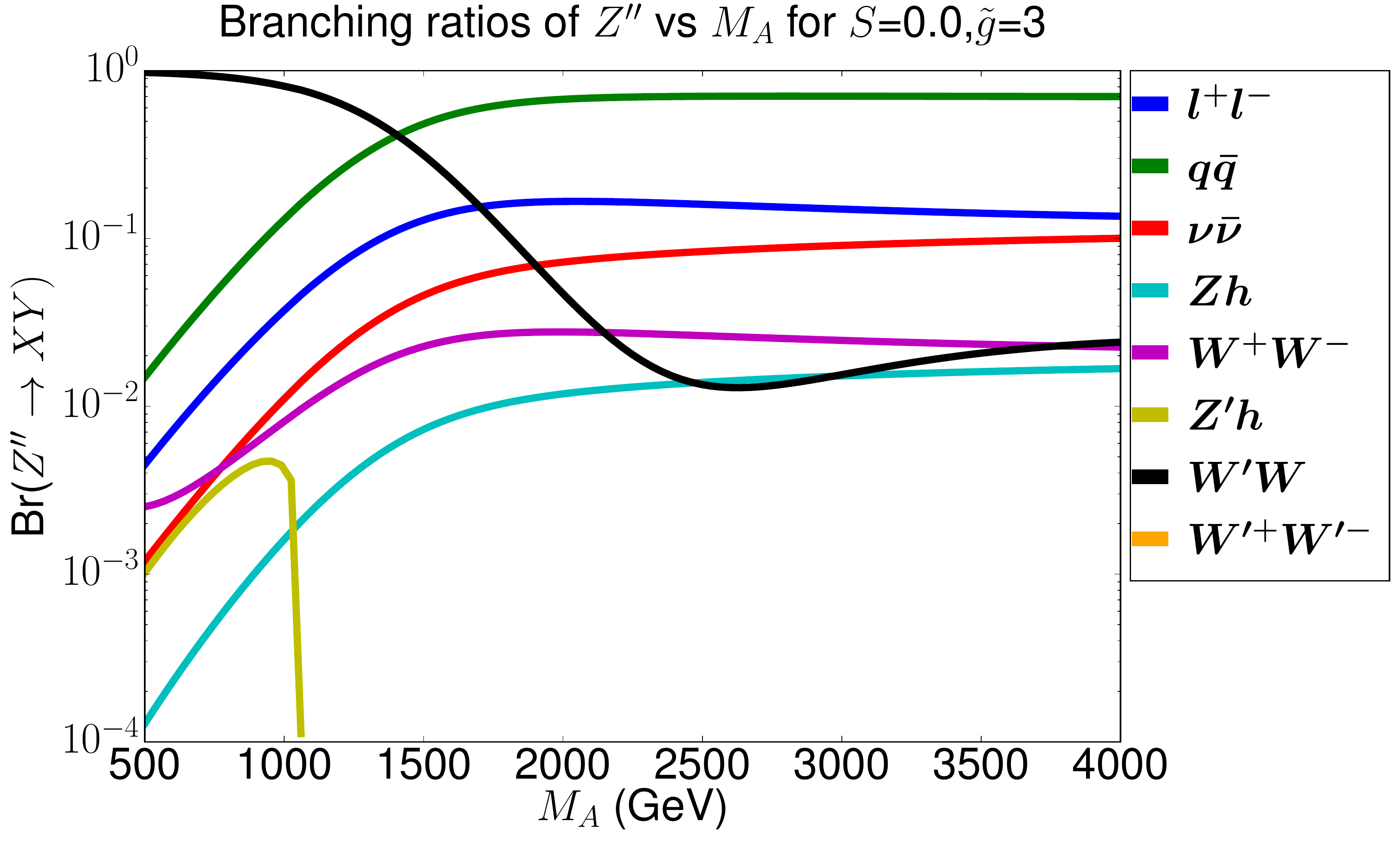}}\\
\subfigure[]{\includegraphics[type=pdf,ext=.pdf,read=.pdf,width=0.5\textwidth]{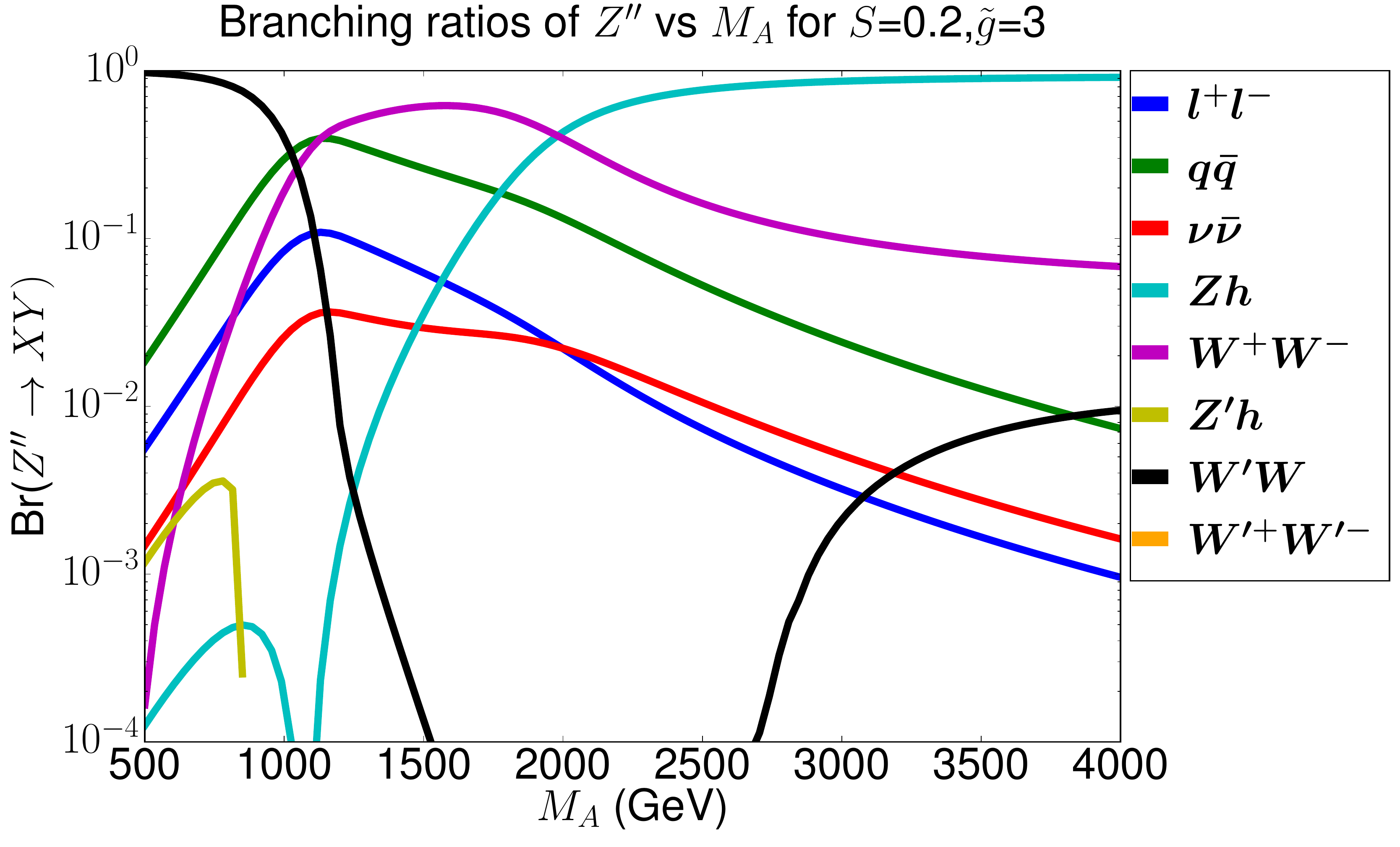}}%
\subfigure[]{\includegraphics[type=pdf,ext=.pdf,read=.pdf,width=0.5\textwidth]{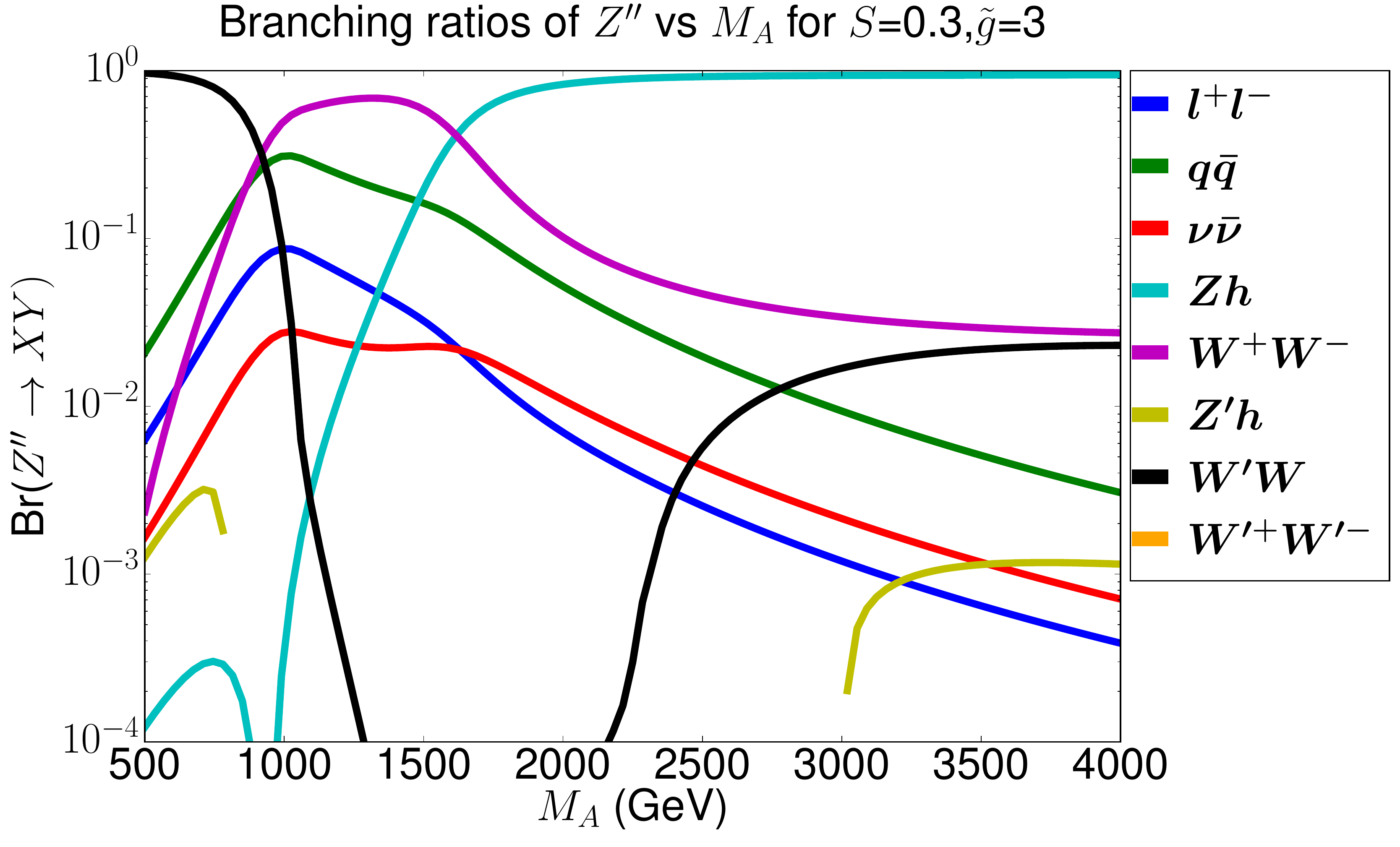}}
\caption{\label{fig:mzpp-allbr-gt3-with-s} $Br(Z'')$ for all decay channels  as a function of $M_A$ at  fixed value of 
$\tilde{g}=3$ for
$S=-0.1$ (a), $S=0.0$ (b), $S=0.2$ (c),  $S=0.3$ (d) respectively  
$S=-0.1$ (a), $S=0.0$ (b), $S=0.2$ (c),  $S=0.3$ (d) respectively}
\end{figure}

\begin{figure}[htb]
\subfigure[]{\includegraphics[type=pdf,ext=.pdf,read=.pdf,width=0.5\textwidth]{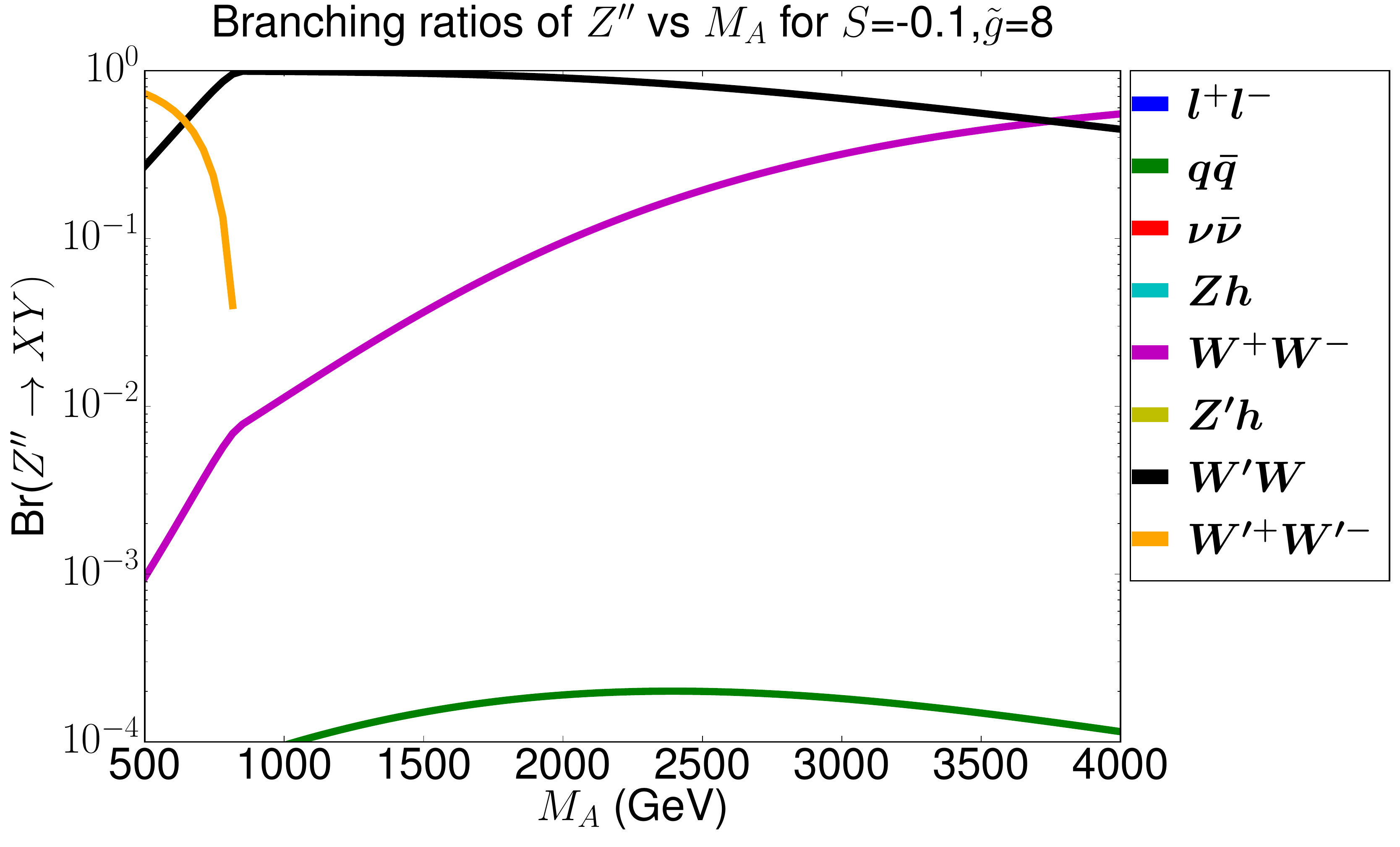}}%
\subfigure[]{\includegraphics[type=pdf,ext=.pdf,read=.pdf,width=0.5\textwidth]{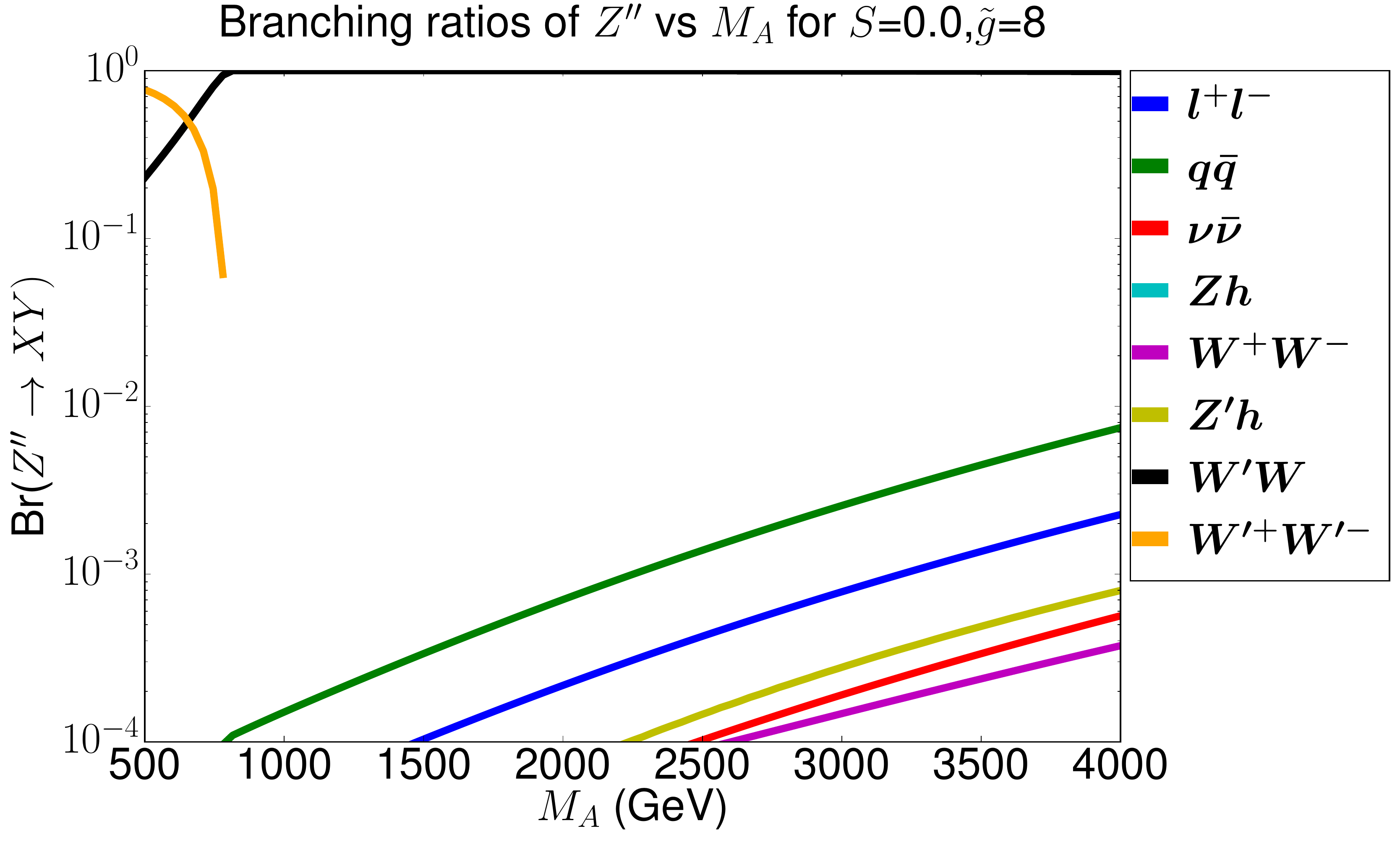}}\\
\subfigure[]{\includegraphics[type=pdf,ext=.pdf,read=.pdf,width=0.5\textwidth]{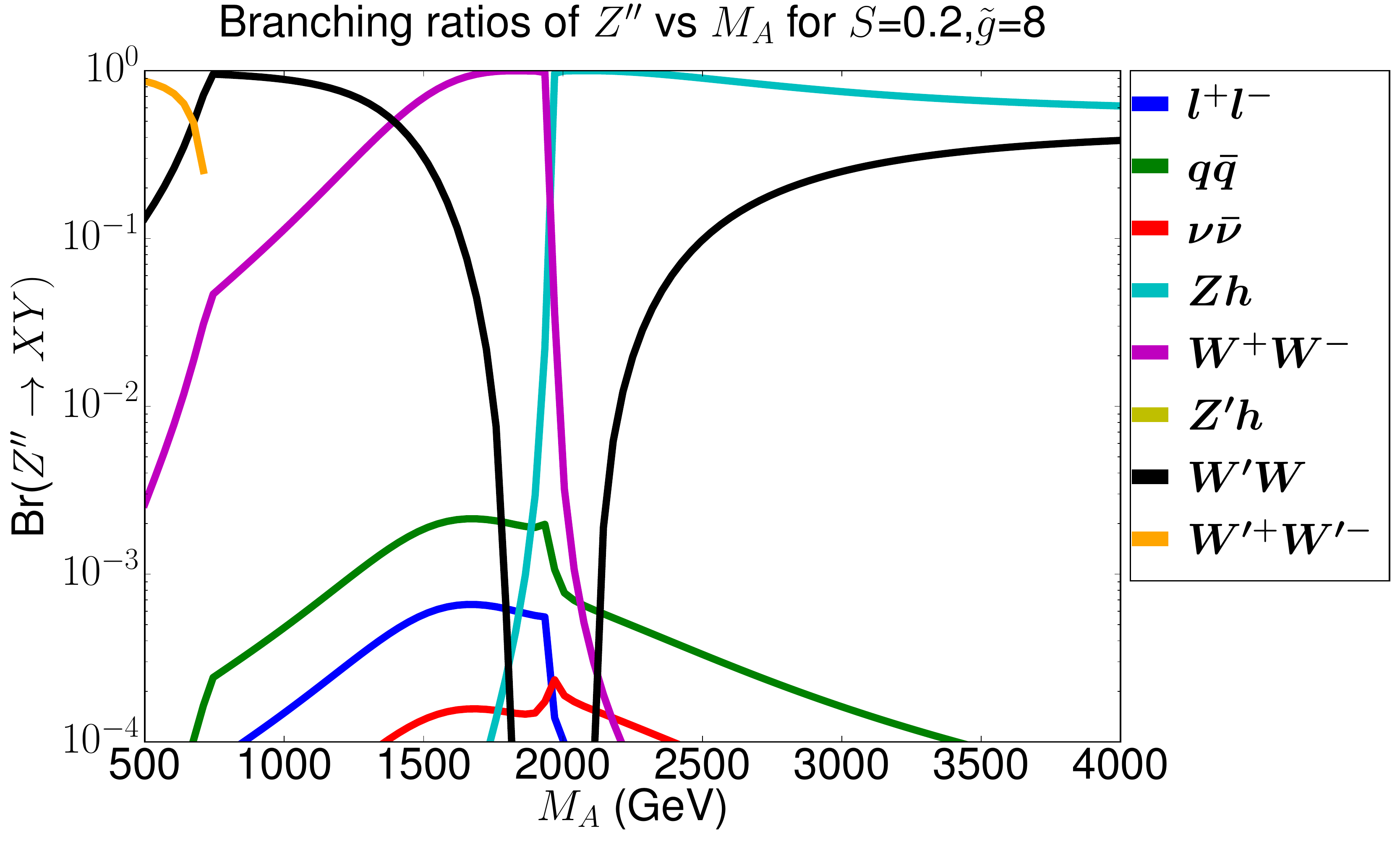}}%
\subfigure[]{\includegraphics[type=pdf,ext=.pdf,read=.pdf,width=0.5\textwidth]{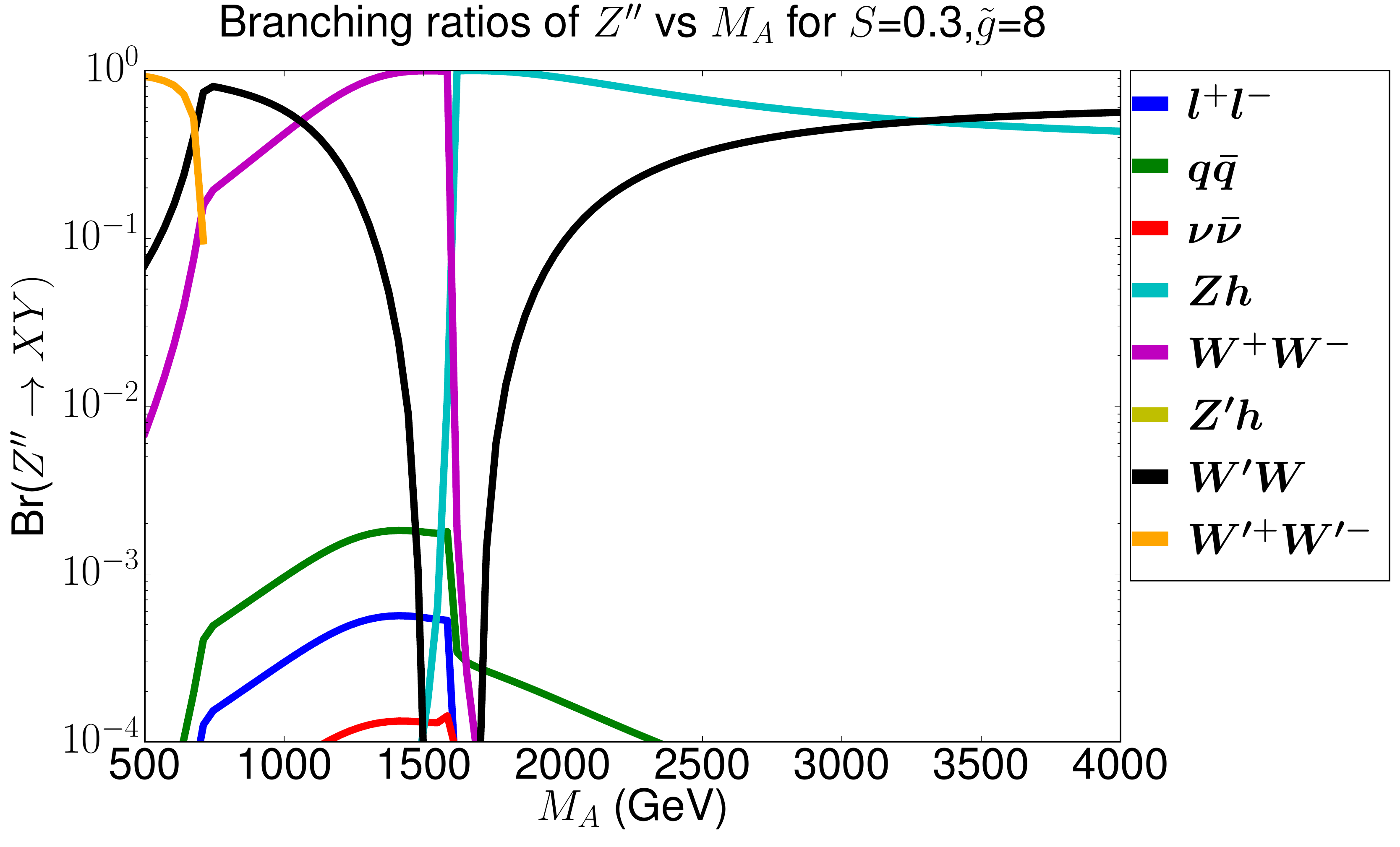}}
\caption{\label{fig:mzpp-allbr-gt8-with-s} $Br(Z'')$ for all decay channels  as a function of $M_A$ at  fixed value of 
$\tilde{g}=8$ for
$S=-0.1$ (a), $S=0.0$ (b), $S=0.2$ (c),  $S=0.3$ (d) respectively  
$S=-0.1$ (a), $S=0.0$ (b), $S=0.2$ (c),  $S=0.3$ (d) respectively}
\end{figure}

 The branching ratio spectra for the $Z^{\prime}$ with $\tilde{g}=3,8$ 
 is presented in Figures \ref{fig:mzp-allbr-gt3-with-s},\ref{fig:mzp-allbr-gt8-with-s}), 
 and for the $Z^{\prime\prime}$ with $\tilde{g}=3,8$ --- in Figures \ref{fig:mzpp-allbr-gt3-with-s}, \ref{fig:mzpp-allbr-gt8-with-s} for various values $S$. The features of the branching ratio spectra such as the dips in the $VV/Vh$ channels are discussed in section \ref{subsec:width-and-br}, and again we note that the $Z^{\prime\prime}\rightarrow W^{\prime +}W^{\prime -}$ channel is opened at low $M_{A}$, high $\tilde{g}$ at all values of $S$. Also note that for the $Z^{\prime}$, at $S=0$ where the resonance is very narrow, the dilepton and diquark branching ratios are boosted and are the dominant decay channels across the whole ($M_{A}$, $\tilde{g}$) parameter space. 

Again, the mass inversion point can also be identified as the point at which the $W^{+}W^{-}$ and $Zh$ branching ratios have a crossing point, hence the lack of crossing point at $S=-0.1,0$.   

% Branching Z'->XY, gt=3

% Branching Z'->XY, gt=8

% Branching Z''->XY, gt=3

% Branching Z''->XY, gt=8

\clearpage

\subsubsection{Cross sections}
\label{appendix:cs}

\begin{figure}[htb]
\subfigure[]{\includegraphics[type=pdf,ext=.pdf,read=.pdf,width=0.5\textwidth]{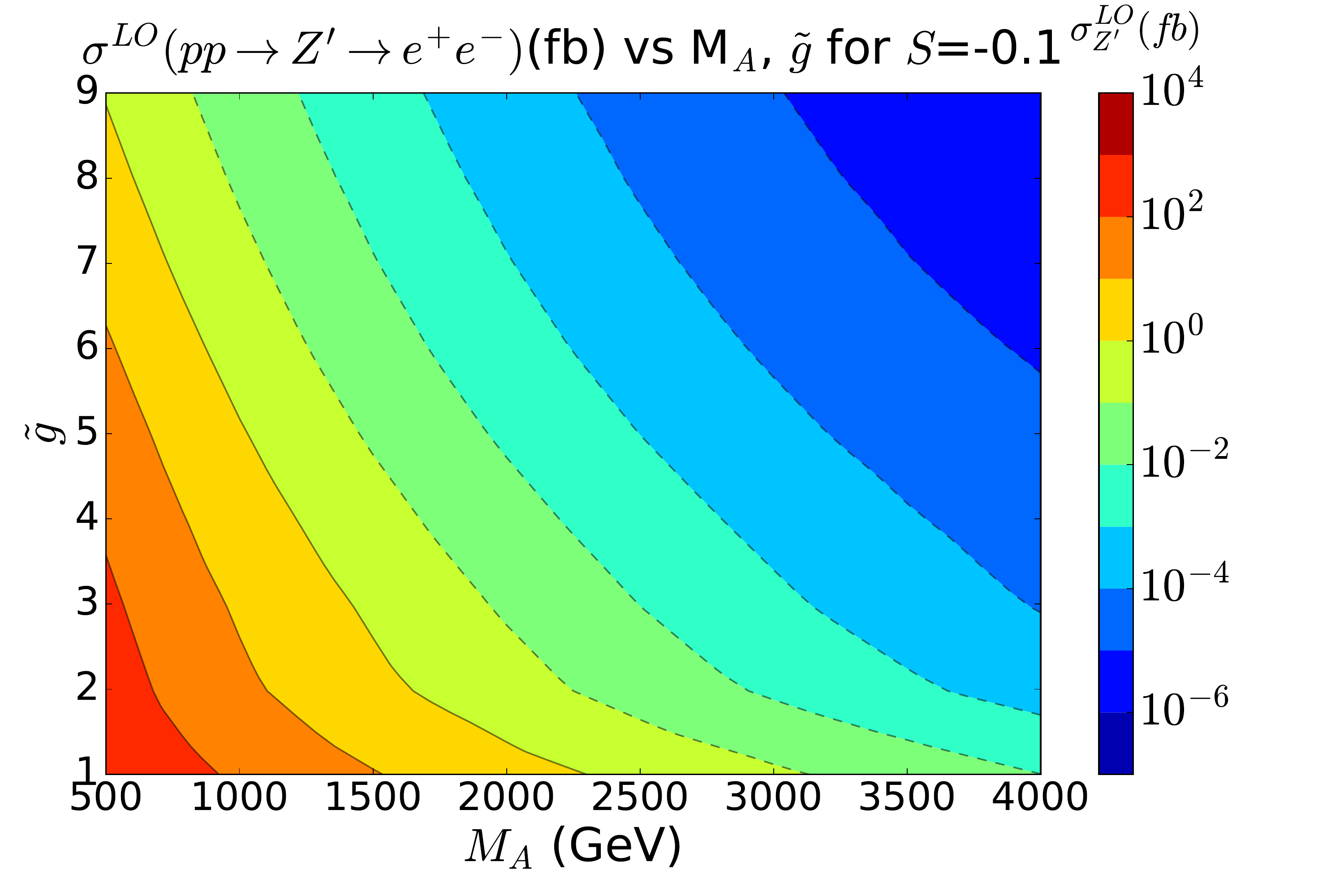}}%
\subfigure[]{\includegraphics[type=pdf,ext=.pdf,read=.pdf,width=0.5\textwidth]{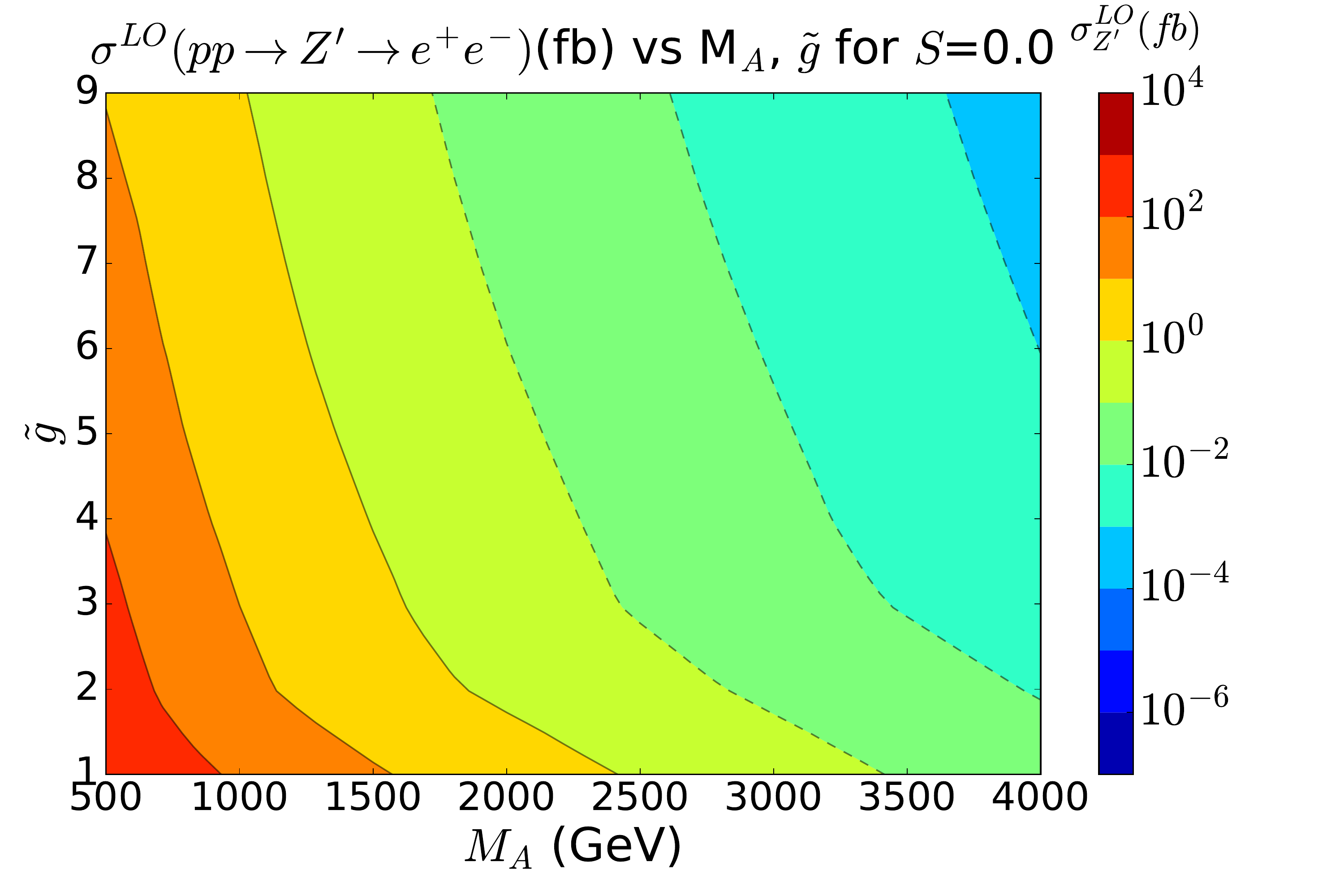}}\\
\subfigure[]{\includegraphics[type=pdf,ext=.pdf,read=.pdf,width=0.5\textwidth]{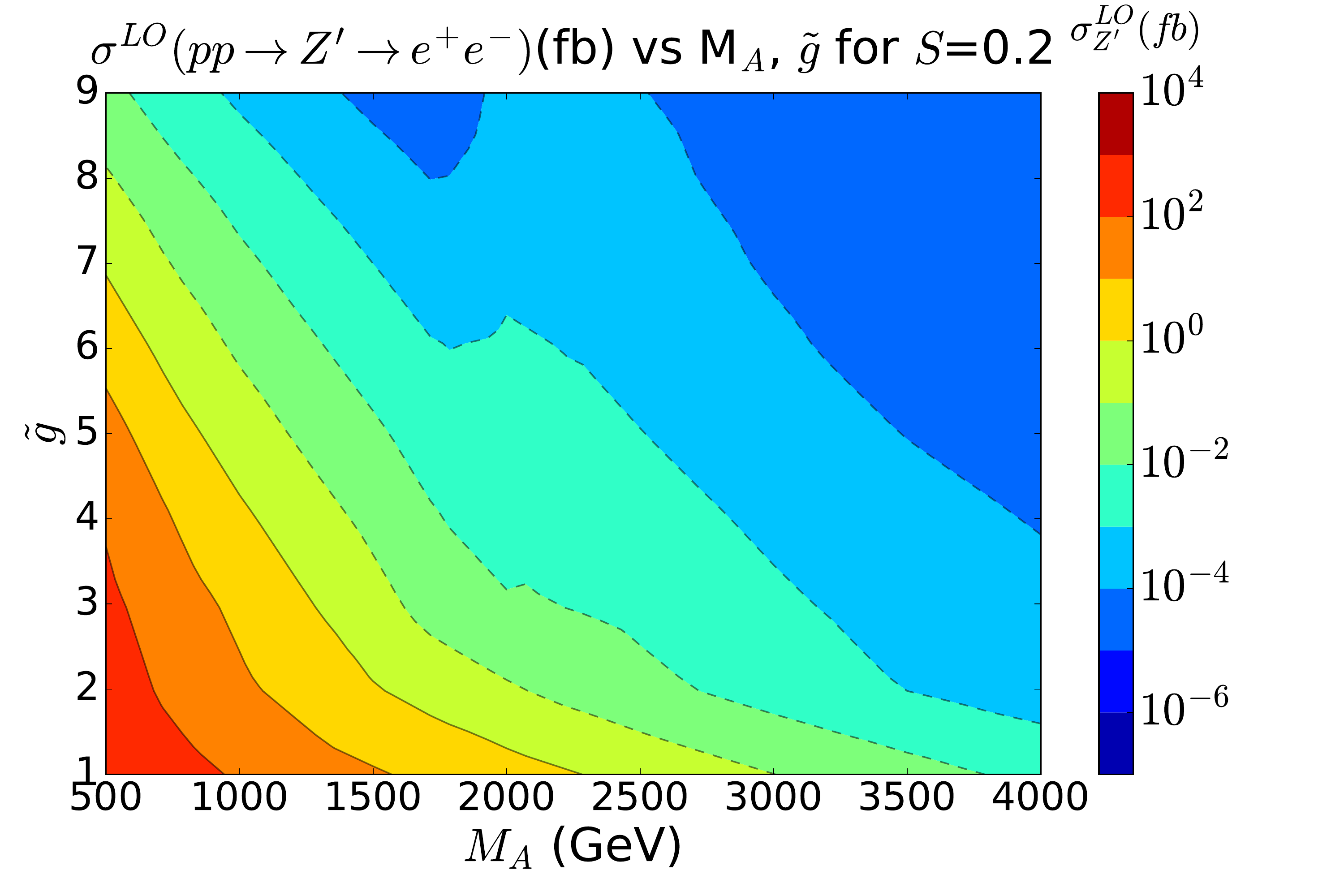}}%
\subfigure[]{\includegraphics[type=pdf,ext=.pdf,read=.pdf,width=0.5\textwidth]{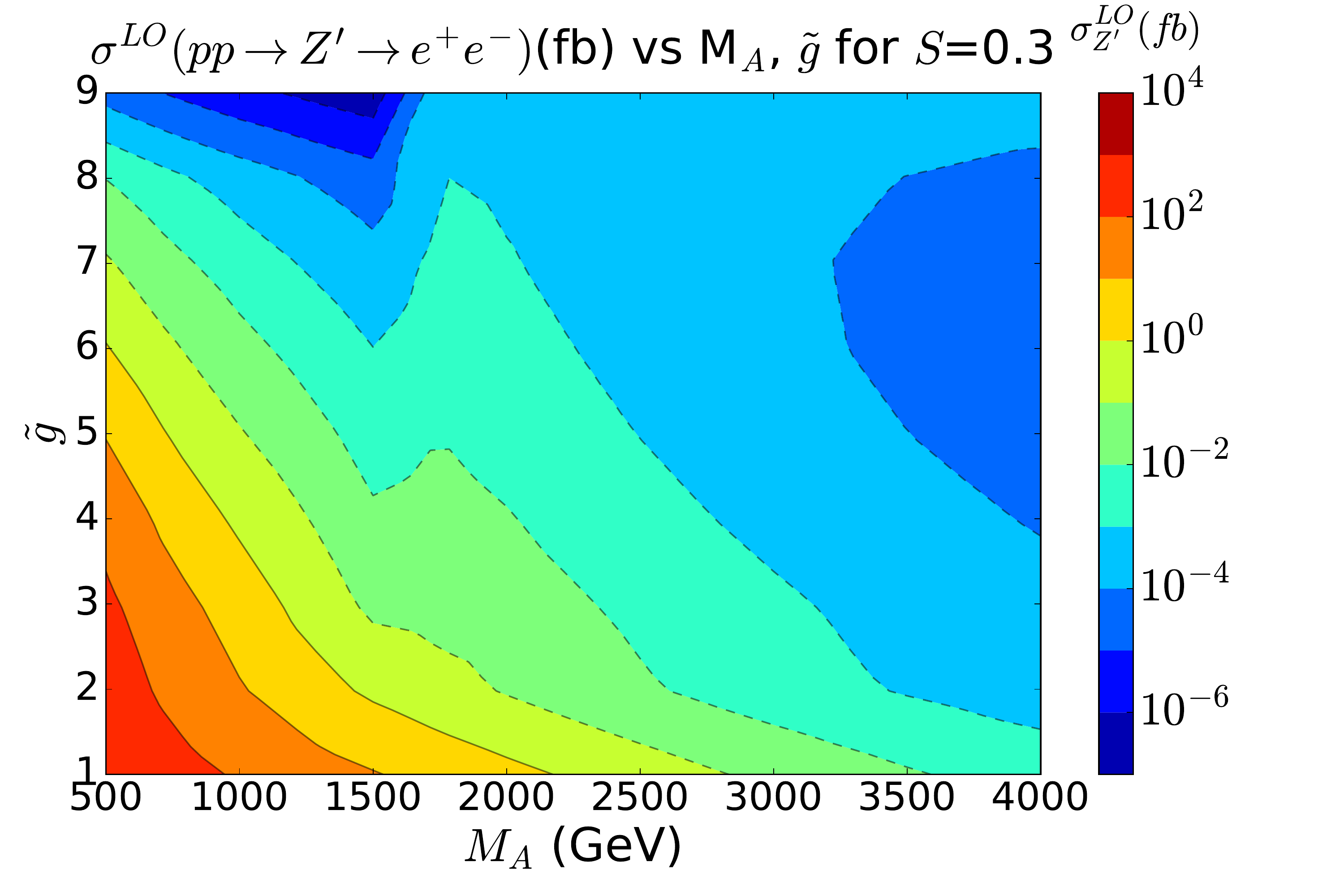}}
\caption{\label{fig:zp-theory-cs-with-s}
DY production cross sections at LO  for $pp\to Z'\to e^+ e^-$ for
$S=-0.1$ (a), $S=0.0$ (b), $S=0.2$ (c),  $S=0.3$ (d) respectively}
\end{figure}

\begin{figure}[htb]
\subfigure[]{\includegraphics[type=pdf,ext=.pdf,read=.pdf,width=0.5\textwidth]{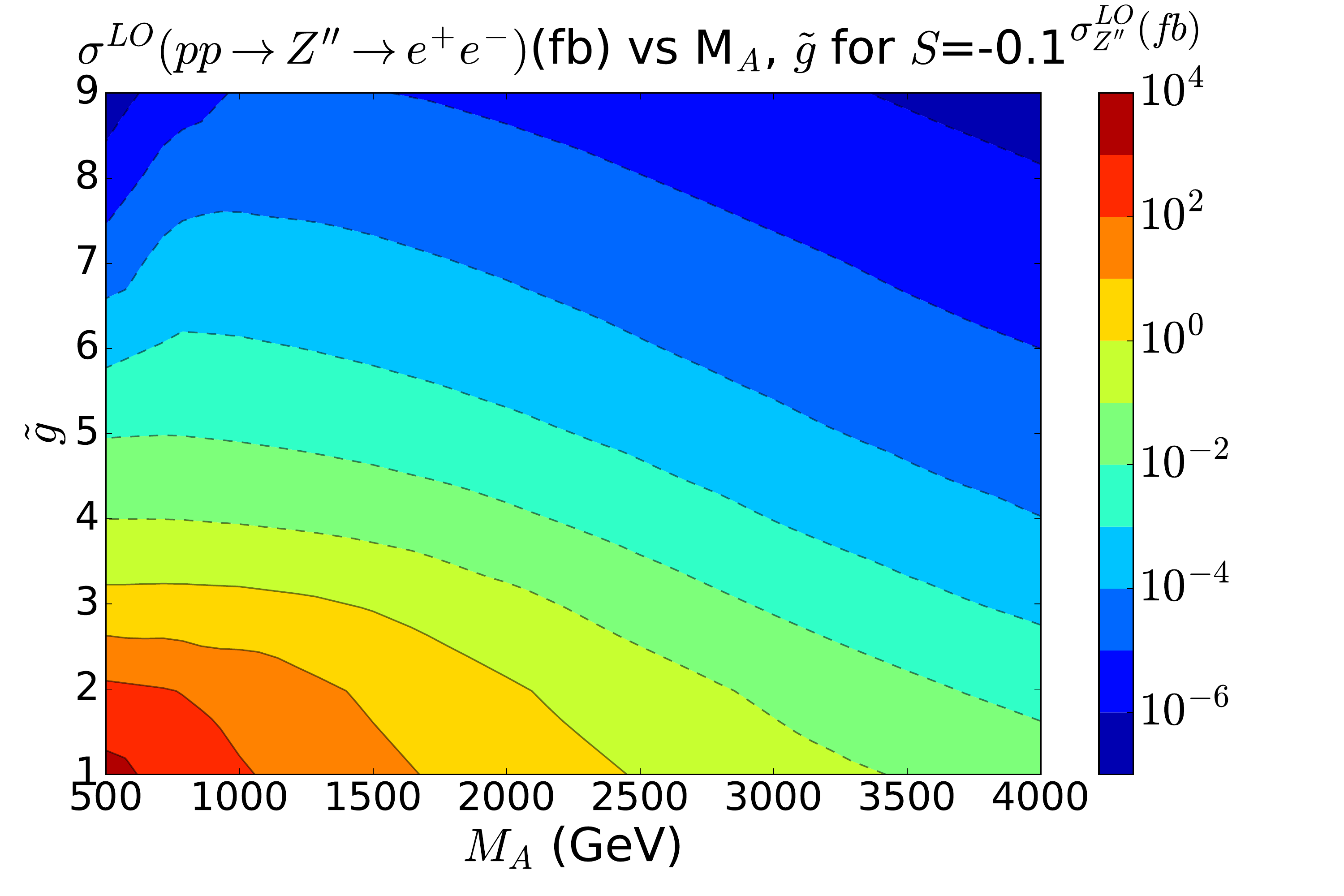}}%
\subfigure[]{\includegraphics[type=pdf,ext=.pdf,read=.pdf,width=0.5\textwidth]{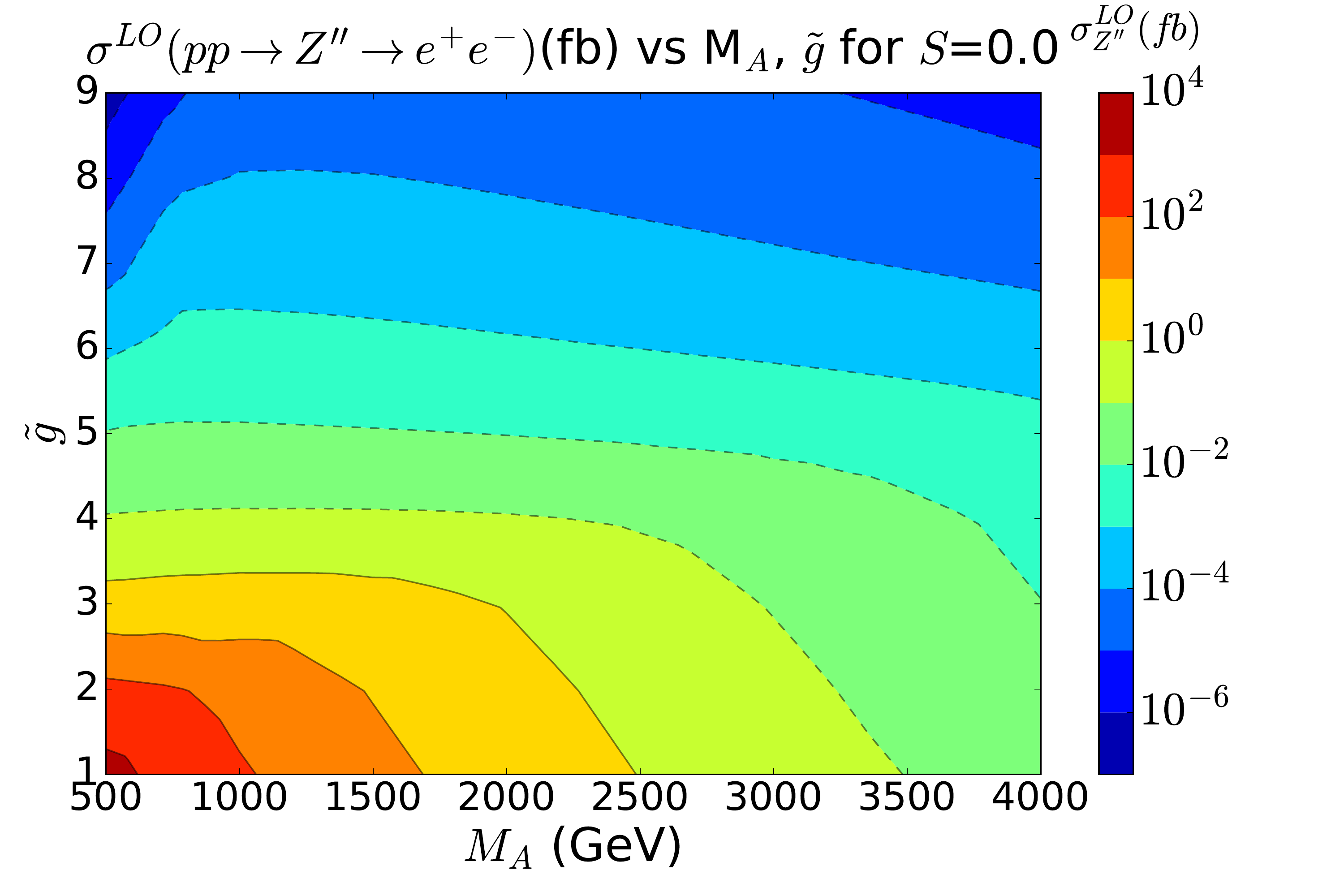}}\\
\subfigure[]{\includegraphics[type=pdf,ext=.pdf,read=.pdf,width=0.5\textwidth]{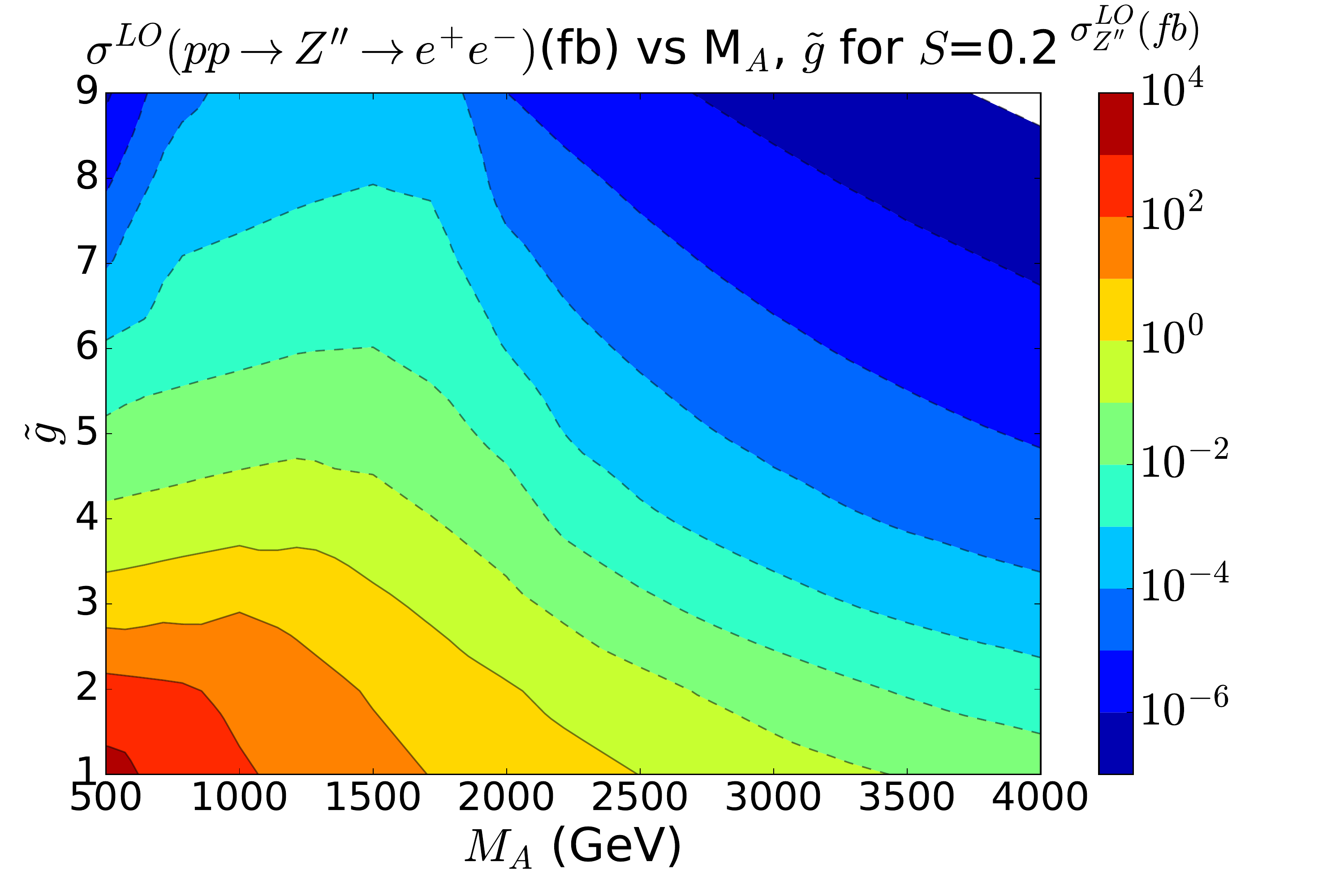}}%
\subfigure[]{\includegraphics[type=pdf,ext=.pdf,read=.pdf,width=0.5\textwidth]{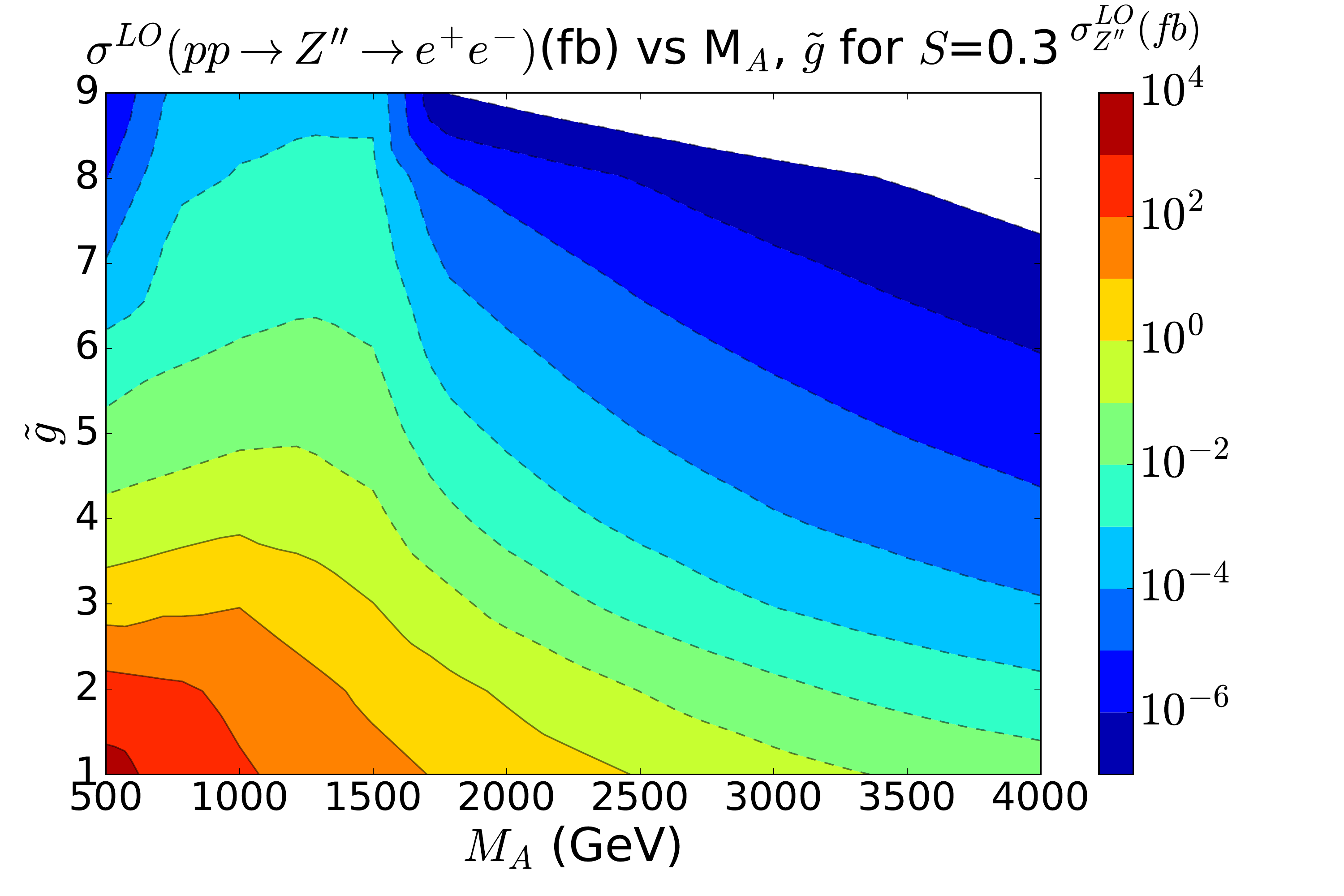}}%
\caption{\label{fig:zpp-theory-cs-with-s}
DY production cross sections at LO  for $pp\to Z''\to e^+ e^-$ for
$S=-0.1$ (a), $S=0.0$ (b), $S=0.2$ (c),  $S=0.3$ (d) respectively}
\end{figure}

The DY production cross sections at LO  for $pp\to Z'\to e^+ e^-$   $pp\to Z''\to e^+ e^-$ processes
are presented in Fig.~\ref{fig:zp-theory-cs-with-s}  and  Fig.~\ref{fig:zpp-theory-cs-with-s} respectively
as contour levels of the cross section in $(M_A,\tilde g)$ space  for different $S$.

\clearpage

\subsection{Effect of $S$ on Parameter Space Exclusions}
\label{subsec:exclusions-with-s}

\begin{figure}[htb]
\subfigure[]{\includegraphics[width=0.5\textwidth]{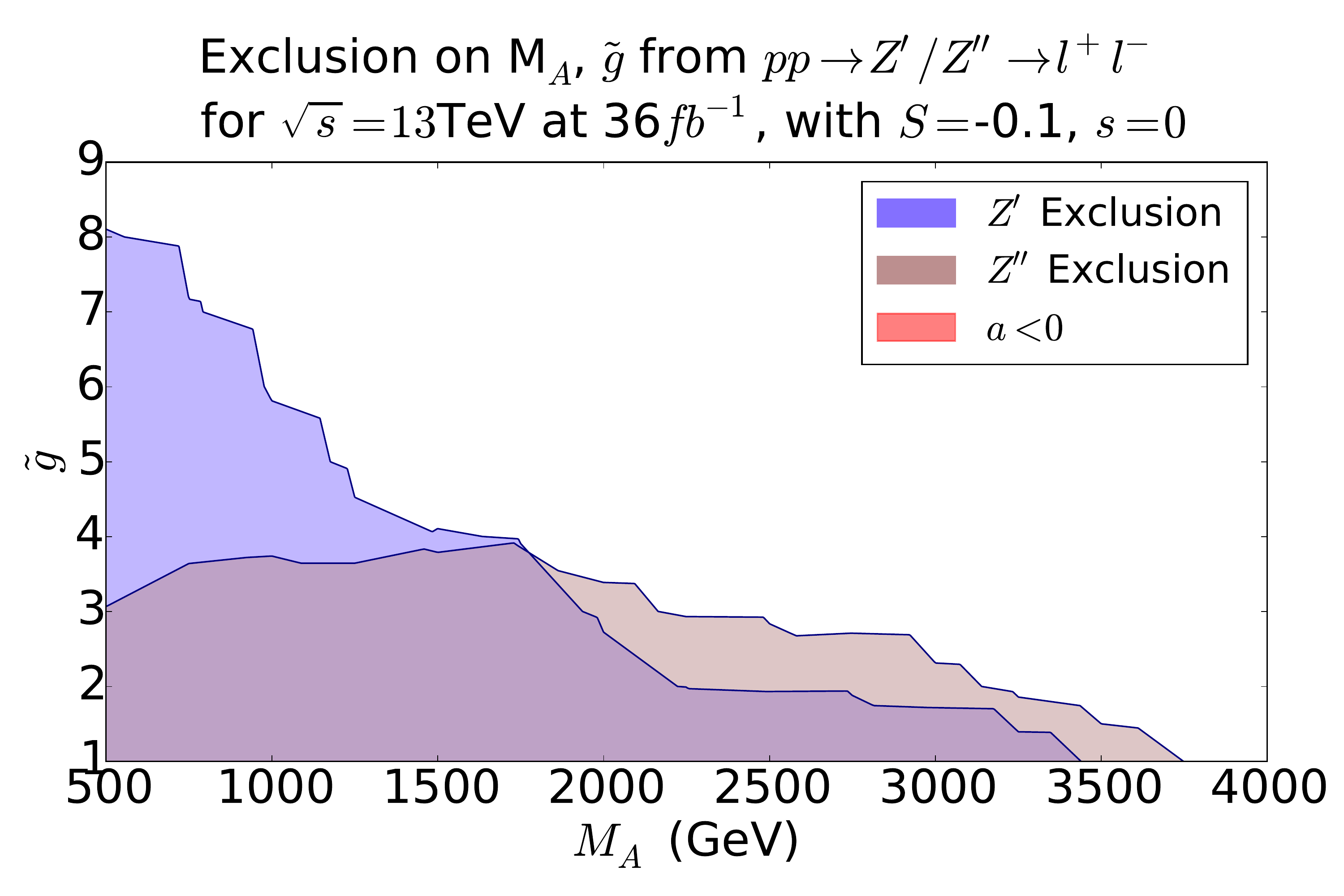}}%
\subfigure[]{\includegraphics[width=0.5\textwidth]{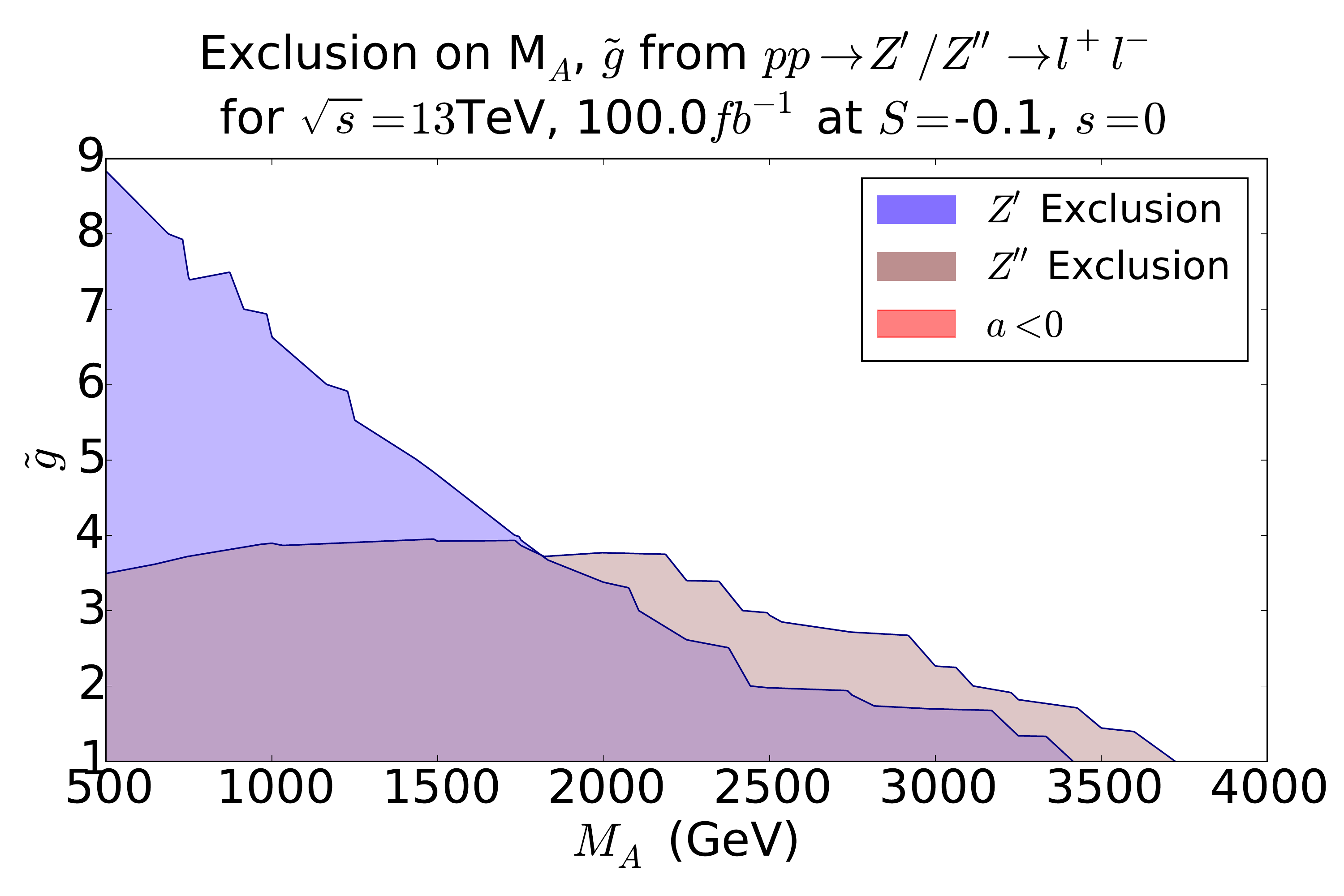}}\\
\subfigure[]{\includegraphics[width=0.5\textwidth]{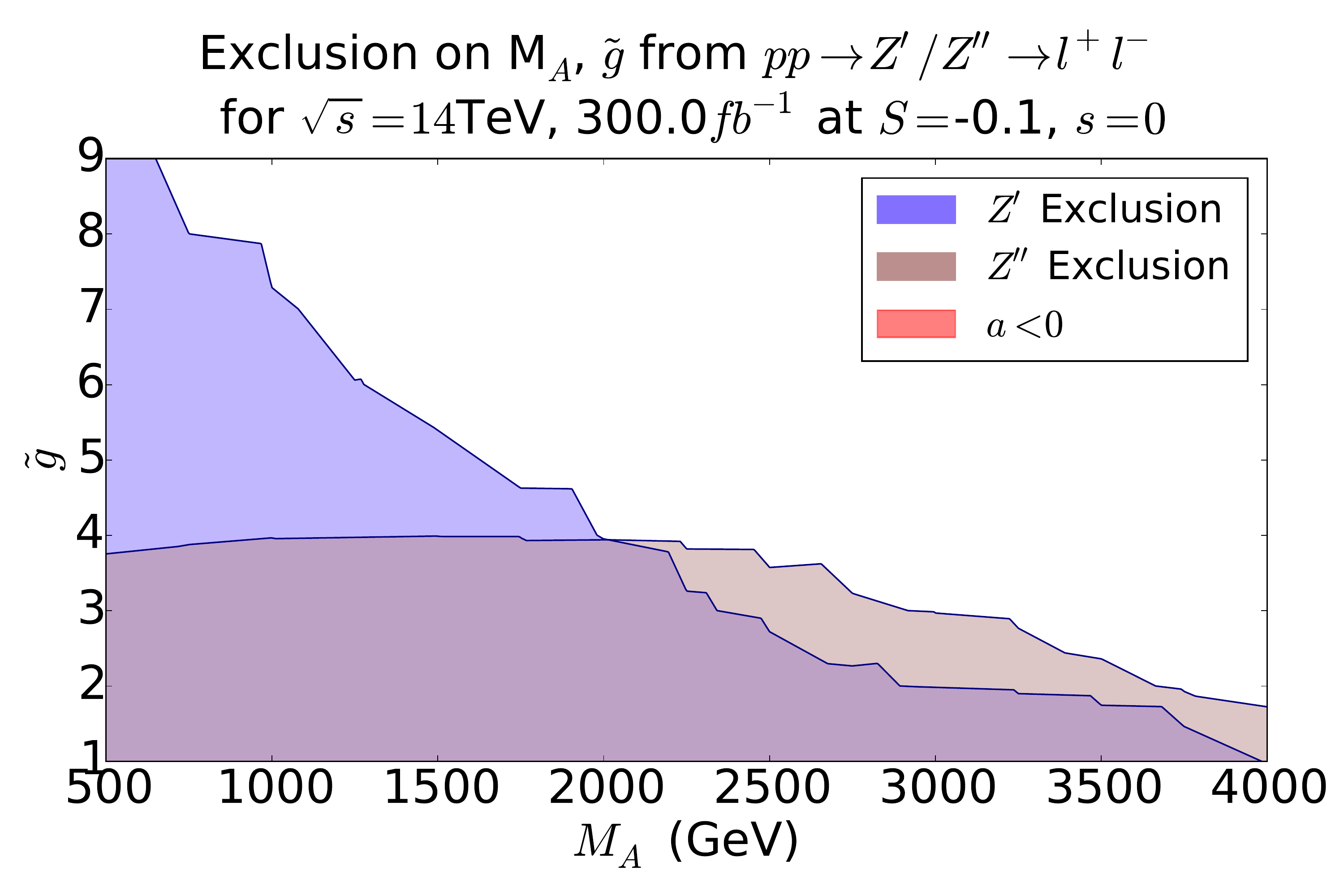}}%
\centering\subfigure[]{\includegraphics[width=0.5\textwidth]{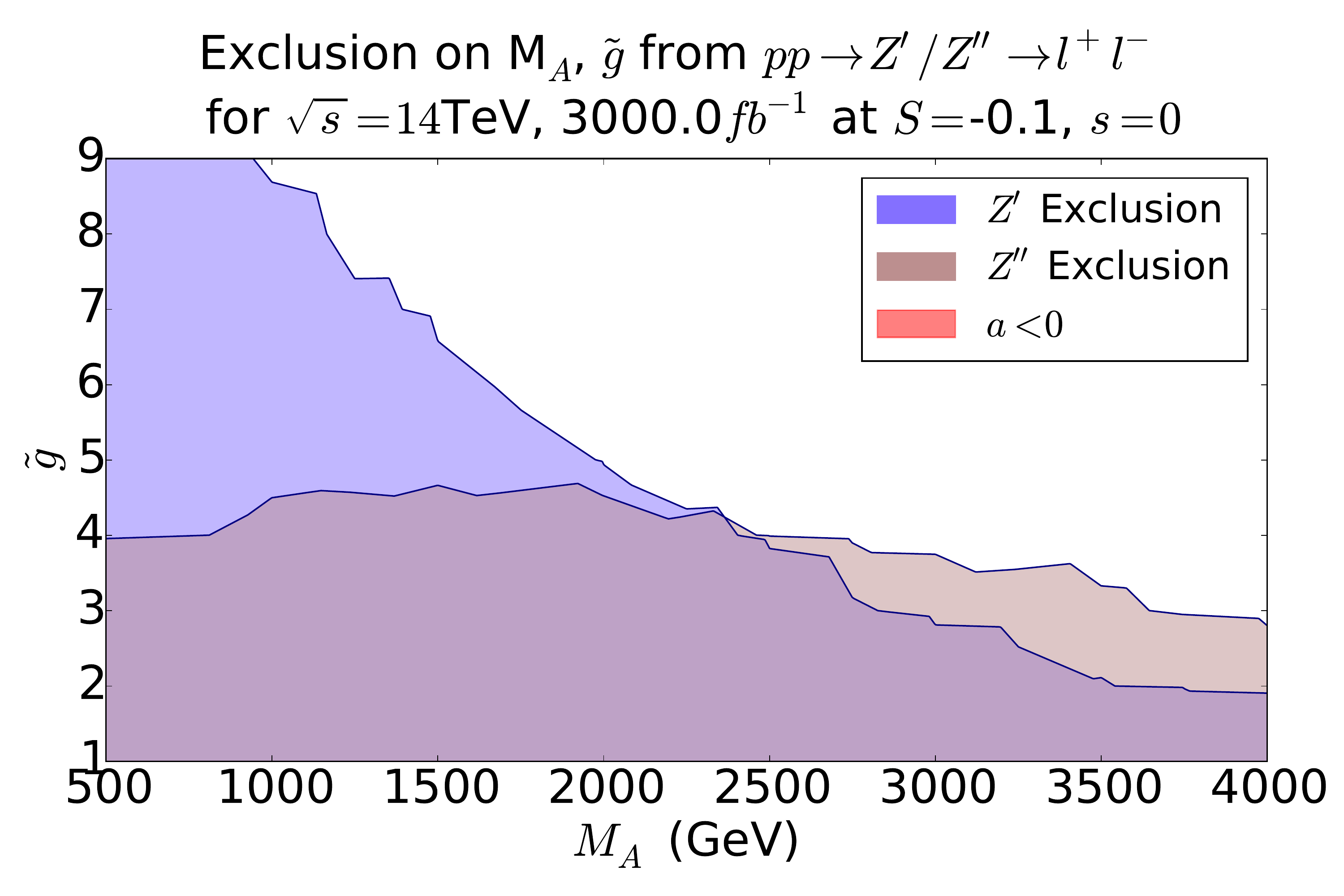}}
\caption{\label{fig:exclusions-s-minus0.1} Exclusion of the $M_A$,$\tilde{g}$ parameter space from $Z^{\prime}$ and $Z^{\prime\prime}$ DY processes at $\sqrt{s}=13$TeV and luminosity of $36$fb$^{-1}$(a); Predicted exclusion regions for the NMWT parameter space at (a) $\sqrt{s}=13$TeV and $\mathcal{L}=100fb^{-1}$, (b)$\sqrt{s}=14$TeV and $\mathcal{L}=300fb^{-1}$, (c) $\sqrt{s}=14$TeV and $\mathcal{L}=3000fb^{-1}$}
\end{figure}

\begin{figure}[htb]
\subfigure[]{\includegraphics[width=0.5\textwidth]{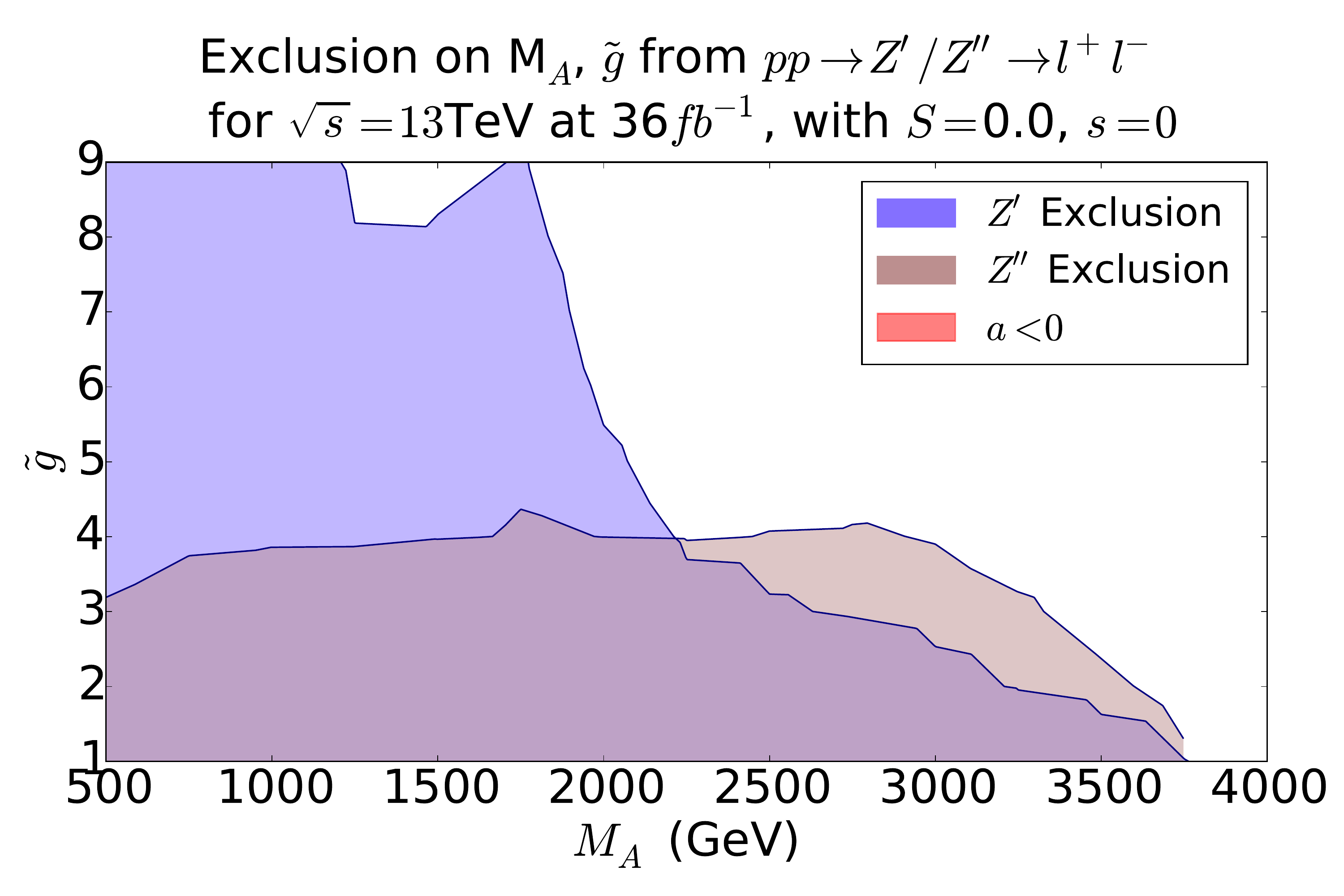}}%
\subfigure[]{\includegraphics[width=0.5\textwidth]{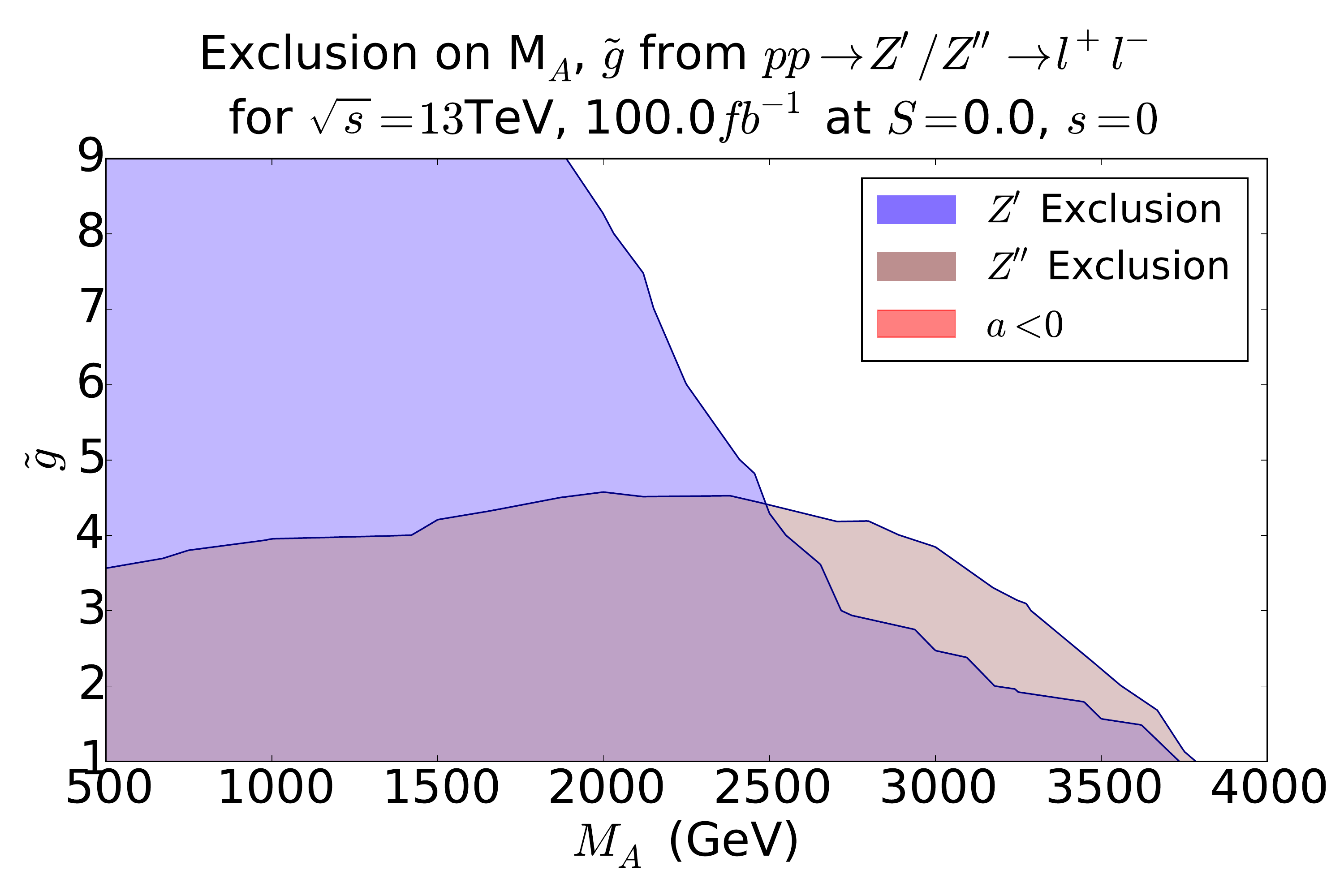}}\\
\subfigure[]{\includegraphics[width=0.5\textwidth]{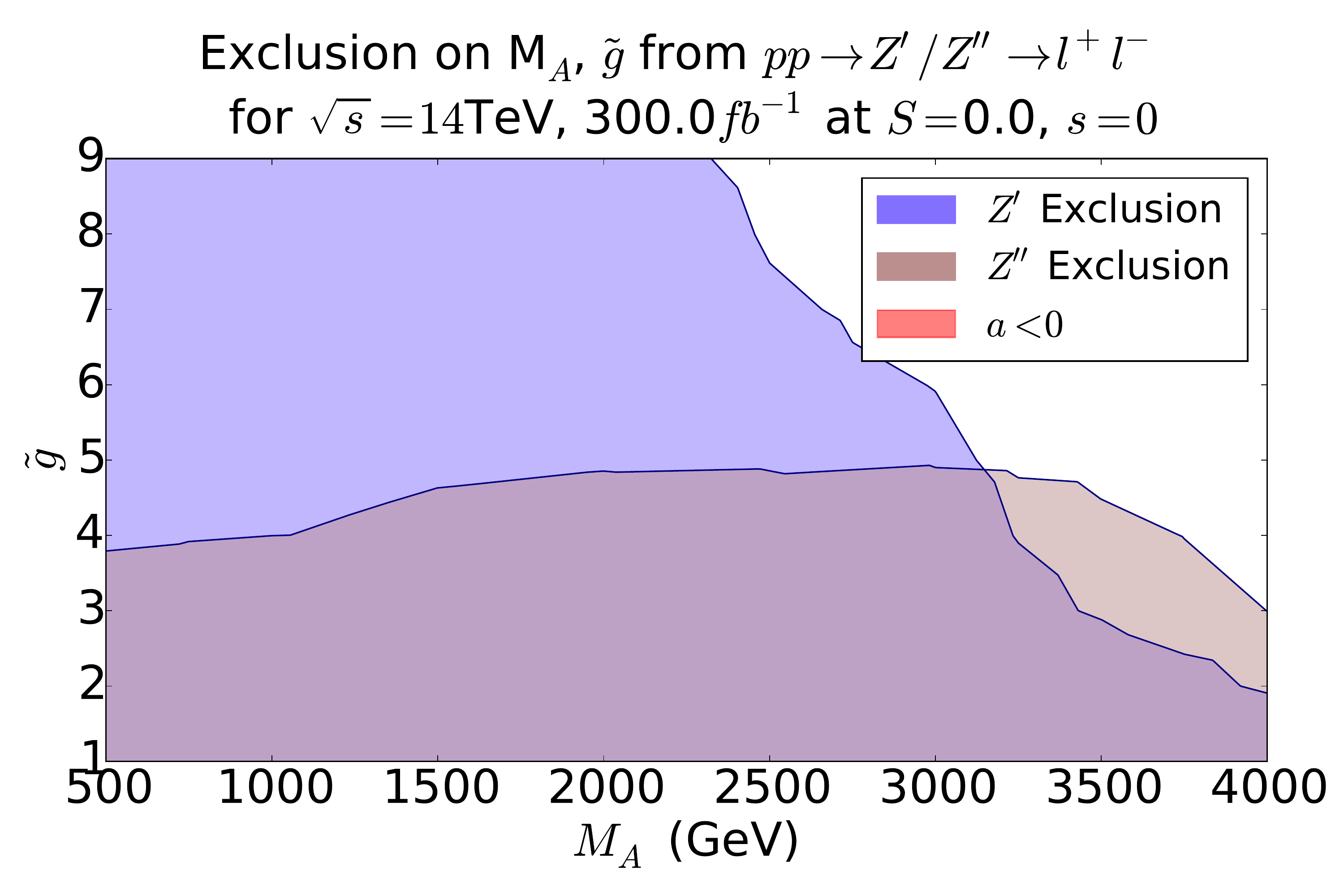}}%
\centering\subfigure[]{\includegraphics[width=0.5\textwidth]{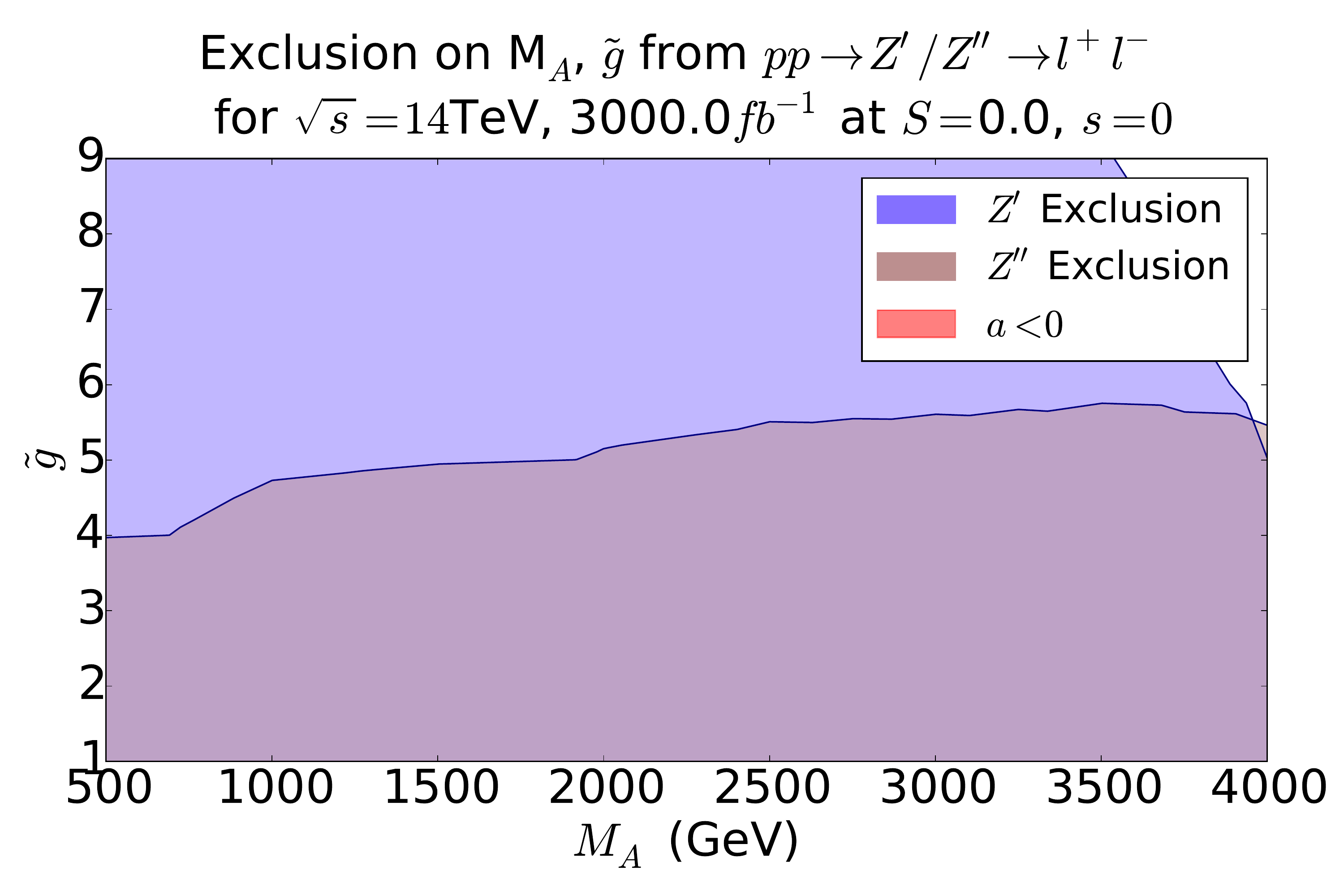}}
\caption{\label{fig:exclusions-s-0} Exclusion of the $M_A$,$\tilde{g}$ parameter space from $Z^{\prime}$ and $Z^{\prime\prime}$ DY processes at $\sqrt{s}=13$TeV and luminosity of $36$fb$^{-1}$(a); Predicted exclusion regions for the NMWT parameter space at (a) $\sqrt{s}=13$TeV and $\mathcal{L}=100fb^{-1}$, (b)$\sqrt{s}=14$TeV and $\mathcal{L}=300fb^{-1}$, (c) $\sqrt{s}=14$TeV and $\mathcal{L}=3000fb^{-1}$}
\end{figure}

\begin{figure}[htb]
\subfigure[]{\includegraphics[width=0.5\textwidth]{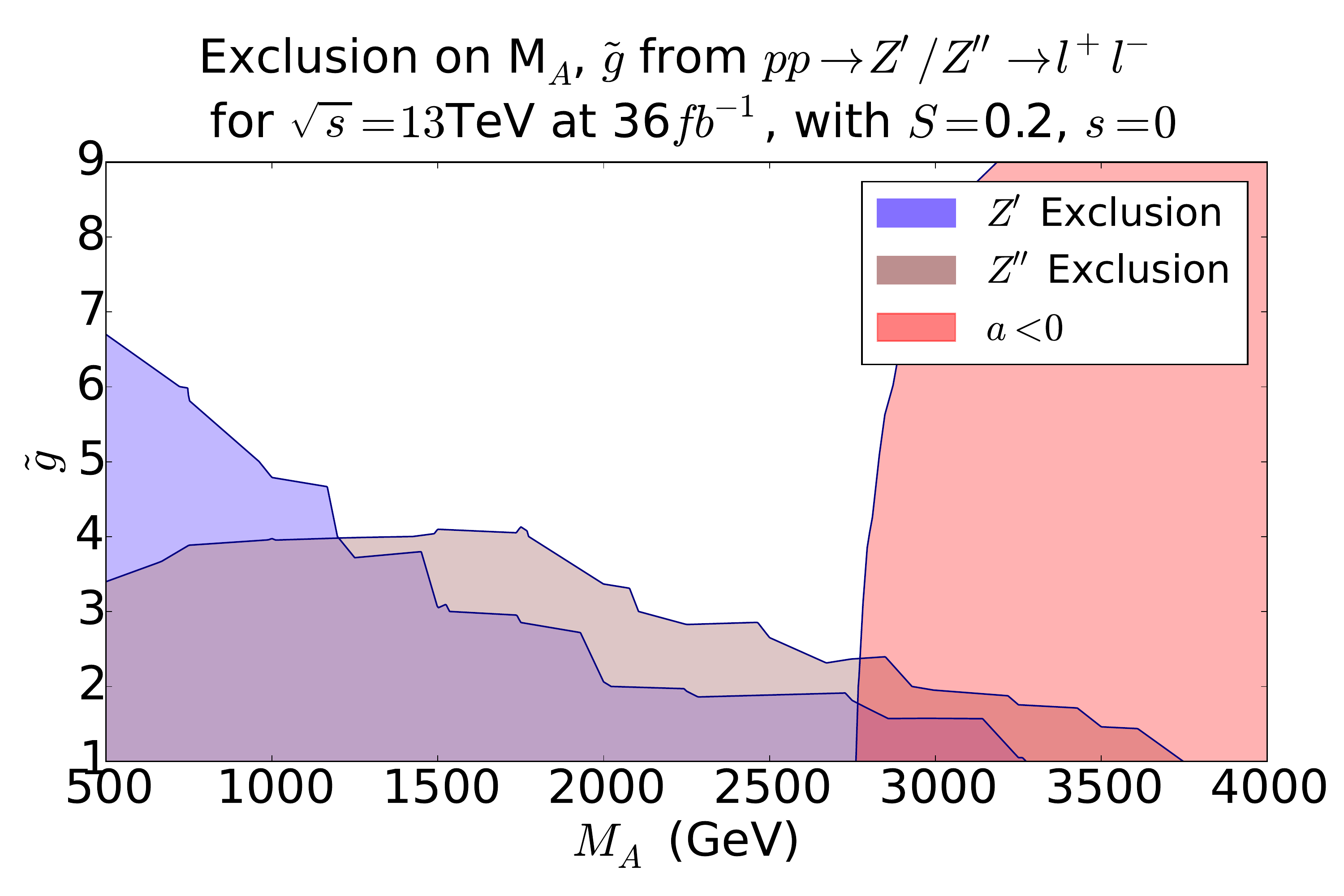}}%
\subfigure[]{\includegraphics[width=0.5\textwidth]{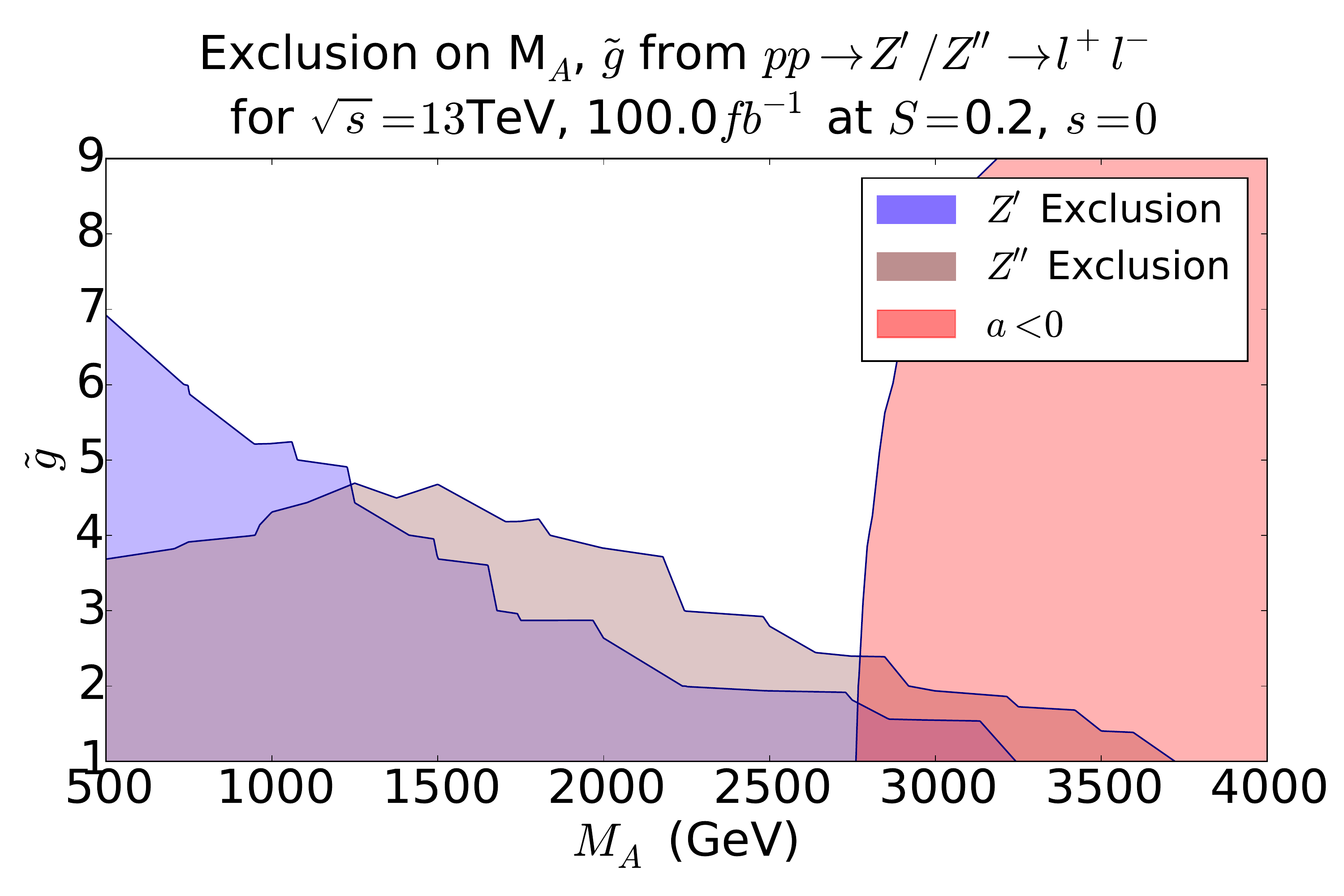}}\\
\subfigure[]{\includegraphics[width=0.5\textwidth]{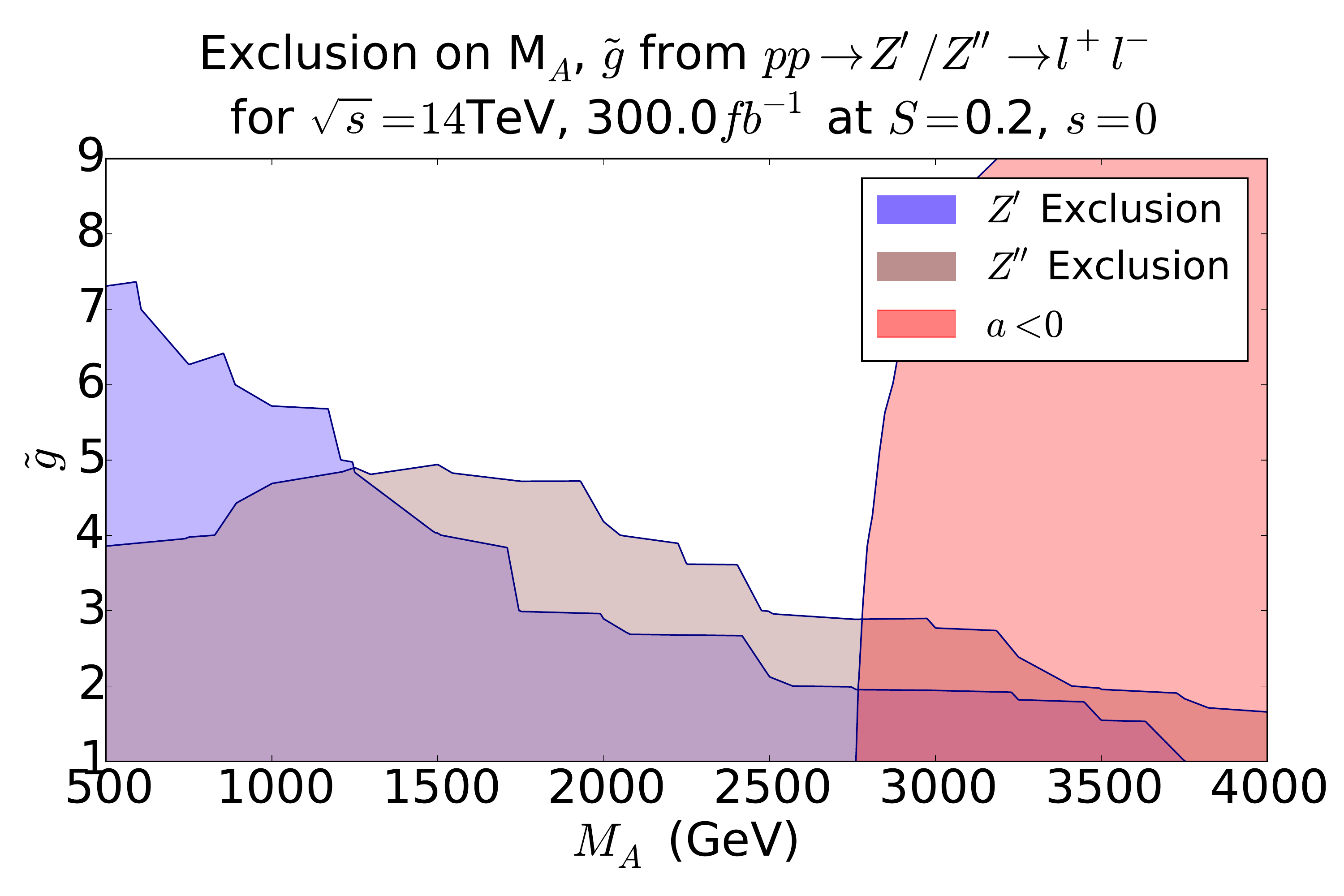}}%
\centering\subfigure[]{\includegraphics[width=0.5\textwidth]{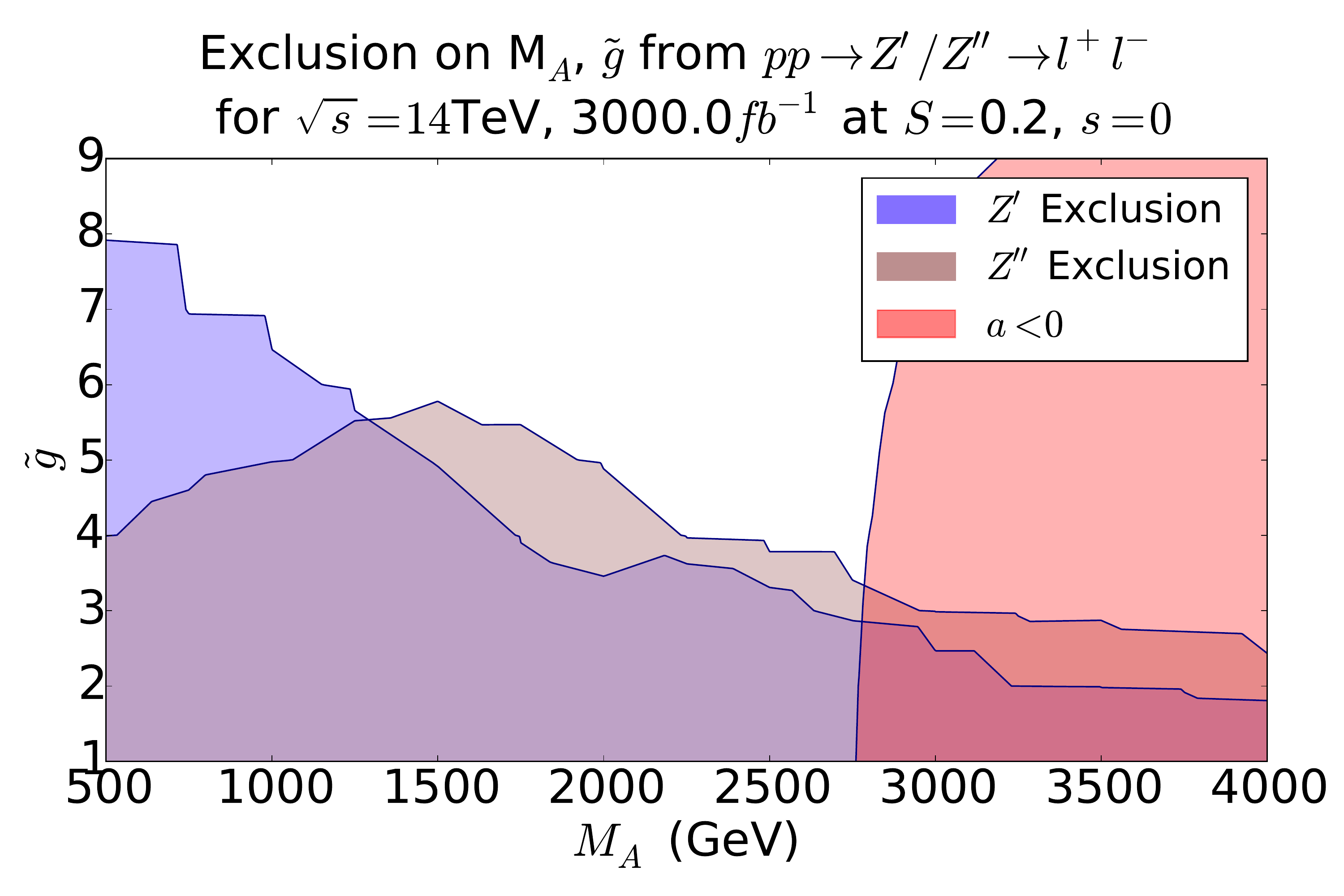}}
\caption{\label{fig:exclusions-s-0.2} Exclusion of the $M_A$,$\tilde{g}$ parameter space from $Z^{\prime}$ and $Z^{\prime\prime}$ DY processes at $\sqrt{s}=13$TeV and luminosity of $36$fb$^{-1}$(a); Predicted exclusion regions for the NMWT parameter space at (a) $\sqrt{s}=13$TeV and $\mathcal{L}=100fb^{-1}$, (b)$\sqrt{s}=14$TeV and $\mathcal{L}=300fb^{-1}$, (c) $\sqrt{s}=14$TeV and $\mathcal{L}=3000fb^{-1}$}
\end{figure}

\begin{figure}[htb]
\subfigure[]{\includegraphics[width=0.5\textwidth]{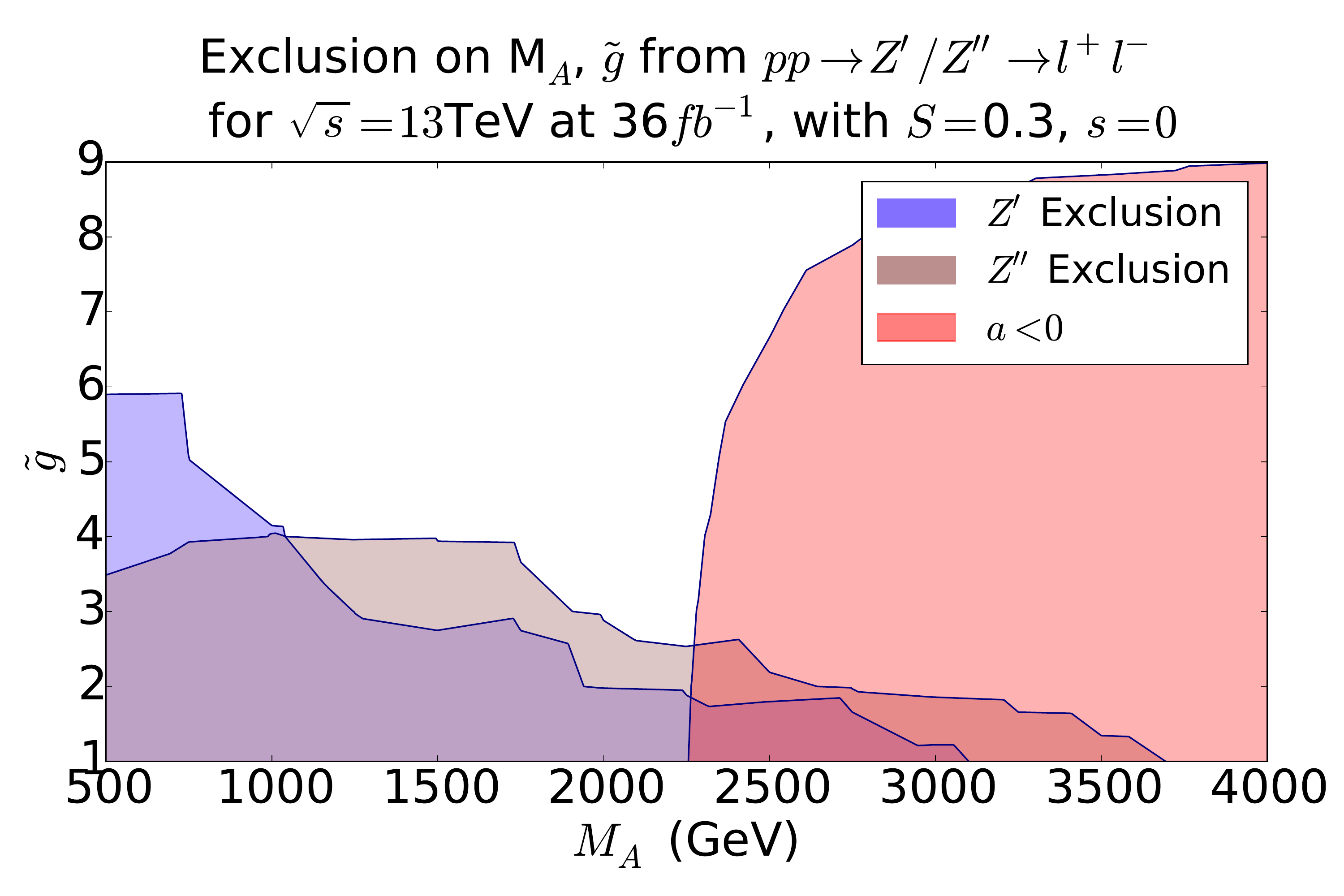}}%
\subfigure[]{\includegraphics[width=0.5\textwidth]{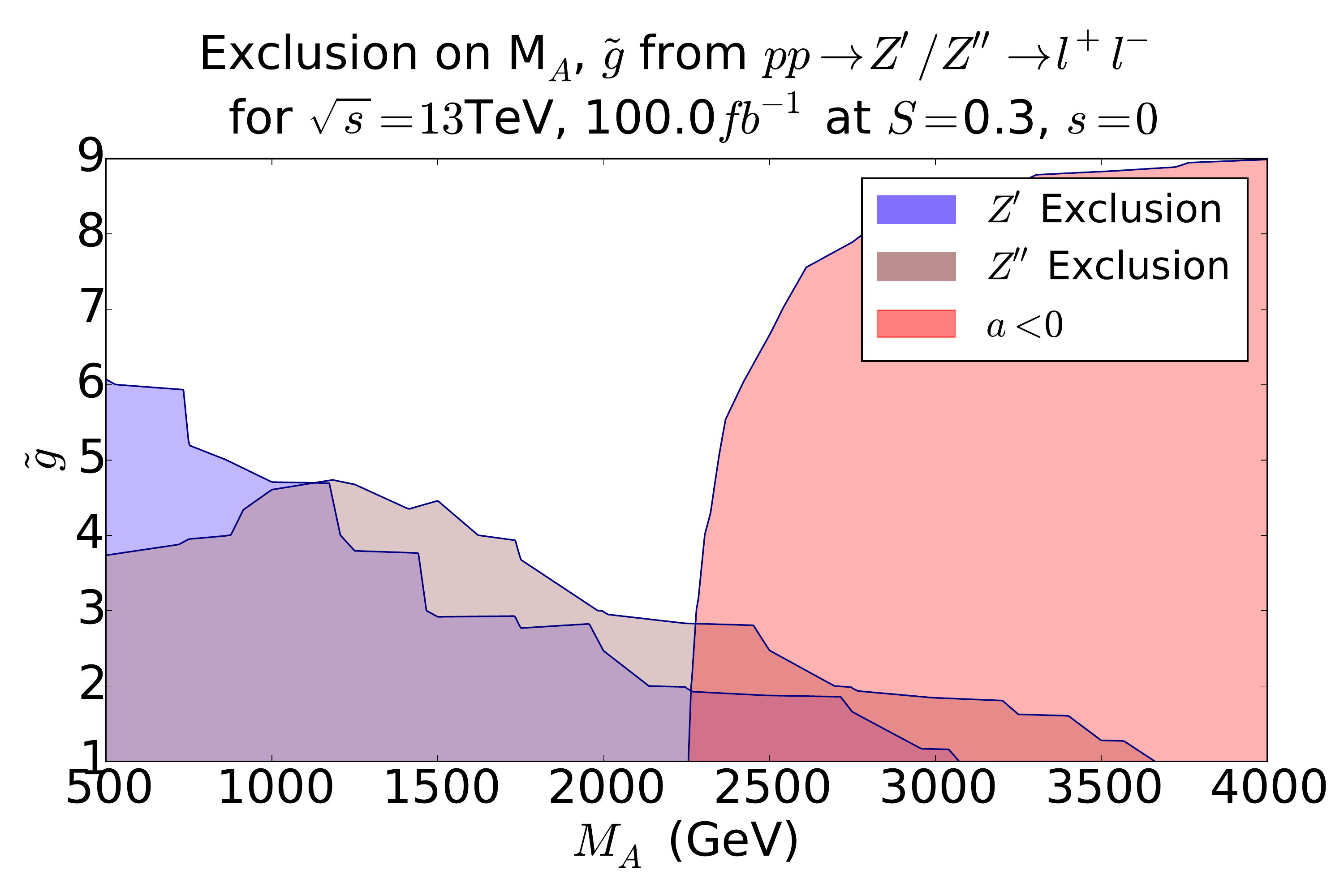}}\\
\subfigure[]{\includegraphics[width=0.5\textwidth]{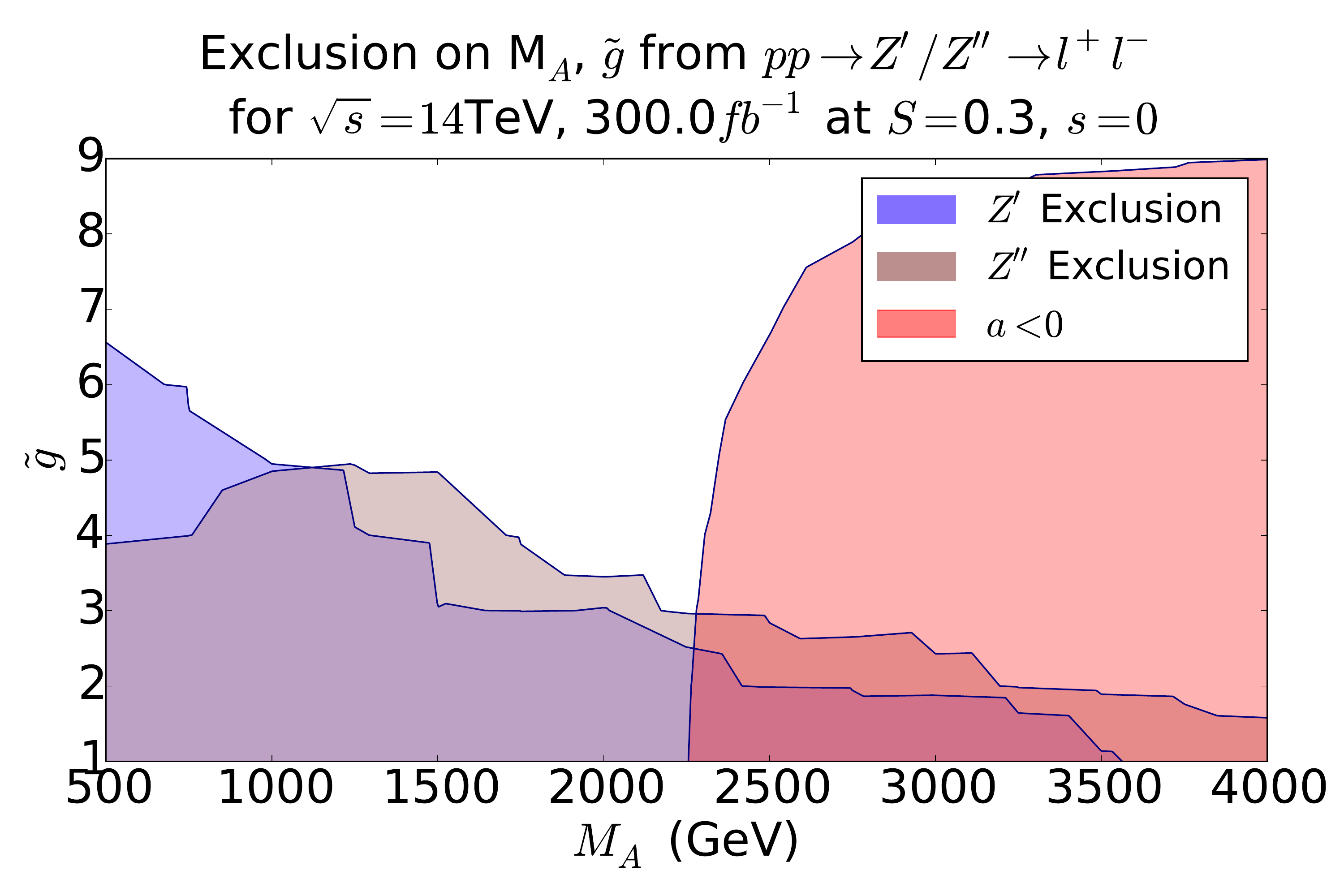}}%
\centering\subfigure[]{\includegraphics[width=0.5\textwidth]{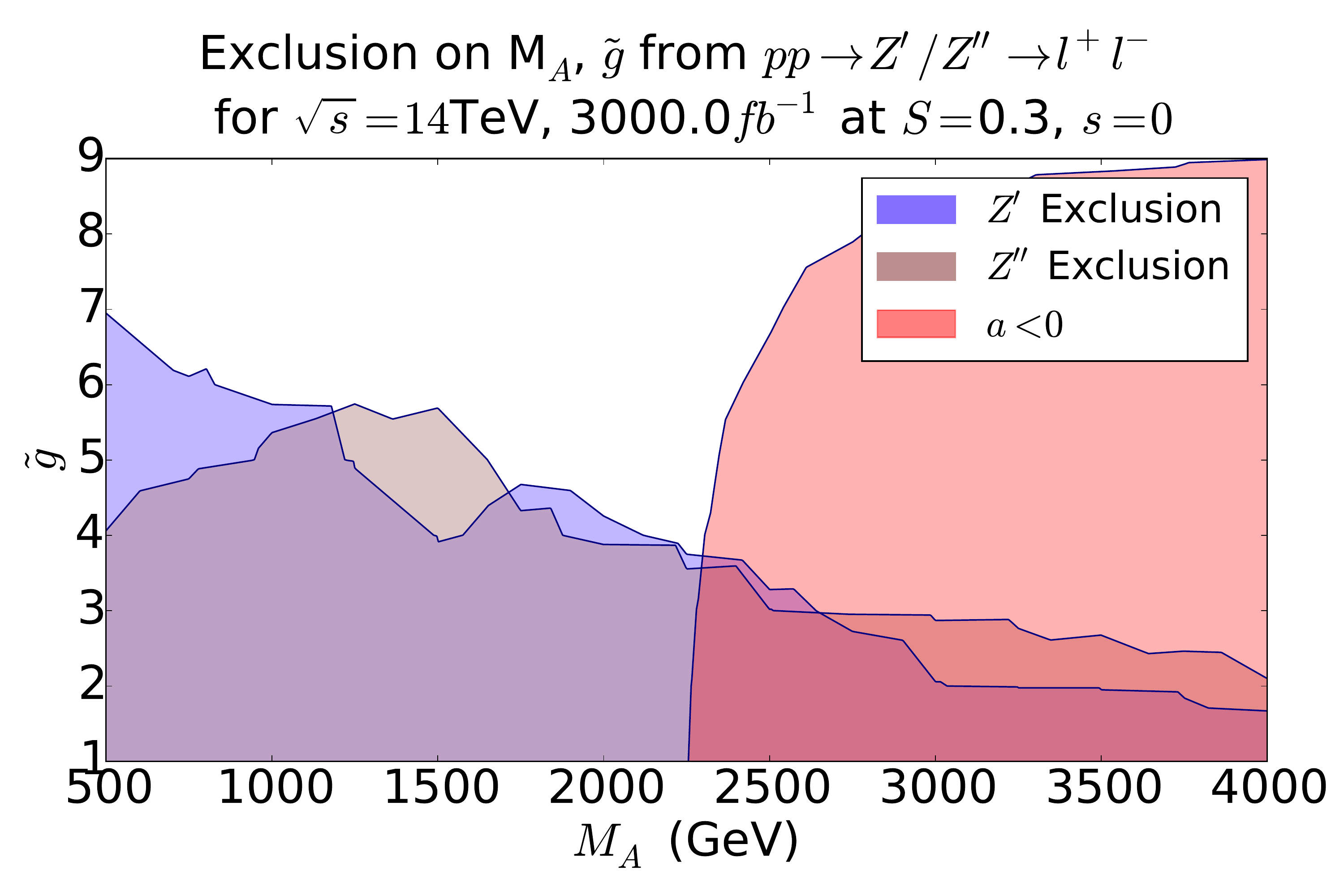}}
\caption{\label{fig:exclusions-s-0.3} Exclusion of the $M_A$,$\tilde{g}$ parameter space from $Z^{\prime}$ and $Z^{\prime\prime}$ DY processes at $\sqrt{s}=13$TeV and luminosity of $36$fb$^{-1}$(a); Predicted exclusion regions for the NMWT parameter space at (a) $\sqrt{s}=13$TeV and $\mathcal{L}=100fb^{-1}$, (b)$\sqrt{s}=14$TeV and $\mathcal{L}=300fb^{-1}$, (c) $\sqrt{s}=14$TeV and $\mathcal{L}=3000fb^{-1}$}
\end{figure}

As noted in section \ref{sec:conclusions}, the $S$ parameter could be of great importance in determining the excluded region of WTC parameter space. As such, we present a set of figures for each discrete $S$ in which we show the current and future limits on the WTC parameter space for fixed $S$. This is for direct comparison to the exclusions quoted and discussed in  section \ref{subsec:13tev-exclusions}. Figures \ref{fig:exclusions-s-minus0.1}, \ref{fig:exclusions-s-0}, \ref{fig:exclusions-s-0.2}, \ref{fig:exclusions-s-0.3} show the excluded regions of $M_{A}$, $\tilde{g}$ for $S=-0.1, 0, 0.2, 0.3$ respectively. 

The projected limits depend strongly on the $S$ parameter, and for large $S$ , the limit from dilepton searches at the LHC covers less of the parameter space, while  the theoretical limit requiring $a>0$ excludes a large portion of the $M_A$ parameter space from above. 

\end{document}